\pdfoutput=1

\documentclass[11pt]{article}

\usepackage{setspace}
\usepackage{jheppub}

\usepackage{amsmath}
\usepackage{amssymb}
\usepackage{mathtools}
\usepackage{graphicx,xcolor,slashed}
\usepackage{enumerate}
\usepackage{ifpdf}
\usepackage[english]{babel}
\usepackage{url}
\usepackage{booktabs}
\usepackage{makecell}

\notoc

\renewcommand{\baselinestretch}{1.18}

\definecolor{darkgreen}{rgb}{0.0, 0.26, 0.15}
\definecolor{darkred}{rgb}{0.65,0.15,0}

\usepackage{float}

\hypersetup{bookmarksnumbered,bookmarksdepth=3}

\allowdisplaybreaks[1]

\DeclareFontFamily{U}{mathx}{\hyphenchar\font45}
\DeclareFontShape{U}{mathx}{m}{n}{
      <5> <6> <7> <8> <9> <10>
      <10.95> <12> <14.4> <17.28> <20.74> <24.88>
      mathx10
      }{}
\DeclareSymbolFont{mathx}{U}{mathx}{m}{n}
\DeclareFontSubstitution{U}{mathx}{m}{n}
\DeclareMathAccent{\widecheck}{0}{mathx}{"71}

\font\tenshuffle=shuffle10 \font\sevenshuffle=shuffle7 \font\fiveshuffle=shuffle7 at 5pt
\def\shuffle{{%
\def\Dshuffle{\mathbin{\hbox{\tenshuffle\char'001}}}%
\def\Sshuffle{\mathbin{\hbox{\sevenshuffle\char'001}}}%
\def\SSshuffle{\mathbin{\hbox{\fiveshuffle\char'001}}}%
\mathchoice{\Dshuffle}{\Dshuffle}{\Sshuffle}{\SSshuffle}}}

\restylefloat{figure}
\definecolor{dgreen}{rgb}{0,0.70,0.30}
\definecolor{gold}{rgb}{0.85,.66,0}
\definecolor{purple}{rgb}{1.0,0.3,0.6}

\usepackage{tikz}
\usetikzlibrary{calc} \usetikzlibrary{patterns} \usetikzlibrary{decorations.pathreplacing} \usetikzlibrary{decorations.markings} \usetikzlibrary{decorations.pathmorphing} \usetikzlibrary{positioning} \usetikzlibrary{bending}

\def\beq{\begin{equation}}
\def\eeq{\end{equation}}

\newcommand{\ccb}{\left(\begin{array}{cc}}
\newcommand{\cce}{\end{array}\right)}

\newcommand{\EBR}[3]{{\cal E}\! \left[\begin{smallmatrix}#1\\#2\end{smallmatrix};#3\right]}
\newcommand{\EBRno}[2]{{\cal E}\! \left[\begin{smallmatrix}#1\\#2\end{smallmatrix}\right]}
\newcommand{\EsvBR}[3]{{\cal E}^{\rm sv}\!  \left[\begin{smallmatrix}#1\\#2\end{smallmatrix};#3\right]}
\newcommand{\EsvBRno}[2]{{\cal E}^{\rm sv}\! \left[\begin{smallmatrix}#1\\#2\end{smallmatrix}\right]}
\newcommand{\EsvBRmin}[3]{{\cal E}_{\rm min}^{\rm sv}\!  \left[\begin{smallmatrix}#1\\#2\end{smallmatrix};#3\right]}
\newcommand{\EsvBRminno}[2]{{\cal E}_{\rm min}^{\rm sv}\! \left[\begin{smallmatrix}#1\\#2\end{smallmatrix}\right]}
\newcommand{\Esv}{{\cal E}^{\rm sv}}
\newcommand{\esv}[1]{{\cal E}^{\rm sv}\! \left[\begin{smallmatrix}#1\end{smallmatrix}\right]}
\newcommand{\esvtau}[1]{{\cal E}^{\rm sv}\! \left[\begin{smallmatrix}#1\end{smallmatrix};\tau\right]}
\newcommand{\bsv}{\beta^{\rm sv}}
\newcommand{\betasv}[1]{\beta^{\rm sv}\! \left[\begin{smallmatrix}#1\end{smallmatrix}\right]}
\newcommand{\betasvtau}[1]{\beta^{\rm sv}\! \left[\begin{smallmatrix}#1\end{smallmatrix}; \tau\right]}
\newcommand{\betasvStau}[1]{\beta^{\rm sv}\! \left[\begin{smallmatrix}#1\end{smallmatrix};-\tfrac{1}{\tau}\right]}
\newcommand{\bsvBR}[3]{\beta^{\rm sv} \! \left[\begin{smallmatrix}#1\\#2\end{smallmatrix};#3\right]}
\newcommand{\bsvBRno}[2]{\beta^{\rm sv}\! \left[\begin{smallmatrix}#1\\#2\end{smallmatrix}\right]}
\newcommand{\alphaBR}[3]{\alpha\! \left[\begin{smallmatrix}#1\\#2\end{smallmatrix};#3\right]}
\newcommand{\alphaBRno}[2]{\alpha\! \left[\begin{smallmatrix}#1\\#2\end{smallmatrix}\right]}

\newcommand{\fBR}[3]{f\! \left[\begin{smallmatrix}#1\\#2\end{smallmatrix};#3\right]}
\newcommand{\fBRno}[2]{f\! \left[\begin{smallmatrix}#1\\#2\end{smallmatrix}\right]}

\newcommand{\Yhat}{\widehat Y}

\newcommand{\SG}{\mathcal{S}}
\newcommand{\EE}{{\rm E}}

\newcommand{\BB}{{\rm B}}

\newcommand{\bea}{\begin{eqnarray}}
\newcommand{\eea}{\end{eqnarray}}

\newcommand{\mycomment}[1]{}

\let\Re\relax
\let\Im\relax
\DeclareMathOperator{\Re}{Re}
\DeclareMathOperator{\Im}{Im}

\newcommand{\nn}{\nonumber}

\newcommand{\dd}{\mathrm{d}}
\newcommand{\te}{\textrm}
\newcommand{\ap}{{\alpha'}}
\newcommand{\ep}{{\epsilon}}

\DeclareMathOperator{\KN}{KN}

\newcommand{\cform}[1]{\,{\cal C}\!\left[\protect\begin{smallmatrix}#1\protect\end{smallmatrix}\right]}
\newcommand{\cformtri}[3]{\,{\cal C}\!\left[\protect\begin{smallmatrix}#1\protect\end{smallmatrix}\middle|\protect\begin{smallmatrix}#2\protect\end{smallmatrix}\middle|\protect\begin{smallmatrix}#3\protect\end{smallmatrix}\right]}
\newcommand{\aform}[1]{\,{\cal A}[\protect\begin{smallmatrix}#1\protect\end{smallmatrix}]}

\newcommand{\NN}{\mathbb N}
\newcommand{\ZZ}{\mathbb Z}

\newcommand{\QQ}{\mathbb Q}

\usepackage[format=plain,
            labelfont=bf,
            textfont=it]{caption}

\title{Generating series of all modular graph forms from iterated Eisenstein integrals}

\author[a]{Jan E.\ Gerken,}
\author[a,b]{Axel Kleinschmidt,}
\author[c]{Oliver Schlotterer}

\affiliation[a]{Max-Planck-Institut f\"ur Gravitationsphysik,
Albert-Einstein-Institut,
DE-14476 Potsdam, Germany}
\affiliation[b]{International Solvay Institutes ULB--Campus Plaine CP231, BE-1050 Brussels, Belgium}
\affiliation[c]{Department of Physics and Astronomy, Uppsala University, SE-75108 Uppsala, Sweden}

\emailAdd{jan.gerken@aei.mpg.de}
\emailAdd{axel.kleinschmidt@aei.mpg.de}
\emailAdd{oliver.schlotterer@physics.uu.se}

\date{\today}

\abstract{We study generating series of torus integrals that contain all so-called modular graph forms relevant for massless one-loop closed-string amplitudes. By analysing the differential equation of the generating series we construct a solution for their low-energy expansion to all orders in the inverse string tension $\alpha'$. Our solution is expressed through initial data involving multiple zeta values and certain real-analytic functions of the modular parameter of the torus. These functions are built from real and imaginary parts of holomorphic iterated Eisenstein integrals and should be closely related to Brown's recent construction of real-analytic modular forms. We study the properties of our real-analytic objects in detail and give explicit examples  to a fixed order in the $\ap$-expansion. In particular, our solution allows for a counting of linearly independent modular graph forms at a given weight, confirming previous partial results and giving predictions for higher, hitherto unexplored weights. It also sheds new light on the topic of uniform transcendentality of the $\alpha'$-expansion.}

\preprint{UUITP--09/20}

\begin{document}

\maketitle{}

\newpage

\setcounter{page}{1}
\pagenumbering{roman}

\setcounter{tocdepth}{2}

\tableofcontents
\vspace{1em}
\hrule

\numberwithin{equation}{section}


\section{Introduction}
\label{sec:intro}

\setcounter{page}{1}
\pagenumbering{arabic}

Closed-string scattering amplitudes at perturbative one-loop order are formulated as integrals over the 
complex-structure parameter $\tau$ of the torus worldsheet. The function of $\tau$ in the integrand has to 
be modular invariant under the group ${\rm SL}_2(\mathbb{Z})$ of large diffeomorphisms of the torus and 
arises from integrating a conformal field theory (CFT) correlator over the punctures $z_i$ of the torus. 
This work is dedicated to performing the integrals over torus punctures in a low-energy expansion 
in powers of Mandelstam variables $s_{ij}$ (in units of the inverse string tension $\ap$).

The families of modular invariants and more generally modular forms that can arise in this low-energy expansion have
been studied from various perspectives \cite{Green:1999pv, Green:2008uj, Green:2013bza, DHoker:2015gmr, DHoker:2015sve, Basu:2015ayg, Zerbini:2015rss, DHoker:2015wxz, DHoker:2016mwo, Basu:2016xrt, Basu:2016kli, Basu:2016mmk, DHoker:2016quv, Kleinschmidt:2017ege, DHoker:2017zhq, Basu:2017nhs, Broedel:2018izr, Ahlen:2018wng,Zerbini:2018sox,Gerken:2018zcy, Gerken:2018jrq, DHoker:2019txf, Dorigoni:2019yoq, DHoker:2019xef, DHoker:2019mib, DHoker:2019blr, Basu:2019idd, Zagier:2019eus, Berg:2019jhh, Gerken:2019cxz, Hohenegger:2019tii}\footnote{See \cite{DHoker:2013fcx, DHoker:2014oxd, Pioline:2015qha, DHoker:2017pvk, DHoker:2018mys, Basu:2018bde} for higher-genus incarnations of modular graph forms.}, and they are now known as \textit{modular graph forms} (MGFs). The name MGF refers to the fact that they can be characterised by (decorated) Feynman-like graphs on the torus where the vertices of the graphs 
correspond to the integrated punctures in the CFT correlator. 
Moreover, MGFs have a definite modular behaviour under ${\rm SL}_2(\mathbb{Z})$ acting on $\tau$. 

On the one hand, it is straightforward to obtain MGFs as nested lattice sums over discrete loop 
momenta on the torus by Fourier transformation of the underlying CFT correlators. On the other hand,
many crucial properties of MGFs, including their behaviour at the cusp
$\tau \rightarrow i\infty$, are laborious to extract from their lattice-sum representations. In particular,
the lattice-sum representation does not manifest that MGFs obey an intricate web of relations over 
rational numbers and multiple zeta values (MZVs).
The last years have witnessed tremendous progress in performing basis reductions of individual MGFs \cite{DHoker:2015gmr, DHoker:2015sve, DHoker:2016mwo, 
DHoker:2016quv, Gerken:2018zcy}, mostly through the differential equations they satisfy. Still, the workload in simplifying 
the low-energy expansion of torus integrals grows drastically with the order in the $\ap$-expansion.

In this work, we study generating series of torus integrals and derive
an all-order formula for their $\ap$-expansion as our main result, that also exposes all relations among MGFs.
These generating series are conjectured to contain all MGFs that are relevant to closed-string 
one-loop amplitudes of type-II, heterotic and bosonic string theories. The advantage of working 
with generating series is that their differential equations in $\tau$, derived in our previous work~\cite{Gerken:2019cxz}, are valid to all orders in
$\ap$ and take a simple form for any number $n$ of punctures. 

Similar types of generating series have been constructed for one-loop open-string amplitudes,
i.e.\ for a conjectural basis of integrals over punctures on the boundary of a cylinder or M\"obius-strip 
worldsheet \cite{Mafra:2019ddf,Mafra:2019xms}. Their differential equations
have been solved to yield explicitly known combinations of iterated integrals over holomorphic
Eisenstein series ${\rm G}_k$ at all orders of the open-string $\ap$-expansions.\footnote{The $\ap$-expansion of the cylinder- and M\"obius-strip integrals in the
simplest one-loop open-string amplitudes is known to be expressible in terms
of iterated Eisenstein integrals from earlier work \cite{Broedel:2014vla, Broedel:2015hia, Broedel:2017jdo, Broedel:2019vjc}. An alternative method to determine all-order $\ap$-expansions of open-string
integrals from differential equations in auxiliary punctures has been introduced in \cite{Broedel:2019gba}.} We shall here exploit 
that the first-order differential equations of closed-string generating series have the same structure 
as their open-string counterparts \cite{Gerken:2019cxz}: Our main result is a solution of 
the closed-string differential equations that pinpoints a systematic parametrization of arbitrary MGFs in terms of iterated Eisenstein integrals and their complex conjugates.\footnote{The 
relation between MGFs and iterated Eisenstein integrals
has already been established for certain classes of 
examples \cite{Ganglzagier, DHoker:2015wxz, 
  Brown:2017qwo, Broedel:2018izr}.} The existence of such parametrizations is implied by the constructive proof announced in talks by Panzer, cf.\ e.g.\  \cite{Panzertalk}. Our generating series also provide 
a new angle on the problem of constructing bases of MGFs at given modular weights and reducing the topology of graphs one needs to consider. 

The results of this work provide a link to recent developments in the mathematics literature:
Brown constructed a class of non-holomorphic modular forms from iterated Eisenstein integrals 
and their complex conjugates which share the algebraic and differential properties of MGFs 
\cite{Brown:mmv,Brown:2017qwo,Brown:2017qwo2}. We expect the combinations of iterated Eisenstein 
integrals in our parametrization of MGFs to occur in Brown's 
generating series of single-valued iterated Eisenstein integrals that drive his construction of
modular forms: At the level of the respective generating series, single-valued iterated Eisenstein 
integrals and closed-string integrals both obey differential equations of Knizhnik--Zamolodchikov--Bernard-type in $\tau$.
Moreover, both constructions give rise to modular forms with an identical counting of
independent representatives, which is governed by holomorphic integration kernels
$\tau^j {\rm G}_k(\tau)$ with $0\leq j \leq k{-}2$ and Tsunogai's derivation algebra~\cite{Tsunogai}.

In order to generate MGFs from first-order differential equations of closed-string integrals,
we need to supplement initial values at the cusp $\tau \rightarrow i\infty$. Our generating series
at $n$ points is believed to degenerate to genus-zero integrals over moduli spaces of $(n{+}2)$-punctured spheres
similar to those in closed-string tree amplitudes. The appearance of sphere integrals will be made explicit at $n=2$
and is under investigation at $n\geq3$ \cite{yinprogress}, i.e.\ conjectural at the time of writing. 
Once the degeneration to sphere integrals is fully established at $n$ points, the initial values in our $\ap$-expansions at genus one are series in single-valued MZVs\footnote{Single-valued MZVs are obtained from evaluating single-valued polylogarithms
at unit argument \cite{Schnetz:2013hqa, Brown:2013gia}.} which arise in the $\ap$-expansion of sphere integrals \cite{Schlotterer:2012ny, Stieberger:2013wea, Stieberger:2014hba, Schlotterer:2018abc, Brown:2018omk, Vanhove:2018elu, Brown:2019wna}. Hence, the formalism in this work should reduce all MGFs to 
single-valued MZVs and real-analytic combinations of iterated Eisenstein integrals. Our results can thus be viewed as a concrete step towards genus-one relations between closed strings and single-valued open-string amplitudes 
as pioneered in \cite{Zerbini:2015rss, DHoker:2015wxz, Broedel:2018izr, Gerken:2018jrq, Zagier:2019eus}.

The present work concerns MGFs that are the building blocks of closed-string scattering amplitudes. In order to obtain the actual scattering amplitude one still has to perform the integral over the modular parameter $\tau$. While our methods do not directly give new insights into this final step, we note that a parametrisation in terms of iterated Eisenstein integrals can help in view of recent progress in representing these in terms of Poincar\'e series~\cite{Ahlen:2018wng,Dorigoni:2019yoq,Dorigoni:2020oon}. Poincar\'e-series representations of modular-invariant functions feature crucially in the Rankin--Selberg--Zagier method for integrals over $\tau$~\cite{MR656029,Bump:2005,Green:1999pv,Angelantonj:2011br,Angelantonj:2012gw} and related work in the context of MGFs can be found in~\cite{DHoker:2015gmr,DHoker:2019txf,DHoker:2019mib}.


\subsection{Summary of results}
\label{sec1.1}

The generating series of MGFs that is central to the present paper can be written in the schematic form
\begin{align}
  Y^\tau_{\vec{\eta}}(\sigma|\rho) = (\tau{-}\bar \tau)^{n-1}& \int \Big( \prod_{j=2}^n \frac{ \dd^2 z_j  }{\Im \tau}\Big)   \exp \Big( \sum_{1\leq i<j}^n s_{ij} G(z_{i}{-}z_{j},\tau) \Big)
  \label{eq1.2}  \\
  &  \times  \sigma \big[ \overline{ \varphi^\tau(z_j,\eta_j,\bar\eta_j) } \big] \rho\big[ \varphi^\tau(z_j,(\tau{-}\bar \tau)\eta_j,\bar\eta_j) \big] \, , \notag 
\end{align}
where the $n$ punctures $z_j$ are integrated over a torus of modular parameter $\tau$ (after fixing $z_1=0$ by translation invariance) and the $\eta_j$ and $\bar\eta_j$ are the formal variables of the generating series. Expanding with respect to these and the dimensionless Mandelstam variables 
\begin{align}
  s_{ij} = -\frac{\ap}{2} k_i \cdot k_j \, , \ \ \ \ \ \
  1\leq i<j\leq n
  \label{eq1.1}
\end{align}
generates MGFs. The integrand of (\ref{eq1.2})
involves doubly-periodic functions $ \varphi^\tau(z_j,\ldots) = \varphi^\tau(z_j{+}1,\ldots) =\varphi^\tau(z_j{+}\tau,\ldots) $ that 
will be spelled out below in~\eqref{eq2.13}. The asymmetric rescaling of the holomorphic bookkeeping 
variables $\eta_j$ and $\bar \eta_j$ in the last factor 
$\varphi^\tau(z_j,(\tau{-}\bar \tau)\eta_j,\bar\eta_j)$ is chosen in view of the modular properties of the generating series.
The integrals $Y^\tau_{\vec{\eta}}$ in (\ref{eq1.2}) are indexed
by permutations $\sigma,\rho \in {\cal S}_{n-1}$ that act on the subscripts $2,3,\ldots,n$
of the $\{z_j,\eta_j\}$ variables and leave $z_1$ inert. Finally, the permutation-invariant 
exponent in (\ref{eq1.2})
features the closed-string Green function $G(z,\tau)$ on the torus that will be reviewed in section~\ref{sec2} below,
where we also comment on the role of $\sigma,\rho$ in the open string \cite{Mafra:2019ddf, Mafra:2019xms, Gerken:2019cxz}.

We have conjectured in \cite{Gerken:2019cxz} that one can use integration by parts and 
Fay identities such that all basis integrals appearing in
torus amplitudes in various string theories are contained 
in the generating series $Y^\tau_{\vec\eta}$. 
This is true for all examples studied thus far (see
e.g.\ appendix D of~\cite{Gerken:2018jrq}), and it would be interesting to find a general proof,
for instance by computing the dimension of the underlying twisted cohomology as done at
tree level by Aomoto \cite{Aomoto87}.\footnote{See for instance \cite{Mizera:2017cqs, Mizera:2019gea} for a discussion in a physics context.}

It was shown in \cite{Gerken:2019cxz} that the integrals $Y^\tau_{\vec{\eta}}$ in (\ref{eq1.2})
obey a first-order differential equation in $\tau$ of schematic form
\begin{align}
 \partial_\tau Y^\tau_{\vec{\eta}}(\sigma|\rho) &= \frac{ 1 }{(\tau{-}\bar \tau)^2}\sum_{j=2}^n \bar \eta_j \partial_{\eta_j} Y^\tau_{\vec{\eta}}(\sigma|\rho) + \sum_{\alpha \in {\cal S}_{n-1}}
  {\cal D}^\tau_{\vec{\eta}}(\rho|\alpha) Y^\tau_{\vec{\eta}}(\sigma|\alpha) \,.
  \label{eq1.3}
  \end{align}
The $(n{-}1)! \times (n{-}1)!$ matrix ${\cal D}^\tau_{\vec{\eta}}(\rho|\alpha) $ comprises
second derivatives in $\eta_j$, Weierstra\ss{} functions of $\eta_j,\tau$ and has a pole in $(\tau{-}\bar\tau)^{-2}$. It is closely related to analogous operators
$D^\tau_{\vec{\eta}}$ in differential equations of open-string integrals
\cite{Mafra:2019ddf, Mafra:2019xms}.

One of the key steps for presenting the solution of~\eqref{eq1.3} in terms of iterated integrals is a 
redefinition of the generating series by the exponentiated action of a differential operator 
$R_{\vec\eta}(\epsilon_0)$ that is related to Tsunogai's derivation algebra~\cite{Tsunogai}.
The redefinition 
\beq
\Yhat^\tau_{\vec{\eta}}(\sigma|\rho)  =
\sum_{\alpha \in {\cal S}_{n-1}}
 \exp \Big( { -}\frac{R_{\vec{\eta}}(\epsilon_0) }{2\pi i (\tau {-} \bar \tau)} \Big) \, \!_{\rho}{}^\alpha  Y^\tau_{\vec{\eta}}(\sigma|\alpha)
 \label{eq1.4}
\eeq
streamlines the differential equation (\ref{eq1.3}) and in particular removes the 
poles $\sim (\tau{-}\bar \tau)^{-2}$ in both terms on the right-hand side.
This results in a differential equation of the form
\begin{align}
\partial_\tau \Yhat^\tau_{\vec{\eta}}(\sigma|\rho)  &= \sum_{k=4}^{\infty} \sum_{j=0}^{k-2} 
 (\tau{-}\bar\tau)^{j}   {\rm G}_k(\tau) 
 \sum_{\alpha \in {\cal S}_{n-1}}
 {\cal O}_{\vec{\eta}, j , k} (\rho|\alpha) 
 \Yhat^\tau_{\vec{\eta}}(\sigma|\alpha)   \, .
\label{eq1.5}
\end{align}
The $ {\cal O}_{\vec{\eta}, j , k} (\rho|\alpha) $ are matrix-valued operators  
that importantly do not depend on $\tau$ and involve at least one power in $s_{ij}$
and therefore $\ap$. Hence, one can solve (\ref{eq1.5}) perturbatively by
iterated integrals over holomorphic Eisenstein series ${\rm G}_k(\tau) $ and
thereby build up the $\ap$-expansion of (\ref{eq1.4}). The
range of the accompanying powers $(\tau{-}\bar\tau)^j , \ j \in \{0,1,\ldots,k{-}2\}$ ties
in with Brown's iterated Eisenstein integrals~\cite{Brown:mmv,Brown:2017qwo,Brown:2017qwo2}. 
More specifically, (\ref{eq1.5}) will be shown to admit an all-order solution for the original integrals (\ref{eq1.2})
\begin{align}
&Y^\tau_{\vec{\eta}}(\sigma|\rho) =
\sum_{\ell=0}^\infty
(-1)^\ell
 \sum_{\substack{k_1,k_2,\ldots,k_\ell \\=4,6,8,\ldots}}
\sum_{j_1=0}^{k_1-2} \sum_{j_2=0}^{k_2-2} \ldots \sum_{j_\ell=0}^{k_\ell-2} (2\pi i)^{-\ell+\sum_{i=1}^\ell(k_i-j_i)}
   \bsvBR{j_1 &j_2 &\ldots &j_\ell}{k_1 &k_2 &\ldots &k_\ell}{\tau} \label{eq1.6} \\
&\hspace{10mm} \times  \sum_{\alpha,\beta \in {\cal S}_{n-1}} \big[ {\cal O}_{\vec{\eta}, j_\ell , k_\ell} \cdot \ldots
 \cdot  {\cal O}_{\vec{\eta}, j_2 , k_2}  \cdot  {\cal O}_{\vec{\eta}, j_1 , k_1} \big] (\rho|\alpha) 
\exp \Big(  \frac{R_{\vec{\eta}}(\epsilon_0) }{2\pi i (\tau {-} \bar \tau)} \Big)\, \!_{\alpha}{}^\beta
\Yhat^{i\infty}_{\vec{\eta}} (\sigma|\beta)\, ,
\notag
\end{align}
see~\eqref{eq3.21} for the exact expression. The $ \bsvBRno{j_1 &j_2 &\ldots &j_\ell}{k_1 &k_2 &\ldots &k_\ell}$
are the central objects in this paper and expressible in terms of holomorphic iterated Eisenstein integrals and
their complex conjugates.
In case of a single column with entries $k\geq 4$ and $0 \leq j \leq k{-}2$, they are related to
non-holomorphic Eisenstein series, their derivatives and single-valued MZVs,
and we expect general $\bsv$ to occur in Brown's generating series of single-valued iterated 
Eisenstein integrals. 
The right-hand side of (\ref{eq1.6}) also features matrix products $\ldots {\cal O}_{\vec{\eta}, j_2 , k_2}
  \cdot  {\cal O}_{\vec{\eta}, j_1 , k_1} $ of the operators in (\ref{eq1.5}), and these operators will be seen 
to be related to Tsunogai's derivation algebra. Moreover, the
degeneration $\Yhat^{i\infty}_{\vec{\eta}} (\sigma|\beta)$ of the integrals (\ref{eq1.4})
at the cusp $\tau \rightarrow i\infty$ is a series in $s_{ij},\eta_j,\bar\eta_j$ and is conjectured\footnote{This
is a stronger form of Zerbini's conjecture \cite{Zerbini:2015rss, Zerbini:2018sox} that the expansion of modular graph {\it functions} around the cusp contains only single-valued MZVs.}
to contain only single-valued MZVs $\zeta_{k_{1},\dots,k_{r}}^{\rm sv}$ in its coefficients.
For ordinary MZVs of depth $r$ (see e.g.\ \cite{Blumlein:2009cf} for their relations over $\QQ$)
\begin{align}
\label{eq:MZV}
  \zeta_{k_{1}, k_2,\dots,k_{r}}=
  \sum_{0<n_{1}<\dots<n_{r}}^{\infty}
  n_1^{-k_1}n_2^{-k_2} \ldots n_r^{-k_r}\, ,
  \qquad k_{1},k_2,\dots, k_{r}\in\NN\,,\qquad k_{r}\geq2
\end{align}
the single-valued map\footnote{Strictly speaking, the single-valued map as in (\ref{svdepth1}) 
is only defined to exist in passing to motivic MZVs.} \cite{Schnetz:2013hqa, Brown:2013gia} at $r=1$ 
only retains cases of odd weight $k_1$,
\beq
\zeta_{2k+1}^{\rm sv} = 2 \zeta_{2k+1} \, , \ \ \ \ \ \ \zeta_{2k}^{\rm sv} = 0 \, .
\label{svdepth1}
\eeq
The degeneration limit $\Yhat^{i\infty}_{\vec{\eta}} (\sigma|\beta)$ at $n=2$ points will be explicitly
reduced to $\zeta_{k}^{\rm sv}$ via~\eqref{eq4.1} which proves our claim in this case, and the degenerations at higher multiplicities $n\geq 3$ are under investigation~\cite{yinprogress}.

We shall investigate the modular and reality properties of the $\bsv$ appearing in~\eqref{eq1.6} and
express them in terms of iterated Eisenstein integrals and their complex conjugates.
These representations $\bsv$ will follow solely based on their derivative w.r.t.\ $\tau$ together 
with the reality properties of the generating series $Y^\tau_{\vec\eta}$.
In particular, the antiholomorphic constituents of $ \bsvBRno{j_1 &j_2}{k_1 &k_2}$ up to
$k_1{+}k_2\leq 10$ turn out to involve only $\zeta_{k}^{\rm sv}$ as follows from 
detailed studies of the two- and three-point generating series.

The relation of the $\bsv$ to the derivation algebra imply that not all combinations of $\bsv$ 
can actually appear independently in the generating series. Together with the conjecture that 
$Y^\tau_{\vec\eta}$ contains all possible MGFs, this allows us to give a precise count and 
determination of the relations between MGFs beyond the weights that have been studied to date.

By exploiting also the reality properties of the $\bsv$, the counting allows us to distinguish between real and imaginary MGFs in the basis. Since imaginary MGFs are cuspidal \cite{DHoker:2019txf}, we can hence also identify the number of imaginary non-holomorphic cusp forms in the spectrum of MGFs with our method. In particular, we show that at modular weight $(5,5)$ three imaginary cusp forms are necessary for a basis of MGFs of arbitrary topology, extending the analysis of two-loop graphs in \cite{DHoker:2019txf} by one new cusp form. 
The total number of independent MGFs of modular weight $(w,w)$ with $w\leq 8$ and the number of
imaginary cusp forms contained in these is given by
\begin{center}
  \vspace{0.5em}
  \begin{tabular}{cccccccccc}
    \toprule
    mod. weight $(w,\bar{w})$ & $(0,0)$ & $(1,1)$ & $(2,2)$ & $(3,3)$ & $(4,4)$ & $(5,5)$ & $(6,6)$ & $(7,7)$ & $(8,8)$\\
    \midrule
    \# MGFs & 1 & 0 & 1 & 1 & 4 & 7 & 19 & 43 & 108 \\
    \# imag.\ cusp forms & 0 & 0 & 0 & 0 & 0 & 3 & 5 & 19 & 42 \\
    \bottomrule
  \end{tabular}
  \vspace{0.5em}
\end{center}
This counting includes products of MGFs but excludes products involving MZVs. In this table, we have focused on cases with $w=\bar{w}$ that can be turned into modular invariant functions by multiplying by $(\Im \tau)^w$ and these cases include not only the modular graph \textit{functions} originally studied in~\cite{DHoker:2015gmr, DHoker:2015wxz}, but also more general modular invariant objects. A more detailed counting including cases with $w\neq \bar{w}$ will be presented in section~\ref{sec6.2}.


\subsection{Outline}
\label{sec1.2}

We introduce the ingredients of the generating series $Y^\tau_{\vec\eta}$ and its properties in section~\ref{sec2}. In section~\ref{sec3}, we study in detail the transition from~\eqref{eq1.3} to~\eqref{eq1.5} and how this leads to the $\bsv$ together with their relation to iterated Eisenstein integrals. This includes a discussion of integration ambiguities given by antiholomorphic functions and how they can be fixed from reality properties. In section~\ref{sec4}, we implement the general scheme in the simplest two-point case and show how this fixes already a large number of $\bsv$. Further $\bsv$ are then fixed by adding in data from $n=3$ points in section~\ref{sec5}, where we also encounter imaginary cuspidal MGFs and study their properties. Section~\ref{sec6} is devoted to modular transformation properties of general $\bsv$ as well as their implications on the classification of independent MGFs
and the transcendentality properties of closed-string integrals. A summary with some open questions is contained in section~\ref{sec7}. Several appendices collect complementary details and some of the more lengthy expressions for the $\bsv$ and similar objects $\Esv$. 

\medskip

{\bf Note:} Some of the explicit expressions relating MGFs, $\bsv$ and $\Esv$ can be quite lengthy, and the arXiv submission of this paper includes an ancillary Mathematica and data file where these relations and expansions of the generating series $Y^\tau_{\vec\eta}$ at $n=2$ and $n=3$ points up to total order $10$ are available.


\section{Generating series of closed-string integrals}
\label{sec2}

In this section, we will spell out the detailed form of the generating series $Y^\tau_{\vec{\eta}}$ 
in (\ref{eq1.2}) and recall its differential equations derived in \cite{Gerken:2019cxz}. For this we first need to 
introduce the basic building blocks entering $Y^\tau_{\vec{\eta}}$ 
and also review the connection to modular graph forms.


\subsection{Kronecker--Eisenstein integrands and Green function}
\label{sec2.1}

The generating series $Y^\tau_{\vec{\eta}}$ is constructed out of the so-called doubly-periodic Kronecker--Eisenstein series and a Koba--Nielsen factor that involves the scalar Green function on the worldsheet torus.

The torus Kronecker--Eisenstein series in its doubly-periodic form reads~\cite{Kronecker,BrownLev}
 \begin{align}
  \label{eq2.1}
  \Omega(z,\eta,\tau) := \exp\left( 2\pi i \eta \frac{\Im z}{\Im \tau}\right)  \frac{\theta'(0,\tau) \theta(z+\eta,\tau)}{\theta(z,\tau)\theta(\eta,\tau)}
\end{align}
with $\theta(z,\tau)$ the odd Jacobi theta function and $\theta'(z,\tau)$ its derivative in the first argument. 
The function $\Omega(z,\eta,\tau)$ is doubly-periodic in the torus variable $z \cong z+1 \cong z+\tau$ and can be 
Laurent-expanded in the formal variable $\eta$. This expansion yields an infinite tower of 
doubly-periodic functions $f^{(w)}(z,\tau)$ via
\begin{align}
  \label{eq2.2}
  \Omega(z,\eta,\tau) = \sum_{w=0}^{\infty} \eta^{w-1} f^{(w)}(z,\tau)
    \,.
\end{align}
The significance of the functions $f^{(w)}$ is that all correlation functions of one-loop 
massless (and possibly massive) closed-string amplitudes in
bosonic, heterotic and type-II theories are expressible through them~\cite{Gerken:2018jrq}\footnote{More 
specifically, see \cite{Broedel:2014vla, Berg:2016wux} and \cite{Dolan:2007eh} for the appearance of $f^{(w)}(z,\tau)$ in the spin sums
of the RNS formalism and the current algebra of heterotic strings, respectively.}. Their simplest 
instances are
\begin{align}
  f^{(0)}(z,\tau) &=1 \, , \hspace{10mm}
    f^{(1)}(z,\tau) =\partial_z \log \theta(z,\tau) + 2\pi i \frac{\Im z}{\Im \tau} \,,
    \label{eq2.3}
\end{align}
and all the $f^{(w\geq 2)}$ are non-singular on the entire torus. Only $f^{(1)}$ has a simple pole at $z=0$
and in fact at all lattice points $z\in \ZZ + \tau \ZZ$.

The real scalar Green function on the torus is
\begin{align}
  G(z,\tau) = -\log \left| \frac{\theta(z,\tau)}{\eta(\tau)}\right|^2  + \frac{2\pi(\Im z)^2}{\Im \tau}\,,
\label{eq2.4}
\end{align}
where $\eta(\tau)=q^{1/24} \prod_{n=1}^\infty(1-q^n)$ denotes the Dedekind eta-function and $q= e^{2\pi i \tau}$.

Under modular transformations with $\begin{psmallmatrix}\alpha&\beta\\\gamma&\delta\end{psmallmatrix}\in {\rm SL}_{2}(\ZZ)$ the doubly-periodic functions and the Green function obey the following simple transformation laws:
\begin{subequations}
\label{eq2.mod}
\begin{align}
  \Omega\left( \frac{z}{\gamma\tau+\delta} , \frac{\eta}{\gamma\tau+\delta} , \frac{\alpha\tau+\beta}{\gamma\tau+\delta} \right) &= (\gamma\tau+\delta) \Omega(z,\eta,\tau)\,,
  \label{eq2.6}
\\
f^{(w)}\left( \frac{z}{\gamma\tau+\delta} ,  \frac{\alpha\tau+\beta}{\gamma\tau+\delta} \right) &= 
(\gamma\tau+\delta)^w f^{(w)}(z,\tau) \label{eq2.7} \,, \\
G\left( \frac{z}{\gamma\tau+\delta} ,\frac{\alpha\tau+\beta}{\gamma\tau+\delta} \right) &= G(z,\tau) \, .
  \label{eq2.8}
\end{align}
\end{subequations}
Objects that transform with a factor of $(\gamma\tau+\delta)^w (\gamma\bar\tau+\delta)^{\bar{w}}$
under $ {\rm SL}_{2}(\ZZ)$ will be said to carry (holomorphic and antiholomorphic) modular weight $(w,\bar{w})$. 
Thus, one can read off weight $(1,0)$ for $\Omega$, weight $(w,0)$ for $f^{(w)}$ and weight $(0,0)$ for the Green function which is also referred to as modular invariant.

One-loop amplitudes of closed-string states are built from 
$n$-point correlation function of vertex operators on a worldsheet torus.
The plane-wave parts of vertex operators with lightlike external
momenta $k_i$ ($i=1,2,\ldots,n$) always contribute the so-called 
Koba--Nielsen factor~\cite{Green:1987mn}
\begin{align}
  \label{eq2.5}
  \KN^{\tau}_{n} := \prod_{1\leq i<j}^n \exp \left( s_{ij} G(z_{ij}, \tau) \right)\,,
\end{align}
comprising Green functions connecting the various vertex insertions 
and the Mandelstam variables $s_{ij}$ defined in~\eqref{eq1.1}.


\subsection{Generating series and component integrals}
\label{sec2.2}

An $n$-point correlation function of massless vertex operators on a worldsheet torus with fixed 
modular parameter $\tau$ depends on the punctures $z_i$ via Green functions, 
$f^{(w)}$ and $\overline{f^{(w)}}$ \cite{Gerken:2018jrq}. Since the $f^{(w)}$ and $\overline{f^{(w)}}$ are generated by the Kronecker--Eisenstein series $\Omega$ via~\eqref{eq2.2} and the Green functions from the Koba--Nielsen factor $\KN^\tau_n$ via~\eqref{eq2.5}, it is natural to consider generating functions involving these objects.
Moreover, one-loop closed-string amplitudes require integrating the punctures
$z_i\cong z_i {+} m\tau{+}n, \ m,n \in \ZZ$ over the torus, and the modular invariant integration
measure is normalised as $\int \frac{\dd^2z_i}{\Im\tau} =1$.

In order to exhibit our generating series of torus integrals, we begin with the simplest two-point case, where
the above reasoning leads to considering \cite{Gerken:2019cxz}
\beq
Y^\tau_{\eta} =  
(\tau{-}\bar\tau) \int \frac{\dd^2z_2}{\Im \tau}
  \, \overline{\Omega(z_{12} , \eta, \tau)}
\Omega(z_{12} , (\tau{-}\bar \tau) \eta, \tau)
\KN^\tau_2
 \label{eq2.12}
\eeq
with $z_{i j} =z_i - z_j$, and we have used translation invariance on the torus to fix $z_1=0$. 

The $n$-point generalisation is an $(n{-}1)!\times (n{-}1)!$ matrix $Y_{\vec{\eta}}^\tau ( \sigma  | \rho )$
labelled by permutations $\sigma,\rho\in \SG_{n-1}$
and involving $n{-}1$ parameters $\vec\eta=(\eta_2,\eta_3,\ldots,\eta_{n})$ \cite{Gerken:2019cxz},
\begin{align}
Y_{\vec{\eta}}^\tau &( \sigma  | \rho ) = Y_{\vec{\eta}}^\tau \big(1, \sigma(2,\ldots, n) | 1,\rho(2,\ldots,n) \big)  = 
(\tau{-}\bar\tau)^{n-1} \int \Big( \prod_{j=2}^n \frac{\dd^2z_j}{\Im\tau} \Big) \KN_n^\tau \notag \\
&\ \ \times
\sigma\Big[ \overline{\Omega(z_{12} , \eta_{23\ldots n}, \tau)} \,\overline{\Omega(z_{23},\eta_{34\ldots n}, \tau)} \cdots \overline{\Omega(z_{n-2,n-1}, \eta_{n-1,n},\tau) }\,\overline{\Omega(z_{n-1,n} ,\eta_n,\tau)}     \Big]     \label{eq2.13}
\\ 
&\ \ \times  \rho\Big[ \Omega(z_{12} ,(\tau{-}\bar \tau) \eta_{23\ldots n}, \tau) \,\Omega(z_{23}, (\tau{-}\bar \tau)\eta_{34\ldots n}, \tau)\cdots  \Omega(z_{n-1,n} ,(\tau{-}\bar \tau) \eta_n,\tau )  \Big]\, , \notag
\end{align}
where we have used the shorthand $\eta_{i \ldots j} = \eta_i + \ldots + \eta_j$.
The permutations $\sigma,\rho$ act on the subscripts of the generating parameters $\eta_i$ and insertion points $z_i$ and are necessary to obtain homogeneous first-order differential equations in $\tau$ for
the matrix $Y_{\vec{\eta}}^\tau ( \sigma  | \rho )$. 
We will refer to the entries of $Y_{\vec{\eta}}^\tau ( \sigma  | \rho )$ by writing the images of the elements $(2,3,\ldots,n)$ under the permutations $\rho$ and $\sigma$.
Thus, $Y^\tau_{\eta_2,\eta_3}(2{,}3|2{,}3)$ at $n=3$ corresponds to the trivial elements of $\mathcal{S}_2$ 
while $Y^\tau_{\eta_2,\eta_3}(3{,}2|2{,}3)$ represents the non-trivial element $\sigma \in \mathcal{S}_2$
that maps the factor of $\overline{ \Omega(z_{12} ,  \eta_{23}, \tau) } \, \overline{\Omega(z_{23}, \eta_3,\tau)}$
in the integrand to $\overline{ \Omega(z_{13} ,  \eta_{23}, \tau) } \, \overline{\Omega(z_{32}, \eta_2,\tau)}$.

In the open-string versions of the integrals (\ref{eq2.13}), the permutation $\sigma$ refers to an 
integration domain (a cyclic ordering of open-string punctures on a cylinder boundary) in the place of 
the complex conjugate $\overline{\Omega}$ \cite{Mafra:2019ddf, Mafra:2019xms, Gerken:2019cxz}.
  The asymmetric choice of second arguments $(\tau{-}\bar \tau)\eta_j$ and $\bar\eta_j$
  for the $\Omega$ and $\overline{ \Omega}$ in the generating series (\ref{eq2.13}) is motivated by aiming for specific 
  modular weights as we shall discuss below. 
  
We will use extensively the following component integrals
\beq
\label{eq2.14}\
Y^\tau_{(a|b)} = \frac{1}{(2\pi i )^b} Y^{\tau}_{\eta_2} \, \big|_{\eta_2^{a-1} \bar \eta_2^{b-1}}
=  \frac{  (\tau {-} \bar \tau)^a }{ (2\pi i)^{b} } \int  \frac{ \dd^2 z_2 }{\Im \tau}  \, 
\KN_2^\tau  \,  f^{(a)}_{12} \, \overline{f^{(b)}_{12}} 
\eeq
with the shorthand
\beq
f^{(a)}_{ij} = f^{(a)}(z_{i}-z_{j},\tau) \, ,
\label{eq2.short}
\eeq
where the normalising factor $(2\pi i)^{-b}$ was chosen to simplify some relations under complex conjugation below.
The components of the $n$-point generating function~\eqref{eq2.13} are similarly defined as
\begin{align}
Y^\tau_{(a_2,a_3,\ldots,a_n|b_2,b_3,\ldots, b_n)}(\sigma|\rho) &= \frac{1}{(2\pi i )^{b_2+b_3+\ldots +b_n}} Y^{\tau}_{\vec{\eta}}(\sigma|\rho)  \, \big|_{\eta_{23\ldots n}^{a_2 -1}\eta_{3\ldots n}^{a_3 -1}\ldots \eta_{n}^{a_n -1} \bar \eta_{23\ldots n}^{b_2-1}  \bar \eta_{3\ldots n}^{b_3-1} \ldots  \bar \eta_{n}^{b_n-1}} \notag \\
&=  \frac{  (\tau {-} \bar \tau)^{a_2+a_3+\ldots+a_n} }{ (2\pi i)^{b_2+b_3+\ldots+b_n} } \Big( \prod_{j=2}^n \int  \frac{ \dd^2 z_j }{\Im \tau} \Big)\, \KN_n^\tau
\label{eq2.15} \\
&\ \ \ \ \ \ \ \ \ \  \times \rho\big[  f^{(a_2)}_{12} \,  f^{(a_3)}_{23} \ldots  f^{(a_n)}_{n-1,n}\big] 
 \,  \sigma\big[ \overline{f^{(b_2)}_{12}} \, \overline{f^{(b_3)}_{23}} \ldots  \overline{f^{(b_n)}_{n-1,n}}  \big] \, .
 \notag
\end{align}
By the modular transformations (\ref{eq2.mod}) together with
\beq
\Im \left(  \frac{\alpha\tau+\beta}{\gamma\tau+\delta} \right) = \frac{ \Im \tau }{(\gamma\tau+\delta)(\gamma\bar \tau+\delta) }
\label{eq2.16}
\eeq
the modular weights of the component integrals (\ref{eq2.14}) and (\ref{eq2.15}) are
\beq
Y^\tau_{(a|b)} \leftrightarrow \te{weight} \, (0,b{-}a) \, , \ \ \ \ \ \
Y^\tau_{(a_2,\ldots,a_n|b_2,\ldots, b_n)}(\sigma|\rho)
\leftrightarrow \te{weight} \, \Big(0,\sum_{j=2}^n(b_j{-}a_j)\Big)\, .
\label{eq2.17}
\eeq
This property holds at each order in the $\ap$-expansion, so the integrals are generating
functions of modular forms. The fact that the holomorphic modular weight of the component integral vanishes was the reason for making the asymmetric definitions~\eqref{eq2.12} and~\eqref{eq2.13}. These definitions also lead to a more tractable differential equation that we shall analyse in detail in this paper.

We will later make essential use of the following reality properties: Complex conjugation of 
component integrals over $f^{(a)}_{12}  \overline{f^{(b)}_{12}} $ exchanges 
$a\leftrightarrow b$, so we have
\beq
Y^\tau_{(a|b)}  = (4y)^{a-b} \overline{ Y^\tau_{(b|a)} } \, , \ \ \ \ \ \ 
\overline{ Y^\tau_{(a|b)} } = (4y)^{a-b} Y^\tau_{(b|a)} \, ,
\label{eq2.18}
\eeq
where $y = \pi \Im \tau$, and similarly,
\begin{align}
 Y^\tau_{(a_2,\ldots,a_n|b_2,\ldots, b_n)}(\sigma|\rho)  &= 
 \Big(\prod_{j=2}^n (4y)^{a_j -b_j}  \Big)
 \overline{ Y^\tau_{(b_2,\ldots, b_n| a_2,\ldots,a_n)} (\rho|\sigma) } \, .
\label{eq2.19}
\end{align}
%


\subsection{Modular graph forms}
\label{sec2.3}

In a series-expansion w.r.t.\ $\alpha'$, the component integrals (\ref{eq2.14}) and (\ref{eq2.15}) can be 
conveniently performed in Fourier space, see appendix \ref{secA.1}. This leads to nested lattice sums over non-vanishing discrete momenta $p = m \tau + n$ on the torus with $m,n\in \ZZ$.
In the case of the two-point 
component integrals (\ref{eq2.14}), the $z_2$ integration yields for instance expressions of the type
\begin{align}
  \cform{a_1&a_2&\ldots &a_R \protect\\b_1&b_2&\ldots &b_R}\! (\tau) = 
   \sum_{p_1,\ldots,p_R\neq 0} \frac{\delta(p_1+\ldots +p_R)}{p_1^{a_1} \bar{p}_1^{b_1} \cdots p_R^{a_R}\bar{p}_R^{b_R}} 
 \label{eq2.22}
\end{align}
with integer labels $a_i,b_i$. Here,  $\sum_{p\neq 0}$ instructs us to sum over 
all $p = m \tau + n$ with $(m,n) \in \ZZ^2$ and 
$(m,n) \neq (0,0)$, resulting in modular weight $\sum_{i=1}^R (a_i, b_i)$.

\begin{figure}
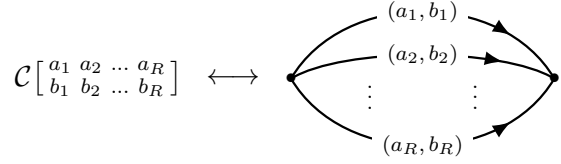

  \begin{center}
    \tikzpicture[scale=0.7,line width=0.3mm,decoration={markings,mark=at position 0.8 with {\arrow[scale=1.3]{latex}}}]
    \draw(-2.7,0)node{$\cform{a_1&a_2&\ldots &a_R \protect\\b_1&b_2&\ldots &b_R}$};
    \draw(-0.1,0)node{$\displaystyle \longleftrightarrow$};
    \draw[postaction={decorate}] (1,0)node{\scriptsize $\bullet$} .. controls (2,1.7) and (5,1.7) .. node[fill=white]{\scriptsize $(a_1,b_1)$} (6,0)node{\scriptsize $\bullet$};
    \draw[postaction={decorate}] (1,0) .. controls (2,0.6) and (5,0.6) .. node[fill=white]{\scriptsize $(a_2,b_2)$} (6,0);
    \draw (2.5,-0.2) node{\scriptsize $\vdots$};
    \draw (4.5,-0.2) node{\scriptsize $\vdots$};
    \draw[postaction={decorate}] (1,0) .. controls (2,-1.7) and (5,-1.7) .. node[fill=white]{\scriptsize $(a_R,b_R)$} (6,0);
    \endtikzpicture
    \caption{\it Dihedral graph with decorated edges and notation for modular graph form.}
    \label{fig:diMGF}
  \end{center}
\end{figure}

Formula~\eqref{eq2.22} is an example of a modular graph form (MGF)~\cite{DHoker:2015wxz, DHoker:2016mwo},
here associated with a dihedral graph topology of lines connecting the two insertion points, see figure~\ref{fig:diMGF}. The momentum-conserving 
delta function obstructs nonzero one-column MGFs, so the simplest examples of dihedral topology are
\begin{align}
  \cform{a&0 \protect\\b &0}\! (\tau) =  \sum_{p\neq 0} \frac{1}{p^{a} \bar{p}^{b} } 
  = \sum_{\substack{(m,n)\in \ZZ^2 \\ (m,n)\neq (0,0) }}  \frac{1}{(m\tau{+}n)^{a} (m\bar{\tau}{+}n)^{b} } \,.
\label{eq2.23}
\end{align}
Special cases are given by non-holomorphic Eisenstein series (convergent for $k\geq 2$)
\beq
{\rm E}_k(\tau) = \Big( \frac{\Im \tau}{\pi} \Big)^k   \cform{k&0 \protect\\k &0}\! (\tau)
= \Big( \frac{\Im \tau}{\pi} \Big)^k \sum_{p\neq 0} \frac{1}{ |p|^{2k}}  \,.
\label{eq2.24}
\eeq
They are real and modular invariant due to the prefactor $(\Im\tau)^k$, see (\ref{eq2.16}).

\medskip

Similar to the MGF~\eqref{eq2.22} associated with the dihedral topology in figure~\ref{fig:diMGF},
one can introduce MGFs for any graph $\Gamma$ with labelled edges~\cite{DHoker:2015wxz, DHoker:2016mwo}. 
As exemplified for the trihedral case in appendix~\ref{secA.2},
the notation ${\cal C}_\Gamma\!\big[\raisebox{.1\height}{\scalebox{.7}{$\begin{array}{c}\mathcal{A}\\[-2mm] \mathcal{B}\end{array}$}}\big]$ for the corresponding MGF has to track
the holomorphic labels $\mathcal{A}$, the antiholomorphic labels $\mathcal{B}$ and the adjacency properties of the edges of the graph $\Gamma$.
We follow the conventions of~\cite{Gerken:2019cxz} for their normalisation where
the modular weight $(w,\bar{w})$ is obtained by summing the labels of all edges,
\begin{align}
{\cal C}_\Gamma\!\big[\raisebox{.1\height}{\scalebox{.7}{$\begin{array}{c}\mathcal{A}\\[-2mm] \mathcal{B}\end{array}$}}\big]
\ \ \ \leftrightarrow \ \ \ \te{modular weight} \
  (w,\bar{w})=\left(\sum_{a\in\mathcal{A}}a,\sum_{b\in\mathcal{B}}b\right)\,,
  \label{eq:1}
\end{align}
cf.\ appendix~\ref{secA.2} for the trihedral case. Even though more complicated graph 
topologies are ubiquitous in the MGF literature, one of the results of the present paper is that, up to total modular 
weight $w{+}\bar{w}=12$, dihedral MGFs are sufficient for providing a basis of all MGFs. This is discussed in more 
detail in section~\ref{sec:esumm}.

\subsubsection{Differential operators and equations} 

The more general case of~\eqref{eq2.23} with $a\neq b$ can be accounted for by using the following derivative operators\footnote{These were called $\nabla_{\text{DG}}$ in~\cite{Gerken:2019cxz} and correspond to $(\Im\tau)$ times Maa\ss{} raising and lowering operators. The normalisation conventions for $\nabla$ are identical to those
in \cite{DHoker:2016mwo, DHoker:2016quv, Broedel:2018izr, Gerken:2018jrq, DHoker:2019txf}.}
\begin{align}
\label{eq2.25}
\nabla:=2i (\Im\tau)^{2}\partial_{\tau} \,,\quad\quad \overline{\nabla}:= - 2i (\Im\tau)^{2}\partial_{\bar\tau} \,.
\end{align}
The operator $\nabla$ has the property that it maps an object of modular weight $(0,\bar{w})$ to $(0,\bar{w}-2)$ and, by~\eqref{eq2.17}, is therefore an appropriate operator for the component integrals~\eqref{eq2.15}. The operator $\overline\nabla$ similarly maps weight $(w,0)$ to $(w{-}2,0)$.

Acting with the differential operator $\nabla$ on the non-holomorphic Eisenstein series 
${\rm E}_k$ in~\eqref{eq2.24} leads to
\beq
\nabla^m {\rm E}_k(\tau)  = \frac{(\Im \tau)^{k+m}}{\pi^k} \frac{(k+m-1)!}{(k-1)!}  \cform{k+m&0 \protect\\k-m &0}\! (\tau)\,.
\label{eq2.26}
\eeq
Cases with $m=k$ yield holomorphic Eisenstein series 
\beq
{\rm G}_k(\tau) = \cform{k&0 \protect\\0 &0}\! (\tau)
= \sum_{p\neq 0} \frac{1}{ p^{k}}  
\label{eq2.27}
\eeq
that converge absolutely for $k\geq 4$ and vanish for odd $k$. Formula~\eqref{eq2.26} specialises in this case to
\beq
(\pi \nabla)^k \mathrm{E}_k(\tau) = \frac{ (2k-1)! }{ (k-1)!} (\Im \tau)^{2k} {\rm G}_{2k}(\tau)\,.
\label{eq2.28}
\eeq
We will encounter the following generalisations that are also real and modular-invariant~\cite{Broedel:2018izr}:$ \! \! \! \! \! $
\begin{subequations}
\label{eq2.com}
\begin{align}
  \mathrm{E}_{2,2} &=\Big(\frac{\Im\tau}{\pi}\Big)^{4}\cform{1&1&2\\1&1&2} -\frac{9}{10}\mathrm{E}_{4} \,,
   \label{eq2.29}\\
  \mathrm{E}_{2,3} &=\Big(\frac{\Im\tau}{\pi}\Big)^{5}\cform{1&1&3\\1&1&3} -\frac{43}{35}\mathrm{E}_{5}\,,
   \label{eq2.30}\\
  \mathrm{E}_{3,3} &=\Big(\frac{\Im\tau}{\pi}\Big)^{6}\big(3\cform{1&2&3\\1&2&3}+\cform{2&2&2\\2&2&2}\big)-\frac{15}{14}\mathrm{E}_{6}\,,
  \label{eq:3}\\
  \mathrm{E}'_{3,3} &=\Big(\frac{\Im\tau}{\pi}\Big)^{6}\big(\cform{1&2&3\\1&2&3}+\frac{17}{60}\cform{2&2&2\\2&2&2}\big)-\frac{59}{140}\mathrm{E}_{6}\,,
  \label{eq:4}\\
  \mathrm{E}_{2,4} &=\Big(\frac{\Im\tau}{\pi}\Big)^{6}\big(9\cform{1&1&4\\1&1&4}+3\cform{1&2&3\\1&2&3}+\cform{2&2&2\\2&2&2}\big)-13\mathrm{E}_{6}\,,
  \label{eq:5}\\
  \mathrm{E}_{2,2,2} &=\Big(\frac{\Im\tau}{\pi}\Big)^{6}\Big(\frac{232}{45}\cform{2&2&2\\2&2&2}+\frac{292}{15}\cform{1&2&3\\1&2&3}+\frac{2}{5}\cform{1&1&4\\1&1&4}-\cform{1&1&2&2\\1&1&2&2}\Big)\nonumber\\
  &\quad+2\mathrm{E}_{3}^{2}+\mathrm{E}_{2}\mathrm{E}_{4}-\frac{466}{45}\mathrm{E}_{6}\,.\label{eq:6}
\end{align}
\end{subequations}
The above MGFs all belong to the dihedral class and arise in $(n{\geq} 2)$-point component integrals. More complicated graph topologies arise for the higher-point component integrals (\ref{eq2.15}), and a brief review of trihedral
modular graph forms can be found in appendix \ref{secA.2}.

The particular choice of combinations in the above expressions simplifies 
the differential equation and delays the occurrence of holomorphic Eisenstein series 
${\rm G}_k$ as much as possible when taking Cauchy--Riemann derivatives.
This leads for instance to the following differential 
equations~\cite{DHoker:2016mwo, Broedel:2018izr}
\begin{align}
(\pi \nabla)^3 \mathrm{E}_{2,2} &= - 6 (\Im \tau)^4 {\rm G}_4 \pi \nabla \mathrm{E}_{2} 
\label{eq2.31}  \,, \\
 (\pi \nabla)^3 \mathrm{E}_{2,3} &= - 2 (\pi \nabla \mathrm{E}_{2}) (\pi \nabla)^2 \mathrm{E}_{3} 
 - 4 (\Im \tau)^4 {\rm G}_4  \pi \nabla  \mathrm{E}_{3} \, .
\notag
 \end{align}

\subsubsection{\texorpdfstring{Examples in $\ap$-expansions}{Examples in alpha'-expansions}  }

As an example of how MGFs occur in the component integrals of the generating series $Y^\tau_{\vec{\eta}}$, we consider the two-point integrals~\eqref{eq2.14}. It can be checked by using identities for modular graph forms~\cite{DHoker:2016mwo} that the first few component integrals have the following $\alpha'$-expansions\footnote{When comparing with the $\ap$-expansions in (2.69) of \cite{Gerken:2019cxz}, note that the component integrals
(\ref{eq2.14}) are related to the $W_{(a|b)}^\tau$ in the reference via $Y_{(a|b)}^\tau
= \frac{ (2i \Im \tau)^a }{(2\pi i )^b} W_{(a|b)}^\tau$.}
 \begin{subequations}
 \label{eq2.all36}
\begin{align}
Y^{\tau}_{(0|0)}&=
1 + \frac{1}{2}s_{12}^2 {\rm E}_2 +\frac{1}{6} s_{12}^3  ({\rm  E}_3 + \zeta_3)
+ s_{12}^4 \Big( {\rm  E}_{2,2} +\frac{ 1}{8} {\rm  E}_2^2 + \frac{ 3}{20} {\rm  E}_4  \Big)
\label{eq2.36}  \\
&\quad + s_{12}^5 \Big(  \frac{1}{2} {\rm  E}_{2,3} + \frac{1}{12} {\rm  E}_2 ({\rm  E}_3 + \zeta_3) + \frac{3}{14} {\rm  E}_5
 + \frac{ 2 \zeta_5}{15} \Big) + {\cal O}(s_{12}^6)\,,
\notag \\
Y^{\tau}_{(2|0)}&=
2 s_{12} \pi \nabla {\rm E}_2   + \frac{2}{3} s_{12}^2 \pi \nabla  {\rm E}_{3}
+ s_{12}^3 \Big( \frac{ 3}{5} \pi \nabla {\rm E}_{4} + 4 \pi \nabla {\rm E}_{2,2} + {\rm E}_2  \pi \nabla {\rm E}_{2} \Big) \label{eq2.37}\\
&\quad +s_{12}^4 \Big(\frac{ 6}{7}  \pi \nabla {\rm E}_{5} + 2  \pi \nabla {\rm E}_{2,3} + \frac{1}{3}  {\rm E}_{2} \pi \nabla {\rm E}_{3}  + \frac{ 1}{3} {\rm E}_{3}  \pi \nabla {\rm E}_{2} + \frac{1}{3} \zeta_3  \pi \nabla {\rm E}_{2}  \Big)+ {\cal O}(s_{12}^5)\,,
\notag \\
Y^{\tau}_{(4|0)}&= - \frac{4}{3}
s_{12}  ( \pi \nabla)^2  {\rm E}_{3}  + s_{12}^2 \Big( {-} \frac{6}{5}   ( \pi \nabla)^2 {\rm E}_{4} 
+ 2    ( \pi \nabla  {\rm E}_{2})^2 \Big)  \label{eq2.38}  \\
&\quad +s_{12}^3 \Big( {-} \frac{12}{7}  ( \pi \nabla)^2 {\rm E}_{5} - 4   ( \pi \nabla)^2 {\rm E}_{2,3}
  - \frac{ 4}{3} (  \pi \nabla  {\rm E}_{2})  ( \pi \nabla  {\rm E}_{3})
 - \frac{ 2}{3}  {\rm E}_{2}   ( \pi \nabla)^2 {\rm E}_{3}  \Big)+ {\cal O}(s_{12}^4)
\notag\,.
 \end{align}
  \end{subequations}

\subsubsection{Laurent polynomials} 
 
The expansion of MGFs around the 
the cusp $\tau \to i \infty$ are expected to take the form
\begin{align}
{\cal C}_\Gamma\!\big[\raisebox{.1\height}{\scalebox{.7}{$\begin{array}{c}\mathcal{A}\\[-2mm] \mathcal{B}\end{array}$}}\big] = \sum_{m,n\geq 0} c_{m,n}( \Im \tau) q^m \bar{q}^n\,,
\label{eqcusp.1}
\end{align}
where $c_{m,n}( \Im \tau)$ are Laurent polynomials in $\Im \tau$, see e.g.\
Theorem 1.4.1 of \cite{Zerbini:2018sox}. An important property of MGFs is the
Laurent polynomial $c_{0,0}( \Im \tau)$ corresponding to the $q$- and $\bar{q}$-independent terms  
in (\ref{eqcusp.1}). As exemplified by (with Bernoulli numbers $B_{2k}$) \cite{Green:2008uj, DHoker:2015gmr}
\begin{subequations}
\label{eqcusp.mzv}
\begin{align}
{\rm E}_k  &= (-1)^{k-1} \frac{ B_{2k} }{(2k)!} (4y)^k + \frac{ 4 (2k{-}3)! }{(k{-}2)!(k{-}1)!} \frac{ \zeta_{2k-1} }{(4y)^{k-1}} + {\cal O}(q,\bar q)\,, \label{eqcusp.mzva} \\
{\rm E}_{2,2} &= - \frac{ y^4}{20250} + \frac{ y \zeta_3}{45} 
+ \frac{ 5 \zeta_5 }{12y} - \frac{ \zeta_3^2 }{4y^2} + {\cal O}(q,\bar q)\,, \label{eqcusp.mzvb} \\
{\rm E}_{2,3} &= - \frac{ 4y^5 }{297675}+ \frac{ 2 y^2 \zeta_3 }{945} - \frac{ \zeta_5 }{180}
+ \frac{ 7 \zeta_7}{16y^2} - \frac{ \zeta_3 \zeta_5 }{2y^3} + {\cal O}(q,\bar q)  \, ,\label{eqcusp.mzvc} 
\end{align}
\end{subequations}
the coefficients in the Laurent polynomials $c_{0,0}( \Im \tau)$
are conjectured\footnote{In the case of modular graph functions with $a_j=b_j$, the coefficients
in the Laurent polynomials $c_{0,0}( \Im \tau)$ are proven to be $\QQ$-linear combinations
of cyclotomic MZVs \cite{Zerbini:2015rss, Zerbini:2018sox}.} to be $\QQ$-linear 
combinations of single-valued MZVs \cite{Zerbini:2015rss, DHoker:2015wxz, Zerbini:2018sox}
when written in terms of $y= \pi \Im \tau$.

\subsubsection{Cusp forms}

Modular graph forms have a simple transformation under complex conjugation that just exchanges the $a_i$ and $b_i$ labels. For dihedral graphs this means
\begin{align}
\overline{  \cform{a_1&a_2&\ldots &a_R \protect\\b_1&b_2&\ldots &b_R}  } =   \cform{b_1&b_2&\ldots &b_R \protect\\a_1&a_2&\ldots &a_R} \,.
\label{eqcusp.2}
\end{align}
We will encounter imaginary combinations of MGFs in the context of three-point $Y^\tau_{\vec{\eta}}$-integrals,
\begin{align}
\label{eqcusp}
 \aform{a_1&a_2&\ldots &a_R \protect\\b_1&b_2&\ldots &b_R} 
= \cform{a_1&a_2&\ldots &a_R \protect\\b_1&b_2&\ldots &b_R}  - \overline{
\cform{a_1&a_2&\ldots &a_R \protect\\b_1&b_2&\ldots &b_R} 
}\,.
\end{align}
Imaginary MGFs of this type have been first studied in \cite{DHoker:2019txf} and were shown to be cusp forms with vanishing Laurent-polynomials
$\sim q^0 \bar q^0$: 
The $\mathcal{A}\left[\cdots\right](\tau)$ in (\ref{eqcusp}) are odd
under $\tau \rightarrow -\bar \tau$ that sends $\Re\tau \to -\Re\tau$ while keeping $\Im\tau$ unchanged. This reflection moreover acts on any modular graph form by\footnote{Here, we make use of the assumption that the entries $a_i$ and $b_i$ of the MGFs are integers~\cite{DHoker:2019txf}.} $\mathcal{C}[\cdots ] (-\bar\tau) = \overline{\mathcal{C}[\cdots ] (\tau)}$ since this operation exchanges holomorphic and antiholomorphic momenta up to a change of summation variable, thus making $\mathcal{A}\left[\cdots\right](\tau)$ an odd function under this reflection. But since $\Im \tau$ and thus the zero mode $c_{0,0}(\Im\tau) - \overline{c_{0,0}(\Im\tau)}$ of $\mathcal{A}\left[\cdots\right](\tau)$ are even this means that the zero mode must vanish. Also real cusp forms occur among MGFs, for instance products of two imaginary cusp forms (\ref{eqcusp}).


\subsection{Differential equation}
\label{sec2.4}

The differential equation of the generating series $Y^\tau_{\vec{\eta}}$ defined in section~\ref{sec2.2} was derived in~\cite{Gerken:2019cxz}. At two points, the integral (\ref{eq2.12}) was shown to obey the homogeneous first-order equation
\begin{align}
2\pi i \partial_\tau Y^\tau_{\eta} &= 
\bigg\{ {-} \frac{ 1 }{(\tau{-}\bar \tau)^2}  R_{\eta}(\epsilon_0) 
+ \sum_{k=4}^{\infty} (1{-}k) (\tau{-}\bar \tau)^{k-2}{\rm G}_k(\tau)  R_{\eta}(\epsilon_k) 
\bigg\}Y^\tau_{\eta} 
\label{eq2.32}
\end{align}
with the following $\eta$- and $\bar\eta$-dependent operators
\beq
R_\eta(\epsilon_0) = s_{12} \Big( \frac{1}{\eta^2} - \frac{1}{2} \partial_\eta^2 \Big) - 2\pi i \bar \eta \partial_{\eta}
 \, , \ \ \ \ \ \
R_\eta(\ep_k) = s_{12} \eta^{k-2}
\, , \ \ \ \ \ \ k\geq 4 \, .
\label{eq2.33}
\eeq
The generalisation to $(n{\geq}3)$ points requires $(n{-}1)! \times (n{-}1)!$ matrix-valued operators
$R_{\vec{\eta}}(\epsilon_k)_{\rho}{}^\alpha$ acting on the indices $\rho$ of the integrals (\ref{eq2.13}). 
The first-order differential equation is 
\begin{align}
2\pi i \partial_\tau Y^\tau_{\vec{\eta}}(\sigma|\rho) &= \sum_{\alpha \in S_{n-1}}
\bigg\{ {-} \frac{ 1 }{(\tau{-}\bar \tau)^2}  R_{\vec{\eta}}(\epsilon_0)_{\rho}{}^\alpha
+ \sum_{k=4}^{\infty} (1{-}k) (\tau{-}\bar \tau)^{k-2}{\rm G}_k(\tau)  R_{\vec{\eta}}(\epsilon_k)_{\rho}{}^\alpha
\bigg\}Y^\tau_{\vec{\eta}} (\sigma|\alpha)\,,
\label{eq2.34}
\end{align}
see also \cite{Gerken:2019cxz} for homogeneous second-order Laplace equations among 
the $ Y^\tau_{\vec{\eta}}(\sigma|\rho)$.
At three points, for instance, the $R_{\vec{\eta}}(\epsilon_k)$ in 
(\ref{eq2.34}) are $2\times 2$ matrices
\begin{align}
R_{\eta_2,\eta_3}(\ep_0) &= \frac{1}{ \eta^2_{23} } \ccb s_{12} &-s_{13} \\-s_{12}&s_{13} \cce
+ \frac{1}{ \eta^{2}_{2} }\ccb 0 &0 \\ s_{12} &s_{12}{+}s_{23} \cce
+ \frac{1}{ \eta^{2}_{3}} \ccb s_{13}{+}s_{23} & s_{13} \\0 &0 \cce
 \notag \\
 & -\ccb 1\,  &\, 0  \\ 0 \, & \, 1  \cce \Big( \frac{1}{2} s_{12} \partial_{\eta_2}^2 + \frac{1}{2} s_{13} \partial_{\eta_3}^2
+ \frac{1}{2} s_{23} (\partial_{\eta_2} {-} \partial_{\eta_3})^2 
+2 \pi i (\bar \eta_2 \partial_{\eta_2} {+}\bar \eta_3 \partial_{\eta_3}  ) \Big) \,,
\label{eq2.35} \\
R_{\eta_2,\eta_3}(\ep_k) &= \eta^{k-2}_{23} \ccb s_{12} &-s_{13} \\-s_{12}&s_{13} \cce
+ \eta^{k-2}_{2} \ccb 0 &0 \\ s_{12} &s_{12}{+}s_{23} \cce
+ \eta^{k-2}_{3} \ccb s_{13}{+}s_{23} & s_{13} \\0 &0 \cce \, , \ \ \ \ k\geq 4\,, \notag
\end{align}
and their higher-multiplicity analogues following 
from \cite{Mafra:2019ddf, Mafra:2019xms, Gerken:2019cxz} are reviewed in appendix~\ref{appderiv}.

The differential equations among the $Y^\tau_{\vec{\eta}}$ are generating series
for differential equations among the component integrals. The simplest two-point
examples are\footnote{When comparing with the differential equations in (3.25) and (3.26) of \cite{Gerken:2019cxz}, 
note that the component integrals are related by $Y_{(a|b)}^\tau
= \frac{ (2i \Im \tau)^a }{(2\pi i )^b} W_{(a|b)}^\tau$. Moreover, the powers of $\Im \tau$
are tailored such that the operators $\nabla^{(w)}$ in the reference can be effectively replaced
by $(\tau{-}\bar \tau)\partial_\tau$.}
\beq
2\pi i \partial_\tau Y^{\tau}_{(0|0)} = \frac{s_{12}}{4(\Im\tau)^2}  Y^{\tau}_{(2|0)} \, , \ \ \ \ \ \ 
2\pi i \partial_\tau Y^{\tau}_{(2|0)} = -\frac{s_{12}}{2(\Im\tau)^2} Y^{\tau}_{(4|0)} + 12  s_{12} (\Im\tau)^2  {\rm G}_4(\tau) Y^{\tau}_{(0|0)}\,,
\label{eq2.36a}
\eeq
and generalise to (we are setting $Y^{\tau}_{(a|-1)} :=0$) 
\begin{align}
2\pi i \partial_\tau Y^{\tau}_{(a|b)} &= {-} \frac{ a}{4 (\Im \tau)^2} Y^{\tau}_{(a+1|b-1)}
+\frac{ (1{-}a)(a{+}2) s_{12} }{8 (\Im \tau)^2 }  Y^{\tau}_{(a+2|b)}
\notag\\
&+ s_{12} \sum_{k=4}^{a+2} (1{-}k) {\rm G}_k(\tau) (2i \Im \tau)^{k-2} Y^{\tau}_{(a+2-k|b)} \, .\label{eq2.36b}
\end{align}
The expansion of the component integrals in terms of MGFs given in~\eqref{eq2.all36} 
together with the differential equations (\ref{eq2.28}) and (\ref{eq2.31}) of the MGFs
can be used to verify (\ref{eq2.36a}) order by order in $\alpha'$. Conversely, one can use~\eqref{eq2.36b} and its generalisations to $n$ points to deduce properties of MGFs.

We also note that $Y^\tau_{\vec\eta}$ satisfies the following equation when differentiated with respect to $\bar\tau$~\cite[Eq.~(6.12)]{Gerken:2019cxz}:
\begin{align}
\label{eqYbartau}
-2\pi i \partial_{\bar \tau} Y^\tau_{\vec{\eta}}(\sigma|\rho) &=  \sum_{\alpha\in\mathcal{S}_{n-1}} \Big\{  2\pi i \sum_{j=2}^n \Big[  2 \eta_j \partial_{\bar \eta_j} + \frac{  \eta_j \partial_{\eta_j} - \bar \eta_j \partial_{\bar \eta_j}  }{\tau - \bar \tau}
\Big] \delta_\alpha^\sigma  -  \overline{R_{\vec\eta}(\epsilon_0)}_\alpha{}^\sigma   \\
&\hspace{40mm}+\sum_{k\geq 4} (1{-}k) \overline{{\rm G}_k(\tau)}\,  \overline{R_{\vec\eta}(\epsilon_k)}_\alpha{}^\sigma \Big\} Y_{\vec\eta}^\tau (\alpha|\rho)\,.\nn
\end{align}
We shall not use this equation extensively but rather the holomorphic $\tau$-derivative~\eqref{eq2.34} 
together with the reality properties~\eqref{eq2.19} of the component integrals. Similar to Brown's construction \cite{Brown:2017qwo,Brown:2017qwo2,brown2017a} of non-holomorphic modular forms, 
the series $Y^\tau_{\vec{\eta}}$
is engineered to simplify the holomorphic derivative (\ref{eq2.34}) at the expense of
the more lengthy expression (\ref{eqYbartau}) for the antiholomorphic one.


\subsection{Derivation algebra}
\label{sec2.5}

The notation $R_{\vec{\eta}}(\epsilon_k)$ for the operators in the differential
equations (\ref{eq2.32}) and (\ref{eq2.34}) was chosen to highlight a connection
with Tsunogai's derivations $\ep_{k}$ \cite{Tsunogai}. They arise in the differential
equation of the elliptic KZB associator and act on its non-commutative variables
\cite{KZB, EnriquezEllAss, Hain}. The derivation
algebra $\{\ep_{k}, \ k\geq 0\}$ is characterised by a variety of relations,
and the operators $R_{\vec{\eta}}(\epsilon_k)$ in this work are believed to 
form matrix representations of these the relations. For instance,
from the absence of $k=2$ in (\ref{eq2.32}) and (\ref{eq2.34}), we have 
$R_{\vec{\eta}}(\epsilon_2)=0$ which is consistent with the general relation
\beq
[\ep_2,\ep_k] = 0 \, .
\label{eq2.39}
\eeq
With the notation
\beq
{\rm ad}^{m}_{\ep_j}(\ep_k) = 
[\underbrace{ \ep_j,[\ep_j,[\ldots [\ep_j,[\ep_j}_{m},\ep_k]]\ldots]]] 
\label{eq2.40}
\eeq
for the repeated adjoint action ${\rm ad}_{\ep_j}(\ep_k) = [\ep_j,\ep_k]$, a crucial
set of relations among Tsunogai's derivations is the (adjoint) nilpotency of $\epsilon_0$
\beq
{\rm ad}^{k-1}_{\ep_0}(\ep_k) =  0 \, , \ \ \ \ \ \ k \geq 2\, , 
\label{eq2.41}
\eeq
for instance $[\ep_0,[\ep_0,[\ep_0,\ep_4]]]=0$.
%
The operators $R_{\vec{\eta}}(\epsilon_k)$ in the differential equations
(\ref{eq2.32}) and (\ref{eq2.34}) are expected to preserve (\ref{eq2.41}),
\beq
R_{\vec{\eta}}\big( {\rm ad}^{k-1}_{\ep_0}(\ep_k)\big) :=   {\rm ad}^{k-1}_{ R_{\vec{\eta}}(\ep_0)}\big(R_{\vec{\eta}}(\ep_k)\big)  =  0 \, , \ \ \ \ \ \ k \geq 2\, ,
\label{eq2.42}
\eeq
as furthermore supported by a variety of explicit checks, also see section 4.5 of~\cite{Mafra:2019xms} 
for supporting arguments in an open-string context.\footnote{Reference~\cite{Mafra:2019xms} deals with open-string amplitudes and features operators $r_{\vec{\eta}}(\epsilon_k)$ that agree with $R_{\vec{\eta}}(\epsilon_k)$ for $k\neq 0$. The difference for $\epsilon_0$ is that $r_{\vec{\eta}}(\epsilon_0)$ contains an additional term proportional to $\zeta_2$ but does not contain the term $\sim \bar \eta_j \partial_{\eta_j}$ of $R_{\vec{\eta}}(\epsilon_0)$.}
There is a variety of further relations in the derivation algebra \cite{LNT, Pollack, Broedel:2015hia}
that are related to the counting of holomorphic cusp forms at
various modular weights~\cite{Pollack}
\begin{subequations}
\label{eq2.43} 
\begin{align}
0 &=[\ep_{10},\ep_4]-3[\ep_{8},\ep_6] \label{eq2.43a}  \,,\\
0&=2 [\ep_{14},\ep_4] - 7[\ep_{12},\ep_6] + 11 [\ep_{10},\ep_8]  \,,\label{eq2.43b} \\
0&=80[\ep_{12},[\ep_4,\ep_{0}]] + 16 [\ep_4,[\ep_{12},\ep_0]] - 250 [\ep_{10},[\ep_6,\ep_0]] \notag \\
& \ \ \ \  - 125 [\ep_6,[\ep_{10},\ep_0]] + 280 [\ep_8,[\ep_8,\ep_0]]- 462 [\ep_4,[\ep_4,\ep_8]] - 1725 [\ep_6,[\ep_6,\ep_4]] \, .
\label{eq2.43c} 
\end{align}
\end{subequations}
We have tested that these relations are preserved
by the $R_{\vec{\eta}}(\ep_k)$, e.g.\
\beq
R_{\vec{\eta}}\big(  [\ep_{10},\ep_4]-3[\ep_{8},\ep_6]\big) 
:= \big[ R_{\vec{\eta}}(\ep_{10}), R_{\vec{\eta}}(\ep_4)\big]
 - 3 \big[R_{\vec{\eta}}(\ep_{8}),R_{\vec{\eta}}(\ep_6) \big] 
= 0 \, .
\label{eq2.44}
\eeq
Similarly, the $R_{\vec{\eta}}(\ep_k)$ at $n\leq 5$ points have been checked
to preserve various generalisations of (\ref{eq2.43}) that can be downloaded from \cite{WWWe}:
\begin{itemize}
\item relations among $[\ep_{k_1},\ep_{k_2}]$ at $k_1{+}k_2\leq 30$ and $n{=}2,3,4$ as well as $k_1{+}k_2\leq 18$ and $n{=}5$,
\item relations among $[\ep_{\ell_1},[\ep_{\ell_2},\ep_{\ell_3}]]$ at $\ell_1{+}\ell_2{+}\ell_3\leq 30$ and $n=2,3,4$,
\item relations among $[\ep_{p_1},[\ep_{p_2},[\ep_{p_3},\ep_{p_4}]]]$ at $p_1{+}p_2{+}p_3{+}p_4\leq 26,\ n=2,3$ as well as (partially relying on numerical methods) $p_1{+}p_2{+}p_3{+}p_4\leq 18,\ n=4$
\end{itemize}
We will see in section \ref{sec6.2} that relations like (\ref{eq2.43}) will play a key role in the 
counting of independent MGFs at given modular weights in their lattice-sum representation
(\ref{eq2.22}), in the same way as they did
for the counting of elliptic MZVs \cite{Broedel:2015hia}. 

Even though the operators $R_{\vec\eta}(\epsilon_k)$ satisfy the derivation-algebra relations (at least to the orders checked), their instances at given multiplicity $n$ are not a faithful representation of the derivation algebra.
In other words, they can also satisfy more relations at fixed $n$. For instance, the two-point example~\eqref{eq2.33} implies that all $R_\eta(\epsilon_k)$ for $k\geq 4$ at $n=2$ commute which is stronger than~\eqref{eq2.43}. As we shall use compositions of the operators $R_{\vec\eta}(\epsilon_k)$ in the rest of the paper to solve~\eqref{eq2.34},
this means that their coefficients only occur in specific linear combinations in low-point results. This will
lead to multiplicity-specific dropouts of MGFs in the $\ap$-expansion of $Y^\tau_{\vec{\eta}}$ at fixed $n$,
in the same way as four-point closed-string tree-level amplitudes do not involve any MZVs of depth $\geq 2$.


\section{Solving differential equations for generating series}
\label{sec3}

The goal of this section is to derive the form of the all-order 
$\alpha'$-expansion of the $Y^\tau_{\vec{\eta}}$ integrals (\ref{eq2.13}) from
their differential equation (\ref{eq2.34}). As a first step we will rewrite the differential equation in a slightly different form using relations in the derivation algebra. This improved differential equation will allow for a formal solution whose properties we discuss in this section. In the next sections we make the formal solution
fully explicit at certain orders by exploiting the reality properties of two- and three-point integrals.


\subsection{Improving the differential equation}
\label{sec3.1}

Given that the differential equation~\eqref{eq2.34} is linear and of first order in $\tau$, it is tempting to solve it (up to antiholomorphic integration ambiguities) formally by line integrals over $\tau$. In particular, the
appearance of $(\tau{-}\bar\tau)^{k-2} {\rm G}_k(\tau)$ on the right-hand side will introduce iterated integrals
over holomorphic Eisenstein series in a formal solution. However, the differential equation features
singular terms $\sim(\tau{-}\bar\tau)^{-2}$ that do not immediately line up with Brown's iterated 
Eisenstein integrals over $\tau^{j} {\rm G}_k(\tau) , \ j=0,1,\ldots,k{-}2$ with well-studied modular
transformations \cite{Brown:mmv}.

Therefore we first strive to remove the singular term $\sim(\tau{-}\bar\tau)^{-2}$
in~\eqref{eq2.34} that does not have any accompanying Eisenstein series ${\rm G}_{k\geq 4}$. 
This can be done by performing the invertible redefinition\footnote{We are grateful to Nils Matthes and 
Erik Panzer for discussions that led to this redefinition.}
\begin{align}
\Yhat^\tau_{\vec{\eta}}  &= \exp \Big( { -}\frac{R_{\vec{\eta}}(\epsilon_0) }{2\pi i (\tau {-} \bar \tau)} \Big)  Y^\tau_{\vec{\eta}}
= \exp \Big(  \frac{R_{\vec{\eta}}(\epsilon_0) }{4y} \Big)  Y^\tau_{\vec{\eta}} 
\hspace{5mm}\Leftrightarrow\hspace{5mm}
Y^\tau_{\vec{\eta}} =  \exp \Big({-} \frac{R_{\vec{\eta}}(\epsilon_0) }{4y} \Big)  \Yhat^\tau_{\vec{\eta}} 
\,,
\label{eq3.1}
\end{align}
where the matrix multiplication w.r.t.\ the second index of 
$Y^\tau_{\vec{\eta}}(\sigma|\rho)$ is suppressed for ease of notation\footnote{More explicitly,
$\Yhat^\tau_{\vec{\eta}}(\sigma|\rho) = \sum_{\alpha \in {\cal S}_{n-1}}
 \exp \big(  \frac{R_{\vec{\eta}}(\epsilon_0) }{4y} \big)\, \!_\rho{}^\alpha  Y^\tau_{\vec{\eta}}(\sigma|\alpha) $.}.
The redefined integrals obey a modified version of (\ref{eq2.34})
\begin{align}
2\pi i \partial_\tau \Yhat^\tau_{\vec{\eta}} &= \sum_{k=4}^{\infty} (1{-}k) {\rm G}_k(\tau) (\tau{-}\bar \tau)^{k-2}
e^{{-}  \frac{ R_{\vec{\eta}}(\epsilon_0) }{2\pi i (\tau-\bar \tau)} }   R_{\vec{\eta}}(\epsilon_k)
e^{ \frac{ R_{\vec{\eta}}(\epsilon_0) }{2\pi i (\tau-\bar \tau)} }  \Yhat^\tau_{\vec{\eta}}  \, ,
\label{eq3.2}
\end{align}
where now the term without holomorphic Eisenstein series is absent and the $R_{\vec{\eta}}(\epsilon_k)$
are conjugated by exponentials of $R_{\vec{\eta}}(\epsilon_0)$. By the relations
(\ref{eq2.42}) in the derivation algebra, the exponentials along with a fixed $R_{\vec{\eta}}(\epsilon_k)$ 
truncate to a finite number of terms,
\beq
e^{{-}  \frac{ R_{\vec{\eta}}(\epsilon_0) }{2\pi i (\tau-\bar \tau)} }   R_{\vec{\eta}}(\epsilon_k)
e^{ \frac{ R_{\vec{\eta}}(\epsilon_0) }{2\pi i (\tau-\bar \tau)} }  = \sum_{j=0}^{k-2}
\frac{1}{j!}\Big( \frac{ -1}{2\pi i (\tau{-}\bar \tau)}  \Big)^j  R_{\vec{\eta}}\big( {\rm ad}_{ \epsilon_0}^j (\epsilon_k) \big) \, ,
\label{eq3.3}
\eeq
where we use the following shorthands here and below
\beq
 R_{\vec{\eta}}\big( {\rm ad}_{ \epsilon_0}^j (\epsilon_k) \big)
:= {\rm ad}_{ R_{\vec{\eta}}(\epsilon_0) }^j R_{\vec{\eta}}(\epsilon_k) \, , \ \ \ \ \ \
 R_{\vec{\eta}}   (\epsilon_{k_1} \epsilon_{k_2})   :=  
 R_{\vec{\eta}} (\epsilon_{k_1})    R_{\vec{\eta}}  (\epsilon_{k_2})
  \, .
\label{eq3.4}
\eeq
Hence, the differential equation (\ref{eq3.2}) simplifies to
\begin{align}
2\pi i \partial_\tau \Yhat^\tau_{\vec{\eta}} &= \sum_{k=4}^{\infty} (1{-}k) {\rm G}_k(\tau)
\sum_{j=0}^{k-2} \frac{1}{j!} \Big( \frac{ -1}{2\pi i} \Big)^j (\tau{-}\bar\tau)^{k-2-j} 
R_{\vec{\eta}}\big( {\rm ad}_{ \epsilon_0}^j (\epsilon_k) \big)
 \Yhat^\tau_{\vec{\eta}}  \, .
\label{eq3.5}
\end{align}
Now, the operator on the right-hand side is manifestly free of singular terms in $(\tau{-}\bar\tau)$,
and the sum over $k$ starts at $k=4$. All the integration kernels in this differential equation are of the form  $(\tau{-}\bar\tau)^{j}  {\rm G}_k(\tau)$ with $k\geq4 $ and $0\leq j \leq k{-}2$. Hence, our kernels line up with
those of Brown's holomorphic and single-valued iterated Eisenstein 
integrals \cite{Brown:mmv, Brown:2017qwo, Brown:2017qwo2}.


\subsection{Formal expansion of the solution}
\label{sec3.3}

The form~\eqref{eq3.5} bodes well for a representation in terms of $\Yhat^\tau_{\vec\eta}$ as an (iterated) line integral from $\tau$ to some reference point that we take to be the cusp at $\tau\to i\infty$. In particular, the differential equation contains no negative powers of $(\tau{-}\bar\tau)$ or $y=\pi\Im\tau$, and this property will propagate to the solution $\Yhat^\tau_{\vec\eta}$, see section \ref{sec3.2} for further details. The original integrals $Y^\tau_{\vec\eta}$, in turn, involve combinations of MGFs with negative powers of $y$ from their Laurent polynomials $c_{m,n}(\Im \tau)$ in (\ref{eqcusp.1}). The absence of negative powers of $y$ in $\Yhat^\tau_{\vec\eta}$ is a crucial difference as compared to $Y^\tau_{\vec\eta}$ and is due to the redefinition~\eqref{eq3.1}. We shall later make this more manifest when we discuss explicit examples obtained from low-point amplitudes.

A formal solution of (\ref{eq3.5}), that also exposes the $\ap$-expansion of the integrals, is given by the series
\begin{align}
\Yhat^{\tau}_{\vec{\eta}} &=
\sum_{\ell=0}^\infty \sum_{\substack{k_1,k_2,\ldots,k_\ell \\ =4,6,8,\ldots}}
\sum_{j_1=0}^{k_1-2} \sum_{j_2=0}^{k_2-2} \ldots \sum_{j_\ell=0}^{k_\ell-2}
\bigg(  \prod_{i=1}^\ell \frac{ (-1)^{j_i} (k_i-1)}{(k_i-j_i-2)!} \bigg)  \EsvBR{j_1 &j_2 &\ldots &j_\ell}{k_1 &k_2 &\ldots &k_\ell}{\tau} \notag \\
&\ \ \ \ \ \times
 R_{\vec{\eta}}\big( {\rm ad}_{ \epsilon_0}^{k_\ell-j_\ell-2} (\epsilon_{k_{\ell}})\ldots 
{\rm ad}_{ \epsilon_0}^{k_2-j_2-2} (\epsilon_{k_2}) {\rm ad}_{ \epsilon_0}^{k_1-j_1-2} (\epsilon_{k_1}) \big)
\Yhat^{i\infty}_{\vec{\eta}} \, ,
\label{eq3.8}
\end{align}
if the $\tau$-dependent constituents solve the initial-value problem
\begin{subequations}
\label{eq3.init}
\begin{align}
2\pi i \partial_\tau \EsvBR{j_1 &j_2 &\ldots &j_\ell}{k_1 &k_2 &\ldots &k_\ell}{\tau}
&= - (2\pi i)^{2-k_\ell+j_\ell} ( \tau - \bar \tau )^{j_\ell} {\rm G}_{k_\ell}(\tau)
\EsvBR{j_1 &j_2 &\ldots &j_{\ell-1}}{k_1 &k_2 &\ldots &k_{\ell-1}}{\tau}\,,
\label{eq3.9}
 \\
 \lim_{\tau \rightarrow i\infty} \EsvBR{j_1 &j_2 &\ldots &j_\ell}{k_1 &k_2 &\ldots &k_\ell}{\tau} &= 0\, .
 \label{eq3.10}
\end{align}
\end{subequations}
The vanishing at the cusp here is understood in terms of a regularised limit that will be discussed in more detail in section~\ref{sec3.2} and is akin to the method of `tangential-base-point regularisation' introduced in~\cite{Brown:mmv}.
Its net effect can be summarised by assigning $\int^\tau_{i\infty} \dd \tau' = \tau$ which
regularises the $\tau\rightarrow i\infty$ limit of all (strictly) positive powers of $\tau$ and $\bar\tau$ to zero (in the absence of negative powers)
and hence $\lim_{\tau \rightarrow i \infty} (\Im \tau)^n = 0$ for all $n>0$.

The parameter $\ell$ in~\eqref{eq3.8} will be referred to as `depth', and
we define at depth zero that $\esvtau{}=1$. Since the sums over the $k_i$ start at 
$k_i=4$, and all the $R_{\vec{\eta}}(\ep_{k\geq 4})$ in (\ref{eq2.33}), (\ref{eq2.35}) and 
appendix \ref{appderiv} are linear in $s_{ij}$, the depth-$\ell$ contributions to~\eqref{eq3.8} involve at least $\ell$ powers of $\ap$. As we will see, any order in the $\alpha'$-expansion of the component integrals (\ref{eq2.15}) can be obtained from a finite number of terms in (\ref{eq3.8}) on the basis of elementary operations.
Like this, the relation~\eqref{eq3.8} reduces the $\ap$-expansion of the generating series $\Yhat^\tau_{\vec\eta}$ to
the way more tractable problem of determining the initial values $\Yhat^{i\infty}_{\vec\eta}$ and the
objects ${\cal E}^{\rm sv}$:
\begin{itemize}
\item The initial values $\Yhat^{i\infty}_{\vec\eta}$ at the cusp are series
in $\eta_i,\bar\eta_i,s_{ij}$ whose coefficients should be $\QQ$-linear combinations of single-valued MZVs
from genus-zero sphere integrals \cite{yinprogress}. We shall give a closed formula at two points in section~\ref{sec4.1}. Given that $\Yhat^{i\infty}_{\vec\eta}$ at higher points are still under 
investigation \cite{yinprogress}, we shall here
use MGF techniques to determine the initial data at three points to certain orders, see section \ref{sec5}.
As will be detailed in section
 \ref{sec3.2}, we exploit the absence of negative powers of $y$ in the expansion of $\Yhat^{\tau}_{\vec\eta}$ 
 around the cusp to extract a well-defined initial value $\Yhat^{i\infty}_{\vec\eta}$.
\item The objects $\Esv$ are partly determined by the differential equations~\eqref{eq3.init} but, since the $\Esv$ are non-holomorphic, the  $\partial_\tau$ derivative is not sufficient to determine them: As will be detailed in section
\ref{sec3.B}, one can add antiholomorphic functions of $\bar\tau$ that vanish at the cusp at every step in their iterative construction. The analysis in~\cite{Gerken:2019cxz} also provides a differential equation for the $\partial_{\bar\tau}$-derivative of $Y$, see~\eqref{eqYbartau}. However, we shall be able to determine the $\Esv$ from the
reality properties (\ref{eq2.19}) of the component integrals, i.e.\ without making recourse to the differential equation with respect to $\partial_{\bar\tau}$.
\end{itemize}
As we shall see in the next section, it turns out to be useful for expressing the $Y^\tau_{\vec\eta}$ rather than the $\Yhat^\tau_{\vec\eta}$ to redefine the $\Esv$ into specific linear combinations that satisfy differential equations that are advantageous for the analysis.
The notation $\Esv$ is chosen due to the similarity to holomorphic and single-valued iterated Eisenstein 
integrals defined by Brown and obeying similar first-order differential 
equations \cite{Brown:mmv, Brown:2017qwo, Brown:2017qwo2}.
 Explicit expressions for the $\Esv$ in terms of holomorphic iterated Eisenstein integrals and their complex conjugates will be given in section~\ref{sec3.B} below, where we also address the issue of the integration constants.


\subsection{Solution for the original integrals}
\label{sec3.4}

Our original goal was to expand the $Y^\tau_{\vec{\eta}}$-integrals (\ref{eq2.13}) in
$\alpha'$. In order to translate the formal solution~\eqref{eq3.8} for the redefined integrals $\Yhat^\tau_{\vec\eta}$ to the original ones $Y^\tau_{\vec{\eta}}$ we have to invert the exponentials in (\ref{eq3.1}). 
In the first place, this introduces the exponential in the second line of the $\ap$-expansion~\eqref{eq3.8}
\begin{align}
Y^{\tau}_{\vec{\eta}} &=
\sum_{\ell=0}^\infty \sum_{\substack{k_1,k_2,\ldots,k_\ell \\=4,6,8,\ldots}}
\sum_{j_1=0}^{k_1-2} \sum_{j_2=0}^{k_2-2} \ldots \sum_{j_\ell=0}^{k_\ell-2}
\bigg(  \prod_{i=1}^\ell \frac{ (-1)^{j_i} (k_i-1)}{(k_i-j_i-2)!} \bigg)   \EsvBR{j_1 &j_2 &\ldots &j_\ell}{k_1 &k_2 &\ldots &k_\ell}{\tau} \label{eq3.18}  \\
&\quad \times \exp \Big(  {-} \frac{R_{\vec{\eta}}(\ep_0) }{ 4 y} \Big)
 R_{\vec{\eta}}\big( {\rm ad}_{ \epsilon_0}^{k_\ell-j_\ell-2} (\epsilon_{k_{\ell}})\ldots 
{\rm ad}_{ \epsilon_0}^{k_2-j_2-2} (\epsilon_{k_2}) {\rm ad}_{ \epsilon_0}^{k_1-j_1-2} (\epsilon_{k_1}) \big)
\Yhat^{i\infty}_{\vec{\eta}}  
\notag
\end{align}
that we then commute through adjoint derivation operators to act on the value $\Yhat^{i\infty}_{\vec\eta}$ at the cusp. 
This amounts to conjugating the $ {\rm ad}_{ \epsilon_0}^{k_i-j_i-2} (\epsilon_{k_{i}})$ via
\begin{align}
\exp \Big(  {-} \frac{ R_{\vec{\eta}}(\ep_0) }{ 4 y} \Big) R_{\vec{\eta}}\big({\rm ad}_{ \epsilon_0}^{k-j-2} (\epsilon_{k}) \big)  \exp \Big(   \frac{ R_{\vec{\eta}}(\ep_0) }{ 4 y} \Big) = \sum_{p=0}^{j} \frac{1}{p!} \Big({-}\frac{1}{4y} \Big)^p  R_{\vec{\eta}} \big( {\rm ad}_{ \epsilon_0}^{k-j+p-2} (\epsilon_{k}) \big)\, .
 \label{eq3.19}
\end{align}
The modified powers of $ {\rm ad}_{ \epsilon_0}$ regroup
the ${\cal E}^{\rm sv}$ into the combination
\begin{align}
\bsvBR{j_1 &j_2 &\ldots &j_\ell}{k_1 &k_2 &\ldots &k_\ell}{\tau} &= \sum_{p_1=0}^{k_1- j_1-2}\sum_{p_2=0}^{k_2- j_2-2} \ldots \sum_{p_\ell=0}^{k_\ell- j_\ell-2}
 \binom{k_1{-}j_1{-}2}{p_1}\binom{k_2{-}j_2{-}2}{p_2}\cdots \binom{k_\ell{-}j_\ell{-}2}{p_\ell}\notag \\
 &\quad \times  \Big( \frac{1}{4y} \Big)^{p_1+p_2+\ldots +p_\ell}
\EsvBR{j_1+p_1 & j_2 + p_2 &\ldots & j_\ell + p_\ell}{k_1 & k_2 &\ldots &k_\ell}{\tau}
 \label{eq3.20}
\end{align}
with $0{\leq}j_i{\leq }k_i{-}2$ and $\bsvBR{}{}{\tau}=1$,
i.e.\ the $\ap$-expansion (\ref{eq3.18}) can be compactly rewritten as
\begin{align}
Y^\tau_{\vec{\eta}} &=
\sum_{\ell=0}^\infty \sum_{\substack{k_1,k_2,\ldots,k_\ell \\ =4,6,8,\ldots}}
\sum_{j_1=0}^{k_1-2} \sum_{j_2=0}^{k_2-2} \ldots \sum_{j_\ell=0}^{k_\ell-2}
\bigg(  \prod_{i=1}^\ell \frac{ (-1)^{j_i} (k_i-1)}{(k_i-j_i-2)!} \bigg)   \bsvBR{j_1 &j_2 &\ldots &j_\ell}{k_1 &k_2 &\ldots &k_\ell}{\tau} \label{eq3.21} \\
& \ \ \ \ \ \times 
 R_{\vec{\eta}}\big( {\rm ad}_{ \epsilon_0}^{k_\ell-j_\ell-2} (\epsilon_{k_{\ell}})\ldots 
{\rm ad}_{ \epsilon_0}^{k_2-j_2-2} (\epsilon_{k_2}) {\rm ad}_{ \epsilon_0}^{k_1-j_1-2} (\epsilon_{k_1}) \big)
\exp \Big(  {-} \frac{ R_{\vec{\eta}}(\ep_0) }{ 4 y} \Big)
\Yhat^{i\infty}_{\vec{\eta}} \, .
\notag
\end{align}
This is the formal solution of the $\ap$-expansion of the generating series $Y^\tau_{\vec\eta}$ of worldsheet integrals. As we reviewed in section~\ref{sec2.3}, the component integrals appearing in the Laurent expansion of $Y^\tau_{\vec\eta}$ with respect to the $\eta_i$ variables can be represented in terms of MGFs. Hence, (\ref{eq3.21}) results in a representation of arbitrary MGFs in terms of $\bsv$ and the ingredients of $\exp (  {-} \frac{ R_{\vec{\eta}}(\ep_0) }{ 4 y} )
\Yhat^{i\infty}_{\vec{\eta}}$ --- conjecturally $\QQ[y^{-1}]$-linear combinations of single-valued MZVs.
We stress that by this, all the relations among MGFs \cite{DHoker:2015gmr, DHoker:2015sve, DHoker:2016mwo, DHoker:2016quv} will be automatically exposed in view of the linear-independence result on
holomorphic iterated Eisenstein integrals of \cite{Nilsnewarticle}.

\subsubsection{\texorpdfstring{Properties of $\beta^{\rm sv}$}{Properties of betasv}}

The simplest examples of the relation (\ref{eq3.20}) at depths one and two read
\begin{align}
\bsvBR{j_1 }{k_1 }{\tau} &= \sum_{p_1=0}^{k_1- j_1-2} 
\binom{ k_1{-}j_1{-}2 }{p_1}
\Big( \frac{1}{4y} \Big)^{p_1}
\EsvBR{j_1+p_1}{k_1}{\tau}\,,
 \label{eq3.22} \\
\bsvBR{j_1 &j_2}{k_1 &k_2}{\tau} &= \sum_{p_1=0}^{k_1- j_1-2}\sum_{p_2=0}^{k_2- j_2-2}
\binom{ k_1{-}j_1{-}2 }{p_1}\binom{ k_2{-}j_2{-}2 }{p_2}
  \Big( \frac{1}{4y} \Big)^{p_1+p_2}
\EsvBR{j_1+p_1 & j_2 + p_2}{k_1 & k_2}{\tau}\, .
\notag
\end{align}
One can straightforwardly invert the map between ${\cal E}^{\rm sv}$ and $\beta^{\rm sv}$ at any depth $\ell$
\begin{align}
\EsvBR{j_1 &j_2 &\ldots &j_\ell}{k_1 &k_2 &\ldots &k_\ell}{\tau} &= \sum_{p_1=0}^{k_1- j_1-2}\sum_{p_2=0}^{k_2- j_2-2} \ldots \sum_{p_\ell=0}^{k_\ell- j_\ell-2}
\binom{ k_1{-}j_1{-}2 }{p_1}\binom{ k_2{-}j_2{-}2 }{p_2}\cdots\binom{ k_\ell{-}j_\ell{-}2 }{p_\ell}
 \notag \\
 &\quad \times  \Big({-} \frac{1}{4y} \Big)^{p_1+p_2+\ldots +p_\ell}
\bsvBR{j_1+p_1 & j_2 + p_2 &\ldots & j_\ell + p_\ell}{k_1 & k_2 &\ldots &k_\ell}{\tau}\, .
 \label{eq3.23}
\end{align}
From (\ref{eq3.20}) and~\eqref{eq3.init} one can check 
that the differential equations obeyed by the $\beta^{\rm sv}$ is
\begin{align}
-4\pi \nabla
\bsvBR{j_1 &j_2&\ldots &j_\ell}{k_1 &k_2 &\ldots &k_\ell}{\tau} &=
\sum_{i=1}^\ell
 (k_i{-}j_i{-}2) \bsvBR{j_1 &j_2 &\ldots  &j_{i-1} &j_i +1 &j_{i+1} &\ldots &j_\ell}{k_1 &k_2 &\ldots &k_{i-1} &k_i  &k_{i+1} &\ldots &k_\ell}{\tau} \notag \\
 &\ \ \ \ \ \ - \delta_{j_\ell,k_\ell-2} (\tau {-} \bar \tau)^{k_\ell} {\rm G}_{k_\ell}(\tau) \bsvBR{j_1 &j_2 &\ldots &j_{\ell-1}}{k_1 &k_2 &\ldots &k_{\ell-1}}{\tau} \,,
 \label{eq3.24}
\end{align}
where we have used the differential operator $\nabla$ defined in~\eqref{eq2.25} as it has a nice action on the MGFs appearing in the component integrals. Compared to~\eqref{eq3.init}, the differential equation produces holomorphic Eisenstein series only when the last pair $(j_\ell,k_\ell)$ of $\bsv$ obeys $j_\ell=k_\ell-2$. As the $\bsv$ are linear combinations of the $\Esv$, the boundary condition for the $\bsv$ is still that
\begin{align}
 \lim_{\tau \rightarrow i\infty} \betasvtau{j_1 &j_2 &\ldots &j_\ell\\k_1 &k_2 &\ldots &k_\ell} &= 0\, ,
\end{align}
again in the sense of a regularised limit. 
As was the case for the $\Esv$, the $\partial_\tau$ derivative~\eqref{eq3.24} and the boundary 
condition are not sufficient to determine the $\bsv$ but the reality properties of the component $Y$-integrals will resolve the integration ambiguities.

At depths one and two the differential equation~\eqref{eq3.24} specialises as follows
\begin{subequations}
\begin{align}
\label{eq:Dbsv1}
-4\pi \nabla \bsvBR{j_1 }{k_1}{\tau} &= (k_1{-}j_1{-}2) \bsvBR{j_1 +1}{k_1}{\tau}
- \delta_{j_1,k_1-2} (\tau {-} \bar \tau)^{k_1} {\rm G}_{k_1}(\tau)\,,
\\*
-4\pi \nabla \bsvBR{j_1 &j_2}{k_1 &k_2}{\tau} &=
 (k_1{-}j_1{-}2) \bsvBR{j_1 +1 &j_2}{k_1 &k_2}{\tau}
 +  (k_2{-}j_2{-}2) \bsvBR{j_1  &j_2+1}{k_1 &k_2}{\tau} \notag \\
 &\ \ \ \ \ \ - \delta_{j_2,k_2-2} (\tau {-} \bar \tau)^{k_2} {\rm G}_{k_2}(\tau) \bsvBR{j_1 }{k_1}{\tau} \, .
 \label{eq3.25}
\end{align}
\end{subequations}

\subsubsection{Constraints from the derivation algebra}

The relations of the derivation algebra such as~\eqref{eq2.44} imply that, starting from $\sum_i k_i\geq 14$, not all $\Esv$ and $\bsv$ appear individually in the expansion of the generating series $Y^\tau_{\vec\eta}$ and $\Yhat^\tau_{\vec\eta}$ but only certain linear combinations can arise. We currently do not have an independent definition of all $\Esv,\bsv$ and such a definition is not needed for the component integrals in this paper that conjecturally cover all closed-string one-loop amplitudes.

The simplest instance where the derivation-algebra relation~\eqref{eq2.44} yields all-multiplicity
dropouts of certain $\bsv$ is in the weight-$14$ part of the expansion~\eqref{eq3.21}
\begin{align}
Y^\tau_{\vec\eta} = \bigg[\ldots &+ 27 \betasv{ 8 & 2 \\ 10 & 4} R_{\vec\eta}(\epsilon_4 \epsilon_{10}) + 27\betasv{ 2 & 8 \\ 4 & 10} R_{\vec\eta}(\epsilon_{10} \epsilon_4 )\nn\\
&+ 35 \betasv{ 6 & 4 \\ 8 & 6}R_{\vec\eta}( \epsilon_6 \epsilon_8 )+35 \betasv{ 4 & 6 \\ 6 & 8}R_{\vec\eta}( \epsilon_8 \epsilon_6 )+\ldots \bigg] \exp\left( -\frac{R_{\vec\eta}(\epsilon_0)}{4y}\right) \Yhat^{i \infty}_{\vec\eta}\label{eq:dropout}\\
=\bigg[\ldots &+ \Big\{ 27  \betasv{ 8 & 2 \\ 10 & 4} +27 \betasv{ 2 & 8 \\ 4 & 10}  \Big\} R_{\vec\eta}(\epsilon_4 \epsilon_{10} )
+ \Big\{ 35 \betasv{ 6 & 4 \\ 8 & 6} -81 \betasv{ 2 & 8 \\ 4 & 10} \Big\}R_{\vec\eta}(\epsilon_{6} \epsilon_8 ) \nn\\
&+ \Big\{ 35 \betasv{ 4 & 6 \\ 6 & 8} +81 \betasv{ 2 & 8 \\ 4 & 10} \Big\}R_{\vec\eta}(\epsilon_{8} \epsilon_6 ) + \ldots \bigg]\exp\left( -\frac{R_{\vec\eta}(\epsilon_0)}{4y}\right) \Yhat^{i \infty}_{\vec\eta}\,,
\nn
\end{align}
where we have solved~\eqref{eq2.44} for $R_{\vec{\eta}}(\epsilon_{10}\epsilon_4)$ in the second step. This shows that only a three-dimensional subspace of the four-dimensional span $\big\langle\betasv{ 8 &2\\10&4},\betasv{ 2 &8\\4&10}, \betasv{6 & 4\\ 8&6}, \betasv{4 & 6\\6&8}\big\rangle$ is realised by the generating series~\eqref{eq3.21}.
The manipulations in (\ref{eq:dropout}) can be repeated for
contributions to $Y^\tau_{\vec{\eta}}$ with ${\rm ad}_{\ep_0}^j$ acting on the $R_{\vec{\eta}}(\epsilon_{4})$,
$R_{\vec{\eta}}(\epsilon_{10})$, $R_{\vec{\eta}}(\epsilon_{6})$, $R_{\vec{\eta}}(\epsilon_{8})$. This
implies similar dropouts among the $\betasv{  j_1&j_2\\10&4},\betasv{ j_1&j_2\\4&10}, \betasv{ j_1&j_2\\ 8&6}, \betasv{j_1&j_2 \\6&8}$ at all values of $j_1{+}j_2\leq10$ and modifies the counting of MGFs at various modular weights, see section \ref{sec:e10e4} for details.


\subsection{Improved initial data and consistent truncations}
\label{sec3.2}

In this section, we illustrate the usefulness of the redefinition~\eqref{eq3.1} from $Y^\tau_{\vec\eta}$ to $\Yhat^\tau_{\vec\eta}$ further by discussing how it acts on and improves the initial data at the cusp $\tau\to i\infty$
that is contained in the Laurent polynomial defined in (\ref{eqcusp.1}). In this context, we also discuss practical aspects of extracting information on the component integrals 
by truncating the series $Y^\tau_{\vec\eta}$ and $\Yhat^\tau_{\vec\eta}$ to specific orders in $s_{ij}$, $\eta_i$ and $\bar\eta_i$.
 
\subsubsection{Behaviour of generating series near the cusp} 
 
Another virtue of the redefinition (\ref{eq3.1}) is that $\Yhat^\tau_{\vec\eta}$ is 
better behaved at the cusp than $Y^\tau_{\vec\eta}$.
While the Laurent polynomial of $Y^\tau_{\vec\eta}$ is known to feature both positive and negative powers 
of $y= \pi \Im \tau$ (see e.g.\ \eqref{eqcusp.mzv} for Laurent polynomials of MGFs in the $\ap$-expansion), we shall see that the
Laurent polynomial of $\Yhat^\tau_{\vec\eta}$ only has non-negative powers.
In order to define an initial value supplementing the differential equations, 
we will take a regularised limit of the Laurent polynomial with the convention to discard strictly positive powers 
of $y$ as $\tau \rightarrow i \infty$.\footnote{One can think of this regularised limit
as realising the $\tau \rightarrow i \infty$
limit of integrals $\int^\tau_{i\infty}$ that remove strictly positive powers of $\Im \tau$ through their tangential-base-point regularisation \cite{Brown:mmv}.}

However, such a regularised limit leads to inconsistencies with products such as $1 = y^n \cdot y^{-n}, \ n>0$ 
when both positive and negative powers are present. This problem is relevant to $Y^\tau_{\vec\eta}$ but not to
$\Yhat^\tau_{\vec\eta}$, where negative powers of $y$ are absent. 
While their absence is not immediately obvious from the redefinition, we have already remarked above that it can be understood from the differential equations as we shall now explain in more detail.

The differential equation (\ref{eq3.5}) relates $\partial_\tau \Yhat^\tau_{\vec\eta}$ to products 
${\rm G}_k(\tau)(\tau{-}\bar \tau)^{k-2-j}R_{\vec{\eta}}\big( {\rm ad}_{ \epsilon_0}^j (\epsilon_k) \big) 
\Yhat^\tau_{\vec{\eta}}$ with $k\geq 4$ and $j\leq k{-}2$. The lowest explicit power of $y=\pi \Im\tau$ is therefore $y^0$ and in general only non-negative powers arise since ${\rm G}_k$ is holomorphic in $\tau$ and the derivations $R_{\vec{\eta}}\big( {\rm ad}_{ \epsilon_0}^j (\epsilon_k) \big)$ do not depend on $\tau$ at all. The differential equation is therefore consistent with $\Yhat^\tau_{\vec{\eta}}$ having only non-negative powers of $y$. 

We note that the differential operator on the left-hand side of~\eqref{eq3.5} lowers the $y$-power  via $\partial_\tau y^{-m} = - m y^{-m-1}$ and therefore the presence of any negative power $y^{-m}$ in $\Yhat^\tau_{\vec\eta}$ requires the presence of even more negative powers by the differential equation.\footnote{Since the derivative of a constant vanishes, this argument does not connect positive to negative powers.} This is even true at any fixed order in the Mandelstam variables $s_{ij}$ and the parameters~$\bar{\eta}_i$ since any operator on the right-hand side of~\eqref{eq3.5} is either linear in $s_{ij}$ or in $\bar{\eta}_i$ by looking at the expressions in section~\ref{sec2.4}.

From the argument above we could still allow for an infinite series of negative powers in $y$ appearing in $\Yhat^\tau_{\vec\eta}$. To rule this out we consider the component integrals arising in the original generating series $Y^\tau_{\vec\eta}$ defined in~\eqref{eq2.13}.
The integrands of the $n$-point component integrals (\ref{eq2.15}) have negative powers of $y$  bounded by $y^{ \geq -(a+b)}$ at the order of $s_{ij}^a \bar \eta_i^{b-n+1}$.
This bound follows from the fact that Green functions and $f^{(k)}$ or $\overline{f^{(k)} }$ 
contribute at most $y^{-1}$ and $y^{-k}$, respectively, as can for instance be seen from their 
lattice-sum representations (\ref{eq2.lat}).\footnote{In terms of the lattice sums, factors of $p=m\tau+n$ or $\bar{p}=m\bar\tau+n$ both count as a factor of $y$ when approaching the cusp as $\Re\tau$ does not matter there. Inspecting the powers of $p$ and $\bar{p}$ in the lattice-sum representations (\ref{eq2.lat}) of the Green function and $f^{(k)}$ or $\overline{f^{(k)}}$ leads to the claim. An alternative way of seeing this for $f^{(k)}$ is to note from~\eqref{eq2.1} that the cuspidal behaviour receives contributions from the exponential prefactor $\exp(2\pi i \eta \frac{\Im z}{\Im \tau})$, and the order of $\eta^{k-1}$ is thus accompanied by up to $k$ inverse powers of $y$. For the Green function one may also inspect its explicit Laurent polynomial given for example in~\cite[Eq.~(2.15)]{Green:2008uj}.}
Moreover, this bound is uniformly valid at all orders 
in $\eta_j$ since the latter are introduced in the combinations $(\tau{-}\bar\tau)\eta_i$ by the 
Kronecker--Eisenstein integrands in (\ref{eq2.13}). Finally, the bound of $y^{ \geq -(a+b)}$ at the order 
of $s_{ij}^a \bar \eta_i^{b-n+1}$ can be transferred from  $Y^\tau_{\vec\eta}$ to $\Yhat^\tau_{\vec\eta}$ since
they are related by the exponential of $\frac{R_{\vec{\eta}}(\epsilon_{0})}{y}$ and the derivation in
the numerator is linear in $(s_{ij},\bar \eta_i)$. Therefore, we conclude that $\Yhat^\tau_{\vec\eta}$ does not contain any negative powers of $y$ at any order in its $\ap$-expansion.

On these grounds, we define the initial value by the regularised limit,
\beq
 \Yhat^{i\infty}_{\vec{\eta}} = \Yhat^{\tau}_{\vec{\eta}} \, \big|_{q^0 \bar q^0} \, \big|_{y^0} \, ,
 \label{eq3.6}
 \eeq
which does not suffer from inconsistencies caused by products involving negative powers of $y$.

\subsubsection{Expansion and truncation of initial data}

The absence of negative powers of $y$ in the Laurent polynomials of $\Yhat^\tau_{\vec\eta}$ can
also be verified explicitly in examples at fixed order in the expansion variables. In practice, this is done by imposing cutoffs on the powers of
$s_{ij}, \eta_j,\bar\eta_j$ in the expansion of $Y^\tau_{\vec\eta}$ or $\Yhat^\tau_{\vec\eta}$. 
We will make our scheme of cutoffs more transparent by defining the \textit{order} of a series in 
$\eta_i$ $\bar\eta_i$ and $s_{ij}$ through the assignment
\beq
\te{order}(\eta_i) = 1 \, , \ \ \ \ \ \ 
\te{order}(\bar \eta_i) = 1 \, , \ \ \ \ \ \ 
\te{order}(s_{ij}) = 2 \, . 
\label{deforder}
\eeq
More precisely, the order of $Y^\tau_{\vec\eta}$ and $\Yhat^\tau_{\vec\eta}$ is counted relative to
the most singular term of homogeneity degree $\eta_j^{1-n} \bar \eta_j^{1-n}$ to make sure that the $\alpha'\rightarrow0$ limit
of the plain Koba--Nielsen integrals $Y^\tau_{(0,\ldots,0|0,\ldots,0)}(\sigma|\rho) = (\prod_{j=2}^n\int \frac{ \dd^2 z_j}{\Im \tau}) {\rm KN}^\tau_{n} = 1 + {\cal O}(\ap^2)$ has order zero.
The assignment \eqref{deforder} is consistent with a counting of (inverse) lattice momenta: every factor of $\eta_i$ or $\bar\eta_i$ corresponds to an inverse momentum according to~\eqref{eq2.lat} while $s_{ij}$ always appears together with a Green function that contains two inverse momenta.

The notion of order in (\ref{deforder}) ensures that $\te{order}(R_{\vec{\eta}}(\epsilon_0))=0$ by inspection of its explicit form in section~\ref{sec2.4}, i.e.\ that the operator $\frac{R_{\vec{\eta}}(\epsilon_0) }{4y}$ in the exponential preserves the order of an expression. 
As we have shown above, $\Yhat^{\tau}_{\vec\eta}$ does not have any negative powers of $y$ at any order in the sense of~\eqref{deforder} and, at the same time, it has bounded positive powers of $y$ at each order by inspection of the component integrals. Since $Y^\tau_{\vec\eta}$ has a bounded negative power of $y$ at any order, this implies that the exponential $\exp  (\frac{R_{\vec{\eta}}(\epsilon_0) }{4y} )$ entering the redefinition~\eqref{eq3.1} terminates to a polynomial at any fixed order.

For instance, in the case of two points, (\ref{eq3.6}) results in the following initial data 
\begin{align}
&\Yhat^{i\infty}_{\eta} = \frac{1}{\bar \eta} \Big\{ \frac{1}{\eta} \Big[1+ 
\frac{1}{6} s_{12}^3 \zeta_3  + \frac{43}{360} s_{12}^5 \zeta_5 
\Big]
+   \eta   \Big[   {-} 2  s_{12} \zeta_3  - \frac{5}{3}   s_{12}^3 \zeta_5 
  - \frac{ 1}{3}  s_{12}^4 \zeta_3^2 
   \Big]
\notag\\
 & \ \ \ \ 
  +  \eta^3    \Big[ {-}  2   s_{12} \zeta_5 +  2   s_{12}^2 \zeta_3^2   - \frac{ 7}{2}   s_{12}^3 \zeta_7 
  \Big]
  +   \eta^5  \big[{-} 2   s_{12} \zeta_7 + 4   s_{12}^2 \zeta_3 \zeta_5 
  \big]
 +  \eta^7   \big[ {-} 2  s_{12} \zeta_9  \big]   \Big\} \notag \\
&\ \ \ \ + (2\pi i) \Big\{ {-}\Big[ \frac{1}{s_{12}} + \frac{ s_{12}^2 }{6} \zeta_3 + \frac{ 43 s_{12}^4}{360} \zeta_5 \Big]
+ \eta^2 \Big[
 2  \zeta_3 + \frac{ 5}{3}s_{12}^2 \zeta_5 +   \frac{ s_{12}^3}{3} \zeta_3^2   \Big] \notag \\
& \ \ \ \ \ \ \ \ + \eta^4 \Big[
   2 \zeta_5   - 2 s_{12} \zeta_3^2 + \frac{  7}{2} s_{12}^2 \zeta_7    \Big]
 + \eta^6 \big[   2 \zeta_7   - 4  s_{12} \zeta_3  \zeta_5  \big]
+  \eta^8 \big[ 2 \zeta_9\big] \Big\} \label{eq3.7}\\
&\ \ \ \ + (2\pi i)^2 \bar \eta \Big\{ \frac{1}{ \eta} \Big[ \frac{ s_{12}^3 \zeta_3}{60} \Big] 
- \eta  \Big[ \frac{  s_{12}^3 \zeta_5}{30}  \Big] \Big\}
- (2\pi i)^3 \bar \eta^2 \Big\{   \Big[ \frac{ s_{12}^2 \zeta_3 }{60} \Big] 
+ \eta^2    \Big[ \frac{  s_{12}^2 \zeta_5 }{30}   \Big]  \Big\} \notag \\
&\ \ \ \ - (2\pi i)^4 \bar \eta^3 \Big\{ \frac{ 1 }{ \eta} \Big[ \frac{ s_{12}^3 \zeta_3 }{1512 }
\Big]\Big\}
+(2\pi i)^5 \bar \eta^4 \Big\{  \Big[ \frac{ s_{12}^2 \zeta_3}{1512}  
\Big] \Big\}   +\text{(terms of order $\geq 12$)}
\,, \notag
\end{align}
where we have given all terms of the form $\eta^{a-1} \bar{\eta}^{b-1} s_{12}^{c}$ up to the order $a+b+2c\leq 10$. 
The above expression has been obtained from a general formula for the two-point Laurent polynomial 
that we shall present in (\ref{eq4.1}) below and the expansion (\ref{eq3.7}) is also available in 
machine-readable form in an 
ancillary file within the arXiv submission of this paper. When disregarding the $(2\pi i )^{k} \bar \eta^{k-1}$,
the all-order expansion of $\Yhat^{i\infty}_{\eta}$ features no MZVs other than $\zeta^{\rm sv}_k$,
in agreement with the results of \cite{DHoker:2019xef, Zagier:2019eus} on the terms $\sim \eta^{-1}\bar \eta^{-1}$.

At low orders, (\ref{eq3.7}) can be crosschecked by
analysing the MGFs in $Y^\tau_\eta$, inserting their Laurent polynomials in (\ref{eqcusp.mzv})
and extracting the initial value according to~\eqref{eq3.6}.
In both approaches, the redefinition~\eqref{eq3.1} has been performed, 
the exponential $\frac{R_{\eta}(\ep_0)}{4y}$ truncates to a polynomial
and one can verify order by order that all negative $y$-powers are eliminated from the Laurent polynomials.

With the assignments in (\ref{deforder}), the operators $R_{\vec\eta}\big( {\rm ad}_{\ep_0}^j(\epsilon_k) \big)$ 
in (\ref{eq3.4}) have order $k$ for any value of $j=0,1,\ldots,k{-}2$. This is evident from the explicit examples 
given in section~\ref{sec2.4} as well as appendix \ref{appderiv} and this is important for truncations of the formal solution 
of the differential equation 
to a fixed order. The expansions of $Y^\tau_{\vec\eta}$ and $\Yhat^\tau_{\vec\eta}$ to the $m$-th order
can be related by the truncation of the exponential $\exp  ({\pm} \frac{R_{\vec{\eta}}(\epsilon_0) }{4y} ) \rightarrow \sum_{r=0}^m 
\frac{1}{r!} ({\pm} \frac{R_{\vec{\eta}}(\epsilon_0) }{4y} )^r$ when acting on the expansions.

Once the contributions of the operators $R_{\vec{\eta}}({\rm ad}_{\ep_0}^{j_\ell} \ep_{k_\ell} \ldots
{\rm ad}_{\ep_0}^{j_1} \ep_{k_1} )$ in (\ref{eq3.21}) are computed
to the order of $k_1+\ldots + k_\ell = m$, one can access the component integrals 
$Y^\tau_{(a_2,\ldots,a_n|b_2,\ldots,b_n)}(\sigma|\rho)$ up to
and including homogeneity degree $\frac{1}{2}m- \frac{1}{2} \sum_{j=2}^n(a_j{+}b_j)$ in $s_{ij}$. 
Conversely, the $\bsvBRno{j_1 &\ldots &j_\ell}{k_1 &\ldots &k_\ell}$ 
appearing at homogeneity degree $s_{ij}^w$ in the $\ap$-expansion of 
$Y^\tau_{(a_2,\ldots,a_n|b_2,\ldots,b_n)}(\sigma|\rho)$ are bounded to feature
$k_1+\ldots + k_\ell \leq 2w+ \sum_{j=2}^n(a_j{+}b_j)$. 

The above bounds rely on the fact that, at $n$ points, the order of the series $\Yhat^{i \infty}_{\vec\eta}$ 
is bounded by the most singular term $\eta_j^{1-n} \bar \eta_j^{1-n}$ exposed by the 
Kronecker--Eisenstein integrand in (\ref{eq2.13}). At two points, for instance, the bound is
saturated by the terms $\Yhat^{i\infty}_{\eta} \rightarrow \frac{1}{\eta \bar \eta} - \frac{ 2\pi i }{s_{12}}$ 
without $\zeta_{2k+1}$ in (\ref{eq3.7}). Their three-point analogues are given by
\begin{align}
\Yhat^{i\infty}_{\eta_2,\eta_3}(2,3|2,3) &= \frac{ 1}{\eta_{23} \eta_3 \bar \eta_{23} \bar \eta_3} 
- \frac{2\pi i }{\eta_3 \bar \eta_3 s_{12}}- \frac{2\pi i }{\eta_{23} \bar \eta_{23} s_{23}}
+ \Big( \frac{1}{s_{12}} + \frac{1}{s_{23}} \Big) \frac{(2\pi i)^2}{s_{123}} + \ldots\,,
\notag \\*
\Yhat^{i\infty}_{\eta_2,\eta_3}(2,3|3,2) &= \frac{ 1}{\eta_{23} \eta_2 \bar \eta_{23} \bar \eta_3} 
+ \frac{2\pi i }{\eta_{23} \bar \eta_{23} s_{23}}
-  \frac{(2\pi i)^2}{ s_{23} s_{123}} + \ldots
\label{3ptlow}
\end{align}
and permutations in $2\leftrightarrow 3$, where $s_{123} = s_{12}{+}s_{13}{+}s_{23}$, and 
all the terms in the ellipsis comprise MZVs and are higher order in the sense of (\ref{deforder}).
At four points, conjectural expressions for the MZV-free part of the 
$\Yhat^{i\infty}_{\eta_2,\eta_3,\eta_4}(2,3,4|\rho),\ \rho \in {\cal S}_3$ can be found in appendix~\ref{app:n4}.


\subsection{Real-analytic combinations of iterated Eisenstein integrals}
\label{sec3.B}

In this section, we relate the objects we called $\Esv$ in~\eqref{eq3.8} to iterated integrals over holomorphic Eisenstein series. We will use Brown's holomorphic iterated Eisenstein integrals 
subject to tangential-base-point regularisation \cite{Brown:mmv},
\begin{align}
\EBR{j_1 &j_2 &\ldots &j_\ell}{k_1 &k_2 &\ldots &k_\ell}{\tau}
&= - (2\pi i)^{1+j_\ell-k_\ell} \int^\tau_{i \infty} \dd \tau' \, (\tau')^{j_{\ell}} {\rm G}_{k_\ell}(\tau')
\EBR{j_1 &j_2 &\ldots &j_{\ell-1}}{k_1 &k_2 &\ldots &k_{\ell-1}}{\tau'} \notag \\
&\hspace{-15mm}= (-1)^{\ell} \Big(  \prod_{i=1}^\ell (2\pi i)^{1+j_i-k_i} \Big)
\int^\tau_{i \infty} \dd \tau_\ell \, (\tau_\ell)^{j_{\ell}} {\rm G}_{k_\ell}(\tau_\ell)
\int^{\tau_\ell}_{i \infty} \dd \tau_{\ell-1} \, (\tau_{\ell-1})^{j_{\ell-1}} {\rm G}_{k_{\ell-1}}(\tau_{\ell-1})
\ldots \notag \\
&\ \ \ \ \ \ \ \ldots \int^{\tau_3}_{i \infty} \dd \tau_{2} \, (\tau_{2})^{j_{2}} {\rm G}_{k_{2}}(\tau_{2})
\int^{\tau_2}_{i \infty} \dd \tau_{1} \, (\tau_{1})^{j_{1}} {\rm G}_{k_{1}}(\tau_{1})\, ,
 \label{eq3.11}
\end{align}
which can be expressed straightforwardly in terms of the iterated Eisenstein integrals
$\gamma_0(\ldots)$ or ${\cal E}_0(\ldots)$ seen in the $\ap$-expansion of open-string integrals
\cite{Broedel:2015hia, Broedel:2018izr, Broedel:2019vjc, Mafra:2019ddf, Mafra:2019xms}, cf.\ appendix~\ref{appC}. 
The holomorphic iterated integrals (\ref{eq3.11}) obey the following differential equations and initial conditions
\begin{subequations}
\label{eq3.hol}
\begin{align}
2\pi i \partial_\tau \EBR{j_1 &j_2 &\ldots &j_\ell}{k_1 &k_2 &\ldots &k_\ell}{\tau}
&= - (2\pi i)^{2-k_\ell+j_\ell}  \tau^{j_\ell} {\rm G}_{k_\ell}(\tau)
\EBR{j_1 &j_2 &\ldots &j_{\ell-1}}{k_1 &k_2 &\ldots &k_{\ell-1}}{\tau}\,,
\label{eq3.12}
 \\
 \lim_{\tau \rightarrow i\infty} \EBR{j_1 &j_2 &\ldots &j_\ell}{k_1 &k_2 &\ldots &k_\ell}{\tau} &= 0\, .
 \label{eq3.13}
\end{align}
\end{subequations}
These equations are similar to those of ${\cal E}^{\rm sv}$ in~\eqref{eq3.init} but feature $ \tau  ^{j_\ell} {\rm G}_{k_\ell}(\tau)$ in the place of $( \tau {-} \bar \tau )^{j_\ell} {\rm G}_{k_\ell}(\tau)$. The holomorphic iterated Eisenstein integrals~\eqref{eq3.11} obey the standard shuffle identities
\beq
{\cal E}[A_1,A_2,\ldots,A_\ell;\tau] \, {\cal E}[B_1,B_2,\ldots,B_{m};\tau]
= {\cal E}[ (A_1,A_2,\ldots,A_\ell) \shuffle (B_1,B_2,\ldots,B_{m});\tau]
\label{eq3.shuff}
\eeq
with respect to the combined letters $A_i = \begin{smallmatrix} j_i \\ k_i \end{smallmatrix}$, e.g.\ $\EBR{j_1}{k_1}{\tau} \EBR{j_2 }{k_2}{\tau} = \EBR{j_1 &j_2 }{k_1 &k_2}{\tau} + \EBR{j_2 &j_1 }{k_2 &k_1}{\tau}$ and where $\shuffle$ denotes the standard shuffle product of ordered sequences. 
There are no linear relations among the ${\cal E}$ with different entries \cite{Nilsnewarticle}.

It is tempting to define a solution to our differential equations (\ref{eq3.init}) by starting from \eqref{eq3.11} 
and simply replacing the holomorphic integration kernels $\tau  ^{j_\ell} {\rm G}_{k_\ell}(\tau)$ 
by the non-holomorphic expressions $( \tau {-} \bar \tau )^{j_\ell} {\rm G}_{k_\ell}(\tau)$: 
\begin{align}
\EsvBRmin{j_1 &j_2 &\ldots &j_\ell}{k_1 &k_2 &\ldots &k_\ell}{\tau}
&=  (-1)^{\ell} \Big(  \prod_{i=1}^\ell (2\pi i)^{1+j_i-k_i} \Big)
\int^\tau_{i \infty} \dd \tau_\ell \, (\tau_\ell{-}\bar \tau)^{j_{\ell}} {\rm G}_{k_\ell}(\tau_\ell)
 \label{eq3.14} \\
&\quad \int^{\tau_\ell}_{i \infty} \dd \tau_{\ell-1} \, (\tau_{\ell-1}{-}\bar \tau)^{j_{\ell-1}} {\rm G}_{k_{\ell-1}}(\tau_{\ell-1})
\cdots \int^{\tau_2}_{i \infty} \dd \tau_{1} \, (\tau_{1}{-}\bar \tau)^{j_{1}} {\rm G}_{k_{1}}(\tau_{1}) \, .
 \notag
\end{align}
Since $\bar \tau$ is not the complex conjugate of the integration variables $\tau_j$,
these integrals are homotopy invariant.
We call (\ref{eq3.14}) the {\it minimal solution} of (\ref{eq3.init}), 
and it also obeys the standard shuffle relations (\ref{eq3.shuff})
with ${\cal E}^{\rm sv}_{\rm min}$ in the place of ${\cal E}$.
Binomial expansion of the integration kernels straightforwardly relates this minimal solution
to the holomorphic iterated Eisenstein integrals~(\ref{eq3.11})

\begin{align}
\EsvBRmin{j_1 &j_2 &\ldots &j_\ell}{k_1 &k_2 &\ldots &k_\ell}{\tau} &= 
\sum_{r_1=0}^{j_1}\sum_{r_2=0}^{j_2} \ldots \sum_{r_\ell=0}^{j_\ell}
\binom{j_1}{r_1}\binom{j_2}{r_2}\cdots\binom{j_\ell}{r_\ell}
 \label{eq3.15} \\
 & \quad\quad \times (-2\pi i \bar\tau)^{r_1+r_2+\ldots+r_\ell}  \EBR{j_1-r_1 &j_2-r_2 &\ldots &j_\ell-r_\ell}{k_1  &k_2 &\ldots &k_\ell}{\tau} \, .
\notag 
\end{align}
However, (\ref{eq3.8}) is supposed to generate real-analytic modular forms
such as ${\rm E}_k(\tau)$ in (\ref{eq2.24}) and its Cauchy--Riemann derivatives, as e.g.\
(\ref{eq2.all36}). Hence, the minimal solutions (\ref{eq3.14}) need to be augmented
by antiholomorphic functions $\overline{\fBR{j_1 &j_2&\ldots &j_\ell}{k_1 &k_2&\ldots &k_\ell}{\tau}}$
that vanish at the cusp, and we shall solve (\ref{eq3.init}) via
  \begin{align}
    \EsvBR{j_1 &j_2 &\ldots &j_\ell}{k_1 &k_2 &\ldots &k_\ell}{\tau}&= \sum_{i=0}^{\ell}\overline{\fBR{j_1 & j_2&\ldots &j_{i}}{k_1 &k_2&\ldots &k_{i}}{\tau}}\EsvBRmin{j_{i+1}&\ldots &j_\ell}{k_{i+1} &\ldots &k_\ell}{\tau}\label{eq:2}
  \end{align}
 with $ \overline{\fBR{ }{ }{\tau}}=\EsvBRmin{}{}{\tau}=1$.
 The functions $\overline{\fBR{j_1 &\ldots &j_\ell}{k_1 &\ldots &k_\ell}{\tau}}$ will be 
 determined systematically by extracting the $\ap$-expansion of the component 
 integrals (\ref{eq2.15}) and imposing their reality properties (\ref{eq2.19}). 
 In particular, these reality properties imply that the
 $\overline{f}$ must be expressible in terms of antiholomorphic iterated Eisenstein 
 integrals (with $\QQ$-linear combinations of MZVs and powers of $\bar \tau$ in its coefficients): Referring back to~\eqref{eqYbartau}, we see that the antiholomorphic derivative $\partial_{\bar\tau} Y^{\tau}_{\vec{\eta}}$ contains only $(\tau{-}\bar \tau)^{-1}$ and the kernels $\overline{{\rm G}}_k$ of antiholomorphic iterated Eisenstein integrals, thus excluding any other objects in $\overline{f}$.

\subsubsection{Depth one}

As will be derived in detail in section \ref{sec4.3}, the appropriate choice of integration
constants at depth $\ell=1$ is given by the purely antiholomorphic expression
\beq
 \overline{\protect\fBR{j_1}{k_1}{\tau}}= \sum_{r_1=0}^{j_1} (-2\pi i \bar\tau)^{r_1}\binom{j_1}{r_1}
 (-1)^{j_1-r_1} \overline{\protect\EBR{j_1-r_1}{k_1}{\tau} }\,.
\label{eq:ff1}
\eeq
Hence, for $\Esv$ at depth one, we obtain,
\beq
\EsvBR{j_1}{k_1}{\tau} = \sum_{r_1=0}^{j_1} (-2\pi i \bar\tau)^{r_1} \binom{j_1}{r_1}
\Big( \EBR{j_1-r_1}{k_1}{\tau}
+(-1)^{j_1-r_1} \overline{  \EBR{j_1-r_1}{k_1}{\tau} } \Big)\, ,
\label{eq3.16}
\eeq
where the contributions $\sim (-2\pi i \bar\tau)^{r_1} \EBR{j_1-r_1}{k_1}{\tau}$ match the
minimal solution (\ref{eq3.15}) while the additional terms are due to \eqref{eq:ff1}. 
Such expressions should 
be contained in Brown's generating series of single-valued iterated Eisenstein 
integrals \cite{Brown:mmv, Brown:2017qwo,Brown:2017qwo2}, 
and similar objects have been discussed in \cite{unpub} as building 
blocks for a single-valued map at depth one.
The reality properties of our component integrals yield an independent construction
of (\ref{eq3.16}) that will be detailed in section \ref{sec4.3}.

The simplest instances of~\eqref{eq3.16} are given by
\begin{align}
\esvtau{0\\4} &= \EBR{0}{4}{\tau}  + \overline{\EBR{0}{4}{\tau}}\,,
\notag\\
\esvtau{1\\4} &= \EBR{1}{4}{\tau}  - \overline{\EBR{1}{4}{\tau}}  + (-2\pi i \bar\tau)  \left(\EBR{0}{4}{\tau}  + \overline{\EBR{0}{4}{\tau}}\right)  \,,\label{more.01}
\\
\esvtau{2\\4} &= \EBR{2}{4}{\tau}  + \overline{\EBR{2}{4}{\tau}}  + 2(-2\pi i \bar\tau) \left(\EBR{1}{4}{\tau}  - \overline{\EBR{1}{4}{\tau}}\right)  + (-2\pi i \bar\tau)^2  \left(\EBR{0}{4}{\tau}  + \overline{\EBR{0}{4}{\tau}}\right) \, ,
\notag
\end{align}
and it is easy to check from~\eqref{eq3.hol} that $\esv{j\\4}$ at $j=0,1,2$ satisfy~\eqref{eq3.init}. As the 
holomorphic iterated Eisenstein integrals (and their complex conjugates) are homotopy-invariant, these 
expressions represent well-defined real-analytic functions, and one can straightforwardly obtain their 
$(q,\bar q)$-expansion from the methods of appendix \ref{secC.0}.

\subsubsection{Depth two}

We next elaborate on the general form of the depth-two $\Esv$. Starting from the minimal solution~\eqref{eq3.15}, the reality properties of the component integrals dictate the following integration constant at depth $\ell=2$
\beq
\overline{\protect\fBR{j_1 &j_2}{k_1 &k_2}{\tau}}=\sum_{r_1=0}^{j_1}\sum_{r_2=0}^{j_2} 
(2\pi i \bar\tau)^{r_1+r_2} (-1)^{j_1+j_2}
\binom{j_1}{r_1}\binom{j_2}{r_2}
\overline{\protect\EBR{j_2-r_2 &j_1-r_1}{k_2 &k_1}{\tau} }+\overline{\protect\alphaBR{j_1 &j_2}{k_1 &k_2}{\tau}}\,,
\label{eq:ff2}
\eeq
where $\overline{ \EBR{j_2-r_2 &j_1-r_1}{k_2 &k_1}{\tau} }$
and $\overline{\alphaBR{j_1 &j_2}{k_1 &k_2}{\tau}}$ are purely antiholomorphic and
individually vanish at the cusp in the regularised limit $\tau\to i \infty$. 
Together with the depth-one expression \eqref{eq:ff1}, the decomposition \eqref{eq:2} into ${\cal E}_{\rm min}^{\rm sv}$ then implies
\begin{align}
\EsvBR{j_1 &j_2}{k_1 &k_2}{\tau} &=  \sum_{r_1=0}^{j_1}\sum_{r_2=0}^{j_2} (-2\pi i \bar\tau)^{r_1+r_2}
\binom{j_1}{r_1}\binom{j_2}{r_2}
\label{eq3.17} 
\Bigg\{ \EBR{j_1 -r_1&j_2-r_2}{k_1 &k_2}{\tau}\notag\\
&\hspace{10mm} +(-1)^{j_1-r_1} \overline{ \EBR{j_1-r_1}{k_1}{\tau} }  \EBR{j_2-r_2}{k_2}{\tau} 
+(-1)^{j_1+j_2-r_1-r_2} \overline{ \EBR{j_2-r_2 &j_1-r_1}{k_2 &k_1}{\tau} }  
\Bigg\}   \notag\\
&\hspace{5mm}+\overline{\alphaBR{j_1 &j_2}{k_1 &k_2}{\tau}} \, ,
\end{align}
where the first term is the minimal solution~\eqref{eq3.15}.
We expect similar expressions to follow from
Brown's generating series of single-valued iterated Eisenstein 
integrals \cite{Brown:mmv, Brown:2017qwo,Brown:2017qwo2}.
Moreover, the first two lines of (\ref{eq3.17}) with lower-depth corrected versions of ${\cal E}$,
$\overline{{\cal E}}$ and the need for further antiholomorphic corrections have featured
in discussions about finding an explicit form of a single-valued map at depth two \cite{unpub}.
As we shall see in sections \ref{sec4.3} and \ref{sec5.8}, the reality properties of our component integrals yield an independent construction
of (\ref{eq3.17}).

We have separated the two terms in (\ref{eq:ff2}) for the following reasons:
\begin{itemize}
\item
The $\overline{ \EBR{j_2-r_2 &j_1-r_1}{k_2 &k_1}{\tau}}$ exhaust the antiholomorphic iterated Eisenstein
integrals at depth two within $\EsvBR{j_1 &j_2}{k_1 &k_2}{\tau}$ which are necessary 
to satisfy the required reality properties.
The $\overline{\alphaBR{j_1 &j_2}{k_1 &k_2}{\tau}} $ in turn conjecturally comprise $\zeta_{2k+1}$
and antiholomorphic iterated Eisenstein integrals of depth one. They are determined
on a case-by-case basis for $(k_1,k_2)=(4,4),(6,4),(4,6)$ in this paper, see (\ref{eq4.24})
and (\ref{G4G6alpha}), and we leave a general
discussion for the future. 

We also note that, since the derivation-algebra relations such as~\eqref{eq2.43} imply that at higher weight only certain linear combinations of the $\Esv$ arise in the solution of $Y^\tau_{\vec\eta}$, not all integration constants can be determined individually from the component integrals. For instance, (\ref{eq2.44}) implies that certain
linear combinations of $\overline{\alphaBR{j_1 &j_2}{k_1 &k_2}{\tau}}$ with $(k_1,k_2) \in \{ 
(10,4),(4,10),(8,6),(6,8)\}$ and $j_i\leq k_i{-}2$ do not occur in the expansion of $Y^\tau_{\vec{\eta}}$ and
are inaccessible with the methods of this work.
\item
Even in absence of $\overline{\alphaBR{j_1 &j_2}{k_1 &k_2}{\tau}} $, the right-hand side of (\ref{eq3.17}) is invariant under the modular $T$-transformation
$\tau \rightarrow \tau + 1$. As will be argued in section \ref{sec6.1} the ${\cal E}^{\rm sv}$ must be $T$-invariant
as well, so the unknown $\overline{\alphaBR{j_1 &j_2}{k_1 &k_2}{\tau}} $ need to be
individually $T$-invariant (on top of being antiholomorphic and vanishing at the cusp).
\end{itemize}
An exemplary expression resulting from~\eqref{eq3.17} is
\begin{align}
\label{eq:Esv44}
\esvtau{2&0\\4&4} &= \EBR{2 &0}{4 &4}{\tau} + \overline{\EBR{2}{4}{\tau}}\EBR{0}{4}{\tau} + \overline{\EBR{0&2}{4&4}{\tau}}\nn\\
&\quad  +2 (-2\pi i \bar\tau)\left\{\EBR{1 &0}{4 &4}{\tau}  -  \overline{\EBR{1}{4}{\tau}}\EBR{0}{4}{\tau}  - \overline{\EBR{0&1}{4&4}{\tau}}\right\}\nn\\
&\quad + (-2\pi i \bar\tau)^2\left\{ \EBR{0 &0}{4 &4}{\tau}   +  \overline{\EBR{0}{4}{\tau}}\EBR{0}{4}{\tau}  + \overline{\EBR{0&0}{4&4}{\tau}}\right\}\nn\\
&\quad +  \frac{2 \zeta_3}{3}  \Big( \overline{  \EBR{0}{4}{\tau} }  - \frac{ i \pi \bar \tau}{360} \Big) \,,
\end{align}
where the last line corresponds to $\overline{\alphaBR{2 & 0}{4&4}{\tau}}$ that will be determined in~\eqref{eq4.24}.

\subsubsection{Higher depth and shuffle}

The ${\cal E}^{\rm sv}$ at depth $\ell \geq 3$ will introduce additional
antiholomorphic $\overline{\fBR{j_1 &j_2 &\ldots &j_\ell}{k_1 &k_2 &\ldots &k_\ell}{\tau}}$
that vanish at the cusp. These antiholomorphic integration constants will preserve
the shuffle relations
\beq
{\cal E}^{\rm sv}[A_1,A_2,\ldots,A_\ell;\tau] \, {\cal E}^{\rm sv}[B_1,B_2,\ldots,B_{m};\tau]
= {\cal E}^{\rm sv}[ (A_1,A_2,\ldots,A_\ell) \shuffle (B_1,B_2,\ldots,B_{m});\tau]
\label{eq3.shuff1}
\eeq
analogous to those of the holomorphic counterparts (\ref{eq3.shuff}). At depth two,
the last terms of (\ref{eq3.17}) are then constrained to obey $ \overline{\alphaBR{j_1 &j_2}{k_1 &k_2}{\tau}}+ \overline{\alphaBR{j_2 &j_1}{k_2 &k_1}{\tau}}=0$. We expect that the decomposition (\ref{eq:2}) 
of ${\cal E}^{\rm sv}$ is related to Brown's construction of single-valued iterated
Eisenstein integrals \cite{Brown:mmv, Brown:2017qwo, Brown:2017qwo2} by composing holomorphic and 
antiholomorphic generating series. A discussion of depth-$(\ell \geq 3)$ instances
and more detailed connections with the work of Brown are left to the future.

Given the expressions (\ref{eq3.16}) and (\ref{eq3.17}) for the simplest ${\cal E}^{\rm sv}$,
also the $\beta^{\rm sv}$ at depth $\ell \leq 2$ can be reduced to iterated Eisenstein integrals
via (\ref{eq3.20}). More specifically, this completely determines the $\beta^{\rm sv}$ at depth one 
and fixes their depth-two examples up to the antiholomorphic and $T$-invariant 
$\overline{\alphaBR{j_1 &j_2}{k_1 &k_2}{\tau}} $ in (\ref{eq3.17}). The latter will later be exemplified 
to comprise antiholomorphic iterated Eisenstein integrals at depth one and powers of $\bar \tau$. Note that the relation (\ref{eq3.20}) between
${\cal E}^{\rm sv}$ and $\beta^{\rm sv}$ preserves the shuffle property and therefore
\beq
\beta^{\rm sv}[A_1,A_2,\ldots,A_\ell;\tau] \, \beta^{\rm sv}[B_1,B_2,\ldots,B_{m};\tau]
= \beta^{\rm sv}[ (A_1,A_2,\ldots,A_\ell) \shuffle (B_1,B_2,\ldots,B_{m});\tau] \, .
\label{eq3.shuff2}
\eeq

\subsubsection{Expansion around the cusp}

The expansion of the above ${\cal E}^{\rm sv}$ around the cusp takes the same form
as that of MGFs in (\ref{eqcusp.1}). Tangential-base-point regularisation  of the
holomorphic iterated Eisenstein integrals leads to the behaviour \cite{Brown:mmv}
\begin{align}
\EBR{j_1}{k_1}{\tau} &=  \frac{ B_{k_1} }{k_1!} \frac{ (2 \pi i \tau)^{j_1+1} }{j_1+1} + {\cal O}(q)\,,
\label{more.02} \\
\EBR{j_1 &j_2}{k_1 &k_2}{\tau} &=  \frac{ B_{k_1} B_{k_2} }{k_1! k_2!} \frac{ (2 \pi i \tau)^{j_1+j_2+2} }{(j_1+1)(j_1+j_2+2)} + {\cal O}(q)
\notag
\end{align}
with Bernoulli numbers $B_{k_i}$. As a consequence of \eqref{eq3.16} and~\eqref{eq3.17},
the Laurent monomial at the order of $q^0 \bar q^0$ in ${\cal E}^{\rm sv}$ at depth $\leq 2$ 
can be given in closed form,
\begin{align}
\esvtau{j_1\\k_1}  &=  \frac{ B_{k_1} }{k_1!} \frac{ (-4y)^{j_1+1} }{j_1+1} + {\cal O}(q,\bar q) \,,
\label{more.03}\\
\EsvBR{j_1 &j_2}{k_1 &k_2}{\tau}   &= \frac{ B_{k_1} B_{k_2} }{k_1! k_2!} \frac{ (-4y)^{j_1+j_2+2} }{(j_1+1)(j_1+j_2+2)}  + {\cal O}(q,\bar q)  \, .
\notag
\end{align}
The $\overline{\alphaBR{j_1 &j_2}{k_1 &k_2}{\tau}}$ which are currently unknown at
$k_1+ k_2\geq12$ cannot contribute to the Laurent monomial since they need to be antiholomorphic,
$T$-invariant and vanishing at the cusp. Note that the regime (\ref{more.03}) of ${\cal E}^{\rm sv}$ 
can be formally obtained from (\ref{more.02}) for ${\cal E}$ by replacing $\tau \rightarrow \tau - \bar \tau$, 
in line with the proposal for an elliptic single-valued map in \cite{Broedel:2018izr}.

The Laurent monomials of the $\beta^{\rm sv}$ at depth $\leq 2$ resulting from (\ref{eq3.22}) 
and (\ref{more.03}) read
\begin{subequations}
\label{more.bsv}
\begin{align}
\bsvBR{j_1 }{k_1 }{\tau}  &=  \frac{ B_{k_1}  j_1! (k_1{-}2{-}j_1)! (-4y)^{j_1+1} }{k_1! \, (k_1{-}1)!}+ {\cal O}(q,\bar q) \,,
\label{more.bsva}\\
\bsvBR{j_1 &j_2}{k_1 &k_2}{\tau}   &= \frac{ B_{k_1} B_{k_2}  (j_1{+}j_2{+}1)!   (k_2{-}2{-}j_2)!  (-4y)^{j_1+j_2+2}  }{ (j_1{+}1) k_1! k_2!  (k_2{+}j_1)!} \label{more.bsvb} \\
&\ \ \ \ \times
 \, _3F_2\Big[
\begin{smallmatrix}
1 {+} j_1,\ 2 {+} j_1 {+} j_2, \  2 {+} j_1 {-} k_1 \\
2 {+} j_1,\ 1 {+} j_1 {+} k_2 
\end{smallmatrix} ; 1
\Big] + {\cal O}(q,\bar q) \notag
\, .
\end{align}
\end{subequations}
%


\section{Explicit forms at two points}
\label{sec4}

In this section, we evaluate explicitly the generating function $Y^\tau_\eta$ at two points up to order $10$ and use this to determine several $\bsv$ and $\Esv$ that were introduced in the previous section. The starting point is an explicit determination of the Laurent polynomial to obtain the initial data $\Yhat^{i\infty}_\eta$ for equation~\eqref{eq3.21} where we present an all-order result for two points. By exploiting the reality properties of the resulting two-point component integrals, we can find the integration constants in various $\bsv$ and $\Esv$.


\subsection{Laurent polynomials and initial data}
\label{sec4.1}

The general idea is to obtain the initial data at $n$ points by reducing the one-loop calculation in the degeneration limit $\tau\to i\infty$ of the torus to an $(n{+}2)$-point tree-level calculation on the sphere.\footnote{For open-string integrals
over the $A$-cycle of the torus, the $\tau\to i\infty$ limit at $n$ points has been reduced to explicitly
known combinations of $(n{+}2)$-point disk integrals in \cite{Mafra:2019ddf, Mafra:2019xms}.} At $n=2$, 
mild generalisations
of the techniques of \cite{DHoker:2019xef, Zagier:2019eus} lead to a closed formula involving the usual Virasoro--Shapiro four-point amplitude on the sphere,
\beq
 \frac{ \Gamma(1{-}a) \Gamma(1{-}b) \Gamma(1{-}c)  }{ \Gamma(1{+}a)\Gamma(1{+}b)\Gamma(1{+}c)} 
 = \exp \Big( 2 \sum_{k=1}^{\infty} \frac{ \zeta_{2k+1} }{2k{+}1} \big[ a^{2k+1} + b^{2k+1}+c^{2k+1}\big] \Big) \, , \ \ \ \ \ \ 
 a+b+c=0\,.
   \label{eq4.2}
\eeq
Its specific combinations that generate the two-point Laurent polynomial of (\ref{eq2.12})
can be written in the following form \cite{yinprogress}, using the shorthand $\xi= \frac{ i \pi \bar \eta}{2y}$, 
\begin{align}
&Y^\tau_{\eta} \, \big|_{q^0 \bar q^0}=
i \pi \exp\Big(  \frac{ s_{12} y }{3} \Big)   \label{eq4.1}   
\bigg\{ \big[  \cot( 2i\eta y ) - i \big] \big[  \cot( \pi \bar \eta ) + i \big]
 \notag\\
 & \ \ \ \ \ \ \times \exp \Big( \frac{ s_{12} }{8y}  \partial_{\eta}^2 \Big) \frac{1}{s_{12}{+}2\eta{+}2\xi} \bigg[ 
 \frac{ \Gamma(1{+}\tfrac{ s_{12}}{2} {+} \eta{+}\xi) \Gamma(1{-}s_{12}) \Gamma(1{+}\tfrac{ s_{12}}{2} {-} \eta{-}\xi)   }{\Gamma(1 {-} \tfrac{ s_{12}}{2} {+} \eta{+}\xi)  \Gamma(1{+}s_{12}) \Gamma(1 {-} \tfrac{ s_{12}}{2} {-} \eta{-}\xi)}
  - e^{-y( s_{12} +2 \eta+2\xi) }\bigg] \notag \\
 & \ \ + \big[  \cot( 2i\eta y ) + i \big] \big[  \cot( \pi \bar \eta ) - i \big]
 \notag \\
 & \ \ \ \ \  \times \exp \Big( \frac{ s_{12} }{8y}  \partial_{\eta}^2 \Big) 
 \frac{1}{s_{12}{-}2\eta{-}2\xi} \bigg[ 
 \frac{
\Gamma(1{+}\tfrac{ s_{12}}{2} {+} \eta{+}\xi) \Gamma(1{-}s_{12}) \Gamma(1{+}\tfrac{ s_{12}}{2} {-} \eta{-}\xi)   }{\Gamma(1 {-} \tfrac{ s_{12}}{2} {+} \eta{+}\xi)  \Gamma(1{+}s_{12}) \Gamma(1 {-} \tfrac{ s_{12}}{2} {-} \eta{-}\xi)}
  - e^{-y(s_{12} -2 \eta-2\xi) }\bigg] \notag \\
  & \ \ - \frac{2}{s_{12}}  \exp \Big( \frac{ s_{12} }{8y}  \partial_{\eta}^2 \Big)   \frac{
\Gamma(1{+}\tfrac{ s_{12}}{2} {+} \eta{+}\xi) \Gamma(1{-}s_{12}) \Gamma(1{+}\tfrac{ s_{12}}{2} {-} \eta{-}\xi)   }{\Gamma(1 {-} \tfrac{ s_{12}}{2} {+} \eta{+}\xi)  \Gamma(1{+}s_{12}) \Gamma(1 {-} \tfrac{ s_{12}}{2} {-} \eta{-}\xi)}
\bigg\} \, .
\end{align}
By tracking the coefficients of $\eta^{a-1} \bar \eta^{b-1}$, this results in the Laurent polynomials of the component integrals $Y_{(a|b)}^\tau$ defined in~\eqref{eq2.14}. Some exemplary instances are
\begin{subequations}
  \label{eq4.ex}
\begin{align}
Y^\tau_{(0|0)} \, \big|_{q^0 \bar q^0} &=
1 + s_{12}^2 \Big(  \frac{ y^2}{90} + \frac{  \zeta_3}{2 y}   \Big) + 
 s_{12}^3 \Big( \frac{ y^3}{2835} + \frac{  \zeta_3}{6} + \frac{  \zeta_5}{8 y^2}   \Big) \nn\\
  &\quad + s_{12}^4 \Big( \frac{ y^4}{22680} +   \frac{  y \zeta_3 }{36}
 + \frac{     5 \zeta_5}{12 y} - \frac{  \zeta_3^2}{8 y^2} + \frac{ 3 \zeta_7}{32 y^3}  \Big)  
  \label{eq4.3}\\
 & \quad + s_{12}^5 \Big( \frac{ y^5}{561330} + \frac{  y^2 \zeta_3 }{  324 } 
  + \frac{ 19 \zeta_5}{144} + \frac{  \zeta_3^2}{12 y} 
 + \frac{ 7 \zeta_7}{ 32 y^2}  - \frac{ 3 \zeta_3 \zeta_5}{16 y^3} 
 + \frac{ 15 \zeta_9}{128 y^4}  \Big) + {\cal O}(s_{12}^6)
\nn  \,,\\
Y^\tau_{(2|0)} \, \big|_{q^0 \bar q^0} &=
s_{12} \Big(  \frac{ 4 y^3}{45} - 2 \zeta_3  \Big) + s_{12}^2 \Big( \frac{ 4 y^4}{945} - \frac{  \zeta_5}{y}  \Big) + 
 s_{12}^3 \Big( \frac{ 2 y^5}{2835} + \frac{ y^2 \zeta_3}{  9} 
 - \frac{ 5 \zeta_5}{ 3 } + \frac{  \zeta_3^2}{y}  - \frac{ 9 \zeta_7}{8 y^2}  \Big) \notag \\
 &\quad + s_{12}^4 \Big( \frac{ 2 y^6}{56133} + \frac{ 2 y^3 \zeta_3}{ 81} - \frac{  \zeta_3^2}{3} 
 - \frac{ 7 \zeta_7}{4 y}   + \frac{  9 \zeta_3 \zeta_5}{4 y^2} - \frac{  15 \zeta_9}{8 y^3}  \Big)
 + {\cal O}(s_{12}^5)\,,
  \label{eq4.4}
  \\
  %
  %
Y^\tau_{(4|2)} \, \big|_{q^0 \bar q^0} &=
-\frac{ 8 y^4}{945} + \frac{ 2 \zeta_5}{y} + 
 s_{12} \Big({-}\frac{ 8 y^5}{14175} + \frac{ 2 y^2 \zeta_3  }{45} - \frac{ 2 \zeta_3^2}{y} 
 + \frac{  45 \zeta_7}{8 y^2} \Big) \notag \\
 & \quad + 
 s_{12}^2 \Big({-} \frac{ 4 y^6}{22275} - \frac{  y \zeta_5 }{ 30 }
  + \frac{ 7 \zeta_7}{2 y}  - \frac{ 45 \zeta_3 \zeta_5}{ 4 y^2} 
+ \frac{ 135 \zeta_9}{8 y^3} \Big)
 + {\cal O}(s_{12}^3)\,,
  \label{eq4.6}
\end{align}
\end{subequations}
see appendix \ref{appB.1} for similar expressions for the Laurent polynomials of
$Y^\tau_{(0|2)} ,Y^\tau_{(4|0)}$ and $Y^\tau_{(3|5)}$.
These expressions have been consistently expanded up to total order $10$:
According to the discussion around (\ref{deforder}), two-point component 
integrals $Y^\tau_{(a|b)}$ are said to be expanded 
to the order $2k$ if the coefficients up to and including $s_{12}^{k-(a+b)/2}$ are worked out.

As one can see clearly, the Laurent polynomials (\ref{eq4.ex}) of the $Y^\tau_{\eta}$-integrals contain negative powers of $y=\pi\Im\tau$. Passing to $\Yhat^\tau_{\eta} $ via the redefinition (\ref{eq3.1}), the negative powers of $y$ disappear, and we extract the initial value already given in (\ref{eq3.7}) from the zeroth power in~$y$.


\subsection{\texorpdfstring{Component integrals in terms of $\bsv$}{Component integrals in terms of betasv}}
\label{sec4.2}

Having obtained the initial value (\ref{eq3.7}), we now need to apply the series of operators in (\ref{eq3.21}) and 
extract the coefficients of $\eta^{a-1} \bar \eta^{b-1}$ to identify the component integrals $Y^\tau_{(a|b)}$ defined in \eqref{eq2.14}. The two-point representation (\ref{eq2.33}) of the derivation algebra
is not faithful and realises fewer linear combinations of $\bsvBR{\ldots}{\ldots}{\tau}$
as compared to the $R_{\vec{\eta}}(\ep_k)$ at $(n\geq 3)$ points:
Since the operators $R_{\eta}(\ep_{k\geq 4})$ at two points are multiplicative ($\partial_\eta$
only occurs in $R_{\eta}(\ep_{0})$), all the commutators
$[R_{\eta}(\ep_{k_1}), R_{\eta}(\ep_{k_2})]$ with $k_1,k_2\geq 4$ vanish. 

Given that $[R_{\eta}(\ep_{4}), R_{\eta}(\ep_{6})]=0$, for instance, only a restricted set
of $\bsvBR{j_1& j_2}{4& 6}{\tau}$ and $\bsvBR{j_1& j_2}{6& 4}{\tau}$
can be found in (\ref{eq3.21}). In particular, $\bsvBR{2& 4}{4& 6}{\tau}$ and
$\bsvBR{4& 2}{6& 4}{\tau}$ do not show up individually but always appear in the
symmetric combination $\bsvBR{2& 4}{4& 6}{\tau}+\bsvBR{4& 2}{6& 4}{\tau}
= \bsvBR{4}{6}{\tau} \bsvBR{2}{4}{\tau}$. In order to determine all the $\bsvBR{j_1& j_2}{4& 6}{\tau}$ individually, we shall study three-point integrals and their reality properties in section~\ref{sec5}.

Applying the operators in~\eqref{eq3.21}, we extract for example the following expressions for the
simplest component integrals in terms of the initial data following from (\ref{eq4.1}) and the $\bsv$:
\vspace{-1ex}
\begin{subequations}
\label{eqYctp2pt}
\begin{align}
Y^{\tau}_{(0|0)}&=
1 + s_{12}^2 \Big( {-}3 \bsvBR{1}{4}{\tau} + \frac{ \zeta_3}{2 y}  \Big)
+s_{12}^3 \Big( {-}5 \bsvBR{2}{6}{\tau} + \frac{ \zeta_3}{6} + \frac{  \zeta_5}{8 y^2}  \Big) \notag \\
&\quad+s_{12}^4 \Big( {-}21 \bsvBR{3}{8}{\tau} + 9  \bsvBR{1& 1}{4& 4}{\tau} - 18 \bsvBR{2& 0}{4& 4}{\tau} \notag \\
&\ \ \ \ \quad+ 
 12  \zeta_3 \bsvBR{0}{4}{\tau}  - \frac{ 3 \zeta_3}{2 y}  \bsvBR{1}{4}{\tau}  - \frac{  \zeta_3^2}{8 y^2}
  + \frac{ 5 \zeta_5}{12 y} + \frac{ 3 \zeta_7}{ 32 y^3}   \Big) \notag \\
 &\quad +s_{12}^5 \Big( {-}135 \bsvBR{4}{10}{\tau} - 60 \bsvBR{3& 0}{6& 4}{\tau} + 
 15 \bsvBR{1& 2}{4& 6}{\tau} + 15 \bsvBR{2& 1}{6& 4}{\tau} - 
 60 \bsvBR{2& 1}{4& 6}{\tau}  \notag \\
 &\quad\ \ \ \ - \frac{ 1}{2} \zeta_3 \bsvBR{1}{4}{\tau}  
 + \frac{ 6  \zeta_5}{y}  \bsvBR{0}{4}{\tau} - \frac{3  \zeta_5}{8 y^2} \bsvBR{1}{4}{\tau} 
  + 40 \zeta_3 \bsvBR{1}{6}{\tau}  - \frac{5  \zeta_3}{2 y}  \bsvBR{2}{6}{\tau}  \notag \\
  &\quad\ \ \ \
 + \frac{43 \zeta_5}{360}  + \frac{ \zeta_3^2}{12 y} + \frac{7 \zeta_7}{32 y^2}  - \frac{3 \zeta_3 \zeta_5}{ 16 y^3}
 + \frac{15 \zeta_9}{128 y^4} \Big) + {\cal O}(s_{12}^6)\,,
 \label{eq4.11}  \\
Y^{\tau}_{(2|0)}&=
s_{12} ( 3 \bsvBR{2}{4}{\tau} - 2 \zeta_3 ) + s_{12}^2 \Big( 10 \bsvBR{3}{6}{\tau} - \frac{\zeta_5}{y}  \Big) \notag \\*
&\quad +s_{12}^3 \Big( 63 \bsvBR{4}{8}{\tau} - 9 \bsvBR{1& 2}{4& 4}{\tau} + 27 \bsvBR{2& 1}{4& 4}{\tau} \notag \\*
&\quad \ \ \ \ - 
 18 \zeta_3 \bsvBR{1}{4}{\tau}  + \frac{3  \zeta_3}{2 y}   \bsvBR{2}{4}{\tau} 
- \frac{5 \zeta_5}{3} + \frac{\zeta_3^2}{y}  - \frac{9 \zeta_7}{8 y^2}  \Big) \notag \\
 &\quad  +s_{12}^4 \Big( 540 \bsvBR{5}{10}{\tau} - 30 \bsvBR{1& 3}{4& 6}{\tau} + 
 165 \bsvBR{2& 2}{4& 6}{\tau} - 15 \bsvBR{2& 2}{6& 4}{\tau} 
  \notag \\*
 &\quad \ \ \ \  + 90 \bsvBR{3& 1}{6& 4}{\tau} + 60 \bsvBR{4& 0}{6& 4}{\tau} + \frac{1}{2}  \zeta_3 \bsvBR{2}{4}{\tau}
 - 24 \zeta_5 \bsvBR{0}{4}{\tau}  \notag \\*
 &\quad \ \ \ \ - \frac{9 \zeta_5}{y}  \bsvBR{1}{4}{\tau} 
 + \frac{3  \zeta_5}{8 y^2}  \bsvBR{2}{4}{\tau}
  - 110 \zeta_3 \bsvBR{2}{6}{\tau}   + \frac{5  \zeta_3}{y}  \bsvBR{3}{6}{\tau} \notag \\
  &\quad \ \ \ \   -  \frac{ \zeta_3^2}{3} - \frac{7 \zeta_7}{4 y} + \frac{9 \zeta_3 \zeta_5}{4 y^2} 
   - \frac{15 \zeta_9}{8 y^3} \Big) + {\cal O}(s_{12}^5)   \,,
 \label{eq4.12}   \\
 Y^\tau_{(0|2)} &=
s_{12} \Big(3 \bsvBR{0}{4}{\tau} -  \frac{ \zeta_3}{8 y^2} \Big) + 
 s_{12}^2 \Big(10 \bsvBR{1}{6}{\tau} -  \frac{\zeta_5}{16 y^3} \Big) \notag \\*
 &\quad  + 
 s_{12}^3 \Big(63 \bsvBR{2}{8}{\tau} - 9 \bsvBR{0& 1}{4& 4}{\tau} + 
    27 \bsvBR{1& 0}{4& 4}{\tau}  - \frac{  9  \zeta_3}{2 y} \bsvBR{0}{4}{\tau} 
      \notag \\*
    &\quad \ \ \ \ + \frac{3  \zeta_3}{ 8 y^2} \bsvBR{1}{4}{\tau}+ \frac{\zeta_3}{60} - \frac{ 5 \zeta_5}{48 y^2}
      + \frac{  \zeta_3^2}{16 y^3} - \frac{  9 \zeta_7}{128 y^4} \Big) \notag \\*
 &\quad  + s_{12}^4 \Big(540 \bsvBR{3}{10}{\tau} - 15 \bsvBR{0& 2}{4& 6}{\tau} + 
    90 \bsvBR{1& 1}{4& 6}{\tau} - 30 \bsvBR{1& 1}{6& 4}{\tau}   \notag \\*
    &\quad \ \ \ \ +    60 \bsvBR{2& 0}{4& 6}{\tau} + 165 \bsvBR{2& 0}{6& 4}{\tau} + 
    \frac{  \zeta_3 }{2} \bsvBR{0}{4}{\tau}  - 40  \zeta_3  \bsvBR{0}{6}{\tau}  \notag \\*
    &\quad \ \ \ \
    - \frac{15  \zeta_3}{y}  \bsvBR{1}{6}{\tau} + \frac{5  \zeta_3}{8 y^2} \bsvBR{2}{6}{\tau} 
     - \frac{ 33  \zeta_5}{8 y^2} \bsvBR{0}{4}{\tau} 
   + \frac{ 3  \zeta_5}{16 y^3} \bsvBR{1}{4}{\tau}  \notag \\*
   & \quad \ \ \ \ + \frac{ \zeta_5}{120 y} - \frac{ \zeta_3^2}{48 y^2} - \frac{7 \zeta_7}{64 y^3}
    + \frac{9 \zeta_3 \zeta_5}{ 64 y^4} - \frac{15 \zeta_9}{128 y^5} \Big) + {\cal O}(s_{12}^5) \, .
 \label{eq4.13}  
\end{align}
\end{subequations}
 \vspace{-1ex}
Further expansions of component integrals to order 10 can be found in appendix \ref{appB.1}.

 
\subsection{\texorpdfstring{$\bsv$ versus modular graph forms}{betasv versus modular graph forms}}
\label{sec4.2b}
 
As exemplified by (\ref{eq2.all36}), the $\ap$-expansion of component integrals $Y_{(a|b)}^{\tau}$ 
is expressible in terms of MGFs. By comparing the expansion of various component integrals in
terms of MGFs with those in terms of the $\bsv$ as derived above, we arrive at a dictionary between
the two types of objects. More specifically, the two-point component integrals $Y_{(a|b)}^{\tau}$ 
are sufficient to express all $\bsv$ at depth one and all depth-two $\bsv$ with $(k_1,k_2)=(4,4)$
in terms of MGFs. Depth-two instances with $(k_1,k_2)=(6,4)$ or $(k_1,k_2)=(4,6)$
are not individually accessible at two points as explained at the beginning of section~\ref{sec4.2} 
and will be fixed from three-point considerations in section~\ref{sec5}.
 
The resulting expressions one obtains in this way at depth one are\footnote{We will no longer
spell out the argument $\tau$ of $\bsv[\ldots]$ in (\ref{2ptbetasv}) and later equations unless the argument
is transformed. The same notation applies to ${\cal E}^{\rm sv}[\ldots]$ (which is real-analytic
like the $\bsv[\ldots]$) and the holomorphic quantities ${\cal E}[\ldots], \alpha[\ldots] ,f[\ldots]$.}
 \begin{subequations}
 \label{2ptbetasv}
   \begin{align}
\begin{array}{ll} \displaystyle &\displaystyle \Bigg. \bsvBRno{0}{6}  = -  \frac{(  \pi \overline{ \nabla} )^2 {\rm E}_3}{960 y^4} +  \frac{  \zeta_5}{640 y^4}\,,
\\
\displaystyle \bsvBRno{0}{4} = \frac{  \pi \overline{ \nabla} {\rm E}_2 }{24 y^2} +  \frac{  \zeta_3}{24 y^2} \, , 
 &\displaystyle \bsvBRno{1}{6}  =  \frac{  \pi \overline{ \nabla} {\rm E}_3 }{240 y^2} +  \frac{  \zeta_5}{160 y^3}\,,
\\
\displaystyle \bsvBRno{1}{4} = - \frac{1}{6}{\rm E}_2  +  \frac{ \zeta_3}{6 y} \, , \ \ \ \ \ \ \ \  \ \ \ \ \ \ \ \ 
 &\displaystyle \Bigg.  \bsvBRno{2}{6} = - \frac{1}{30} {\rm E}_3 + \frac{ \zeta_5}{40 y^2}\,,
 \\
\displaystyle \bsvBRno{2}{4} = \frac{2}{3} \pi \nabla {\rm E}_2 + \frac{2 \zeta_3 }{3} \, ,
 &\displaystyle   \bsvBRno{3}{6} = \frac{1}{15} \pi \nabla {\rm E}_3 + \frac{\zeta_5}{10 y}\,,
 \\
 &\displaystyle \Bigg.  \bsvBRno{4}{6} = -\frac{4}{15} ( \pi \nabla)^2 {\rm E}_3 +\frac{ 2 \zeta_5 }{5}
 \end{array}
  \label{eq4.18} 
 \end{align}
as well as
 \begin{align}
 \bsvBRno{3}{8} = -\frac{1}{140} {\rm E}_4 + \frac{ \zeta_7}{224 y^3} \, , \ \ \ \ \ \
 \bsvBRno{4}{10} = - \frac{1}{630} {\rm E}_5 + \frac{ \zeta_9}{1152 y^4} \, ,
 \label{eq4.19} 
\end{align}
and similar expressions for the remaining $\bsvBRno{j}{8} ,\bsvBRno{j}{10} $ in terms of Cauchy--Riemann derivatives of ${\rm E}_4, {\rm E}_5$ can be found in appendix \ref{appBd1}.

At depth two, we find the modular graph function ${\rm E}_{2,2}$ in (\ref{eq2.29}) and its derivatives:
\begin{align}
\bsvBRno{0&0}{4&4} &= \frac{(\pi \overline{\nabla}{\rm E}_2)^2}{1152y^4} + \frac{\zeta_3  \pi \overline{\nabla}{\rm E}_2}{576y^4} + \frac{\zeta_3^2}{1152 y^4} = \frac12  \big( \, \bsvBRno{0 }{4 } \,\big)^2 \,,  \notag\\
\bsvBRno{0&1}{4&4}&=-\frac{\pi\overline{\nabla}{\rm E}_{2, 2}}{144y^2}-\frac{{\rm E}_{2}\pi\overline{\nabla}{\rm E}_{2}}{144y^2}-\frac{\zeta_3 {\rm E}_{2}}{144y^2}+\frac{\zeta_3}{2160} -\frac{5\zeta_5}{1728y^2}+\frac{\zeta_3^2}{288y^3} \,, \notag\\
\bsvBRno{0&2}{4&4}&=\frac{{\rm E}_{2, 2}}{18}+\frac{(\pi\overline{\nabla}{\rm E}_{2} )\pi \nabla{\rm E}_{2}}{36y^2}+\frac{\zeta_3 \pi\nabla{\rm E}_{2}}{36y^2}-\frac{5\zeta_5}{216y} +\frac{\zeta_3^2}{72y^2} \,, \notag\\
\bsvBRno{1&0}{4&4}&=\frac{\pi\overline{\nabla}{\rm E}_{2, 2}}{144y^2}
+\frac{\zeta_3 \pi\overline{\nabla}{\rm E}_{2}}{144y^3}
-\frac{\zeta_3}{2160} +\frac{5\zeta_5}{1728y^2}+\frac{\zeta_3^2}{288y^3} \,, \notag\\
\bsvBRno{1&1}{4&4} &= \frac{{\rm E}_2^2}{72} - \frac{\zeta_3 {\rm E}_2}{36y}  + \frac{\zeta_3^2}{72 y^2}= \frac12 \big( \, \bsvBRno{1}{4 } \, \big)^2\,,  \label{eq4.27}\\
\bsvBRno{1&2}{4&4}&=
-\frac{\pi\nabla{\rm E}_{2, 2}}{9}
-\frac{{\rm E}_{2}\pi\nabla{\rm E}_{2}}{9}+\frac{\zeta_3 \pi\nabla{\rm E}_{2}}{9y} -\frac{5\zeta_5}{108}+\frac{\zeta_3^2}{18y}  \,,\notag \\
\bsvBRno{2&0}{4&4}&=-\frac{{\rm E}_{2, 2}}{18}+\frac{\zeta_3 \pi\overline{\nabla}{\rm E}_{2}}{36y^2}
+\frac{5\zeta_5}{216y}+\frac{\zeta_3^2}{72y^2}\,, \notag\\
\bsvBRno{2&1}{4&4}&= \frac{\pi\nabla{\rm E}_{2, 2}}{9}-\frac{\zeta_3 {\rm E}_{2}}{9} +\frac{5\zeta_5}{108}+\frac{\zeta_3^2}{18y}\,, \notag\\
\bsvBRno{2&2}{4&4}&= \frac{2 (\pi \nabla {\rm E}_2)^2 }{9}+ \frac{4\zeta_3 \pi \nabla{\rm E}_2}{9} + \frac{2\zeta_3^2}{9}= \frac12  \big( \, \bsvBRno{2 }{4 }\, \big)^2 \, .
 \notag
  \end{align}
\end{subequations}
Similar expressions arise when the associated ${\cal E}^{\rm sv}$ 
are expressed in terms of MGFs via (\ref{eq3.23}), see appendix \ref{appEsv}. From the expressions above one can verify the shuffle property~\eqref{eq3.shuff2} of the $\bsv$ in a straightforward manner, e.g.
\begin{align}
\betasv{0&2\\4&4} + \betasv{2&0\\4&4} = \betasv{0\\4} \betasv{2\\4}\,.
\end{align}

\subsubsection{\texorpdfstring{Modular graph forms in terms of $\bsv$}{Modular graph forms in terms of betasv}}

These relations can also be inverted to obtain expressions for the modular graph forms in terms of the $\bsv$. At depth one they are
 \begin{subequations}
 \label{2ptMGFBSV}
\begin{align}
&&\frac{(\pi \overline{\nabla} )^2  {\rm E}_{3} }{y^4} &=
-960 \betasv{0\\6} + \frac{3 \zeta_5}{2 y^4}\,,
\notag \\
\frac{\pi \overline{\nabla}   {\rm E}_{2} }{y^2} &=
24 \betasv{0\\4} - \frac{ \zeta_3}{y^2} \, ,
&\frac{\pi \overline{\nabla}   {\rm E}_{3} }{y^2} &=
240 \betasv{1\\6} - \frac{3 \zeta_5}{2 y^3}\,,
\notag \\
 {\rm E}_{2} &=
-6 \betasv{1\\4} +\frac{ \zeta_3}{y}\, ,
 &{\rm E}_{3} &=
-30 \betasv{2\\6} + \frac{3 \zeta_5}{4 y^2}\,,
 \label{eq4.16}   \\
\pi \nabla {\rm E}_{2} &=
\frac{3}{2} \betasv{2\\4} - \zeta_3\, ,
&\pi \nabla {\rm E}_{3} &=
15 \betasv{3\\6} - \frac{3 \zeta_5}{2 y}\,,
\notag \\
&&(\pi \nabla)^2 {\rm E}_{3} &=
-\frac{15}{4} \betasv{4\\6} + \frac{3 \zeta_5}{2}
\notag 
\end{align}
as well as
\begin{align}
 {\rm E}_{4} &=
-140 \betasv{3\\8} + \frac{5 \zeta_7}{8 y^3} \, ,
 &{\rm E}_{5} &=
-630 \betasv{4\\10} + \frac{35 \zeta_9}{64 y^4}
 \label{eq4.17} 
\end{align}
and similar expressions for the Cauchy--Riemann derivatives of
${\rm E}_4$ and ${\rm E}_5$ given in appendix~\ref{appBd1}.
Inverting the depth-two relations (\ref{eq4.27}) leads to the
shuffle-irreducible MGFs
\begin{align}
\frac{\pi \overline{\nabla}   {\rm E}_{2,2} }{y^2} &=144 \betasv{1&0\\4&4} 
- \frac{ 24  \zeta_3}{y} \betasv{0\\4} + \frac{ \zeta_3}{15} 
- \frac{5 \zeta_5}{ 12 y^2}+ \frac{ \zeta_3^2}{2 y^3} \,,
\notag \\
{\rm E}_{2,2} &=
-18 \betasv{2&0\\4&4} + 12 \zeta_3 \betasv{0\\4} 
+ \frac{5 \zeta_5}{12 y} - \frac{ \zeta_3^2}{4 y^2} \,,  \label{eq4.26} \\
 \pi \nabla {\rm E}_{2,2} &= 
9 \betasv{2&1\\4&4} - 6 \zeta_3  \betasv{1\\4} - \frac{5 \zeta_5}{12}  + \frac{ \zeta_3^2}{ 2 y}\, . \notag
\end{align}
At two points, one can still derive expressions for the modular graph function
${\rm E}_{2,3}$ in (\ref{eq2.30}) and its Cauchy--Riemann derivatives:
\begin{align}
\frac{(\pi \overline{\nabla})^2  {\rm E}_{2,3} }{y^4} &=
-3840 \betasv{0&1\\4&6} - 7680 \betasv{1&0\\4&6} - 
 11520 \betasv{1&0\\6&4} \notag \\
 &  + 
\frac{ 1280 \zeta_3}{y} \betasv{0\\6}  + \frac{160  \zeta_3}{y^2} \betasv{1\\6}
+ \frac{ 72  \zeta_5}{y^3} \betasv{0\\4}
+ \frac{8 \zeta_3}{189} - \frac{2 \zeta_5}{15 y^2} 
 + \frac{ 7 \zeta_7}{8 y^4} - \frac{3 \zeta_3 \zeta_5}{y^5} \,,\notag \\
\frac{\pi \overline{\nabla}   {\rm E}_{2,3} }{y^2} &= 960 \betasv{1&1\\4&6} + 480 \betasv{2&0\\4&6} + 
 1440 \betasv{2&0\\6&4} \notag \\
 & - 320  \zeta_3 \betasv{0\\6} - 
\frac{ 160  \zeta_3}{y} \betasv{1\\6}
- \frac{ 36  \zeta_5}{y^2}  \betasv{0\\4}
+ \frac{ \zeta_5}{15 y}  - \frac{ 7 \zeta_7}{8 y^3}  + \frac{3 \zeta_3 \zeta_5}{2 y^4} \,,\notag  \\
{\rm E}_{2,3} &=
-120 \betasv{2&1\\4&6} - 120 \betasv{3&0\\6&4} + \frac{ 12 \zeta_5}{y}  \betasv{0\\4} + 
 80  \zeta_3 \betasv{1\\6} - \frac{ \zeta_5}{36} 
+ \frac{ 7 \zeta_7}{16 y^2} - \frac{\zeta_3 \zeta_5}{2 y^3} \,,
\label{exE23}
 \\
\pi \nabla {\rm E}_{2,3} &= 
90 \betasv{2&2\\4&6} + 60 \betasv{3&1\\6&4} + 
 30 \betasv{4&0\\6&4} \notag \\
 & - 60  \zeta_3 \betasv{2\\6} - 
 12  \zeta_5 \betasv{0\\4} - \frac{6 \zeta_5}{y}   \betasv{1\\4}
 - \frac{7 \zeta_7}{8 y}+ \frac{ 3 \zeta_3 \zeta_5}{2 y^2} \,,
\notag \\
(\pi \nabla)^2 {\rm E}_{2,3} &= 
-45 \betasv{2&3\\4&6} - 15 \betasv{3&2\\6&4} - 
 30 \betasv{4&1\\6&4}  \notag \\
 &+ 30  \zeta_3 \betasv{3\\6} + 
 12 \zeta_5  \betasv{1\\4} + \frac{3  \zeta_5}{ 2 y}  \betasv{2\\4}
  + \frac{7 \zeta_7}{8} - \frac{3 \zeta_3 \zeta_5}{y} \, .
 \notag
\end{align}
\end{subequations}
However, we will need three-point input to solve for the individual $\beta^{\rm sv}$ in
terms of MGFs. In particular, we will fix the antiholomorphic integration constants 
$\overline{\alphaBRno{j_1 &j_2}{6 &4}}$ or
$\overline{\alphaBRno{j_1 &j_2}{4 &6}}$ in section \ref{sec5}.

\subsubsection{Closed formulae at depth one}

As detailed in appendix \ref{closedform}, one can compute the $(s_{12} \rightarrow 0)$-limit
of the component integrals $Y^\tau_{(a|b)}$ with $a{+}b \geq 4$ in closed form.
By comparing the leading order of $Y^\tau_{(k|k)}$ resulting from (\ref{eq3.21}) with
the lattice-sum representations (\ref{eq2.24}) of non-holomorphic Eisenstein series, one obtains
\beq
{\rm E}_k = \frac{ (2k{-}1)! }{ [(k{-}1)!]^2} \bigg\{ 
{-} \betasv{ k-1\\ 2k} + \frac{ 2 \zeta_{2k-1} }{(2k{-}1) (4y)^{k-1} }
\bigg\} \, .
\label{eq:eksol.2}
\eeq
Similarly, the lattice-sum representations (\ref{eq2.26})
of their Cauchy--Riemann derivatives arise at the
$s_{12}^0$ order of $Y^\tau_{(a|b)}$ with $a\neq b$,
and comparison with (\ref{eq3.21}) implies ($0\leq m \leq k{-}1$)
\begin{subequations}
\label{eq:bsv0}
\begin{align}
(\pi \nabla)^m {\rm E}_k &= \Big( {-}\frac{1}{4} \Big)^{m} \frac{ (2k{-}1)! }{(k{-}1)! (k{-}1{-}m)!} 
\bigg\{
{-} \betasv{ k-1+m\\ 2k} + \frac{ 2 \zeta_{2k-1} }{(2k{-}1) (4y)^{k-1-m} }
\bigg\}\,,
\\
\frac{(\pi \overline{\nabla})^m {\rm E}_k}{y^{2m}} &=  \frac{({-}4)^{m} (2k{-}1)! }{(k{-}1)! (k{-}1{-}m)!} 
\bigg\{
{-} \betasv{ k-1-m\\ 2k} + \frac{ 2 \zeta_{2k-1} }{(2k{-}1) (4y)^{k-1+m} }
\bigg\} \, .
\end{align}
\end{subequations}
By solving these relations for the $\bsv$, one arrives at
\begin{align}
\betasv{ k-1 \\2k} = -\frac{[(k{-}1)!]^2}{(2k{-}1)!} {\rm E}_k + \frac{2\zeta_{2k-1}}{(2k{-}1) (4y)^{k-1}}
\label{eq:eksol.1}
\end{align}
as well as ($0\leq m \leq k{-}1$)
\begin{subequations}
\label{eq:bsv1}
\begin{align}
\betasv{ k-1+m \\ 2k} &=  - \frac{(-4)^m (k{-}1)!\,(k{-}1{-}m)! \, (\pi \nabla)^m {\rm E}_k}{(2k{-}1)!} + \frac{2\zeta_{2k-1}}{(2k{-}1) (4y)^{k-1-m}}\,,\\
\betasv{ k-1-m \\ 2k} &=  - \frac{ (k{-}1)!\,(k{-}1{-}m)! \, (\pi \overline\nabla)^m {\rm E}_k}{(-4)^m(2k-1)! y^{2m}} + \frac{2\zeta_{2k-1}}{(2k{-}1) (4y)^{k-1+m}}\,.
\end{align}
\end{subequations}
%

 
\subsection{Simplifying modular graph forms}
\label{sec4.2banana}

By the linear-independence result on iterated Eisenstein integrals \cite{Nilsnewarticle}, the
$\bsv$ are suitable for obtaining relations between MGFs which are hard to see from their lattice-sum representation.
In the following, we will illustrate this with the relation
\begin{align}
  {\rm D}_3 &=\EE_3 + \zeta_3\label{eq4.11da}
\end{align}
due to Zagier, where ${\rm D}_{3}$ is the two-loop instance of the `banana' graph functions, the coefficients in the $\alpha'$-expansion of the component integral $Y^{\tau}_{(0|0)}$, defined by \cite{Green:2008uj, DHoker:2015gmr}
\begin{align}
Y^{\tau}_{(0|0)} = \sum_{n=0}^{\infty} \frac{1}{n!} (s_{12})^n {\rm D}_n(\tau) \, , \ \ \ \  
{\rm D}_n(\tau) = \int \frac{ \dd^2 z}{\Im \tau} \big( G(z,\tau) \big)^n
=  \Big( \frac{ \Im \tau}{\pi} \Big)^n
{\cal C}[ \underbrace{\begin{smallmatrix} 1 &1 &\ldots &1 \\1&1&\ldots &1 \end{smallmatrix} }_{n}](\tau)\,.
 \label{eq4.11a} 
\end{align} 
The lattice sums ${\cal C}[\ldots]$ are defined in (\ref{eq2.22}) and immediately pinpoint the simplest non-trivial
banana graph function ${\rm D}_2 = \EE_2$. (One has ${\rm D}_0=1$ and ${\rm D}_1=0$.)

The identity \eqref{eq4.11da} was first proven by explicitly performing one of the sums in ${\rm D}_{3}$. To prove \eqref{eq4.11da} independently using the $\bsv$, we have to identify both MGFs in the relation as coefficients in the $\ap$-expansion of component integrals $Y^{\tau}_{(a|b)}$ which we can write in terms of $\bsv$ using \eqref{eq3.21}. Hence, \eqref{eq4.11da} follows from comparing
\begin{align}
  {\rm D}_{3}&=6Y^{\tau}_{(0|0)}\Big|_{s_{12}^{3}}=-30\bsvBRno{2}{6}+\zeta_{3}+\frac{3\zeta_{5}}{4y^{2}}\\
  \EE_{3}&=Y^{\tau}_{(3|3)}\Big|_{s_{12}^{0}}=-30\bsvBRno{2}{6}+\frac{3\zeta_{5}}{4y^{2}}\,.
\end{align}
Higher-loop generalisations of \eqref{eq4.11da} are known from MGF techniques \cite{DHoker:2015gmr, DHoker:2015sve, DHoker:2016mwo, DHoker:2016quv}, 
\begin{subequations}
  \label{banana5a}
  \begin{align}
    {\rm D}_4 &= 24\EE_{2,2} +3 \EE_2^2 +\frac{18}{5} \EE_4  \label{eq4.11ba} \\
    {\rm D}_5 &=  60 \EE_{2,3} +10 \EE_3\EE_2 + \frac{180}{7} \EE_5 + 10\zeta_3  \EE_2 +16\zeta_5\,,
    \label{eq4.11za}
  \end{align}
\end{subequations}
see also \cite{DHoker:2019xef, Zagier:2019eus} for all-order results on the Laurent polynomials of
banana graph functions. Relations among MGFs like (\ref{banana5a}) can be proven in the same way as \eqref{eq4.11da}: They become manifest once all MGFs in the relation are identified as $\alpha'$ coefficients of component integrals and expressed in terms of $\bsv$ via \eqref{eq3.21}. This will in fact expose all the relations among
MGFs since the $\bsvBRno{j_1 &j_2 &\ldots &j_\ell}{k_1 &k_2 &\ldots &k_\ell}$ with different entries $j_i,k_i$ are linearly independent.

Of course, the reach of this procedure depends on the multiplicity of the $Y^\tau_{\vec{\eta}}$-integrals
under consideration. For instance, $\EE_{2,2}$ and $\EE_{2,3}$ in \eqref{banana5a} contain the two-loop graphs $\cform{1&1&2\\1&1&2}$ and $\cform{1&1&3\\1&1&3}$ which do not appear in any two-point component integral $Y^{\tau}_{(a|b)}$.\footnote{The $\alpha'$-expansion of two-point component integral $Y^{\tau}_{(a|b)}$ only involves the lattice
  sums ${\cal C}[\begin{smallmatrix}  a &0 \\ 0 &b\end{smallmatrix} \begin{smallmatrix} 1 &1 &\ldots &1 \\1&1&\ldots &1 \end{smallmatrix}]$ \cite{Gerken:2019cxz}.}  Instead, the
$\cform{1&1&k\\1&1&k}$ first appear as the coefficient of $s_{23}^2$ in the three-point component integral
$Y^\tau_{(k,0|k,0)}(2,3|3,2)$ discussed in section~\ref{sec5}.

As a reference, we express the lowest-loop banana graphs ${\rm D}_{n}$ in terms of $\bsv$, by comparing \eqref{eq4.11a} with \eqref{eq4.11}, yielding
\begin{subequations}
\label{banana5}
\begin{align}
{\rm D}_2 &= -6  \bsvBRno{1}{4}  + \frac{ \zeta_3}{y} \,,\label{eq4.11c}
\\
  {\rm D}_3&= -30   \bsvBRno{2}{6}  +  \zeta_3 + \frac{ 3 \zeta_5}{4 y^2}\,,\label{eq4.11d}\\
 {\rm D}_4&= 216 \bsvBRno{1& 1}{4& 4} - 
 432 \bsvBRno{2& 0}{4& 4}  -504 \bsvBRno{3}{8}  \label{eq4.11b} \\
 &\quad + 288 \zeta_3 \bsvBRno{0}{4} 
   - \frac{ 36  \zeta_3 }{y} \bsvBRno{1}{4} 
   + \frac{ 10  \zeta_5}{y} - \frac{ 3  \zeta_3^2}{y^2} 
    + \frac{ 9  \zeta_7}{4 y^3}\,,\notag  \\
 {\rm D}_5 &=  1800 \bsvBRno{1& 2}{4& 6} - 
 7200 \bsvBRno{2& 1}{4& 6} + 1800 \bsvBRno{2& 1}{6& 4} - 
 7200 \bsvBRno{3& 0}{6& 4} \notag \\
 &\quad -16200 \bsvBRno{4}{10}  - 60  \zeta_3 \bsvBRno{1}{4}  + 
 4800  \zeta_3 \bsvBRno{1}{6}  
  - \frac{ 300  \zeta_3 }{y} \bsvBRno{2}{6}  
\label{eq4.11e}\\
  &\quad + \frac{720  \zeta_5 }{y}  \bsvBRno{0}{4} 
  - \frac{ 45  \zeta_5 }{y^2} \bsvBRno{1}{4} 
    + \frac{ 43  \zeta_5}{3}  + \frac{10  \zeta_3^2}{y} 
   + \frac{105  \zeta_7}{4 y^2} - \frac{45  \zeta_3  \zeta_5}{2 y^3}
     + \frac{225  \zeta_9}{16 y^4}\,,\notag 
     \end{align}
     \end{subequations}
see appendix \ref{appB} for similar $\beta^{\rm sv}$-representations of $ {\rm D}_6$ and $ {\rm D}_7 $. As expected, these expressions satisfy the relations \eqref{banana5a} if we plug in the $\bsv$ representations \eqref{2ptMGFBSV} of the modular graph functions on the right-hand sides.


\subsection{\texorpdfstring{Explicit $\beta^{\rm sv}$ from reality properties at two points}{Explicit betasv from reality properties at two points}}
\label{sec4.3}

In this section, we derive the antiholomorphic integration constants in 
certain instances of ${\cal E}^{\rm sv}$ in (\ref{eq3.16}) and (\ref{eq3.17})
from reality properties (\ref{eq2.18}) of two-point component integrals.
This will make the iterated-Eisenstein-integral representation of the associated 
$\bsv$ and MGFs fully explicit.
 

\subsubsection{Depth one}
\label{sec4.3.1}

From reality of $y= \pi \Im \tau$ and $Y^{\tau}_{(0|0)}$, the orders $s_{12}^2$
and $s_{12}^3$ of its $\ap$-expansion (\ref{eq4.11}) immediately imply that $ \bsvBRno{1}{4}$
and  $\bsvBRno{2}{6}$ are real. Similarly, from the instance
$Y^{\tau}_{(2|0)}= 16y^2 \overline{Y^{\tau}_{(0|2)}}$ of (\ref{eq2.18}), 
the $s_{12}$ and $s_{12}^2$ orders of (\ref{eq4.12}) and (\ref{eq4.13})
imply that
\beq
\overline{ \bsvBRno{2}{4} } = (4y)^{2} \bsvBRno{0}{4} \, , \ \ \ \ \ \ 
\overline{ \bsvBRno{3}{6} } = (4y)^{2} \bsvBRno{1}{6} \, .
\eeq
By combining (\ref{eq2.18}) with the $s_{ij}^0$-order of general
$Y^{\tau}_{(a|b)}$ with $a+b \geq 4$ derived in appendix \ref{closedform},
one arrives at the closed depth-one formula
\beq
\overline{ \bsvBRno{j}{k} } = (4y)^{2+2j-k} \bsvBRno{k-2-j}{k}  \, .
 \label{eq4.14} 
\eeq
By (\ref{eq3.23}), this also determines the complex conjugation properties of ${\cal E}^{\rm sv}$
\beq
\overline{ \EsvBRno{j}{k} } =  (-1)^{j}  \sum_{p=0}^j \binom{j}{p}
(4y)^p \EsvBRno{j-p}{k}  \, .
 \label{eq4.15} 
\eeq
This is crucial extra information beyond the initial-value problem (\ref{eq3.init}): The latter
only determines $\EsvBRno{j_1 &j_2&\ldots &j_\ell}{k_1 &k_2 &\ldots &k_\ell} $ up to 
antiholomorphic integration constants, denoted
by $\overline{\fBRno{j_1 &j_2&\ldots &j_\ell}{k_1 &k_2 &\ldots &k_\ell}}$ in \eqref{eq:2}, that vanish at the cusp.
The complex-conjugation property (\ref{eq4.15}) in turn relates these integration constants
to the holomorphic ingredients ${\cal E}^{\rm sv}_{\rm min}$ that are fixed by their differential equation and
can be read off from its minimal solution (\ref{eq3.15}). At $k=4$, for instance, (\ref{eq4.15}) reads
\begin{align}
\overline{ \EsvBRno{0}{4}  } &=  \EsvBRno{0}{4}  \,,\notag \\
\overline{ \EsvBRno{1}{4}  } &= -4y \EsvBRno{0}{4}  - \EsvBRno{1}{4} \label{more.08}  \,,\\
\overline{ \EsvBRno{2}{4}  } &= 16 y^2 \EsvBRno{0}{4} + 8 y \EsvBRno{1}{4} + \EsvBRno{2}{4} \notag
\end{align}
and selects the antiholomorphic completion in 
$\EsvBRno{j}{4}= \EsvBRminno{j}{4}+\overline{\fBRno{j}{4}}$: By inserting the expansion~\eqref{eq3.15} of the minimal $\mathcal{E}^{\text{sv}}_{\text{min}}$ in terms of \textit{holomorphic} iterated integrals (\ref{eq3.11})
\begin{align}
\EsvBRminno{0}{4} &= \EBRno{0}{4} \,, \notag \\
\EsvBRminno{1}{4} &= \EBRno{1}{4}  - 2\pi i \bar \tau \EBRno{0}{4} \,, \label{more.09} \\
\EsvBRminno{2}{4} &= \EBRno{2}{4} - 4\pi i \bar \tau \EBRno{1}{4} +(2\pi i \bar \tau)^2 \EBRno{0}{4} \notag 
\end{align}
into (\ref{more.08}) and isolating the purely antiholomorphic terms, one is uniquely led to
\begin{align}
\overline{\fBRno{0}{4}}&= \overline{\EBRno{0}{4}}\,,  \notag \\
\overline{\fBRno{1}{4}}&= {-} \overline{\EBRno{1}{4}}  - 2\pi i \bar \tau \overline{\EBRno{0}{4}}  \,,\label{more.XX} \\
\overline{\fBRno{2}{4}}&= \overline{\EBRno{2}{4}} + 4\pi i \bar \tau \overline{ \EBRno{1}{4}} 
+(2\pi i \bar \tau)^2 \overline{\EBRno{0}{4}} \, . \notag 
\end{align}
This reasoning results in the expressions (\ref{more.01}) for $\EsvBRno{j}{4} $ and 
can be straightforwardly repeated 
at $k \geq 6$: The reality properties (\ref{eq4.15}) completely fix the 
$\overline{\EBRno{j-r}{k}}$ in (\ref{eq3.16}) and uniquely determine 
$\EsvBRno{j}{k} $ in terms of iterated Eisenstein integrals and their complex conjugates.

By combining the expression (\ref{eq3.16}) for $\EsvBRno{j}{k}$ with the
dictionaries (\ref{eq3.22}) and (\ref{2ptMGFBSV}) to $\bsvBRno{j}{k}$ and MGFs,
both ${\rm E}_k$ and their Cauchy--Riemann derivatives can then be reduced to
holomorphic iterated Eisenstein integrals and their complex conjugates, e.g.
\begin{align} 
\pi \nabla {\rm E}_2 &=  \frac{3}{2}  \EsvBRno{2}{4}  - \zeta_3
\label{eq4.71} 
 \\
&= - 12 \pi^2 \bar \tau^2 \Re \EBRno{0}{4}+ 12 \pi \bar \tau \Im \EBRno{1}{4}  + 3  \Re \EBRno{2}{4} -
\zeta_3\,,
\notag\\
{\rm E}_2 &= -6  \EsvBRno{1}{4} - \frac{3  \EsvBRno{2}{4}}{2 y} + \frac{ \zeta_3}{y} 
\label{eq4.72}  \\
&= \frac{   
 12 \pi^2 \tau \bar \tau   \Re \EBRno{0}{4}  -6  \pi ( \tau {+} \bar \tau )  \Im \EBRno{1}{4} - 3   \Re \EBRno{2}{4} + \zeta_3 }{y}\,,
\notag\\
\pi \overline \nabla {\rm E}_2 &= 
24 y^2  \EsvBRno{0}{4} + 12 y  \EsvBRno{1}{4} + \frac{3}{2}  \EsvBRno{2}{4} - \zeta_3   
\label{eq4.73} \\
 &= - 12 \pi^2 \tau^2   \Re \EBRno{0}{4}  +12  \pi \tau  \Im \EBRno{1}{4} + 3   \Re \EBRno{2}{4} - \zeta_3\, .
 \notag
\end{align}
At depth one, these iterated-Eisenstein-integral representations of ${\rm E}_k$ are well-known
\cite{Ganglzagier, DHoker:2015wxz} and serve as a cross-check for the expansion methods of this 
work. At higher depth, however, 
only a small number of MGFs has been expressed in terms of iterated Eisenstein 
integrals \cite{Brown:2017qwo, Broedel:2018izr}, and we will
later provide new representations for non-holomorphic imaginary cusp forms. Most importantly, the
reality properties of component integrals determine the integration constants in higher-depth
${\cal E}^{\rm sv}$ and $\beta^{\rm sv}$ without referring to the MGFs in the $\ap$-expansion.


\subsubsection{Depth two}
\label{sec4.4}

Based on the $\ap$-expansions (\ref{eqYctp2pt}) of two-point component integrals,
the $s_{12}^4$-order of $Y^{\tau}_{(0|0)}=  \overline{Y^{\tau}_{(0|0)}}$
and the $s_{12}^3$-order $Y^{\tau}_{(2|0)}= 16y^2 \overline{Y^{\tau}_{(0|2)}}$
imply
\begin{subequations}
 \label{eq4.g4g4} 
\begin{align}
\overline {  \bsvBRno{1& 1}{4& 4} }&= \bsvBRno{1& 1}{4& 4} \, , \ \ \ \ \ \
\overline {  \bsvBRno{0& 0}{4& 4} }=  \frac{  \bsvBRno{2& 2}{4& 4}}{256 y^4}\,,
 \label{eq4.20}  \\
\overline {  \bsvBRno{2& 0}{4& 4} }&=
 \bsvBRno{2& 0}{4& 4} - \frac{ 2 \zeta_3}{3} \bsvBRno{0}{4}  + \frac{  \zeta_3}{24 y^2} \bsvBRno{2}{4}\,,
 \label{eq4.21}   \\
\overline {  \bsvBRno{0& 2}{4&4} } &=
\bsvBRno{0& 2}{4& 4} + \frac{ 2  \zeta_3}{3} \bsvBRno{0}{4}
 - \frac{  \zeta_3 }{24 y^2} \bsvBRno{2}{4} \,,
  \label{eq4.21a} \\ 
\overline {  \bsvBRno{1& 0}{4& 4} }&= \frac{  \bsvBRno{2& 1}{4& 4}}{16 y^2} 
 - \frac{   \zeta_3}{24 y^2} \bsvBRno{1}{4} + \frac{  \zeta_3}{ 96 y^3} \bsvBRno{2}{4} - \frac{ \zeta_3}{2160}\,,
 \label{eq4.22}  \\
\overline {  \bsvBRno{0& 1} {4& 4}} &=
 \frac{\bsvBRno{1& 2}{4&4} }{16 y^2}  + \frac{ \zeta_3}{24 y^2} \bsvBRno{1}{4} 
  - \frac{  \zeta_3}{  96 y^3} \bsvBRno{2}{4} +  \frac{\zeta_3}{2160}  \, . \label{eq4.22a}
   \end{align}
   \end{subequations}
These simplest depth-two examples illustrate that $ \overline{ \bsvBRno{j_1& j_2}{k_1& k_2} }$ 
introduce admixtures of single-valued MZVs and $\bsv$ of lower depth. There is no depth-one
analogue of this feature in the expression (\ref{eq4.14}) for $ \overline{ \bsvBRno{j}{k} }$.
In section \ref{sec5.8}, three-point $\ap$-expansions will be used to extract
similar complex-conjugation properties for all the individual 
$  \bsvBRno{j_1& j_2}{4& 6} $ and $  \bsvBRno{j_1& j_2}{6& 4} $. Our examples
will line up with the conjectural closed depth-two formula
\beq
\overline{ \bsvBRno{j_1 &j_2}{k_1 &k_2} } = (4y)^{4+2j_1+2j_2-k_1-k_2} \bsvBRno{k_2-2-j_2 &k_1-2-j_1}{k_2 &k_1} \ \te{mod depth $<$ 2} 
 \label{eq4.23}
\eeq
which translates as follows to the ${\cal E}^{\rm sv}$
\beq
\overline{ \EsvBRno{j_1 &j_2}{k_1 &k_2} } =  (-1)^{j_1+j_2}  \sum_{p_1=0}^{j_1}
\sum_{p_2=0}^{j_2} \binom{j_1}{p_1}\binom{j_2}{p_2}
(4y)^{p_1+p_2} \EsvBRno{j_2-p_2 &j_1-p_1}{k_2 &k_1} \ \te{mod depth $<$ 2} \, .
 \label{eq4.51}
\eeq
These complex-conjugation properties of $\Esv$ are consistent with the general depth-two expression~\eqref{eq3.17}, assuming the conjecture that the iterated Eisenstein integrals
in $\overline{\alpha[\cdots]}$ have depth one and zero.

The $\zeta_3$-admixtures in (\ref{eq4.g4g4}) propagate to the following 
shuffle-inequivalent $\overline{ \EsvBRno{j_1 &j_2}{4 &4} } $,
\begin{align}
\overline{ \EsvBRno{1 &0}{4 &4} }  &=  -4 y \EsvBRno{0& 0}{4& 4} - \EsvBRno{0& 1}{4& 4}\,,
\notag\\
\overline{ \EsvBRno{2 &0}{4 &4} }  &=  16 y^2 \EsvBRno{0& 0}{4& 4} + 8 y \EsvBRno{0& 1}{4& 4} + 
 \EsvBRno{0& 2}{4& 4} - \frac{ y \zeta_3 }{270} +  \frac{ 2\zeta_3}{3} \EsvBRno{0}{4}  \,,
 \label{eq4.52} \\
\overline{ \EsvBRno{2 &1}{4 &4} }  &= -64 y^3 \EsvBRno{0& 0}{4& 4} - 32 y^2 \EsvBRno{0& 1}{4& 4} - 
 4 y \EsvBRno{0& 2}{4& 4} - 16 y^2 \EsvBRno{1& 0}{4& 4} \notag \\
 & \quad\quad- 8 y \EsvBRno{1& 1}{4& 4} - \EsvBRno{1& 2}{4& 4} + \frac{ y^2 \zeta_3 }{135} - 
 \frac{ 8y \zeta_3}{3}  \EsvBRno{0}{4}  -  \frac{ 2 \zeta_3}{3} \EsvBRno{1}{4} \, . \notag
\end{align}
These equations uniquely fix all the integration constants $\overline{\alphaBRno{j_1& j_2}{4& 4} }$: 
One has to first express the ${\cal E}^{\rm sv}$ in terms of holomorphic iterated Eisenstein
integrals ${\cal E}$ and their complex conjugates via (\ref{eq3.16}) and (\ref{eq3.17}). Then 
by comparing the purely holomorphic terms 
$\sim   \tau, {\cal E}, \alphaBRno{j_1& j_2}{4& 4} $ on 
the two sides of (\ref{eq4.52}), one can read off
\begin{align}
 \alphaBRno{1& 0}{4& 4} &=  \alphaBRno{0& 1}{4& 4} = 0  \,,\notag \\
  \alphaBRno{2& 0}{4& 4} &=  \frac{2 \zeta_3}{3}  \Big(   \EBRno{0}{4} 
 + \frac{ i \pi   \tau}{360} \Big) 
 = -  \alphaBRno{0& 2}{4& 4}  \,, \label{eq4.24} \\
  \alphaBRno{2& 1}{4& 4}  &=  \frac{2 \zeta_3}{3}  \Big(  2 \pi i \tau   \EBRno{0}{4}  -    \EBRno{1}{4} - \frac{ \pi^2  \tau^2}{360} \Big)
= -    \alphaBRno{1& 2}{4& 4} 
 \, . \notag
\end{align}
The expressions for $\overline{\alphaBRno{j_1& j_2}{4& 4} }$ that enter the
actual $ \EsvBRno{j_1&j_2}{4 &4}$ follow from complex conjugation, and we have used the shuffle 
relations (\ref{eq3.shuff1}) to infer the $\alphaBRno{j_1& j_2}{4& 4} $ with $j_1<j_2$. Moreover,
as exhibited in appendix \ref{appC}, the $\overline{ \alphaBRno{j_1& j_2}{4& 4} }$
in (\ref{eq4.24}) are invariant under
the modular $T: \tau \rightarrow \tau +1$ transformation,
in line with the discussion in section \ref{sec3.B}.


\section{Explicit forms at three points}
\label{sec5}

An analysis similar to the one of section~\ref{sec4} can be done at three points. 
Unlike formula~\eqref{eq4.1} we do not have a closed expression for the all-order 
Laurent polynomial at three points to obtain the initial data directly. 
For this reason, we first discuss a basis of modular graph forms that we use for expanding the component integrals $Y^\tau_{\vec\eta}$. From this expansion and the knowledge of the Laurent polynomials of the modular graph forms as given in~\cite{DHoker:2015gmr} we can construct the initial data $\Yhat^{i\infty}_{\vec\eta}$ and solve for the remaining $\bsv$ at depth two having $(k_1,k_2)=(4,6)$ or $(k_1,k_2)=(6,4)$. 
The expansion of the initial data to order 10 is available in machine-readable form in an 
ancillary file within the arXiv submission of this paper.
As a consistency check of our procedure the instances of $\bsv$ that were determined from the two-point analysis in section~\ref{sec4} are consequences of the three-point considerations.

The detailed discussion of $\betasv{j_1&j_2\\4&6},\, \betasv{j_1&j_2\\6&4} $ and the associated MGFs 
in this section is motivated as follows: The depth-two integrals $\betasv{j_1&j_2\\4&4}$ have been
described in terms of real MGFs ${\rm E}_2$ and ${\rm E}_{2,2}$, see (\ref{eq4.27}) and (\ref{eq4.26}),
and their reality is a particularity of having the same Eisenstein series ${\rm G}_4$
in both integration kernels. Generic $\betasv{j_1&j_2\\k_1&k_2}$ with $k_1\neq k_2$, by contrast, 
introduce complex MGFs. The irreducible MGFs besides products of depth-one quantities can be 
organised in terms of real MGFs such as ${\rm E}_{2,3}$ and imaginary MGFs such as the non-holomorphic 
cusp forms (\ref{eqcusp}) which have been discussed first in \cite{DHoker:2019txf}.
Hence, the $\betasv{j_1&j_2\\4&6},\, \betasv{j_1&j_2\\6&4} $ in this
section are the simplest non-trivial window into the generic properties of depth-two MGFs.


\subsection{Bases of modular graph forms up to order 10}
\label{sec5.1}

At two points, the expansion of any component integral $Y^{\tau}_{(a|b)}$ to order 10
is entirely expressible in terms of the modular graph functions ${\rm E}_{k\leq 5}, {\rm E}_{2,2}, {\rm E}_{2,3}$
as well as their Cauchy--Riemann derivatives, cf.\ (\ref{eqYctp2pt}) and appendix \ref{appBd1}. At
three points, this is no longer the case: The $\ap$-expansion of various component 
integrals (\ref{eq2.15}) introduces additional MGFs that are not expressible in terms
of the real quantities ${\rm E}_{k\leq 5}, {\rm E}_{2,2}$ and ${\rm E}_{2,3}$. This resonates with the
comments in early section \ref{sec4.2} that the operators $R_{\eta}(\ep_k)$ in the
two-point differential equations obey relations that no longer hold for their three-point analogues
$R_{\eta_2,\eta_3}(\ep_k)$ in (\ref{eq2.35}).

The additional MGFs that start appearing at three points can be understood from the perspective of
lattice sums. 
Expanding three-point component integrals $Y^{\tau}_{(a_2,a_3|b_2,b_3)}$ 
to order 10 introduces a large variety of dihedral and trihedral MGFs\footnote{See appendix
\ref{secA.2} for trihedral MGFs.} whose
modular weight adds up to $\leq 10$. In the notation $\cform{a_1&a_2&\ldots &a_R \protect\\b_1&b_2&\ldots &b_R}$ 
for the dihedral case in (\ref{eq2.22}), this amounts to holomorphic and antiholomorphic
modular weights $w=\sum_{j=1}^R a_j$ and $\bar w= \sum_{j=1}^R b_j$ 
subject to $w+\bar w\leq 10$.

Already the two-loop graphs 
$\cform{a_1&a_2&a_3 \protect\\b_1&b_2&b_3}$ with $w+\bar w= 10$
were found in \cite{DHoker:2019txf} to introduce irreducible cusp forms 
$\aform{a_1&a_2&a_3 \protect\\b_1&b_2&b_3}$
as defined in (\ref{eqcusp}) with vanishing Laurent polynomials. 
The known types of relations among dihedral and trihedral MGFs
\cite{DHoker:2016mwo, DHoker:2016quv, Gerken:2018zcy} -- see \cite{package} for a Mathematica
implementation and data mine -- leave three independent
cusp forms built from $\cform{a_1&a_2&\ldots &a_R \protect\\b_1&b_2&\ldots &b_R}$
with $w = \bar w =5$. One of them is expressible as the antisymmetrised
product
\beq
\frac{ (\nabla\mathrm{E}_{2})\overline{\nabla}\mathrm{E}_{3} -(\overline{\nabla}\mathrm{E}_{2})\nabla\mathrm{E}_{3} }{( \Im \tau )^2} = 6 \Big( \frac{ \Im \tau}{\pi} \Big)^5 \Big\{ 
 \cform{3&0 \protect\\1 &0}  \cform{2&0 \protect\\ 4&0} 
 -  \cform{1&0 \protect\\3 &0}  \cform{4&0 \protect\\2 &0}
 \Big\} \, ,
 \label{redcusp}
\eeq
and we additionally have two irreducible cusp forms that can be taken to be $ \aform{0&2&3\\3&0&2}$
and $\aform{0&1&2&2\\1&1&0&3}$. While (\ref{redcusp}) and $ \aform{0&2&3\\3&0&2}$ have been
discussed in \cite{DHoker:2019txf}, the cusp form $\aform{0&1&2&2\\1&1&0&3}$ exceeds the loop orders
studied in the reference.

For lattice sums $\cform{a_1&a_2&\ldots &a_R \protect\\b_1&b_2&\ldots &b_R}$ 
with different holomorphic and antiholomorphic modular weights
$w \neq \bar w$, one can construct basis elements
from Cauchy--Riemann derivatives of modular invariants. As detailed in  
table~\ref{tab:Ebasis}, the bases\footnote{We are not counting combinations of MGFs that evaluate
to MZVs or products involving MZVs or holomorphic Eisenstein series in the table. This is why we did 
not include $\zeta_3=( \frac{ \Im \tau}{\pi})^3 ( \cform{1&1&1 \protect\\1&1&1}- \cform{3 &0 \protect\\3 &0} )$ 
at $(w,\bar w )=(3,3)$
or any of $\zeta_5 , \, \zeta_3 {\rm E}_2, \, {\rm G}_4 \overline \nabla^2 {\rm E}_3$ 
at $(w,\bar w )=(5,5)$.} for $w +\bar w \leq 8$
can be assembled from Cauchy--Riemann derivatives of ${\rm E}_{k\leq 4}$
and ${\rm E}_{2,2}$ (including products of ${\rm E}_{2},\, \nabla {\rm E}_{2}$ 
and $\overline\nabla {\rm E}_{2}$). For $w+ \bar w = 10$ in turn,
one needs to adjoin combinations of ${\rm E}_{5},\, {\rm E}_{2,3},\, \aform{0&2&3\\3&0&2} ,\, \aform{0&1&2&2\\1&1&0&3}$ and their Cauchy--Riemann derivatives
to obtain complete lattice-sum bases.

\begin{table}
  \begin{center}
    {\setstretch{1.0}
      \begin{tabular}{cc}
      \toprule
      weight $(w,\bar{w})$ & basis elements\\
      \midrule
      (2,2) & $\mathrm{E}_{2}$\\
      (3,1) & $\nabla \mathrm{E}_{2}$\\
      \midrule
      (3,3) & $\mathrm{E}_{3}$\\
      (4,2) & $\nabla \mathrm{E}_{3}$ \\
      (5,1) & $\nabla^{2} \mathrm{E}_{3}$ \\
      \midrule
      (4,4) & $\mathrm{E}_{4},\ \mathrm{E}_{2,2},\ \mathrm{E}_{2}^{2},
      \ (\Im\tau)^{-2} \nabla \mathrm{E}_{2}\overline{\nabla} \mathrm{E}_{2}$\\
      (5,3) & $\nabla \mathrm{E}_{4},\ \nabla \mathrm{E}_{2,2},
      \ \mathrm{E}_{2}\nabla \mathrm{E}_{2}$ \\
      (6,2) & $\nabla^{2} \mathrm{E}_{4},\ (\nabla \mathrm{E}_{2})^{2}$ \\
      (7,1) & $\nabla^{3}\mathrm{E}_{4}$\\
      \midrule
      (5,5) & $\mathrm{E}_{5},\ \mathrm{E}_{2,3},\ \mathrm{E}_{2}\mathrm{E}_{3},
      \ (\Im\tau)^{-2} \overline{\nabla}\mathrm{E}_{2}\nabla \mathrm{E}_{3},\ (\Im\tau)^{-2} \nabla\mathrm{E}_{2}\overline{\nabla} \mathrm{E}_{3},\
   \BB_{2,3},\ \BB'_{2,3}$ \\   
      (6,4) & $\nabla\mathrm{E}_{5},\ \nabla \mathrm{E}_{2,3},\ \mathrm{E}_{3}\nabla\mathrm{E}_{2},
      \ \mathrm{E}_{2}\nabla\mathrm{E}_{3},\ (\Im\tau)^{-2} \overline{\nabla}\mathrm{E}_{2}\nabla^{2}\mathrm{E}_{3},
\ \nabla\BB_{2,3},\ \nabla \BB'_{2,3}$\\
      (7,3) & $\nabla^{2}\mathrm{E}_{5},\ \nabla^{2}\mathrm{E}_{2,3},
      \ \nabla \mathrm{E}_{2}\nabla \mathrm{E}_{3},\ \mathrm{E}_{2} \nabla^{2}\mathrm{E}_{3},
\ \nabla^2 \BB'_{2,3}$\\
      (8,2) & $\nabla^{3}\mathrm{E}_{5},\ \nabla\mathrm{E}_{2}\nabla^{2}\mathrm{E}_{3},
\ \nabla^3 \BB'_{2,3}$\\
      (9,1) & $\nabla^{4}\mathrm{E}_{5}$\\
      \bottomrule
    \end{tabular}
  }
  \caption{Basis of MGFs of modular weight
    $w+\bar{w}\leq10$ in terms of $\nabla^{k}\mathrm{E}_{\dots}$ and imaginary cusp forms and their derivatives. There is a similar basis with $\bar{w}>w$ where $\nabla$ is replaced by $\overline\nabla$ and $\BB_{\dots}$ by $\overline{\BB}_{\dots}$. 
The modular weights $(w,\bar w)$ refer to the lattice sums ${\cal C}[\ldots]$ after 
stripping off overall suitable factors of $(\Im \tau)^k$.
For instance, ${\rm E}_k$ is counted as $(w,\bar w)=(k,k)$ from the lattice-sum contribution
 $\cform{k&0 \protect\\k &0}= \sum_{p\neq 0} |p|^{-2k}$ to (\ref{eq2.24}). Additional factors of $(\Im\tau)^{-2}$ have been added explicitly in the table whenever there is an $\overline\nabla$ to ensure that all entries in one row have the same modular properties. }
    \label{tab:Ebasis}
  \end{center}
\end{table}

In order to obtain simple expressions for the full range of
$\betasv{j_1&j_2\\4&6}$ and $\betasv{j_1&j_2\\6&4} $ in terms of lattice 
sums, it is convenient to delay the appearance of holomorphic Eisenstein series in the 
Cauchy--Riemann equations. This can be achieved by taking the combinations
\begin{subequations}
\label{bforms}
\begin{align}
  \BB_{2,3}&=\left(\frac{\Im\tau}{\pi}\right)^5\aform{0&1&2&2\\1&1&0&3}
  +\frac{(\nabla\mathrm{E}_{2})\overline{\nabla}\mathrm{E}_{3} -(\overline{\nabla}\mathrm{E}_{2})\nabla\mathrm{E}_{3}}{6(\Im \tau)^2} \,,\\
\BB'_{2,3}&=\BB_{2,3}+\frac{1}{2}\left(\frac{\Im\tau}{\pi}\right)^5\aform{0&2&3\\3&0&2}- \frac{21}{4}\mathrm{E}_{2,3}- \frac{1}{2} \zeta_{3}\mathrm{E}_{2}
\end{align}
\end{subequations}
as basis elements for the lattice sums with $w = \bar w = 5$, where the rescaling by $\Im\tau/\pi$
is analogous to (\ref{eq2.24}) or (\ref{eq2.com}) and renders $\BB_{2,3} , \BB_{2,3}'$ modular invariant.
The lowest-order Cauchy--Riemann derivatives that contain holomorphic Eisenstein series are
\begin{subequations}
\label{G4G6s}
\begin{align}
(\pi \nabla)^2 {\rm B}_{2,3} &=
+ \frac{3}{2} (\pi \nabla)^2 {\rm E}_{2,3}+ \frac{2}{7} (\pi \nabla)^2 {\rm B}'_{2,3}
- \frac{3}{2} {\rm E}_2  (\pi \nabla)^2 {\rm E}_3\notag \\
&\quad+ \frac{3}{2} (\pi \nabla {\rm E}_2) (\pi \nabla {\rm E}_3)
+  (\Im \tau)^4  {\rm G}_4  (9 {\rm E}_3 + 3 \zeta_3)  \,,
\\
(\pi \nabla)^4 {\rm B}'_{2,3} &= 1260(\Im \tau)^6  {\rm G}_6 \pi \nabla {\rm E}_2  \, .
\end{align}
\end{subequations}
While the modular graph form $\BB_{2,3}$ is also an imaginary cusp form, the second form $\BB'_{2,3}$ is neither real nor a cusp form, and its Laurent polynomial is determined by the known Laurent polynomials (\ref{eqcusp.mzv}) of 
$\EE_{2,3}$ and $\EE_2$,  
\beq
 \BB_{2,3} = {\cal O}(q,\bar q) \, , \ \ \ \ \ \ 
  \BB'_{2,3}= \frac{y^5}{14175} - \frac{y^2 \zeta_3}{45 }
  + \frac{ 7 \zeta_5}{240} 
  - \frac{ \zeta_3^2}{2 y} 
     - \frac{147 \zeta_7}{64 y^2}
  + \frac{21 \zeta_3 \zeta_5}{8 y^3}
 + {\cal O}(q,\bar q)\, .
\eeq
Accordingly, the complex-conjugation properties are
\beq
\overline{\BB}_{2,3}= - \BB_{2,3} \, , \ \ \ \ \ \ \overline{\BB}'_{2,3}= - \BB'_{2,3} - \frac{21}{2}\mathrm{E}_{2,3}- 
 \zeta_{3}\mathrm{E}_{2} \, .
\eeq
The imaginary cusp forms with $w=\bar w=5$ studied in~\cite{DHoker:2019txf} were denoted by
$\mathcal{A}_{1,2;5}$ and $\mathcal{A}_{1,4;5}$ there and can be rewritten in our basis 
as follows\footnote{Note that the normalisation conventions of \cite{DHoker:2019txf} for
$\cform{a_1&a_2&\ldots &a_R \protect\\b_1&b_2&\ldots &b_R}$ and $ \aform{a_1&a_2&\ldots &a_R \protect\\b_1&b_2&\ldots &b_R}$ differ from ours in (\ref{eq2.22}) and (\ref{eqcusp}) by an additional factor of 
$( \frac{ \Im \tau }{\pi} )^{\frac{1}{2} \sum_{j=1}^R(a_j+b_j)}$.} 
\begin{subequations}
\label{DHKcusp}
\begin{align}
\mathcal{A}_{1,2;5} &= \frac13  \left(\frac{\Im\tau}{\pi}\right)^5 \aform{0&2&3\\3&0&2} = \frac23 \left(\BB'_{2,3} -\BB_{2,3} + \frac{21}4 \EE_{2,3} +\frac{\zeta_3}2 \EE_2\right)\,,\\
\mathcal{A}_{1,4;5} &= \frac{(\nabla\mathrm{E}_{2})\overline{\nabla}\mathrm{E}_{3} -(\overline{\nabla}\mathrm{E}_{2})\nabla\mathrm{E}_{3}}{6(\Im \tau)^2} \,.
\end{align}
\end{subequations}
The extra cusp form $\aform{0&1&2&2\\1&1&0&3}$ entering the definition of $\BB'_{2,3}$ did not arise in~\cite{DHoker:2019txf} as its lattice-sum representation requires three-loop graphs on the worldsheet.

The reason for defining the particular combinations~\eqref{bforms} is that their derivatives 
$\nabla \BB_{2,3}$, $\nabla \BB'_{2,3}$, $\nabla^2 \BB'_{2,3}$ and $\nabla^3 \BB'_{2,3}$ 
do not contain any explicit holomorphic Eisenstein series. The collection of MGFs in table~\ref{tab:Ebasis}
forms a basis for MGFs of total modular weight $w+\bar w =10$, excluding factors of ${\rm G}_k$
or MZVs. First, the techniques in the literature have been used to decompose all dihedral and trihedral MGFs
with $w+\bar w =10$ in the basis of the table \cite{package}. 
Second, the counting of basis elements matches the number of $\bsv$ that can enter the
$\ap$-expansion of $Y^\tau_{\vec{\eta}}$ at the relevant order. As will be detailed in the following sections, see in particular (\ref{excusp.2}) and (\ref{excusp.4}) to (\ref{excusp.6}) or (\ref{bsvwt10}), 
the correspondence between lattice-sum and iterated-integral bases is
\begin{align}
\begin{array}{c} \mathrm{E}_{2,3},\ \mathrm{E}_{2}\mathrm{E}_{3},
      \ (\Im \tau)^{-2} \overline{\nabla}\mathrm{E}_{2}\nabla \mathrm{E}_{3} \\
     (\Im \tau)^{-2}  \nabla\mathrm{E}_{2}\overline{\nabla} \mathrm{E}_{3},\
   \BB_{2,3},\ \BB'_{2,3} \end{array} \bigg\}
   &\leftrightarrow \bigg\{ \begin{array}{c} \betasv{0 &3\\4&6} , \ \betasv{1 &2\\4&6}, \ \betasv{2 &1\\4&6} \\
   \betasv{3 &0\\6&4} , \ \betasv{2 &1\\6&4}, \ \betasv{1 &2\\6&4} \end{array}
 \notag  \\
  \begin{array}{c}  \nabla \mathrm{E}_{2,3},\ \mathrm{E}_{3}\nabla\mathrm{E}_{2},
      \ \mathrm{E}_{2}\nabla\mathrm{E}_{3} \\
     (\Im \tau)^{-2}  \overline{\nabla}\mathrm{E}_{2}\nabla^{2}\mathrm{E}_{3},
\ \nabla\BB_{2,3},\ \nabla \BB'_{2,3} \end{array} \bigg\}
&\leftrightarrow \bigg\{ \begin{array}{c} \betasv{0 &4\\4&6} , \ \betasv{1 &3\\4&6}, \ \betasv{2 &2\\4&6} \\
   \betasv{4 &0\\6&4} , \ \betasv{3 &1\\6&4}, \ \betasv{2 &2\\6&4} \end{array}
\label{corresp} \\
 \nabla^{2}\mathrm{E}_{2,3},   \ \nabla \mathrm{E}_{2}\nabla \mathrm{E}_{3},\ \mathrm{E}_{2} \nabla^{2}\mathrm{E}_{3},\ \nabla^2 \BB'_{2,3}
 &\leftrightarrow \betasv{1 &4\\4&6} , \ \betasv{2 &3\\4&6}, \
   \betasv{4 &1\\6&4} , \ \betasv{3 &2\\6&4}
 \notag   \\
  \nabla\mathrm{E}_{2}\nabla^{2}\mathrm{E}_{3},\ \nabla^3 \BB'_{2,3}
  &\leftrightarrow \betasv{2 &4\\4&6} , \   \betasv{4 &2\\6&4} \notag \, ,
  \end{align}
where the powers of $\Im \tau$ were inserted to harmonise the modular weights.
Similarly, we have $\nabla^m {\rm E}_5 \leftrightarrow \betasv{4+m\\10}, \ (\Im \tau)^{-2m}\overline{\nabla}^m {\rm E}_5 \leftrightarrow \betasv{4-m\\10}$ with $m\leq 4$ according to (\ref{eq4.19app}).
All the $\bsv$ in (\ref{corresp}) are understood to carry admixtures of lower depth analogous
to the terms involving $\zeta_k$ in (\ref{2ptMGFBSV}). Generalisations of~\eqref{corresp} to higher weight will be discussed in section~\ref{sec6.2}.


\subsection{Three-point component integrals and cusp forms}
\label{sec5.9}

Based on the generating series (\ref{eq3.21}), we have expanded all three-point component 
integrals $Y^{\tau}_{(a_2,a_3|b_2,b_3)}(\sigma|\rho)$ to order 10. Similar to the two-point case, 
the leading orders of the simplest cases $Y^{\tau}_{(0,0|0,0)}(\sigma|\rho)$ or
$Y^{\tau}_{(1,0|1,0)}(\sigma|\rho), Y^{\tau}_{(1,0|0,1)}(\sigma|\rho)$ are still expressible
in terms of ${\rm E}_{k}$ and ${\rm E}_{p,q}$. The $\ap$-expansion of $Y^{\tau}_{(0,0|0,0)}(\sigma|\rho)$ involves
trihedral modular graph functions ${\rm D}_{a,b,c}(\tau)$ that are discussed
in appendix \ref{secA.2}. They serve as consistency checks of the expansion method
(\ref{eq3.21}) and as another showcase of how the $\bsv$ expose the relations among MGFs.

The simplest instances of cusp forms or the alternative basis elements in (\ref{bforms})
occur in the $\ap$-expansion of the following component integrals:
\vspace{-1.5ex}
\begin{subequations}
\label{excusp}
\begin{align}
Y_{(2,0|0,2)}^\tau(2,3|2,3) &- Y_{(0,2|2,0)}^\tau(2,3|2,3) =
s_{13} (  s_{23} - s_{12}) (s_{12} + s_{13} + s_{23}) \left(\frac{\Im\tau}{\pi}\right)^5\aform{0&1&2&2\\1&1&0&3} \notag \\*
&\ \ \  + (  s_{23} - s_{12}) (2 s_{13}^2 + s_{12} s_{23})  \frac{  (\nabla\mathrm{E}_{2})\overline{\nabla}\mathrm{E}_{3} -(\overline{\nabla}\mathrm{E}_{2})\nabla\mathrm{E}_{3} }{12 (\Im\tau)^2} + {\cal O}(s_{ij}^4)  \notag \\*
&=
(s_{12} {-} s_{23}) (2 s_{12} s_{13} {-} s_{12} s_{23} {+} 2 s_{13} s_{23}) 
 \frac{ (\nabla\mathrm{E}_{2})\overline{\nabla}\mathrm{E}_{3} -(\overline{\nabla}\mathrm{E}_{2})\nabla\mathrm{E}_{3} }{12 (\Im \tau)^2} \notag \\*
&\ \ \ + s_{13} ( s_{23} {-} s_{12} ) (s_{12} {+} s_{13} {+} s_{23}) {\rm B}_{2,3} + {\cal O}(s_{ij}^4)\,, \\
 Y_{(2,1|3,0)}^\tau(2,3|2,3) &- Y_{(3,0|2,1)}^\tau(2,3|2,3) = 
  s_{12} s_{13}  \frac{  (\nabla\mathrm{E}_{2})\overline{\nabla}\mathrm{E}_{3} -(\overline{\nabla}\mathrm{E}_{2})\nabla\mathrm{E}_{3} }{4 (\Im\tau)^2} 
  \notag \\*
 &\ \ \  - s_{13} ( s_{13} {+} s_{23} ) \left(\frac{\Im\tau}{\pi}\right)^5\aform{0&1&2&2\\1&1&0&3}+ \frac{2 }{3} s_{12} s_{13}  \left(\frac{\Im\tau}{\pi}\right)^5\aform{0&2&3\\3&0&2}   + {\cal O}(s_{ij}^3)
 \notag \\
 &=s_{12} s_{13} \Big( \frac{4 {\rm B}_{2,3}'}{3} + 7 {\rm E}_{2,3} 
 + \frac{2}{3} \zeta_3 {\rm E}_2 \Big) - \frac{1}{3} s_{13} (4 s_{12} {+} 3 s_{13} {+} 3 s_{23}) {\rm B}_{2,3}
\notag \\*
&\ \ \  + s_{13} (3 s_{12} {+} 2 s_{13} {+} 2 s_{23}) \frac{ (\nabla\mathrm{E}_{2})\overline{\nabla}\mathrm{E}_{3} {-}(\overline{\nabla}\mathrm{E}_{2})\nabla\mathrm{E}_{3}  }{12 (\Im\tau)^2}  + {\cal O}(s_{ij}^3) \,.
\end{align}
\end{subequations}
The expressions in (\ref{excusp}) have been obtained by integrating over
$z_2,z_3$ in Fourier space as reviewed in appendix \ref{secA.1} and simplifying
the lattice sums via known MGF techniques. By matching these results with
the $\ap$-expansions due to (\ref{eq3.21}),
\begin{subequations}
\label{excusp.1}
\begin{align}
& Y_{(2,1|3,0)}^\tau(2,3|2,3) - Y_{(3,0|2,1)}^\tau(2,3|2,3)\, \Big|_{s_{13}^2} = 
-60 \betasv{0& 3\\4& 6} + 270 \betasv{1& 2\\4& 6} + 
 60 \betasv{1& 2\\6& 4} \notag \\
 &\ \ \ \ - 390 \betasv{2& 1\\4& 6} - 
 270 \betasv{2& 1\\6& 4} + 390 \betasv{3& 0\\6& 4} + 
 3 \zeta_3 \betasv{1\\4}   + 260  \zeta_3 \betasv{1\\6} 
  -  \frac{45  \zeta_3}{y} \betasv{2\\6}  \notag \\
  &\ \ \ \  + \frac{ 5  \zeta_3  }{ 2 y^2} \betasv{3\\6}
   -  \frac{39  \zeta_5 }{y} \betasv{0\\4} 
    + \frac{ 27  \zeta_5  }{4 y^2} \betasv{1\\4} 
     - \frac{ 3  \zeta_5  }{8 y^3} \betasv{2\\4} 
      + \frac{13  \zeta_5  }{120 }
 \\
 & Y_{(2,1|3,0)}^\tau(2,3|2,3) - Y_{(3,0|2,1)}^\tau(2,3|2,3)\, \Big|_{s_{12}s_{13}} = 
-90 \betasv{0& 3\\4& 6} + 360 \betasv{1& 2\\4& 6} + 
 90 \betasv{1& 2\\6& 4}  \notag \\
 &\ \ \ \ + 330 \betasv{2& 1\\4& 6} - 
 360 \betasv{2& 1\\6& 4} - 330 \betasv{3& 0\\6& 4} - 
 220  \zeta_3 \betasv{1\\6}  - \frac{60  \zeta_3}{y} \betasv{2\\6} 
  + \frac{ 15  \zeta_3 }{4y^2} \betasv{3\\6}  \notag \\
  &\ \ \ \  
   + \frac{ 33   \zeta_5}{y} \betasv{0\\4}  + \frac{ 9   \zeta_5 }{y^2} \betasv{1\\4}  
   - \frac{ 9  \zeta_5}{ 16 y^3}  \betasv{2\\4}  - \frac{ \zeta_5 }{90}
\end{align}
\end{subequations}
one can extract the following $\bsv$-representation of ${\rm B}_{2,3}$ and ${\rm B}'_{2,3}$
\begin{subequations}
\label{excusp.2}
\begin{align}
{\rm B}_{2,3}  &=
 450 \betasv{2&1\\4&6}- 450 \betasv{3&0\\6&4}
 +270 \betasv{2&1\\6&4} -270 \betasv{1&2\\4&6}    \label{eq:7}\\*
 & \quad\quad - 
 3 \zeta_3\betasv{1\\4} - 300 \zeta_3 \betasv{1\\6}
 + \frac{ 45\zeta_3 \betasv{2\\6} }{y} 
 + \frac{ 45 \zeta_5 \betasv{0\\4} }{y} - \frac{27\zeta_5 \betasv{1\\4}}{ 4 y^2}
 - \frac{13 \zeta_5}{120} \,,
\notag \\
{\rm B}'_{2,3} &= 1260 \betasv{2&1\\4&6} - 840 \zeta_3\betasv{1\\6} 
+ \frac{7 \zeta_5}{240}  - \frac{ \zeta_3^2}{2 y} 
- \frac{147 \zeta_7}{64 y^2} + \frac{21 \zeta_3 \zeta_5}{ 8 y^3}  \,.
\end{align}
\end{subequations}
Similarly, (\ref{bforms}) implies $\beta^{\rm sv}$-representations of the cusp forms
\begin{subequations}
\label{excusp.3}
\begin{align}
\left(\frac{\Im\tau}{\pi}\right)^5\aform{0&1&2&2\\1&1&0&3} &= 
60 \betasv{0& 3\\4& 6}  -  60 \betasv{1& 2\\6& 4}   + 
 270 \betasv{2& 1\\6& 4} - 270 \betasv{1& 2\\4& 6} + 390 \betasv{2& 1\\4& 6} \notag \\
& \quad - 390 \betasv{3& 0\\6& 4} - 
 3  \zeta_3 \betasv{1\\4}   - 260  \zeta_3 \betasv{1\\6} 
  + \frac{45  \zeta_3}{y} \betasv{2\\6}   - \frac{ 5  \zeta_3 }{2y^2}  \betasv{3\\6} 
   \notag \\
   &\quad + \frac{39  \zeta_5}{y} \betasv{0\\4} 
   - \frac{ 27  \zeta_5 }{4y^2} \betasv{1\\4} 
    + \frac{ 3  \zeta_5 }{8y^3} \betasv{2\\4} 
    - \frac{13  \zeta_5 }{120}\,,
 \\
 \left(\frac{\Im\tau}{\pi}\right)^5\aform{0&2&3\\3&0&2} &=
 540 \betasv{1& 2\\4& 6} -  540 \betasv{2& 1\\6& 4} + 360 \betasv{2& 1\\4& 6}  - 360 \betasv{3& 0\\6& 4} - 
 240  \zeta_3 \betasv{1\\6}  \notag \\
 &\quad - \frac{90  \zeta_3 }{y} \betasv{2\\6} 
   + \frac{36  \zeta_5}{y} \betasv{0\\4} 
   + \frac{ 27  \zeta_5}{2y^2} \betasv{1\\4}  - \frac{ \zeta_5}{60}\, ,
\end{align}
\end{subequations}
where the vanishing of their Laurent polynomials can be crosschecked
through the asymptotics (\ref{more.bsv}) of the $\beta^{\rm sv}$.
Once we have fixed the antiholomorphic integration constants of the
$\betasv{j_1& j_2\\4& 6}$ and $\betasv{j_1& j_2\\6& 4}$ in section \ref{sec5.8},
one can extract the $q$-expansions of the MGFs from their new representations
(\ref{excusp.2}) and (\ref{excusp.3}), see appendix \ref{appC}.


\subsection{\texorpdfstring{Cauchy--Riemann derivatives of cusp forms and $\beta^{\rm sv}$}{Cauchy--Riemann derivatives of cusp forms and betasv}}
\label{sec5.3}

The above procedure to relate the new basis elements ${\rm B}_{2,3}$ and 
${\rm B}'_{2,3}$ to cusp forms can be repeated based on component
integrals $Y_{(a_2,a_3|b_2,b_3)}^\tau(\sigma|\rho) $ of non-vanishing
modular weight $(0,b_2{+}b_3{-}a_2{-}a_3)$. Their expansion  in terms of $\bsv$ 
to order 10 is available in an 
ancillary file within the arXiv submission of this paper.
On top of (\ref{excusp.2}), we find
\begin{subequations}
 \label{excusp.4}
\begin{align}
\pi \nabla {\rm B}_{2,3}  &= 135 \betasv{1&3\\4&6} - 270 \betasv{2&2\\4&6} - 
 \frac{135}{2} \betasv{2&2\\6&4} + 90 \betasv{3&1\\6&4} + 
 \frac{225}{2} \betasv{4&0\\6&4} + \frac{3 \zeta_3}{4} \betasv{2\\4} \nn\\
 &\quad  + 
 180 \zeta_3  \betasv{2\\6} - \frac{45  \zeta_3}{ 2 y}  \betasv{3\\6}
 - 45 \zeta_5 \betasv{0\\4}  - \frac{ 9 \zeta_5}{y}  \betasv{1\\4}
 + \frac{27  \zeta_5}{ 16 y^2}  \betasv{2\\4}\,,\\
\frac{\pi \overline{\nabla} \,\overline{ {\rm B}}_{2,3} }{y^2} &=  1440 \betasv{1&1\\4&6}-1080 \betasv{0&2\\4&6} + 
 2160 \betasv{1&1\\6&4} + 1800 \betasv{2&0\\4&6} - 
 4320 \betasv{2&0\\6&4} \notag \\
 & \quad  - 12 \zeta_3 \betasv{0\\4}   -  1200 \zeta_3 \betasv{0\\6}  - \frac{ 240 \zeta_3}{y}   \betasv{1\\6}
 + \frac{ 45  \zeta_3}{y^2}  \betasv{2\\6}
 + \frac{ 108  \zeta_5}{y^2} \betasv{0\\4} \nn\\
 &\quad - \frac{27  \zeta_5}{ 2 y^3} \betasv{1\\4}
 - \frac{ \zeta_5}{4 y} \,,
\end{align}
\end{subequations}
as well as
\begin{subequations}
 \label{excusp.5}
\begin{align}
\frac{(\pi \overline{\nabla})^3  \overline{ {\rm B}}'_{2,3} }{y^6} &= 
-483840 \betasv{0&0\\6&4} + \frac{756 \zeta_5}{y^4}  \betasv{0\\4}
 - \frac{8 \zeta_3}{15 y} - \frac{7 \zeta_5}{ 5 y^3} 
 - \frac{ 63 \zeta_3 \zeta_5}{4 y^6}\,,
 \\
\frac{ (\pi \overline{\nabla})^2  \overline{ {\rm B}}'_{2,3} }{y^4} &= 
120960 \betasv{1&0\\6&4}
- \frac{756  \zeta_5}{y^3} \betasv{0\\4}
 - \frac{2 \zeta_3}{15} + \frac{7 \zeta_5}{ 5 y^2} 
 - \frac{147 \zeta_7}{32 y^4} + \frac{ 63 \zeta_3 \zeta_5}{4 y^5}\,,
 \\
\frac{\pi \overline{\nabla} \,\overline{ {\rm B}}'_{2,3} }{y^2} &= 
-15120 \betasv{2&0\\6&4} - 24  \zeta_3 \betasv{0\\4}
 + \frac{ 378  \zeta_5}{y^2}\betasv{0\\4}
- \frac{7 \zeta_5}{10 y} + \frac{ \zeta_3^2}{2 y^2} + \frac{147 \zeta_7}{32 y^3} - \frac{63 \zeta_3 \zeta_5}{ 8 y^4} 
\end{align}
\end{subequations}
and
\begin{subequations}
\label{excusp.6}
\begin{align}
\pi \nabla {\rm B}'_{2,3} &=
-945 \betasv{2&2\\4&6} + 630  \zeta_3 \betasv{2\\6} + 
 \frac{\zeta_3^2}{2}  + \frac{147 \zeta_7}{32 y} - \frac{63 \zeta_3 \zeta_5}{8 y^2}\,,
\\
(\pi \nabla)^2 {\rm B}'_{2,3} &=
\frac{945}{2} \betasv{2&3\\4&6} - 315  \zeta_3 \betasv{3\\6}
 - \frac{147 \zeta_7}{32} + \frac{ 63 \zeta_3 \zeta_5}{4 y}\,,
 \\
(\pi \nabla)^3 {\rm B}'_{2,3} &=
- \frac{945}{8} \betasv{2&4\\4&6} + \frac{ 315  \zeta_3}{4} \betasv{4\\6} - 
\frac{ 63}{4} \zeta_3 \zeta_5\, .
\end{align}
\end{subequations}
Higher derivatives in turn involve holomorphic Eisenstein series, see (\ref{G4G6s}).
These relations can be inverted to express all $\betasv{j_1&j_2\\k_1&k_2}$ with $k_1+k_2=10$ in terms of MGFs. The full expressions are given in appendix~\ref{appbetasv}.


\subsection{\texorpdfstring{Explicit $\beta^{\rm sv}$ from reality properties at three points}{Explicit betasv from reality properties at three points}}
\label{sec5.8}

We shall now outline the computation of the antiholomorphic integration 
constants $\overline{\alphaBRno{j_1& j_2}{6& 4} }$ that
enter the key quantities $\betasv{j_1&j_2\\6&4}$ of this section via 
(\ref{eq3.22}) and (\ref{eq3.17}). Similar to the steps in section
\ref{sec4.3}, we first determine the complex conjugate
$\betasv{j_1&j_2\\6&4}$ from the reality properties 
$ Y^\tau_{(a_2,a_3|b_2, b_3)}(\sigma|\rho)  = 
 (4y)^{a_2+a_3 -b_2-b_3} 
 \overline{ Y^\tau_{(b_2, b_3| a_2,a_3)} (\rho|\sigma) }$ of the component integrals,
\begin{subequations}
\label{ccG6G4}
\begin{align}
\overline{ \betasv{0& 0\\6& 4} } &= 
  \frac{\betasv{2& 4\\4& 6}}{4096 y^6}
  -  \frac{  \zeta_3}{6144 y^6} \betasv{4\\6}
   +  \frac{ \zeta_5}{10240 y^6} \betasv{2\\4}
  -  \frac{ \zeta_3}{907200 y}
  -  \frac{\zeta_5}{345600 y^3}\,,
\\
  \overline{ \betasv{1& 0\\6& 4} } &= 
 \frac{ \betasv{2& 3\\4& 6}}{256 y^4}
  -  \frac{  \zeta_3}{384 y^4}  \betasv{3\\6}
   +  \frac{ \zeta_5}{2560 y^5} \betasv{2\\4}
  +  \frac{ \zeta_3}{907200}
  -  \frac{ \zeta_5}{86400 y^2}\,,
  \\
\overline{ \betasv{0& 1\\6& 4} } &= 
 \frac{ \betasv{1& 4\\4& 6}}{256 y^4}
 -  \frac{  \zeta_3}{1536 y^5} \betasv{4\\6}
   +  \frac{ \zeta_5}{640 y^4} \betasv{1\\4} 
  -  \frac{ \zeta_3}{226800 }
  +  \frac{ \zeta_5}{172800 y^2}\,,
  \\
\overline{ \betasv{2& 0\\6& 4} } &= 
 \frac{ \betasv{2& 2\\4& 6}}{16 y^2 }
 -  \frac{ \zeta_3}{ 10080 y^2 } \betasv{2\\4} 
 -  \frac{ \zeta_3}{24 y^2} \betasv{2\\6}
  +  \frac{ \zeta_5}{640 y^4} \betasv{2\\4}
   -  \frac{ \zeta_5}{  21600 y}\,,
  \\
\overline{ \betasv{1& 1\\6& 4} } &= 
 \frac{ \betasv{1& 3\\4& 6}}{16 y^2}
 +  \frac{ \zeta_3}{  6720 y^2} \betasv{2\\4}
  -  \frac{ \zeta_3}{96 y^3}  \betasv{3\\6}
  +  \frac{ \zeta_5}{160 y^3} \betasv{1\\4}
  +  \frac{ \zeta_5}{43200 y}\,,
  \\
\overline{ \betasv{0& 2\\6& 4} } &= 
 \frac{ \betasv{0& 4\\4& 6}}{16 y^2}
  -  \frac{ \zeta_3}{ 1680 y^2} \betasv{2\\4}
   -  \frac{ \zeta_3}{384 y^4} \betasv{4\\6}
    +  \frac{ \zeta_5}{40 y^2} \betasv{0\\4}
     -  \frac{ \zeta_5}{  21600 y}\,,
  \\ 
 \overline{ \betasv{3& 0\\6& 4} } &= 
 \betasv{2& 1\\4& 6} -  \frac{ \zeta_3}{210 } \betasv{1\\4} 
 -  \frac{ 2 \zeta_3}{3} \betasv{1\\6}  
 +  \frac{ \zeta_5}{160 y^3}  \betasv{2\\4}
   -  \frac{\zeta_5}{5400}\,,
  \\
\overline{ \betasv{2& 1\\6& 4} } &= 
 \betasv{1& 2\\4& 6} +  \frac{\zeta_3 }{315} \betasv{1\\4}
  -  \frac{ \zeta_3}{6 y} \betasv{2\\6}  
   +  \frac{ \zeta_5}{40 y^2} \betasv{1\\4}
  + \frac{ \zeta_5}{10800}\,,
  \\
\overline{ \betasv{1& 2\\6& 4} } &= 
 \betasv{0& 3\\4& 6} 
 -  \frac{ \zeta_3}{210 } \betasv{1\\4}  
 -  \frac{  \zeta_3}{24 y^2}  \betasv{3\\6}
  +  \frac{  \zeta_5}{10 y} \betasv{0\\4}
   -  \frac{ \zeta_5}{5400}\,,
  \\
\overline{ \betasv{4& 0\\6& 4} } &= 
 16 y^2 \betasv{2& 0\\4& 6} - 
  \frac{ 16 \zeta_3 y^2 }{105 } \betasv{0\\4}  - 
 \frac{  32 \zeta_3 y^2 }{3 } \betasv{0\\6}  
  +  \frac{ \zeta_5}{40 y^2} \betasv{2\\4}
  -  \frac{y \zeta_5}{1350 }\,,
  \\
\overline{ \betasv{3& 1\\6& 4} } &= 
 16 y^2 \betasv{1& 1\\4& 6} + 
  \frac{ 4 \zeta_3 y^2 }{105 } \betasv{0\\4}  - 
  \frac{ 8 \zeta_3 y }{3  }\betasv{1\\6}   
  +  \frac{ \zeta_5}{10 y} \betasv{1\\4}
  +  \frac{y \zeta_5}{2700}\,,
  \\
\overline{ \betasv{2& 2\\6& 4} } &= 
 16 y^2 \betasv{0& 2\\4& 6} - 
  \frac{ 8 \zeta_3 y^2}{315 } \betasv{0\\4}  - 
   \frac{2 \zeta_3}{3 }\betasv{2\\6}   
  + \frac{ 2  \zeta_5}{5 }\betasv{0\\4} 
  -  \frac{ \zeta_5 y}{1350}\,,
  \\
 \overline{ \betasv{4& 1\\6& 4} } &= 
 256 y^4 \betasv{1& 0\\4& 6}
   -  \frac{ 128 \zeta_3 y^3}{3 } \betasv{0\\6} 
    +  \frac{  2 \zeta_5}{5 } \betasv{1\\4} 
    -  \frac{16 y^4 \zeta_3}{14175}
     +  \frac{ y^2 \zeta_5 }{675 }\,,
  \\
\overline{ \betasv{3& 2\\6& 4} } &= 
 256 y^4 \betasv{0& 1\\4& 6}
   -   \frac{32 \zeta_3 y^2  }{3} \betasv{1\\6}  + 
  \frac{ 8 \zeta_5 y }{5} \betasv{0\\4}  
  +  \frac{4 y^4 \zeta_3}{14175}
   -  \frac{ 2  \zeta_5 y^2  }{675}\,,
  \\
\overline{ \betasv{4& 2\\6& 4} } &= 
 4096 y^6 \betasv{0& 0\\4& 6}  - 
   \frac{ 512 \zeta_3 y^4 }{3 } \betasv{0\\6}  
   +  \frac{ 32 \zeta_5 y^2}{5 } \betasv{0\\4} 
   -  \frac{64 y^5 \zeta_3}{14175}
   -  \frac{8 y^3 \zeta_5 }{675}\, .
\end{align}
\end{subequations}
We emphasise that this reasoning does not rely on any MGF representation
and can be applied at higher orders $k_1+k_2\geq 12$, where a basis of lattice sums may not be
explicitly available. These results line up with the closed depth-two 
formula (\ref{eq4.23}) modulo admixtures of lower depth and determine
$\overline{\betasv{j_1&j_2\\4&6}}$ via shuffle relations and (\ref{eq4.14}).

In close analogy with (\ref{eq4.52}), one can now solve (\ref{ccG6G4}) for the
$\overline{\EsvBRno{j_1&j_2}{6 &4}}$ and introduce the desired integration
constants via (\ref{eq3.17}). By comparing the purely holomorphic terms,
we arrive at
\begin{subequations}
\label{G4G6alpha}
\begin{align}
\alphaBRno{0& 0}{6& 4}  &=  
\alphaBRno{1& 0}{6& 4}  =
\alphaBRno{0& 1}{6& 4}  = 0 \,,
\\
\alphaBRno{2& 0}{6& 4}  &= 
- \frac{  i  \pi \tau \zeta_3}{226800} - \frac{ \zeta_3}{630 } \EBRno{0}{4} \,,
\\
 \alphaBRno{1& 1}{6& 4}  &=
 \frac{i  \pi \tau \zeta_3}{151200 } + \frac{ \zeta_3}{420 } \EBRno{0}{4} \,,
 \\
\alphaBRno{0& 2}{6& 4}  &=
\frac{i  \pi \tau \zeta_3}{56700 } - \frac{ \zeta_3}{105 } \EBRno{0}{4}  - 
\frac{ 2 \zeta_3}{3 } \EBRno{0}{6} \,,
 \\
\alphaBRno{3& 0}{6& 4}  &=
\frac{ \pi^2 \tau^2 \zeta_3}{75600 } - 
 \frac{ i  \pi \tau \zeta_3 }{105  } \EBRno{0}{4}  + \frac{\zeta_3}{210} \EBRno{1}{4} \,,
 \\
 \alphaBRno{2& 1}{6& 4}  &=
 - \frac{ \pi^2 \tau^2 \zeta_3}{113400} + 
 \frac{2 i  \pi \tau \zeta_3 }{315 } \EBRno{0}{4}  - \frac{ \zeta_3}{315 } \EBRno{1}{4} \,,
 \\
\alphaBRno{1& 2}{6& 4}  &= 
- \frac{ \pi^2 \tau^2 \zeta_3}{32400} - 
 \frac{ i  \pi \tau \zeta_3 }{105 } \EBRno{0}{4}  + 
\frac{ \zeta_3}{210 } \EBRno{1}{4} 
 - \frac{ 4 i  \pi \tau \zeta_3 }{3 } \EBRno{0}{6} 
  + \frac{ 2 \zeta_3}{3 } \EBRno{1}{6} \,,
 \\
\alphaBRno{4& 0}{6& 4}  &=
 \frac{i  \pi^3 \tau^3 \zeta_3}{28350 } + 
\frac{ 4  \pi^2 \tau^2  \zeta_3 }{105 } \EBRno{0}{4} + 
\frac{ 4 i  \pi \tau \zeta_3 }{105 } \EBRno{1}{4}  - 
\frac{ \zeta_3}{105 } \EBRno{2}{4}  + \frac{ i  \pi \tau \zeta_5 }{900 }
 + \frac{ 2 \zeta_5}{5 } \EBRno{0}{4} \,,
 \\
 \alphaBRno{3& 1}{6& 4}  &=
 - \frac{ i  \pi^3 \tau^3 \zeta_3}{113400} - 
  \frac{ \pi^2 \tau^2 \zeta_3 }{105 } \EBRno{0}{4}  - 
 \frac{ i  \pi \tau  \zeta_3}{105 } \EBRno{1}{4}  +  \frac{\zeta_3}{420 } \EBRno{2}{4} \,,
 \\
\alphaBRno{2& 2}{6& 4}  &= 
 - \frac{i  \pi^3 \tau^3 \zeta_3}{18900} + 
\frac{ 2  \pi^2 \tau^2  \zeta_3 }{315 } \EBRno{0}{4}  + 
 \frac{2 i  \pi \tau \zeta_3 }{315 } \EBRno{1}{4}   - 
\frac{ \zeta_3}{630 } \EBRno{2}{4}  \notag \\
&\quad + \frac{ 8  \pi^2 \tau^2 \zeta_3 }{3  }\EBRno{0}{6}+ 
 \frac{8 i  \pi \tau \zeta_3 }{3 } \EBRno{1}{6} 
   - \frac{2 \zeta_3}{3 } \EBRno{2}{6} \,,
 \\
 \alphaBRno{4& 1}{6& 4}  &=
 - \frac{ \pi^2 \tau^2 \zeta_5  }{900 }+ 
 \frac{4 i  \pi \tau \zeta_5 }{5 } \EBRno{0}{4}  - \frac{ 2 \zeta_5}{5 }\EBRno{1}{4} \,,
 \\
 \alphaBRno{3& 2}{6& 4}  &=
 \frac{ \pi^4 \tau^4 \zeta_3}{11340  } + 
 \frac{16 i  \pi^3 \tau^3 \zeta_3 }{3 } \EBRno{0}{6}  - 
 8  \pi^2 \tau^2 \zeta_3 \EBRno{1}{6}  - 
 4 i  \pi \tau \zeta_3 \EBRno{2}{6}  +\frac{ 2 \zeta_3}{3 } \EBRno{3}{6} \,,
 \\
\alphaBRno{4& 2}{6& 4}  &= 
 \frac{2 i  \pi^5 \tau^5 \zeta_3}{14175 } - 
 \frac{32  \pi^4 \tau^4 \zeta_3 }{3 } \EBRno{0}{6}  - 
\frac{ 64 i  \pi^3 \tau^3 \zeta_3 }{3 } \EBRno{1}{6}  + 
 16  \pi^2 \tau^2 \zeta_3 \EBRno{2}{6}  + 
\frac{ 16 i  \pi \tau \zeta_3 }{3 } \EBRno{3}{6} \notag \\
&\quad  - \frac{ 2 \zeta_3}{3 } \EBRno{4}{6}   -  \frac{ i  \pi^3 \tau^3 \zeta_5 }{675 } - 
 \frac{8  \pi^2 \tau^2 \zeta_5 }{5 } \EBRno{0}{4}  - 
\frac{ 8 i  \pi \tau \zeta_5 }{5 } \EBRno{1}{4}  + \frac{ 2 \zeta_5}{5 } \EBRno{2}{4}  \, .
\end{align}
\end{subequations}
Note that shuffle relations determine $ \alphaBRno{j_1& j_2}{4& 6}
 = -  \alphaBRno{j_2& j_1}{6& 4} $, and manifestly $T$-invariant representations
 can be found in (\ref{G4G4app}).


\subsection{Laplace equations of cusp forms}
\label{sec5.2}

In this section, we discuss the Laplace equations of the extra basis MGFs
corresponding to $\betasv{j_1&j_2\\6&4}$ and $\betasv{j_1&j_2\\4&6}$.
Their representatives~\eqref{bforms} satisfy
 \vspace{-1ex}
\begin{subequations}
\label{DeltaCusp}
\begin{align}
(\Delta +2) \BB_{2,3} &= 4 \BB'_{2,3} +21 \EE_{2,3} +\frac{3\left((\nabla\mathrm{E}_{2})\overline{\nabla}\mathrm{E}_{3} -(\overline{\nabla}\mathrm{E}_{2})\nabla\mathrm{E}_{3}\right)}{2(\Im\tau)^2}+2 \zeta_3 \EE_2\,,\\
(\Delta - 16) \BB'_{2,3} &= -14 \BB_{2,3} +\frac{105}{2} \EE_{2,3} + 21 \EE_2 \EE_3 +7\zeta_3\EE_2 -\frac{21}{40}\zeta_5\,,
\end{align}
\end{subequations}
as can be shown by combining their $\bsv$ representations in (\ref{excusp.2}) 
with the differential equations~\eqref{eq3.24} obeyed by the $\bsv$.\footnote{To obtain these Laplace equations, one first expresses $\overline{\nabla}\BB_{2,3}$ and $\overline{\nabla}\BB'_{2,3}$ through a combination of $\bsv$ as in~\eqref{excusp.4} and then acts with $\nabla$. The resulting expression is then converted back into MGFs by using the inverse relations shown in e.g.~\eqref{2ptbetasv} and appendix~\ref{appbetasv}. The same result can also be obtained by acting with the derivatives on the lattice sum representations of $\BB_{2,3}$ and $\BB'_{2,3}$ and decomposing the result into the basis summarised in table~\ref{tab:Ebasis}.}
This system can be diagonalised to
\vspace{-0.5ex}
\begin{subequations}
\begin{align}
(\Delta - 12) (- \BB_{2,3}  +  \BB'_{2,3}) &=  \frac{63}{2} \EE_{2,3} + 21 \EE_2 \EE_3 +5 \zeta_3 \EE_2 - \frac{21}{40}\zeta_5\\*
&\quad \quad  -\frac{3\left((\nabla\mathrm{E}_{2})\overline{\nabla}\mathrm{E}_{3} -(\overline{\nabla}\mathrm{E}_{2})\nabla\mathrm{E}_{3}\right)}{2(\Im\tau)^2}   \,,  \nn\\
(\Delta - 2) (- 7\BB_{2,3}  + 2 \BB'_{2,3}) &=  
-42 \EE_{2,3}+42\EE_2 \EE_3  - \frac{21}{20}\zeta_5 \\
&\quad \quad  -\frac{21\left((\nabla\mathrm{E}_{2})\overline{\nabla}\mathrm{E}_{3} -(\overline{\nabla}\mathrm{E}_{2})\nabla\mathrm{E}_{3}\right)}{2(\Im\tau)^2} \,. \nn
\end{align}
\end{subequations}
It is rewarding to rewrite these Laplace equations in terms of cusp forms, i.e.\ eliminate
${\rm B}_{2,3}'$ in favour of the three-column cusp form ${\cal A}_{1,2;5}$ in the normalisation
conventions of (\ref{DHKcusp}):
\begin{subequations}
\label{newsys}
\begin{align}
(\Delta- 12) {\cal A}_{1,2;5} &=  \frac{ (\nabla\mathrm{E}_{3})\overline{\nabla}\mathrm{E}_{2} -(\overline{\nabla}\mathrm{E}_{3})\nabla\mathrm{E}_{2} }{(\Im\tau)^2}\,,
\label{diaglap}
\\
(\Delta - 2) {\rm B}_{2,3} &= 6 {\cal A}_{1,2;5} - \frac{3\big( (\nabla\mathrm{E}_{3})\overline{\nabla}\mathrm{E}_{2} -(\overline{\nabla}\mathrm{E}_{3})\nabla\mathrm{E}_{2}\big)}{2(\Im\tau)^2}\,.
\label{nondiag}
\end{align}
\end{subequations}
Note that (\ref{diaglap}) is a special case of the Laplace equation among two-loop MGFs studied
in \cite{DHoker:2019txf}. The system (\ref{newsys}) can be diagonalised through the following 
linear combination of cusp forms
\beq
(\Delta - 2)\Big({\rm B}_{2,3} - \frac{3}{5} {\cal A}_{1,2;5}\Big) = - \frac{21 \big((\nabla\mathrm{E}_{3})\overline{\nabla}\mathrm{E}_{2} -(\overline{\nabla}\mathrm{E}_{3})\nabla\mathrm{E}_{2}\big) }{10(\Im\tau)^2}\, .
\eeq
Even though they diagonalise the Laplacian, ${\cal A}_{1,2;5}$ and ${\rm B}_{2,3} - \frac{3}{5} {\cal A}_{1,2;5}$ have not been 
chosen as basis elements in table \ref{tab:Ebasis} since their Cauchy--Riemann derivatives yield 
holomorphic Eisenstein series in earlier steps than ${\rm B}_{2,3}$ and ${\rm B}_{2,3}'$, see (\ref{G4G6s}).


\section{\texorpdfstring{Properties of the $\bsv$ and their generating series $Y^\tau_{\vec{\eta}}$}{Properties of the betasv and their generating series Y}}
\label{sec6}

In this section, we study the central objects $\bsv$ and $Y^\tau_{\vec{\eta}}$ in more detail. 
Based on the modular properties of their generating series $Y^\tau_{\vec{\eta}}$, we will determine
the ${\rm SL}_2(\ZZ)$ transformations of the $\bsv$ and assign a modular weight modulo
corrections by $\bsv$ of lower depth. This will be used 
to infer the counting of independent MGFs at various modular weights
from the entries $j_i,k_i$ of the $\bsvBR{j_1 &j_2 &\ldots &j_\ell}{k_1 &k_2 &\ldots &k_\ell}{\tau}$
that occur in the expansion of $Y^\tau_{\vec{\eta}}$.
Finally, based on the transcendental weights of the $\bsv$ and the accompanying combinations
of $y$ and MZVs, we prove that the $\ap$-expansion of $Y^\tau_{\vec{\eta}}$ is uniformly transcendental
if the initial values $\Yhat^{i\infty}_{\vec{\eta}}$ are.


\subsection{Modular properties}
\label{sec6.1}

We first explore the modular properties of the $\bsv$ that can be written in more compact form than
those of the $\Esv$. The modular $T$- and $S$-transformation of the $\bsv$ will be inferred from
their appearance \eqref{eq3.21} in the generating function $Y^\tau_{\vec\eta}$. The torus-integral
representation \eqref{eq2.13} of $Y^\tau_{\vec\eta}$ and the modular properties (\ref{eq2.mod}) 
of its ingredients imply the ${\rm SL}_2(\ZZ)$ transformation
\beq
Y^{  \frac{\alpha \tau+ \beta}{\gamma \tau + \delta} }_{\vec{\eta}}(\sigma|\rho) \, \Big|^{ \eta_j \rightarrow (\gamma \bar \tau + \delta) \eta_j }_{  \bar \eta_j \rightarrow \frac{ \bar \eta_j }{ (\gamma \bar \tau + \delta)}
} = Y^\tau_{\vec{\eta}}(\sigma|\rho) \, .
\label{modtrfY.1}
\eeq
The asymmetric transformation law for the $\bar \eta_j$ and $\eta_j$ stems from the
different choices of arguments for $\Omega$ and  $\overline{\Omega}$ in
the definition \eqref{eq2.13} of the generating series $Y^\tau_{\vec\eta}$. 
By the series expansion \eqref{eq3.21} of both sides of (\ref{modtrfY.1}) in terms of $\bsvBR{j_1 &j_2 &\ldots &j_\ell}{k_1 &k_2 &\ldots &k_\ell}{\tau}$ and $\bsvBR{j_1 &j_2 &\ldots &j_\ell}{k_1 &k_2 &\ldots &k_\ell}{ \frac{\alpha \tau+ \beta}{\gamma \tau + \delta}}$, respectively, we can aim to infer the ${\rm SL}_2(\ZZ)$-properties of $\bsv$.


\subsubsection{\texorpdfstring{$T$- and $S$-transformations}{T- and S-transformations}}

The $T$-modular transformation $\tau \rightarrow \tau{+}1$ is an invariance of both $Y^\tau_{\vec\eta}$ and the operator
$\exp(-\frac{R_{\vec\eta}(\epsilon_0)}{4y})$ acting on the initial values $\Yhat^{i\infty}_{\vec\eta}$ in \eqref{eq3.21}.
Hence, the $T$-invariance of the closed-string integrals can be transferred to the $\beta^{\rm sv}$,
\beq
\bsvBR{j_1 &j_2 &\ldots &j_\ell}{k_1 &k_2 &\ldots &k_\ell}{\tau{+}1} = \bsvBR{j_1 &j_2 &\ldots &j_\ell}{k_1 &k_2 &\ldots &k_\ell}{\tau} \, ,
\label{Ttrf}
\eeq
as is also evident from the explicit low-depth examples worked out in the previous sections.

Under an $S$-modular transformation $\tau \to -1/\tau$, by contrast, we also have to take into account the
(asymmetric) transformation of the $\eta_j,\bar\eta_j$ and that the imaginary part $\Im\tau=y/\pi$ appears 
explicitly in the operator $\exp(-\frac{R_{\vec\eta}(\epsilon_0)}{4y}) \Yhat^{i\infty}_{\vec\eta}$ in \eqref{eq3.21}.
Hence, the $S$-modular transformations of the $\bsv$ can be obtained by inserting
\begin{align}
Y^{-1/\tau}_{\vec{\eta}} \, \big|^{\eta_j \rightarrow \bar \tau \eta_j}_{\bar \eta_j \rightarrow \bar \eta_j/\bar \tau} &=
\sum_{\ell=0}^\infty \sum_{\substack{k_1,\ldots,k_\ell\\ =4,6,8,\ldots}}
\sum_{j_1=0}^{k_1-2} \sum_{j_2=0}^{k_2-2} \ldots \sum_{j_\ell=0}^{k_\ell-2}
\bigg(  \prod_{i=1}^\ell \frac{ (-1)^{j_i} (k_i-1)}{(k_i-j_i-2)!} \bigg)  
 \bsvBR{j_1 &j_2 &\ldots &j_\ell}{k_1 &k_2 &\ldots &k_\ell}{-\tfrac{1}{\tau}}
\notag \\
&\quad \times 
 R_{\vec{\eta}}\big( {\rm ad}_{ \epsilon_0}^{k_\ell-j_\ell-2} (\epsilon_{k_{\ell}})\ldots 
{\rm ad}_{ \epsilon_0}^{k_2-j_2-2} (\epsilon_{k_2}) {\rm ad}_{ \epsilon_0}^{k_1-j_1-2} (\epsilon_{k_1}) \big)
 \label{Yexpand.7x}  \\
&\quad \times 
\exp \Big(  {-} |\tau|^2 \frac{ R_{\vec{\eta}}(\ep_0) }{ 4 y} \Big)
\Yhat^{i\infty}_{\vec{\eta}} \, \big|^{\eta_j \rightarrow \bar \tau \eta_j}_{\bar \eta_j \rightarrow \bar \eta_j/\bar \tau} 
\notag
\end{align}
into the left-hand side of (\ref{modtrfY.1}),
where the substitution on the $\eta$ variables applies to all occurrences on the right-hand side of (\ref{Yexpand.7x}).  

Once a given instance of $\bsv$ has been expressed in terms of MGFs, its
$S$-modular properties can alternatively be inferred from the well-known transformation laws
of the MGFs. Both approaches lead to the following exemplary transformations of the $\bsv$:
\begin{subequations}
\label{Sexampl}
\begin{align}
\betasvStau{0\\4}&=\bar\tau^{2} \Big\{\betasvtau{0\\4} +\frac{\zeta_3}{24y^2}\big( \tau^2{-}1\big)
\Big\}\,,\\
\betasvStau{1\\4}&=\betasvtau{1\\4}+\frac{\zeta_3 \big(|\tau|^2 {-}1 \big)}{6y}
\,,\\
\betasvStau{2\\4}&=\frac{1}{{\bar\tau}^2} \Big\{ \betasvtau{2\\4} +\frac{2\zeta_3}{3} \big( \bar\tau^2{-}1 \big)
\Big\}\,,\\
\betasvStau{2&0\\4&4}&=
\betasvtau{2&0\\4&4} + \frac{ 2 \zeta_3}{3}  (\bar \tau^2{-}1) \betasvtau{0\\4} 
+\frac{ 5 \zeta_5  \big(|\tau|^2{-}1 \big)}{216 y}
+\frac{ \zeta_3^2 \big(1{-} 2 \bar \tau^2 {+} | \tau |^4 \big) }{72y^2}  \, ,
\\
\betasvStau{1&2\\6&4}&=\betasvtau{1&2\\6&4}
+\frac{\zeta_5 \big( \tau^3 \bar\tau{-}1 \big)}{160y^3}\betasvtau{2\\4}
 -\frac{\zeta_3^2 \big(|\tau|^2{-}1 \big)}{2520y} 
  \nn\\
&\quad -\frac{7\zeta_7 \big( |\tau|^4{-}1\big)}{3840y^2}
+\frac{\zeta_3\zeta_5 \big( |\tau|^6{-}2 \tau^3 \bar \tau{+}1\big)}{480y^3}
\,.
\end{align}
\end{subequations}
Based on the general relations (\ref{eq:eksol.1}) and (\ref{eq:bsv1}) to non-holomorphic Eisenstein series,
modular $S$-transformations at depth one can be given in closed form
\begin{align}
\betasvStau{ j \\ k} &= \bar\tau^{k-2-2j} \betasvtau{ j \\ k} - \frac{2\zeta_{k-1}\bar\tau^{k-2-2j} }{(k{-}1) (4y)^{k-2-j}} +\frac{2\zeta_{k-1}  |\tau|^{2(k-2-j)} }{(k{-}1) (4y)^{k-2-j}}\,,
\label{allSbsv}
\end{align}
and their analogues at depth two and $k_1{+}k_2 \leq 10$ can be found in 
appendix~\ref{appSbeta}.


One important immediate consequence of~\eqref{Yexpand.7x} is that the 
maximal-depth term of any $S$-modular transformation is
\begin{align}
\label{eqSlead}
\bsvBR{j_1 &j_2 &\ldots &j_\ell}{k_1 &k_2 &\ldots &k_\ell}{-\tfrac{1}{\tau}} = \bar\tau^{-2\ell - 2(j_1+j_2+\ldots + j_\ell) +  k_1+k_2+\ldots + k_\ell  } \bsvBR{j_1 &j_2 &\ldots &j_\ell}{k_1 &k_2 &\ldots &k_\ell}{\tau}  
\ \te{mod depth} \ \leq \ell{-}1\,,
\end{align}
where the terms of subleading depth are illustrated by the examples in (\ref{Sexampl}).
This follows from taking the terms without MZVs in the initial values
which determine the maximal-depth contributions and whose two- and three-point
instances $\Yhat^{i\infty}_{\eta} \rightarrow \frac{1}{\eta \bar \eta} - \frac{ 2\pi i }{s_{12}}$  
and (\ref{3ptlow}) are invariant under $\eta_j \rightarrow \bar \tau \eta_j$  and
$\bar \eta_j \rightarrow \frac{ \bar \eta_j}{\bar \tau}$. These terms in the initial values
are annihilated by $R_{\vec{\eta}}(\ep_0)$ and therefore unaffected by its
exponential. Hence, for the analysis of maximal-depth terms, it is sufficient to consider 
the rescaling of the $\eta_j,\bar \eta_j$ 
in the operators $R_{\vec\eta}({\rm ad}_{ \epsilon_0}^{k_i-j_i-2} (\epsilon_{k_i}))$ in
(\ref{Yexpand.7x}) that have finite adjoint powers of $\epsilon_0$. Referring back to section~\ref{sec2.4}, we see that 
$R_{\vec\eta}(\epsilon_k)\sim s_{ij} \eta_j^{k-2}$ picks up a factor of $\bar\tau^{k-2}$ under the transformation $(\eta_j,\bar\eta_k) \to (\bar\tau \eta_j,  \bar\eta_j / \bar\tau)$ of~\eqref{Yexpand.7x}. 
In particular, since $R_{\vec\eta}(\epsilon_0)$ picks up a factor of $\bar \tau^{-2}$, 
the operators $R_{\vec\eta}({\rm ad}_{ \epsilon_0}^{k_i-j_i-2} (\epsilon_{k_i}))$ in (\ref{Yexpand.7x})
transform by $\bar \tau^{2+2j_i-k_i}$.

Demanding the maximal-depth terms in the $S$-transformation of the 
$\beta^{\rm sv}$ to cancel all of these factors or $\bar \tau^{2+2j_i-k_i}$ leads to~\eqref{eqSlead}. The argument
is based on the modular invariance of the terms in $\Yhat^{i\infty}_{\vec{\eta}}$ without 
MZVs which amounts to invariance under $\eta_j \rightarrow \bar \tau \eta_j$  and
$\bar \eta_j \rightarrow \frac{ \bar \eta_j}{\bar \tau}$. This is manifest in the two- and three-point examples and we present a conjecture for the MZV-free part of $\Yhat^{i\infty}_{\vec\eta}$ for $n{=}4$ in appendix~\ref{app:n4}. 
The modular transformation (\ref{eqSlead}) is thus firmly established for the combinations of $\beta^{\rm sv}$
that occur in the $Y^\tau_{\vec{\eta}}$-series at $(n {\leq} 3)$ points. Since the counting
of independent MGFs in the next subsection will rely on (\ref{eqSlead}), we have checked
that all the $\beta^{\rm sv}$ entering the weights under consideration there and admitted by the 
derivation algebra do occur in the three-point $Y^\tau_{\vec{\eta}}$. Our counting of MGFs in this
work is therefore not tied to conjectural properties of $(n{\geq 4})$-point initial values.

As a consequence of (\ref{eqSlead}), 
even though the $\bsv$ are not genuine
modular forms, they can be assigned leading modular weights given by
\begin{align}
\label{bsvweight}
\bsvBR{j_1 &j_2 &\ldots &j_\ell}{k_1 &k_2 &\ldots &k_\ell}{\tau} \ \ \ \longleftrightarrow \ \ \ \text{`modular weight'} \quad \Big(0, -2\ell  + \sum_{i=1}^\ell (k_i {-} 2j_i) \Big)
\ \te{mod depth} \ \leq \ell{-}1
\,,
\end{align}
and these will be the modular weights of MGFs associated with the given $\beta^{\rm sv}$ as their leading-depth 
contributions. In order to compensate for the lower-depth corrections to the transformation (\ref{eqSlead}) and
attain a genuine modular form, expressions like (\ref{exE23}), (\ref{eq:bsv0}) for MGFs comprise 
a tail of $\beta^{\rm sv}$ of lower depth. Note that there are only non-holomorphic weights just as 
for the component integrals in~\eqref{eq2.17} as the generating function $Y^\tau_{\vec\eta}$ 
was rescaled by $\Im \tau$ to absorb all holomorphic modular weights.

\subsubsection{A caveat from the derivation-algebra relations}

An important qualification of the above arguments is that the derivation-algebra relations such as~\eqref{eq2.44} imply that the generating series $Y^\tau_{\vec\eta}$ will not contain each possible $\bsv$ with $j_i\leq k_i{-}2$ individually but certain combinations always appear together. The first instance of this implied by~\eqref{eq2.44} occurs at $\ell=2, \ k_1{+}k_2=14$ and was spelt out in~\eqref{eq:dropout}.
Therefore, even though $Y^\tau_{\vec\eta}$ has a perfectly well-defined modular transformation given by~\eqref{modtrfY.1}, this does not uniquely fix the modular behaviour of all the individual $\bsv$.
Instead, from weight $\sum_i k_i \geq 14$ onward, only the specific combinations of $\bsv$
realised in the $\ap$-expansion of $Y^\tau_{\vec\eta}$ (see for instance \eqref{eq:dropout})
have to obey the modular properties (\ref{Ttrf}) and (\ref{eqSlead}). In principle, there is the freedom for
the individual $\bsv$ in these combinations to depart from the above $T$- and $S$-transformations, as long as
these departures cancel from the $Y^\tau_{\vec\eta}$.

Fortunately, this ambiguity does not affect the closed-string integrals 
or the MGFs in its $\ap$-expansion. For the combinations of $\bsv$ that drop out from $Y^\tau_{\vec\eta}$ 
(and therefore all component integrals) by derivation-algebra relations, we do not need or give 
an independent definition in this work.\footnote{For combinations of $\bsv$ that drop
out from the $\ap$-expansion (\ref{eq3.21}), we cannot determine the antiholomorphic integration constants
from the reality properties (\ref{eq2.19}) of component integrals either.}
Hence, (\ref{eqSlead}) can be used as an effective modular transformation that holds
for all combinations of $\bsv$ relevant to this work.
When studying the implications on MGFs in the next section, the dropouts 
of $\bsv$ at given $\sum_i j_i$ and $\sum_i k_i$ will be taken into account, 
so the counting of MGFs can be safely based on (\ref{eqSlead}) and the relations
in the derivation algebra.


\subsection{Counting of modular graph forms}
\label{sec6.2}

The modular properties (\ref{eqSlead}) of the $\bsv$ can be used to count the number of 
independent MGFs of a given weight. 
This will lend further support to our basis of MGFs in table~\ref{tab:Ebasis}. The modular weights
$(w,\bar{w})$ of general lattice sums 
${\cal C}_\Gamma\!\big[\raisebox{.1\height}{\scalebox{.7}{$\begin{array}{c}\mathcal{A}\\[-2mm] \mathcal{B}\end{array}$}}\big]$ (cf.~\eqref{eq:1}) are related to the entries of the highest-depth terms
 $\bsvBRno{j_1 &j_2 &\ldots &j_\ell}{k_1 &k_2 &\ldots &k_\ell}$
 in their integral representation via
\beq
w + \bar w = \sum_{i=1}^\ell k_i  \, , \ \ \ \ \ \ w - \bar w = 2 \ell + \sum_{i=1}^\ell (2j_i - k_i) \, .
\label{bigclaim}
\eeq 
Note that our convention of modular weights is implied by the lattice-sum conventions (\ref{eq2.22}) and differs by the factor of $ ( \frac{ \Im \tau }{\pi} )^{\frac{1}{2}(w+\bar w)}$ from~\cite{DHoker:2016mwo, DHoker:2016quv, DHoker:2019txf}.

While the second correspondence involving $w {-} \bar w$ is simply a consequence of 
(\ref{bsvweight}), the first one $w {+} \bar w = \sum_{i=1}^\ell k_i$ requires further justification
since $\cform{a&\ldots  \protect\\b &\ldots}$ and $\Im \tau\cform{a{+}1&\ldots  \protect\\b{+}1 &\ldots}$ 
have the same total modular weight. It can be understood by comparing
the integral-representation (\ref{eq2.13}) of $Y^\tau_{\vec{\eta}}$ with its $\ap$-expansion (\ref{eq3.21})
in terms of $\beta^{\rm sv}$.

The integrals can be performed order by order in $\ap$ and
$\eta_j,\bar \eta_j$, where the lattice-sum representations (\ref{eq2.lat}) of $G(z,\tau),f^{(k)}(z,\tau)$
and $\overline{ f^{(k)}(z,\tau)}$ yield MGFs such as 
$\cform{a_1&a_2&\ldots &a_R \protect\\b_1&b_2&\ldots &b_R}$ 
and higher graph topologies \cite{Gerken:2018jrq}. The respective contributions 
to the expansion variables and the modular weights of the lattice sums are
\begin{align}
G(z,\tau) \ \ &\leftrightarrow   \ \ s_{ij} \ \& \ \te{modular weights} \ (1,1) \notag \\
f^{(k)}(z,\tau) \ \ &\leftrightarrow   \ \ (\eta_j)^k \ \& \ \te{modular weights} \ (k,0)
\label{latcount}\\
\overline{ f^{(k)}(z,\tau)} \ \ &\leftrightarrow   \ \ (\bar \eta_j)^k \ \& \ \te{modular weights} \ (0,k)\, . \notag
\end{align}
We are disregarding powers of $\Im \tau$ and overall prefactors $\sim (\eta_j \bar \eta_j)^{1-n}$ of
the $Y^\tau_{\vec{\eta}}$, i.e.\ the modular weight $(1,1)$ of the Green function refers to
its contributions to the entries of $\cform{\ldots  \protect\\\ldots}$.

In the $\ap$-expansion (\ref{eq3.21}), in turn, 
the correlation between powers of $s_{ij}, \eta_j,\bar \eta_j$ and the entries of $\beta^{\rm sv}$
is governed by the derivations. Their homogeneity degrees are 
$R_{\vec{\eta}}(\ep_0)\sim s_{ij}/\eta_j^2 + \bar \eta_j/\eta_j$ and
$R_{\vec{\eta}}(\ep_k)\sim s_{ij} \eta_j^{k-2}$, which correspond to
modular weights $R_{\vec{\eta}}(\ep_0)\leftrightarrow (-1,1)$ and $R_{\vec{\eta}}(\ep_k) \leftrightarrow 
(k{-}1,1)$ from the lattice-sum viewpoint (\ref{latcount}).
Hence, the $(s_{ij}, \eta_j,\bar \eta_j)$-counting of any operator $R_{\vec{\eta}}({\rm ad}_{\ep_0}^j \ep_k)$
is the same as having an extra $w + \bar w=k$ in lattice sums, regardless of the power $j$
of ${\rm ad}_{\ep_0}$. This explains why $w {+} \bar w$ has to grow with $\sum_{i=1}^\ell k_i$.

Finally, the absence of a $k_i$-independent offset $w {+} \bar w- \sum_{i=1}^\ell k_i$ can be checked by 
comparing the overall powers of $(s_{ij}, \eta_j,\bar \eta_j)$ 
in the initial value $\Yhat^{i \infty}_{\vec{\eta}}$ and the integral representation of $Y^\tau_{\vec{\eta}}$.
This is most conveniently done by noting the low-energy limit $Y^\tau_{(0,\ldots,0|0,\ldots,0)}(\sigma|\rho) = 1 + {\cal O}(\ap^2)$ of the simplest component
integral at the leading order $\sim (\eta_j \bar \eta_j)^{1-n}$.

On these grounds, we will perform a counting of independent MGFs on the basis of (\ref{bigclaim})
in the rest of this section. Our counting only refers to MGFs that do not evaluate to MZVs or products involving
MZVs or holomorphic Eisenstein series. We explain our methods in most detail for modular invariant objects, where we also distinguish between real and imaginary invariants, but these methods also cover weights
$(w,\bar w)$ with $w\neq \bar w$ that we list in table~\ref{tab:MGFcount}.
For all the values of $(w,\bar{w})$ where we perform the counting below we have verified explicitly that the relevant action of the operators $R_{\vec\eta}(\epsilon_k)$ on the MZV-free part of $\Yhat^{i\infty}_{\vec\eta}$ at three points does not produce accidental linear dependences. Hence, (\ref{bsvweight}) is firmly established in these cases,
and the counting is accurate.
 
 \subsubsection{\texorpdfstring{Reviewing weight $w+\bar w \leq 8$}{Reviewing weight W+wbar<=8}}

Up to total weight $\sum_i k_i<8$ the only possible basis elements stem from $\betasv{j\\k}$ of depth one. 
At fixed $k$, each choice of $0 \leq j \leq k{-}2$ leads to a different modular weight according to~\eqref{bsvweight}. This is in agreement with table~\ref{tab:Ebasis} featuring only a single basis element for all total weights $w{+}\bar w<8$. For instance, modular invariants are obtained for $\betasv{j\\k}$ whenever $j=\frac{k-2}{2}$, and they are related to the $\EE_{k/2}$ shown in the $(\frac{k}{2},\frac{k}{2})$ rows of table~\ref{tab:Ebasis}, see also the explicit formula~(\ref{eq:eksol.2}).

Starting from lattice sums of total weight $w+\bar{w}=8$, there can also be invariant combinations of depth-two $\bsv$. The condition for modular invariance implied by~\eqref{eqSlead} becomes
\begin{align}
\betasv{j_1&j_2\\4&4} \quad\text{weight $(0,0)$} \quad\Leftrightarrow\quad 2=j_1+j_2 \, , \quad 0 \leq j_1,j_2\leq 2\, ,
\end{align}
and there are three solutions to this condition given by $(j_1,j_2)\in \{ (0,2), (1,1), (2,0) \}$, leading to three additional modular invariants of total weight $8$ besides ${\rm E}_4$. Two linear combinations of such
$\betasv{j_1&j_2\\4&4}\, \big|_{j_1+j_2=2}$ can be realised by the shuffles $\betasv{1\\4}^2$ and
$\betasv{0\\4}\betasv{2\\4}$ which correspond to $\EE_2^2$ and $(\nabla {\rm E}_2)\overline \nabla {\rm E}_2$
by (\ref{eq4.18}).\footnote{For ease of notation, we will suppress here and in the following the overall factors of $(\Im \tau)^{\#}$. They are always implicit and understood to be such that the holomorphic modular weight 
vanishes, cf.\ (\ref{bsvweight}). Therefore,  $(\Im\tau)^{-2}(\nabla {\rm E}_2)\overline \nabla {\rm E}_2$ will be just written as  $(\nabla {\rm E}_2)\overline \nabla {\rm E}_2$.} 
Hence, there is a single shuffle-irreducible modular invariant at depth two which can be 
chosen to be ${\rm E}_{2,2}$, expressed through $\bsv$ in (\ref{eq4.26}). Together with $\EE_4 \leftrightarrow \betasv{3\\8}$ at depth one, this reasoning agrees with the total 
of four entries at weight $(4,4)$ in table~\ref{tab:Ebasis}.

The same counting strategy can be applied at non-zero modular weight. Let us consider the example of $(w,\bar{w})=(5,3)$ in table~\ref{tab:Ebasis} which translates into modular weight $(0,-2)$ after multiplication by $(\Im\tau)^5$. The relevant
$\betasv{j_1&j_2\\4&4}$ at depth two with antiholomorphic weight $-2$ have $j_1{+}j_2=3$ by (\ref{bsvweight}) and this leaves
the two options $(j_1,j_2)\in \{ (1,2), (2,1) \}$. One of them is the shuffle $\EE_2 \nabla\EE_2$,
and the irreducible representative is $ \nabla\EE_{2,2}$, see~\eqref{eq4.26}. The
connection with the irreducible modular invariant ${\rm E}_{2,2}$ can be anticipated by comparing  the differential equation (\ref{eq3.25}) of the $\bsv$ with the equations satisfied by the MGFs. 

In general, the appearance of holomorphic Eisenstein series in Cauchy--Riemann derivatives or relations to shuffles as in $\nabla^2 {\rm E}_{2,2} = -\frac{1}{2} (\nabla {\rm E}_2)^2$ implies
that the number of basis MGFs with weights $(w{+}k,w{-}k)$ decreases with $|k|$. An overview of the
numbers of MGFs and irreducible representatives at $w{+}\bar w\leq 14$ can be found 
in table~\ref{tab:MGFcount} below.

 \subsubsection{\texorpdfstring{Reviewing weight $w+\bar w =10$}{Reviewing weight w+wbar<=10}}

Continuing to total weight $10$, there are now additional possibilities at depth two coming from $(k_1,k_2)=(4,6)$ or $(6,4)$. The condition for modular invariant $\betasv{j_1&j_2\\4&6}$ and $\betasv{j_2&j_1\\6&4}$ 
becomes $j_1{+}j_2=3$ (with $0{\leq} j_1 {\leq }2 $ and $0 {\leq} j_2 {\leq }4$). Both cases lead to three solutions 
each, and thus there is a total of six modular invariants contributing to the lattice sums of weights $(w,\bar w)=(5,5)$ that can be expressed through depth-two $\bsv$. Together with the single contribution $\EE_5 \leftrightarrow \betasv{4\\10}$ from depth one, we find seven modular invariant combinations of $\bsv$ which matches the number of basis elements in the $(5,5)$ sector in table~\ref{tab:Ebasis}.

Three combinations of the modular invariant $\betasv{j_1&j_2\\4&6}$ and $\betasv{j_2&j_1\\6&4}$
can be realised as a shuffle of $\betasv{j\\4}\betasv{3-j\\6}$ with $j=0,1,2$. This translates into modular
invariant products $\EE_2\EE_3, \  (\nabla\EE_2)\overline \nabla \EE_3, \ 
(\overline\nabla\EE_2)\nabla \EE_3$ 
and leaves three irreducible modular invariants at depth two that can be chosen to be ${\rm E}_{2,3}$ and 
${\rm B}_{2,3},{\rm B}'_{2,3}$ in table~\ref{tab:Ebasis}, see (\ref{exE23}) and (\ref{excusp.2}) for
their expressions in terms of $\bsv$.
Alternatively, one can trade ${\rm B}_{2,3},{\rm B}'_{2,3}$ for the imaginary cusp forms $\aform{0&2&3\\3&0&2}, \aform{0&1&2&2\\1&1&0&3}$ and organise the modular invariants according to their reality properties: 
three real basis elements $\EE_2\EE_3, \ {\rm Re}[
(\nabla\EE_2)\overline \nabla \EE_3], \ {\rm E}_{2,3}$ 
(one of them irreducible) and three imaginary basis elements ${\rm Im}[
(\nabla\EE_2)\overline \nabla \EE_3], \ 
\aform{0&2&3\\3&0&2}, \ \aform{0&1&2&2\\1&1&0&3}$ (two of them irreducible).

The counting of real and imaginary forms can also be obtained based on the reality properties~\eqref{eq4.14} and~\eqref{eq4.23} of the $\bsv$: in the modular-invariant case, complex conjugation is only an operation on the labels of the $\bsv$ at leading order in depth. Therefore one has to form combinations of the $\bsv$ that are mapped to themselves or minus themselves under complex conjugation. For instance, since
\begin{align}
\overline{\betasv{2 &1 \\ 4&6}} = \betasv{3&0\\6&4} \quad\text{modulo lower depth,}
\end{align}
the combinations $\betasv{2 &1 \\ 4&6}  \pm \betasv{3&0\\6&4}$ give real and imaginary MGFs modulo lower depth, respectively. See e.g.\ the real $\EE_{2,3}$ in \eqref{exE23} and the imaginary cusp form ${\rm B}_{2,3}$ in \eqref{eq:7}.

The analogous counting of MGFs with $w{+}\bar w = 10$ 
and $w \neq \bar w$ based on the $\bsv$ can be found in table~\ref{tab:MGFcount} below.


 \subsubsection{\texorpdfstring{Predictions for weight $w+\bar w= 12$}{Predictions for weight w+wbar=12}}

For lattice sums of weight $w{+}\bar w=12$, a basis of 19 modular invariants
can be anticipated from $\beta^{\rm sv}$ at depth $\ell=1,2,3$:
\begin{itemize}
\item[(i)] a single depth-one invariant $\EE_6 \leftrightarrow \betasv{5\\12}$
\item[(ii)] 5 depth-two invariants $\betasv{j_1&j_2\\6&6}$ with $j_1+j_2=4$ and $0 \leq j_1,j_2\leq 4$
\item[(iii)] $6$ depth-two invariants $\betasv{j_1&j_2\\4&8}$ \& $\betasv{j_2&j_1\\8&4}$ with 
$j_1{+}j_2=4$ and $0 {\leq} j_1{ \leq} 2$ \& $0{\leq} j_2{\leq} 6$
\item[(iv)] 7 depth-three invariants $\betasv{j_1&j_2 &j_3\\4&4 &4}$ with $j_1+j_2+j_3=3$ and $0 \leq j_i \leq 2$
\end{itemize}
We will analyse the shuffle- and reality properties separately in each sector (ii), (iii), (iv)
and connect with known irreducible modular graph functions.

Sector (ii) comprises three shuffle $\betasv{2\\6}^2, \ \betasv{1\\6}\betasv{3\\6}$ and $\betasv{0\\6}\betasv{4\\6}$ 
that correspond to $\EE_3^2, \ (\nabla {\rm E}_3)\overline \nabla {\rm E}_3$ and 
$(\nabla^2 {\rm E}_3)\overline \nabla^2 {\rm E}_3$ according to~\eqref{2ptbetasv}. This leaves two irreducibles
which can be taken to be the quantities
\begin{align}
{\rm E}_{3,3} &=  450  \betasv{4& 0\\6& 6} - 180 \zeta_5  \betasv{0\\6} + \frac{ \zeta_7}{16 y} - \frac{ 7 \zeta_9}{64 y^3}
 + \frac{ 9 \zeta_5^2}{64 y^4}\,,
 \label{morebsv1}
\\
{\rm E}_{3,3}' &= 120 (   \betasv{4& 0\\6& 6} - \betasv{3& 1\\6& 6}) - 48  \zeta_5 \betasv{0\\6}  + \frac{
 12  \zeta_5}{y}  \betasv{1\\6} + \frac{ 3 \zeta_7}{160 y} -  \frac{7 \zeta_9}{480 y^3}
  \label{morebsv2}
\end{align}
corresponding to the lattice sums \eqref{eq:3} and \eqref{eq:4}. 
The $\beta^{\rm sv}$ representations have been inferred from the differential equations \cite{Broedel:2018izr}
and the Laurent polynomials \cite{DHoker:2016quv} of the real MGFs ${\rm E}_{3,3}$ and ${\rm E}'_{3,3}$.

Sector (iii) also admits three shuffles $\betasv{0\\4}\betasv{4\\8}, \ \betasv{1\\4}\betasv{3\\8}$
and $ \betasv{2\\4}\betasv{2\\8}$
corresponding to $(\overline \nabla {\rm E}_2) \nabla {\rm E}_4, \  {\rm E}_2 {\rm E}_4$ and
$(\nabla {\rm E}_2)\overline \nabla {\rm E}_4$, respectively. Two of them are real ${\rm E}_2 {\rm E}_4$,  
${\rm Re}[ (\nabla {\rm E}_2)\overline \nabla {\rm E}_4]$ whereas a third one ${\rm Im}[ (\nabla {\rm E}_2)\overline \nabla {\rm E}_4]$ is imaginary. The remaining three invariants are shuffle irreducible,
and one real representative
\begin{align}
{\rm E}_{2,4} &= -5670 \betasv{4& 0\\8& 4} - 5670 \betasv{2& 2\\4& 8} + 
 3780 \zeta_3 \betasv{2\\8}  \notag \\
 &\quad + \frac{ 405  \zeta_7}{4 y^2} \betasv{0\\4} - \frac{9 \zeta_7}{80 y} + 
  \frac{25 \zeta_9}{8 y^3} -  \frac{ 135 \zeta_3 \zeta_7}{32 y^4}  
   \label{morebsv3}
\end{align}
corresponds to the lattice sum given in \eqref{eq:5}. As will be argued below, the remaining two shuffle irreducibles can be chosen to be imaginary.

Sector (iv) admits 2+3 shuffles ${\rm E}_2^3, \ {\rm E}_2 (\nabla {\rm E}_2) \overline \nabla {\rm E}_2$
and ${\rm E}_2 {\rm E}_{2,2} , \ (\nabla {\rm E}_2) \overline \nabla {\rm E}_{2,2}$, 
$(\overline\nabla {\rm E}_2) \nabla {\rm E}_{2,2}$. Among the leftover two shuffle-irreducibles,
one real representative
\begin{align}
{\rm E}_{2,2,2} &= -216 \betasv{2& 1& 0\\4& 4& 4} + 144 \zeta_3 \betasv{1& 0\\4& 4} + 10 \zeta_5 \betasv{0\\4} \notag \\
&\quad -\frac{ 12 \zeta_3^2 }{y} \betasv{0\\4}   + \frac{ \zeta_3^2}{ 30} + \frac{ 661 \zeta_7}{1800 y} 
- \frac{ 5 \zeta_3 \zeta_5}{12 y^2} + \frac{ \zeta_3^3}{6 y^3}
 \label{morebsv4}
\end{align}
corresponds to the lattice sum \eqref{eq:6}.
As will be argued below, the second shuffle irreducible is imaginary.

In order to anticipate the number of real and imaginary irreducible modular invariants, the
known types of relations among MGFs have been exhaustively applied to all dihedral and trihedral graph topologies
at weights $(w,\bar w )=(6,6)$ \cite{package}. The solution to the large equation system identifies
14 real and 5 imaginary independent modular invariants, again excluding MZVs and ${\rm G}_k$
from our counting conventions. Given that
modular invariant combinations of the known ${\rm E}_{\ldots}$ already exhaust the
14 real invariants, the remaining shuffle irreducibles must admit imaginary representatives. 
This conclusion lends support to extending the reality properties of the $\bsv$ given in~\eqref{eq4.23}  beyond $k_1{+}k_2>10$, and it is tempting to extrapolate it to arbitrary depth
\beq
\overline{ \bsvBRno{j_1 &j_2 &\ldots &j_\ell}{k_1 &k_2 &\ldots &k_\ell} } = (4y)^{2\ell+ \sum_{i=1}^\ell (2j_i-k_i)} \bsvBRno{k_\ell-2-j_\ell &\ldots &k_2-2-j_2 &k_1-2-j_1}{k_\ell &\ldots &k_2 &k_1} \ \te{mod depth $\leq \ell{-}1$} \, .
 \label{eq4.23ext}
\eeq
This conjecture leads to the same counting of imaginary representatives, and the
power of $4y$ therein vanishes exactly if the modular weight of $\beta^{\rm sv}$ in~\eqref{bsvweight} does.

Hence, the 5 imaginary invariants at $(w,\bar w )=(6,6)$ are ${\rm Im}[ (\nabla {\rm E}_2)\overline \nabla {\rm E}_4]$,
${\rm Im}[ (\nabla {\rm E}_2)\overline \nabla {\rm E}_{2,2}]$, two irreducible cusp forms from
(iii) and one irreducible cusp form from (iv). The paper~\cite{DHoker:2019txf} identified two cusp
forms at $(w,\bar w)=(6,6)$ among the two-loop graphs on the worldsheet. Accordingly, three out
of the five cusp forms in our counting must require lattice sums associated with $(L \geq 3)$-loop
graphs. Indeed, a detailed analysis of the relations between dihedral and trihedral MGFs 
suggests that $\aform{0&2 &4\\ 5 &0 &1}, \aform{0 &1 &2 &3\\ 2 &1 &3 &0}$
and $\aform{0 &2 &2 &2\\ 3 &0 &1 &2}$ qualify as a basis of shuffle-irreducible
cusp forms at $(w,\bar w)=(6,6)$, and ${\rm Im}[ (\nabla {\rm E}_2)\overline \nabla {\rm E}_{2,2}]$ 
also exceeds the two-loop graphs when written in terms of lattice sums. 

In summary, the 19 modular invariant lattice sums of weight $(w,\bar w)=(6,6)$ comprise
11 shuffles (3 from (ii), 3 from (iii) and 5 from (iv)) and 8 shuffle irreducibles. The 
irreducibles admit 5 real representatives known in the literature (${\rm E}_6$ from (i), ${\rm E}_{3,3}, {\rm E}_{3,3}'$ from (ii), ${\rm E}_{2,4}$ from (iii), ${\rm E}_{2,2,2}$ from (iv))
and 3 imaginary cusp forms (two from (iii) and one from (iv)) generalising 
$\aform{0&2&3\\3&0&2}, \ \aform{0&1&2&2\\1&1&0&3}$ described in section \ref{sec5}. 

The analogous counting of MGFs with $w{+}\bar w = 12$ 
and $w \neq \bar w$ can be found in table~\ref{tab:MGFcount} below.

\subsubsection{\texorpdfstring{Weight $w+\bar w= 14$ and the derivation algebra}{Weight w+wbar=14 and the derivation algebra}}
\label{sec:e10e4}

By extending the above counting method to weight $w{+}\bar w=14$, one is na\"ively
led to 44 modular invariants (26 of them shuffles). If all the $\bsv$ were realised 
independently in the expansion (\ref{eq3.21}) of $Y^\tau_{\vec{\eta}}$, the total of 44 
would arise from the following sectors: 
\begin{itemize}
\item[(a)] a single depth-one invariant $\EE_7 \leftrightarrow \betasv{6\\14}$
\item[(b)] 6 depth-two invariants $\betasv{j_1&j_2\\4&10}$ \& $\betasv{j_2&j_1\\10&4}$ with $j_1{+}j_2=5$ and $0 {\leq} j_1 {\leq} 2$ and $0{\leq} j_2{\leq} 8$
\item[(c)] 10 depth-two invariants $\betasv{j_1&j_2\\6&8}$ \& $\betasv{j_2&j_1\\8&6}$ with $j_1{+}j_2=5$ and $0 {\leq} j_1 {\leq} 4$ and $0{\leq} j_2 {\leq} 6$
\item[(d)] 27 depth-three invariants $\betasv{j_1&j_2 &j_3\\6&4 &4}$ with $j_1+j_2+j_3=4$ 
and $0 \leq j_1 \leq 4$ as well as $0 \leq j_2,j_3 \leq 2$ and permutations of $(k_1,k_2,k_3)$
\end{itemize}
However, weight $\sum_{i=1}^\ell k_i = 14$ is the first instance where the derivation algebra exhibits relations 
beyond the nilpotency properties in (\ref{eq2.42}) that we have already used in the derivation of (\ref{eq3.21}). The simplest instance was exhibited in~\eqref{eq:dropout}, see also appendix~\ref{appB} for a representation of 
the MGF ${\rm D}_7$ at $w+\bar{w}=14$ in terms of $\bsv$.

More generally,
the relation (\ref{eq2.44}) implies additional relations under the adjoint $\epsilon_0$ action according to\footnote{We have checked that more general relations of the form
\begin{equation*}
\big( R_{\vec{\eta}}(\ep_0) \big)^{j_1} \Big( \big[ R_{\vec{\eta}}(\ep_{10}), R_{\vec{\eta}}(\ep_4)\big]
 - 3 \big[R_{\vec{\eta}}(\ep_{8}),R_{\vec{\eta}}(\ep_6) \big]  \Big) \big( R_{\vec{\eta}}(\ep_0) \big)^{j_2} = 0\, ,
\end{equation*}
do not yield any further relations among the operators $R_{\vec{\eta}}\big(  
{\rm ad}_{ \epsilon_0}^{j_1} (\epsilon_{k_1}) {\rm ad}_{ \epsilon_0}^{j_2} (\epsilon_{k_2}) \big)$
in the expansion (\ref{eq3.21}) of $Y^\tau_{\vec{\eta}}$.}
\begin{align}
0 &= R_{\vec\eta} \bigg[ {\rm ad}_{\epsilon_0}^{j} \Big( \big[ \ep_{10},\ep_4\big]
 - 3 \big[\ep_{8} , \ep_6 \big]  \Big)\bigg]  
 \label{adjrel}\\
&= \sum_{r=0}^{j} \binom{j}{r}
R_{\vec\eta} \Big( 
 \big[ {\rm ad}_{\epsilon_0}^{r}(\ep_{10}) ,  {\rm ad}_{\epsilon_0}^{j-r}(\ep_4) \big]
-3   \big[ {\rm ad}_{\epsilon_0}^{r}(\ep_{8})  , {\rm ad}_{\epsilon_0}^{j-r}(\ep_6) \big]
 \Big) \, ,
\notag
\end{align}
and similar relations arise at higher weight and depth, see (\ref{eq2.43}) and \cite{LNT, Pollack, Broedel:2015hia}.
In passing to the second line, we have rewritten the relation in terms of the quantities
 $ R_{\vec{\eta}}\big(  {\rm ad}_{ \epsilon_0}^{j_1} (\epsilon_{k_1}) {\rm ad}_{ \epsilon_0}^{j_2} (\epsilon_{k_2}) \big)$
 that occur in the expansion (\ref{eq3.21}) of $Y^\tau_{\vec{\eta}}$ (setting $j {\leq} 10$ in
 (\ref{adjrel}) and using $R_{\vec\eta}({\rm ad}_{ \epsilon_0}^{k-1} (\epsilon_{k}))=0$).
As a consequence, the $\beta^{\rm sv}$ in the sectors (b) 
and (c) cannot all appear independently in the generating series $Y^\tau_{\vec{\eta}}$ of MGFs.

More specifically, (\ref{adjrel}) implies exactly one dropout among the $\betasv{j_1&j_2\\k_1&k_2}$ 
with $k_1{+}k_2=14$ for each value $j=j_1{+}j_2$ with $0\leq j \leq 10$.
At $j=5$, this reduces the total number of independent modular invariants with weight $(w,\bar w)=(7,7)$
by one, leading to $43$ rather than $44$. The commutators in (\ref{adjrel})
imply that this reduction affects the shuffle-irreducible MGFs,
and the dropout at $(w,\bar w)=(7,7)$ concerns an imaginary modular invariant
when the combinations of $\beta^{\rm sv}$ are organised into real and
imaginary ones. Further details and the analogous
counting of forms with $w\neq \bar w$ can be found in table~\ref{tab:MGFcount}.
We have checked that all $\beta^{\rm sv}$ noted in the MGF-column of the table occur in the $Y^\tau_{\vec{\eta}}$-series at three points (without accidental dropouts) and are therefore known to satisfy \eqref{eqSlead} without relying on the conjectural four-point data from appendix~\ref{app:n4}.
The conjectural relation (\ref{eq4.23ext}) implies that the basis at $(w,\bar w)=(7,7)$ can be spanned by
24 real and 19 imaginary invariants.\footnote{As an immediate consequence of (\ref{eq4.23ext}), we
have a single real invariant ${\rm E}_7$ in sector (a) as well as 15 real and 12 imaginary invariants
in sector (d). The sectors (b) and (c) are coupled through the relations (\ref{adjrel}) in the derivation
algebra. It follows from (\ref{eq4.23ext}) that the 15 independent instances 
of $ R_{\vec{\eta}}\big(  {\rm ad}_{ \epsilon_0}^{j_1} (\epsilon_{k_1}) {\rm ad}_{ \epsilon_0}^{j_2} (\epsilon_{k_2}) \big)$
 in (\ref{eq3.21}) are accompanied by 8 real and 7 imaginary linear combinations of 
$\betasv{j_1&j_2\\4&10}, \ \betasv{j_2&j_1\\10&4}, \ \betasv{j_1&j_2\\6&8}, \ \betasv{j_2&j_1\\8&6}$ at $j_1{+}j_2=5$.} It would be interesting to study at the level of the Laurent polynomials if our basis of real MGFs at this weight contains a cusp form.


\begin{table}
  \begin{center}
    {\setstretch{1.0}
      \begin{tabular}{cccccc}
      \toprule
      weight &  $\# \ \bsv$ & $\#$ MGFs & of which shuffle irred. & real MGFs & imag. MGFs\\
      \midrule
      (2,2) & 1 & 1 & 1 &1 &0\\
      (3,1) & 1 & 1 & 1 &-- &--\\
      \midrule
      (3,3) & 1 & 1 & 1 &1 &0\\
      (4,2) & 1 & 1 & 1 &-- &--\\
      (5,1) & 1 & 1 & 1  &-- &--\\
      \midrule
      (4,4) & 4 & 4 & 2 &4 &0\\
      (5,3) & 3 & 3 & 2  &-- &--\\
      (6,2) & 2 & 2 & 1  &-- &--\\
      (7,1) & 1 & 1 & 1 &-- &--\\
      \midrule
      (5,5) & 7 & 7 & 4 &4 &3\\
      (6,4) & 7 & 7 & 4  &-- &--\\
      (7,3) & 5 & 5 & 3  &-- &--\\
      (8,2) & 3 & 3 & 2  &-- &--\\
      (9,1) & 1 & 1 & 1  &-- &--\\
      \midrule
      (6,6) & 19 & 19 & 8 & 14 &5\\
      (7,5) & 17 & 17 & 8  &-- &--\\
      (8,4) & 13 & 13 & 6  &-- &--\\
      (9,3) & 8 & 8 & 4  &-- &--\\
      (10,2) & 4 & 4 & 2  &-- &--\\
      (11,1) & 1 & 1 & 1  &-- &--\\
      \midrule
      (7,7) & 44 & 43 & 17 & 24 &19\\
      (8,6) & 41 & 40 & 16  &-- &--\\
      (9,5) & 33 & 32 & 13  &-- &--\\
      (10,4) & 22 & 21 & 9   &-- &--\\
      (11,3) & 12 & 11 & 5  &-- &--\\
      (12,2) & 5 & 4 & 2  &-- &--\\
      (13,1) & 1 & 1 & 1  &-- &--\\
      \bottomrule
    \end{tabular}
  }
  \caption{Counting of modular graph forms (MGF) up to total weight $w+\bar{w}=14$ based on the number of $\bsv$. The entries list the total number of MGFs, the number of shuffle-irreducible MGFs as well as the number of real and imaginary MGFs in the modular invariant sectors. Up to total weight $w+\bar{w}\leq 12$, the counting has been confirmed by independent methods for dihedral and trihedral MGFs. For $w+\bar{w}=14$, the derivation algebra imposes the additional constraint~\eqref{eq2.44} on the combinations of the $\bsv$ that can appear in the generating function $Y^\tau_{\vec\eta}$, leading to a mismatch of the number of $\bsv$ and MGFs.
 }
    \label{tab:MGFcount}
  \end{center}
\end{table}

\subsubsection{\texorpdfstring{Weight $w+\bar w \geq 16$ and the derivation algebra}{Weight w+wbar=16 and the derivation algebra}}
\label{sec:e12e4}

We have not performed a similarly detailed analysis at higher weight and only offer some general comments. 
At weight $w{+}\bar w = 16$, similar dropouts in the na\"ive count of MGFs via $\beta^{\rm sv}$
arise from the depth-three relation (\ref{eq2.43c}), obstructing for instance the independent
appearance of all the $\betasv{j_1&j_2 &j_3\\8&4 &4}$ with $0 \leq j_1 \leq 6$ and $0 \leq j_2,j_3 \leq 2$. 
In case of modular invariants with $w=\bar{w}=8$, this leads to the dropout of a real MGF, leaving in total $108$ MGFs, out of which $42$ are imaginary cusp forms.

Weight $w{+}\bar w=18$ even allows for three sources of dropouts:
\begin{itemize}
\item the irreducible depth-two relation (\ref{eq2.43b}) involving $(k_i,k_j) \in \{(4,14),(6,12),(8,10) \}$
\item left- and right-multiplication of the $(k_1{+}k_2=14)$-relation (\ref{eq2.43a}) by a single $\ep_4$ and
arbitrary powers of $\ep_0$
\item an irreducible depth-four relation first seen in \cite{Pollack} and available for download at \cite{WWWe}
\end{itemize}
The systematics of relations in the derivation algebra is governed by the counting of holomorphic
cusp forms \cite{Pollack}. The propagation of irreducible relations to higher depth and weight by multiplication
with additional $\ep_k$ has been discussed in detail in \cite{Broedel:2015hia}. The latter reference
is dedicated to classifying relations among elliptic MZVs and counting their irreducible representatives
at various lengths and depths. In this way \cite{Broedel:2015hia} can be viewed as the open-string prototype
of the present counting of MGFs.

\subsubsection{Depth versus graph data}
\label{sec:esumm}

We emphasise that the above counting of MGFs applies to closed-string
integrals of arbitrary multiplicity and therefore to arbitrary graph topologies.
The reason is that the $R_{\vec{\eta}}(\ep_k)$ were assumed to obey no
further relations besides those in the derivation algebra, i.e.\ multiplicity-specific
relations such as commutativity of the $R_\eta(\ep_{k\geq 4})$ at two points were disregarded.

Our bases of MGFs at $w{+}\bar w\leq 10$ and $(w,\bar w)=(6,6)$ as well as $(w,\bar{w})=(7,5)$
were built from dihedral representatives (\ref{eq2.22}). Hence, the results of this section imply
that any MGF at these weights associated with arbitrarily complicated 
graph topologies can be reduced to dihedral MGFs (possibly with $\QQ$-linear combinations
of MZVs in their coefficients), extending the explicit calculations for dihedral and trihedral graphs in \cite{package}.
It would be interesting to determine the first combination
of weights, where the appearance of a trihedral basis MGF is inevitable.

We have not found any general correlation between the loop order of an MGF
and the maximum depth of the associated $\bsv$. On the one hand, one-loop MGFs (\ref{eq2.26})
are still in one-to-one correspondence with $\bsv$ at depth one by (\ref{eq:bsv0}).
On the other hand, a basis of MGFs with $w{+}\bar w=10$ requires at least one 
three-loop graph (e.g.\ $\aform{0&1&2&2\\1&1&0&3}$) while the associated $\bsv$ cannot exceed depth two. Up to $w+\bar{w}=12$ all examples satisfy that the loop order of an irreducible $\bsv$ of depth $\ell$ is 
at least $\ell$, i.e., the depth of an irreducible $\bsv$ appears to be a lower bound for the loop order.


\subsection{Towards uniform transcendentality}
\label{sec6.3}

This section is dedicated to the transcendentality properties of the 
generating series $Y^\tau_{\vec{\eta}}$ that become manifest
from our results. We will show that the
component integrals (\ref{eq2.15}) are uniformly transcendental
provided that the same is true for the initial values\footnote{It will be the main goal of \cite{yinprogress} 
to express the initial values $\Yhat^{i\infty}_{\vec{\eta}}$ in terms of uniformly transcendental 
sphere integrals as done in (\ref{eq4.2}) and (\ref{eq4.1}) for the two-point example.} $\Yhat^{i\infty}_{\vec{\eta}}$.
In other words, the matrix- and operator-valued series
\begin{align}
\Lambda^\tau_{\vec{\eta}} &=
\sum_{\ell=0}^\infty \sum_{\substack{k_1,k_2,\ldots,k_\ell \\=4,6,8,\ldots}}
\sum_{j_1=0}^{k_1-2} \sum_{j_2=0}^{k_2-2} \ldots \sum_{j_\ell=0}^{k_\ell-2}
\bigg(  \prod_{i=1}^\ell \frac{ (-1)^{j_i} (k_i-1)}{(k_i-j_i-2)!} \bigg)   \EsvBR{j_1 &j_2 &\ldots &j_\ell}{k_1 &k_2 &\ldots &k_\ell}{\tau} \label{eq3.21alt}  \\
&\quad \times \exp \Big(  {-} \frac{R_{\vec{\eta}}(\ep_0) }{ 4 y} \Big)
 R_{\vec{\eta}}\big( {\rm ad}_{ \epsilon_0}^{k_\ell-j_\ell-2} (\epsilon_{k_{\ell}})\ldots 
{\rm ad}_{ \epsilon_0}^{k_2-j_2-2} (\epsilon_{k_2}) {\rm ad}_{ \epsilon_0}^{k_1-j_1-2} (\epsilon_{k_1}) \big)
\notag
\end{align}
relating $Y^{\tau}_{\vec{\eta}}= \Lambda^\tau_{\vec{\eta}} \Yhat^{i\infty}_{\vec{\eta}}$ by (\ref{eq3.18}) 
will be demonstrated to enjoy uniform transcendentality.
Our reasoning closely follows the lines of section 7.1 in \cite{Mafra:2019xms}, where the
open-string analogues of the $Y^\tau_{\vec{\eta}}$ are shown to be uniformly transcendental.


\subsubsection{Weight assignments and uniform transcendentality of the generating series}

We assign the following transcendental weights to the \textit{holomorphic} building blocks in 
the $\ap$-expansion of open- and closed-string integrals,
\begin{center}
  \vspace{0.5em}
  {\setlength{\tabcolsep}{12pt}
    \begin{tabular}{ccccc}
      \toprule
      quantity & $\zeta_{n_1,n_2,\ldots,n_r}$ & $\pi$ & $\EBRno{j_1 &j_2 &\ldots &j_\ell}{k_1 &k_2 &\ldots &k_\ell}$ & $\tau$ \\
      \midrule
      \makecell{transcendental\\weight} & $\displaystyle\sum_{j=1}^r n_j$ & $1$ & $\displaystyle\ell + \sum_{i=1}^\ell j_i$ & 0 \\
      \bottomrule
    \end{tabular}
  }
  \vspace{0.5em}
\end{center}
leading to weight 0 for $\nabla$ and for instance weight $j_1{+}1$ for $\EBRno{j_1 }{k_1}$.

Moreover, complex conjugation is taken to preserve the weight, which leads to weight 0 for
 $\bar \tau$, weight 1 for $y$ and weight $\ell + \sum_{i=1}^\ell j_i$ for
$\overline{ \EBRno{j_1 &j_2 &\ldots &j_\ell}{k_1 &k_2 &\ldots &k_\ell} }$.
The weights of the holomorphic iterated Eisenstein integrals are inherited from
those of elliptic MZVs \cite{Broedel:2015hia}.

In order to infer the weight of the real-analytic ${\cal E}^{\rm sv}$ in the $\ap$-expansion (\ref{eq3.21alt}), we
first note that their building blocks $\EsvBRminno{j_1 &j_2 &\ldots &j_\ell}{k_1 &k_2 &\ldots &k_\ell}$
involving only holomorphic ${\cal E}[\ldots]$
have transcendental weight $\ell + \sum_{i=1}^\ell j_i$ as is manifest in their representation (\ref{eq3.15}). 
We will demonstrate in section~\ref{sec:transc-seri-bsv} that this propagates to the
antiholomorphic integration constants $\overline{\fBRno{j_1 &j_2 &\ldots &j_\ell}{k_1 &k_2 &\ldots &k_\ell}}$
in the decomposition (\ref{eq:2}) of $\EsvBRno{j_1 &j_2 &\ldots &j_\ell}{k_1 &k_2 &\ldots &k_\ell}$.

With these definitions we will show that the component integrals carry uniform transcendental weight 
\begin{align}
Y^\tau_{(a_2,\ldots,a_n|b_2,\ldots,b_n)}(\sigma|\rho) \ \te{at order} \ \ap^w \ \ \ \longleftrightarrow \ \ \ 
\te{transcendental weight} \ w+\sum_{j=2}^n a_j\,.
\label{trans.1}
\end{align}
In order to give a uniform transcendental weight to the whole generating series $Y^\tau_{\vec\eta}$ we have to assign
\begin{align}
s_{ij}, \eta_j,\bar \eta_j \ \ \ \longleftrightarrow \ \ \ 
\te{transcendental weight} \ {-}1\ .
\label{trans.4}
\end{align}
With this convention and the inverse factors of $(2\pi i)$ in the definition 
(\ref{eq2.15}) of component integrals, (\ref{trans.1}) is equivalent to having
\beq
\te{claim:} \ \ \ Y^\tau_{\vec{\eta}} \ \ \ \longleftrightarrow \ \ \ 
\te{transcendental weight} \ 2(n{-}1) 
\label{trans.5}
\eeq
for the generating series at $n$ points. This will be shown under the assumption
that the initial data has uniform transcendental weight,
\beq
\te{assumption:} \ \ \ \Yhat^{i\infty}_{\vec{\eta}} \ \ \ \longleftrightarrow \ \ \ 
\te{transcendental weight} \ 2(n{-}1) \, .
\label{trans.5as}
\eeq
%


\subsubsection{\texorpdfstring{Transcendentality of the series in $\bsv$}{Transcendentality of the series in betasv}}
\label{sec:transc-seri-bsv}

We start by inspecting the constituents of the series $\Lambda^\tau_{\vec{\eta}}$ in (\ref{eq3.21alt}).
By the homogeneity degrees $R_{\vec{\eta}}(\ep_0) \sim s_{ij}/\eta_j^2 +2\pi i \bar \eta_j/\eta_j$
and $R_{\vec{\eta}}(\ep_{k\geq 4}) \sim s_{ij}\eta_j^{k-2}$ of the derivations
in section \ref{sec2.4} and appendix \ref{appderiv}, we get
\begin{align}
R_{\vec{\eta}}(\ep_{k}) \ \ \ \longleftrightarrow \ \ \ 
\te{transcendental weight} \ 1{-}k \, , \ \ \ k \geq 0\ .
\label{trans.6}
\end{align}
As an immediate consequence, the operators in the series (\ref{eq3.21alt}) are assigned
\begin{align}
\exp \Big( {-}\frac{ R_{\vec{\eta}}(\ep_{0}) }{4y} \Big) \ \ \ &\longleftrightarrow \ \ \ 
\te{transcendental weight} \ 0  \label{trans.7}\\
 R_{\vec{\eta}}\big({\rm ad}_{\ep_0}^{k-j-2}(\ep_{k})\big)   \ \ \ &\longleftrightarrow \ \ \ 
\te{transcendental weight} \ {-}(j{+}1) 
\ . \notag
\end{align}
Hence, the transcendental weight we have found for the ${\cal E}^{\rm sv}_{\rm min}$
cancels that of the accompanying derivations,
\beq
\EsvBRminno{j_1  &\ldots &j_\ell}{k_1  &\ldots &k_\ell}
 R_{\vec{\eta}}\big( {\rm ad}_{ \epsilon_0}^{k_\ell-j_\ell-2} (\epsilon_{k_{\ell}})\ldots 
 {\rm ad}_{ \epsilon_0}^{k_1-j_1-2} (\epsilon_{k_1}) \big)
\ \ \ \longleftrightarrow \ \ \ 
\te{transcendental weight} \ 0\, .
\label{trans.8alt}
\eeq
We shall now argue that this has to extend to the full ${\cal E}_{\rm min}^{\rm sv} \rightarrow {\cal E}^{\rm sv}$:
By the vanishing transcendental weight of $\exp ( {-}\frac{ R_{\vec{\eta}}(\ep_{0}) }{4y})$, it
follows from (\ref{trans.8alt}) that the ${\cal E}_{\rm min}^{\rm sv}$ contributions
to the series (\ref{eq3.21alt}) have weight zero, i.e.\ $\Lambda^\tau_{\vec{\eta}}$
can only depart from vanishing transcendental weight via ${\cal E}^{\rm sv} - {\cal E}_{\rm min}^{\rm sv} $.
The latter reduce to antiholomorphic integration constants $\overline{\fBRno{j_1 &j_2 &\ldots &j_\ell}{k_1 &k_2 &\ldots &k_\ell}}$, so by our assumption (\ref{trans.5as}) on the initial values, the claims (\ref{trans.5})
on the the series $Y^\tau_{\vec{\eta}}$ and (\ref{trans.1}) on the component integrals can only be violated 
by antiholomorphic quantities.

However, a purely antiholomorphic violation of uniform transcendentality is incompatible with
the reality properties (\ref{eq2.19}) of the component integrals: The contributions from holomorphic 
iterated Eisenstein integrals are uniformly transcendental by (\ref{trans.8alt}), so the same must be
true for those of the antiholomorphic ones. More precisely, by induction in the depth $\ell$
(which can be separated by only considering fixed orders in $\ap$), one can show that 
the $\overline{\fBRno{j_1 &j_2 &\ldots &j_\ell}{k_1 &k_2 &\ldots &k_\ell}}$ must share
the transcendental weights of the ${\cal E}_{\rm min}^{\rm sv}$, i.e.
\begin{align}
&\ \ \ \ \ \EsvBRno{j_1 &j_2 &\ldots &j_\ell}{k_1 &k_2 &\ldots &k_\ell} ,\, 
\bsvBRno{j_1 &j_2 &\ldots &j_\ell}{k_1 &k_2 &\ldots &k_\ell} \ \ \ \longleftrightarrow \ \ \ 
\te{transcendental weight} \ \ell + \sum_{i=1}^\ell j_i 
\label{trans.3} \\
&\EsvBRno{j_1  &\ldots &j_\ell}{k_1  &\ldots &k_\ell}
 R_{\vec{\eta}}\big( {\rm ad}_{ \epsilon_0}^{k_\ell-j_\ell-2} (\epsilon_{k_{\ell}})\ldots 
 {\rm ad}_{ \epsilon_0}^{k_1-j_1-2} (\epsilon_{k_1}) \big)
\ \ \ \longleftrightarrow \ \ \ 
\te{transcendental weight} \ 0 \, . \notag
\end{align}
The matching transcendental weights of ${\cal E}^{\rm sv}$ and $\bsv$ follow from
their relation (\ref{eq3.20}) and $y$ having weight 1. Based on (\ref{trans.3})
and (\ref{trans.7}), each term in the series (\ref{eq3.21alt}) has transcendental weight zero,
and the weight of $Y^\tau_{\vec{\eta}}$ agrees with that of the initial value $\Yhat^{i\infty}_{\vec{\eta}}$.
Hence, the claim (\ref{trans.5}) follows from the assumption (\ref{trans.5as}).

At two points, the initial value following from the
Laurent polynomials (\ref{eq4.1}) has transcendental weight 2, where we again use the vanishing 
weight of $\exp \Big( {-}\frac{ R_{\vec{\eta}}(\ep_{0}) }{4y} \Big)$. This confirms 
the claims (\ref{trans.1}) and (\ref{trans.5}) at $n=2$ since the series in ${\cal E}^{\rm sv}$ 
preserves the weight by (\ref{trans.3}).

At $n\geq 3$ points, the
dictionary between $\Yhat^{i\infty}_{\vec{\eta}}$ and $(n{+}2)$-point sphere integrals is under 
investigation \cite{yinprogress}. From a variety of Laurent-polynomials in $(n{\geq} 3)$-point MGFs
\cite{DHoker:2015gmr, Zerbini:2015rss, DHoker:2016quv} and
preliminary studies of their generating series, there is substantial evidence that the transcendental
weight of $\Yhat^{i\infty}_{\vec{\eta}}$ is $2(n{-}1)$.


\subsubsection{Basis integrals versus one-loop string amplitudes}

We emphasise that the discussion of this section is tailored to the conjectural basis
$Y^\tau_{\vec{\eta}}$ of torus integrals. In order to extract the transcendentality properties
of one-loop string amplitudes, it remains to 
\begin{itemize}
\item express their torus integral in terms of component integrals (\ref{trans.1}), 
where the expansion coefficients\footnote{The reduction of $(n{\geq}4)$-point 
gauge amplitudes of the heterotic string to a basis of
$Y^\tau_{(a_2,\ldots,a_n|b_2,\ldots,b_n)}$ also involves the
modular version
$\widehat {\rm G}_2 = {\rm G}_2- \frac{ \pi }{\Im \tau}$ of
$ {\rm G}_2 = \sum_{n \in \ZZ \setminus \{0 \} } \frac{1}{n^2} + \sum_{m \in \ZZ \setminus \{0 \}}
\sum_{n \in \ZZ} \frac{1}{(m \tau{+}n)^2}$ among the expansion coefficients \cite{Gerken:2018jrq}.}
may involve $\QQ$-linear combinations of ${\rm G}_{k}$ 
\cite{Dolan:2007eh, Broedel:2014vla, Gerken:2018jrq}
\item study the kinematic factors accompanying the component integrals 
\item integrate the modular parameter $\tau$ over the fundamental domain.
\end{itemize}
The subtle interplay of $\tau$-integration with the transcendental weights has
been explored in \cite{DHoker:2019blr} along with a powerful all-order result
for the integrated four-point integral $Y^\tau_{(0,0,0|0,0,0)}$
that was shown to enjoy a natural extension of uniform transcendentality.
In \cite{Basu:2019fim} it was argued that uniform transcendentality is violated starting from two loops.

The kinematic coefficients of $Y^\tau_{(a_2,\ldots,a_n|b_2,\ldots,b_n)}$ or
${\rm G}_k Y^\tau_{(a_2,\ldots,a_n|b_2,\ldots,b_n)}$ may feature
different transcendentality properties, depending on the string theory under investigation.
For the $\tau$-integrands of type-II superstrings, these kinematic factors
should be independent of $\ap$ in a suitable normalisation of the overall one-loop amplitude.
This can for instance be seen from the explicit $(4\leq n \leq 7)$-point results
in \cite{Green:1982sw, Green:2013bza, Mafra:2016nwr, Mafra:2018qqe} and the worldsheet 
supersymmetry in the RNS formalism, even in case of reduced spacetime
supersymmetry \cite{Bianchi:2015vsa, Berg:2016wux}. Hence, the $\tau$-integrands 
of $n$-point type-II amplitudes at one loop are expected to be uniformly transcendental.

Heterotic and bosonic strings in turn are known to involve tachyon poles in their chiral 
halves due to factors like $\partial f^{(k)}_{ij}$ and $f^{(k)}_{ij}f^{(\ell)}_{ij}$ in their CFT correlators. 
They can still be rewritten in terms of $Y^\tau_{(a_2,\ldots,a_n|b_2,\ldots,b_n)}$ via
integration by parts \cite{Gerken:2018jrq}, but the expansion coefficients may involve factors like $(1+s_{ij})^{-1}$
that break uniform transcendentality upon geometric-series expansion. Hence, even if
one-loop amplitudes of heterotic and bosonic strings can be expanded in a uniformly transcendental
integral basis, the overall $\tau$-integrand will generically lose this property through the 
kinematic factors. This effect is well-known from tree-level amplitudes in these theories
\cite{Huang:2016tag, Schlotterer:2016cxa, Azevedo:2018dgo}.


\section{Conclusion and outlook}
\label{sec7}

In this work, we have pinpointed the structure of the $\alpha'$-expansion for the generating
series $Y^\tau_{\vec{\eta}}$ of torus integrals seen in one-loop closed-string amplitudes. 
As main results we have
\begin{itemize}
\item[(i)] exhibited that, for any number of external legs, the polynomial structure in Mandelstam variables is explicitly determined to all orders in $\ap$ by \eqref{eq3.21}. This is based on  conjectural matrix realisations of certain derivations $\ep_k$ dual to Eisenstein series~\cite{Tsunogai}.
\item[(ii)] reduced the torus-puncture integrals to combinations of iterated Eisenstein
integrals with integration kernels $\tau^j {\rm G}_k $ with $k\geq 4$ \& $0\leq j \leq k{-}2$, their complex conjugates and MZVs from their behaviour at the cusp $\tau\to i\infty$.
\item[(iii)] developed methods to count the number of independent modular graph forms (w.r.t.\ relations over $\mathbb{Q[\te{MZV}]}$) that occur at given modular
weight or $\ap$-order in generic one-loop string amplitudes. Our approach exposes all relations between modular graph~forms. 
\end{itemize}
While the main formula (\ref{eq3.21}) of (i) is mainly driven by the holomorphic derivative 
$\partial_\tau Y^\tau_{\vec{\eta}}$, the appearance of antiholomorphic iterated Eisenstein 
integrals in (ii) is inferred from the
complex-conjugation properties of the series. We have presented
non-trivial depth-two examples but leave a detailed study of higher orders for the future. As to
(iii), while the methods of this work determine the counting of modular graph forms at arbitrary 
weight, it is an open problem to condense these mechanisms to a closed formula.

The combinations of (anti-)holomorphic Eisenstein integrals in the $\ap$-expansion of
this work are denoted by $\beta^{\rm sv}$ and we expect them to occur in Brown's generating
series of single-valued iterated Eisenstein integrals \cite{Brown:mmv, Brown:2017qwo, Brown:2017qwo2}.
Both the antiholomorphic constituents of $\beta^{\rm sv}$ and their linear combinations
that yield modular graph forms involve single-valued MZVs. Also Brown's construction
of single-valued iterated Eisenstein integrals involves single-valued MZVs which can be traced back
to the multiple modular values in the $S$-transformation of their holomorphic counterparts.

In our setup, by contrast, all single-valued MZVs descend from the degeneration of the $n$-point $Y^\tau_{\vec{\eta}}$
at the cusp, where $(n{+}2)$-point sphere integrals should be recovered. Following earlier
work of D'Hoker, Green \cite{DHoker:2019xef} and Zagier, Zerbini \cite{Zagier:2019eus},
we have made this fully explicit for the $Y^\tau_{\vec{\eta}}$ series at $n=2$ points, and the 
degenerations at $n\geq 3$ are under investigation \cite{yinprogress}. Hence, by combining the results of the present paper with the evaluation of genus-zero integrals at the cusp,
one can completely determine
\begin{itemize}
\item 
the explicit form of the combinations of holomorphic iterated Eisenstein integrals 
and their complex conjugates that yield non-holomorphic modular forms
\item the generating series of all possible genus-one integrals over closed-string punctures.
\end{itemize}
Finally, the appearance of the $\bsv$ in closed-string one-loop
amplitudes suggests a connection to open strings: At tree level, the sphere integrals in closed-string
amplitudes were identified as single-valued disk integrals occurring in open-string amplitudes
\cite{Schlotterer:2012ny, Stieberger:2013wea, Stieberger:2014hba, Schlotterer:2018abc, Brown:2018omk, 
Vanhove:2018elu, Brown:2019wna}. This calls for an extension of the single-valued map for (motivic) MZVs
\cite{Schnetz:2013hqa, Brown:2013gia} to the holomorphic iterated Eisenstein integrals in the 
$\ap$-expansion of one-loop open-string integrals \cite{Mafra:2019ddf, Mafra:2019xms}.
We hope to report on the relation between one-loop closed-string
integrals and single-valued versions of the open-string integrals 
in the near future.


\section*{Acknowledgements}

We are grateful to Johannes Broedel, Eric D'Hoker, Daniele Dorigoni, Justin Kaidi, Bram Verbeek, Federico Zerbini and in particular Nils Matthes and Erik Panzer for inspiring discussions. 
Moreover, we thank Daniele Dorigoni, Eric D'Hoker and Federico Zerbini for valuable comments on a draft version.
OS is grateful to Johannes Broedel, Erik Panzer and Federico Zerbini for a series of valuable discussions. We thank the organisers for providing a stimulating atmosphere during the ``Elliptic 2019'' conference where early parts of this work were carried out.
JG and AK thank the University of Uppsala for hospitality and JG furthermore thanks Chalmers University of Technology for hospitality during early stages of the project. OS is 
supported by the European Research Council under
ERC-STG-804286 UNISCAMP. JG is supported by the International Max Planck Research School for Mathematical
and Physical Aspects of Gravitation, Cosmology and Quantum Field Theory.

\appendix


\section{Lattice sums}
\label{appA}

This appendix reviews some more background material on modular graph forms.


\subsection{Fourier integrals}
\label{secA.1}

In order to perform the component integrals (\ref{eq2.14}) and (\ref{eq2.15}) 
 order by order in $\alpha'$, we employ the Fourier transformations of its
 doubly-periodic building blocks with respect to the real coordinates $u,v$ of the torus,
\beq
z=u\tau+v \, , \ \ \ \ \ \ u,v\in [0,1] \, ,
\label{eq2.20}
\eeq
namely
\begin{subequations}
\label{eq2.lat}
\begin{align}
  \Omega(z,\eta,\tau) &= \frac{1}{\eta} + \sum_{p \neq 0} \frac{e^{2\pi i \langle p, z\rangle}}{p+\eta}\,,
    \label{eq2.9}
\\
  f^{(w)}(z,\tau) &= (-1)^{w-1} \sum_{p\neq 0}\, \frac{e^{2\pi i \langle p, z\rangle}}{p^{w}}\, ,
    \label{eq2.10}
\\
  G(z,\tau) &=  \frac{\Im\tau}{\pi} \sum_{p\neq 0} \frac{e^{2\pi i \langle p,z \rangle}}{|p|^2}\,.
    \label{eq2.11}
\end{align}
\end{subequations}
The Fourier coefficients are labelled by discrete lattice momenta $p \in \ZZ + \tau \ZZ$ with the notation
\beq
p = m \tau + n \, , \ \ \ \ \ \ m,n\in \ZZ \ \ \ \ \ \
 \Longrightarrow \ \ \ \ \ \ \langle p,z \rangle = mv - nu = \frac{ p \bar z - \bar p z }{\tau - \bar \tau} \, .
\label{eq2.21}
\eeq
Note that the result (\ref{eq2.11}) for the Green function \cite{Green:1999pv} is conditionally convergent,
and (\ref{eq2.10}) only applies to $w\geq 1$. Moreover, the instances of (\ref{eq2.10}) at $w=1,2$
are not absolutely convergent for any $z$ on the torus, though the sum at $w=1$ is formally consistent with
$f^{(1)}(z,\tau) = - \partial_z G(z,\tau)$.


\subsection{Trihedral modular graph forms}
\label{secA.2}
\begin{figure}
  \begin{center}
    \tikzpicture[scale=0.8,line width=0.30mm]
    \scope[shift={(-1,0)},decoration={markings,mark=at position 0.8 with {\arrow[scale=1.3]{latex}}}]
    \fill (0,0) circle [radius=0.07]node[left]{$z_2$};
    \fill (5,0) circle [radius=0.07]node[right]{$z_1$};
    \fill (60:5) circle [radius=0.07]node[above]{$z_3$};
    \draw[postaction={decorate}] (5,0) .. controls (4,0.6) and (1,0.6) ..node[fill=white,transform shape]{$(a_{1},b_{1})$} (0,0);
    \draw (1.5,0.15) node{\scriptsize $\vdots$};
    \draw (3.5,0.15) node{\scriptsize $\vdots$};
    \draw[postaction={decorate}] (5,0) .. controls (4,-0.6) and (1,-0.6) ..node[fill=white,transform shape]{$(a_{Q},b_{Q})$} (0,0);

    \scope[rotate=60]
    \draw[postaction={decorate}] (0,0) .. controls (1,0.6) and (4,0.6) ..node[fill=white,transform shape,sloped]{$(c_{1},d_{1})$} (5,0);
    \draw (1.5,0.15) node[transform shape]{\scriptsize $\vdots$};
    \draw (3.5,0.15) node[transform shape]{\scriptsize $\vdots$};
    \draw[postaction={decorate}] (0,0) .. controls (1,-0.6) and (4,-0.6) ..node[transform shape,fill=white]{$(c_{R},d_{R})$} (5,0);
    \endscope

    \scope[shift={(60:5)},rotate=-60]
    \draw[postaction={decorate}] (0,0) .. controls (1,0.6) and (4,0.6) ..node[fill=white,transform shape]{$(e_{1},f_{1})$} (5,0);
    \draw (1.5,0.15) node[transform shape]{\scriptsize $\vdots$};
    \draw (3.5,0.15) node[transform shape]{\scriptsize $\vdots$};
    \draw[postaction={decorate}] (0,0) .. controls (1,-0.6) and (4,-0.6) ..node[transform shape,fill=white]{$(e_{S},f_{S})$} (5,0);
    \endscope
    \endscope
    \endtikzpicture
    \caption{Depiction of the worldsheet graph associated with the trihedral modular graph form 
      $\cformtri{  a_1  &\ldots &a_{Q} \\ b_1  &\ldots &b_{Q} }{ c_1  &\ldots &c_{R} \\ d_1  &\ldots &d_{R} }{ e_1  &\ldots &e_{S} \\ f_1  &\ldots &f_{S} }$.}
    \label{fig:trihed}
  \end{center}
\end{figure}

The $\ap$-expansion of three-point component integrals (\ref{eq2.15}) introduces MGFs of
trihedral topology, cf.\ figure~\ref{fig:trihed}, \cite{DHoker:2016mwo}
\beq
\cformtri{  a_1  &\ldots &a_{Q} \\ b_1  &\ldots &b_{Q} }
{ c_1  &\ldots &c_{R} \\ d_1  &\ldots &d_{R} }
{ e_1  &\ldots &e_{S} \\ f_1  &\ldots &f_{S} } (\tau) =  \sum_{\substack{ p_1,p_2,\ldots,p_Q \neq 0 
\\ k_1,k_2,\ldots,k_R \neq 0 \\\ell_1,\ell_2,\ldots,\ell_S \neq 0}} \frac{ \delta( \sum_{i=1}^Q p_i -  \sum_{i=1}^R k_i ) \delta( \sum_{i=1}^R k_i - \sum_{i=1}^S \ell_i ) }{ \big( \prod_{j=1}^Q p_j^{a_j} \bar p_j^{b_j} \big)
\big( \prod_{j=1}^R k_j^{c_j} \bar k_j^{d_j} \big) \big( \prod_{j=1}^S \ell_j^{e_j} \bar \ell_j^{f_j} \big)  }
 \, ,
\label{trihlattice}
\eeq
where all of $p_j,k_j,\ell_j$ refer to lattice momenta of the form (\ref{eq2.21}).
Similar to dihedral MGFs (\ref{eq2.22}), the integer exponents $a_j,b_j,\ldots,f_j$ 
lead to holomorphic and antiholomorphic modular weights $\sum_{j=1}^Q a_j + \sum_{j=1}^R c_j + \sum_{j=1}^S e_j $
and $\sum_{j=1}^Q b_j + \sum_{j=1}^R d_j + \sum_{j=1}^S f_j$, respectively.


The $\ap$-expansion of the simplest three-point component integral $Y^{\tau}_{(0,0|0,0)}$
is well-known to be expressible in terms of the banana-graph functions ${\rm D}_{n}$
in (\ref{eq4.11a}) and the trihedral MGFs 
\begin{align}
{\rm D}_{a,b,c}(\tau) &= \int \frac{ \dd^2 z_2}{\Im \tau} \frac{ \dd^2 z_3}{\Im \tau} \big( G(z_{12},\tau) \big)^a
\big( G(z_{23},\tau) \big)^b \big( G(z_{13},\tau) \big)^c \notag \\
&= \Big( \frac{ \Im \tau}{\pi} \Big)^{a+b+c} {\cal C}[ \underbrace{\begin{smallmatrix} 1 &1 &\ldots &1 \\1 &1 &\ldots &1 \end{smallmatrix} }_{a}
|  \underbrace{\begin{smallmatrix} 1 &1&\ldots &1 \\1& 1&\ldots &1 \end{smallmatrix} }_{b}
|  \underbrace{\begin{smallmatrix} 1 &1&\ldots &1 \\1& 1&\ldots &1 \end{smallmatrix} }_{c}](\tau) \, ,
\label{eq5.1}
\end{align}
namely \cite{Green:2008uj}
\begin{align}
Y^{\tau}_{(0,0|0,0)}(2,3|2,3) &= 1 + \sum_{n=2}^{\infty} \frac{1}{n!} (s_{12}^n+s_{13}^n + s_{23}^n) {\rm D}_n(\tau)
+ s_{12}s_{13} s_{23}{\rm D}_{1,1,1}(\tau)
\label{eq5.2} \\
&\hspace{-1.7cm}+ \frac{1}{2} s_{12}s_{13} s_{23}(s_{12} +s_{13}+s_{23}) {\rm D}_{2,1,1}(\tau)
+ \frac{1}{4} (s_{12}^2 s_{13}^2+s_{13}^2s_{23}^2 + s_{12}^2 s_{23}^2)  \big(  {\rm D}_2(\tau) \big)^2 \notag \\
&\hspace{-1.7cm}+ \frac{1}{6} s_{12}s_{13} s_{23}(s^2_{12} +s^2_{13}+s^2_{23}) {\rm D}_{3,1,1}(\tau)
+ \frac{1}{4} s_{12}s_{13} s_{23} (s_{12} s_{13}+s_{13}s_{23} + s_{12} s_{23})  {\rm D}_{2,2,1}(\tau) \notag \\
&\hspace{-1.7cm}+ \frac{1}{12} (s_{12}^2 s_{13}^3+s_{12}^3 s_{13}^2+
s_{13}^2s_{23}^3 +s_{13}^3s_{23}^2 + 
s_{12}^2 s_{23}^3 + s_{12}^3 s_{23}^2){\rm D}_2(\tau) {\rm D}_3(\tau) + {\cal O}(s_{ij}^6)\, .
\notag
\end{align}
Upon comparison with the $\ap$-expansion obtained from (\ref{eq3.21}),
i.e.\ by extracting the coefficient of $ \eta_{23}^{-1}\eta_3^{-1} \bar \eta_{23}^{-1} \bar \eta_3^{-1}$ in the
generating functions, we arrive at the following expressions for the trihedral
modular graph functions (\ref{eq5.1})
\begin{subequations}
\label{eq5.3vertex}
\begin{align}
{\rm D}_{1,1,1}&= -30 \bsvBRno{2}{6} + \frac{3 \zeta_5}{4 y^2}\,,
\label{eq5.3} \\
 {\rm D}_{2,1,1} &= -126 \bsvBRno{3}{8} - 18 \bsvBRno{2& 0}{4& 4} + 
 12  \zeta_3 \bsvBRno{0}{4} + \frac{ 5 \zeta_5}{ 12 y}  - \frac{ \zeta_3^2}{4 y^2}
 + \frac{9 \zeta_7}{16 y^3} \,,\label{eq5.4}  \\
 {\rm D}_{3,1,1} &=  540 \bsvBRno{1& 2}{4& 6} - 
 900 \bsvBRno{2& 1}{4& 6} + 540 \bsvBRno{2& 1}{6& 4} - 
 900 \bsvBRno{3& 0}{6& 4}  \notag \\
 & \quad -1458 \bsvBRno{4}{10} 
 + 600 \zeta_3 \bsvBRno{1}{6}  - \frac{90 \zeta_3}{y} \bsvBRno{2}{6}  \notag \\
 &\quad  + \frac{90 \zeta_5  }{y}  \bsvBRno{0}{4} - \frac{27 \zeta_5}{ 2 y^2}  \bsvBRno{1}{4} 
 -  \frac{\zeta_5}{30}  + \frac{105 \zeta_7}{32 y^2}
  - \frac{3 \zeta_3 \zeta_5}{2 y^3} + \frac{ 81 \zeta_9}{64 y^4}\,,
   \label{eq5.5} \\
 {\rm D}_{2,2,1} &= -1296 \bsvBRno{4}{10} - 240 \bsvBRno{2& 1}{4& 6} - 
 240 \bsvBRno{3& 0}{6& 4}  \notag \\
 &\quad + 160 \zeta_3  \bsvBRno{1}{6} 
 + \frac{24  \zeta_5}{y} \bsvBRno{0}{4}
  + \frac{11 \zeta_5}{45} + \frac{7 \zeta_7}{8 y^2}  
  - \frac{ \zeta_3 \zeta_5}{y^3 }+ \frac{9 \zeta_9}{8 y^4}\, .
 \label{eq5.6} 
 \end{align}
\end{subequations}
The combinations of $\bsv$ in the right-hand side
can be identified with those of ${\rm E}_k$, ${\rm E}_{2,2}$ and ${\rm E}_{2,3}$
in (\ref{2ptMGFBSV}). Hence, the $\bsv$-representations (\ref{eq5.3vertex})
expose the relations among the simplest ${\rm D}_{a,b,c}$ known from
the literature \cite{DHoker:2015gmr, DHoker:2015sve, DHoker:2016mwo, DHoker:2016quv}
\begin{subequations}
\label{5.rels}
\begin{align}
{\rm D}_{1,1,1}&= {\rm E}_3 = {\rm D}_{3} - \zeta_3\,,
 \label{eq5.7} 
\\
{\rm D}_{2,1,1} &= {\rm E}_{2,2} + \frac{ 9}{10} {\rm E}_4\,,
 \label{eq5.8} 
\\
{\rm D}_{3,1,1} &= \frac{15}{2} {\rm E}_{2,3} + 3 {\rm E}_2 {\rm E}_3 + \frac{81 }{35} {\rm E}_5 + \frac{ 7 \zeta_5}{40}\,,
 \label{eq5.9} 
\\ 
{\rm D}_{2,2,1}&= 2 {\rm E}_{2,3} + \frac{72 }{35} {\rm E}_5 + \frac{ 3 \zeta_5 }{10}\, .
 \label{eq5.10} 
\end{align}
\end{subequations}
%


\section{Derivations beyond three points}
\label{appderiv}

This appendix is dedicated to the matrices $R_{\vec{\eta}}(\ep_k)$ 
in the differential equation (\ref{eq2.34}) at multiplicities $n\geq 4$ 
which have been determined in \cite{Mafra:2019ddf, Mafra:2019xms, Gerken:2019cxz}.


\subsection{Four points}
\label{deriv.1}

The $6\times 6$ matrices $R_{\eta_2,\eta_3,\eta_4}(\ep_k) $ at four points are given 
by $R_{\eta_2,\eta_3,\eta_4}(\ep_2) = 0$ and
\begin{align}
&R_{\eta_2,\eta_3,\eta_4}(\ep_k) =  
 \eta_{2}^{k-2}  r (e_{2}) 
 + \eta_{3}^{k-2}  r (e_{3}) 
 + \eta_{4}^{k-2}  r (e_{4})
 + \eta_{23}^{k-2} r (e_{23})
+ \eta_{24}^{k-2}  r(e_{24})
+ \eta_{34}^{k-2}  r (e_{34})
 \notag \\
& \ \ \ \ \ + \eta_{234}^{k-2} r(e_{234})
  - \delta_{k,0}  \Big( \frac{1}{2} \sum_{j=2}^4 s_{1j} \partial_{\eta_j}^2 +  \frac{1}{2}\sum_{2\leq i <j}^4 s_{ij} (\partial_{\eta_i} {-} \partial_{\eta_j})^2 + 2\pi i  \sum_{j=2}^4 \bar \eta_j \partial_{\eta_j} \Big) 1_{6\times 6} 
  \label{der4pt}
\end{align}
for non-negative and even $k\neq2$. The $r(e_{\dots})$ are the $6\times6$ matrices appearing in
equation (4.21) and appendix C.1 of~\cite{Mafra:2019xms} as $r_{\vec\eta}(e_{\dots})$.


\subsection{\texorpdfstring{$n$ points}{n points}}
\label{deriv.2}

The $(n{-}1)!\times (n{-}1)!$ matrices $R_{\vec{\eta}}(\ep_k)$ 
in (\ref{eq2.34}) at $n$ points can be generated from
\cite{Mafra:2019ddf, Mafra:2019xms, Gerken:2019cxz}
\begin{align}
2\pi i \partial_\tau  &Y^\tau_{\vec{\eta}}(\sigma| 1,2,\ldots,n ) =
- \sum_{j=2}^n \wp\big( (\tau{-}\bar \tau) \eta_{j,j+1 \ldots n},\tau\big) 
Y^\tau_{\vec{\eta}}\big(\sigma| S[12 \ldots j{-}1, j(j{+}1) \ldots n] \big)
\label{step0} \\
&+\frac{1}{(\tau{-}\bar \tau)^2} \Big\{ 
2\pi i \sum_{j=2}^n \bar \eta_j \partial_{\eta_j} + \frac{1}{2} \sum_{j=2}^n s_{1j} \partial^2_{\eta_j} + \frac{ 1}{2}
\sum_{2\leq i<j}^n s_{ij}(\partial_{\eta_i}{-}\partial_{\eta_j})^2
\Big\}
Y^\tau_{\vec{\eta}}(\sigma| 1,2,\ldots,n ) \, .
\notag
\end{align}
The entries of $R_{\vec{\eta}}(\ep_k)$ can be read off through the following steps:
\begin{itemize}
\item expand out the $S[A,B]$-map via
\begin{align}
&Y^\tau_{\vec{\eta}}(\sigma| S[a_1 a_2\ldots a_p,b_1 b_2 \ldots b_q])  =  \sum_{i=1}^{p} \sum_{j=1}^{q} (-1)^{i-j+p-1}
s_{a_i b_j}  \label{step1} \\
&\ \ \ \ \ \times Y^\tau_{\vec{\eta}}\big(\sigma| (a_1 a_2\ldots a_{i-1} \shuffle a_{p} a_{p-1}\ldots a_{i+1}),a_i,b_j,
(b_{j-1}\ldots b_2 b_1 \shuffle b_{j+1} \ldots b_{q}) \big) \notag
\end{align}
 \item reduce the integrals to a basis of $Y^\tau_{\vec{\eta}}(\sigma|1,\ldots)$ by means of Kleiss--Kuijf relations
 following from Fay identities \cite{Mafra:2019xms}
\beq
Y^\tau_{\vec{\eta}}(\sigma|a_1, a_2,\ldots ,a_p,1,b_1 ,b_2,\ldots, b_q) = (-1)^{p} Y^\tau_{\vec{\eta}}\big(\sigma|1,(a_p \ldots a_2 a_1 \shuffle b_1 b_2\ldots b_q) \big) 
\label{step2}
\eeq 
\vspace{-6ex}
\item expand the Weierstra\ss\ functions in (\ref{step0}) via
\vspace{-1ex}
\beq
\wp(\eta,\tau) = \frac{1}{\eta^2} + \sum_{k=4}^\infty (k{-}1) \eta^{k-2}{\rm G}_k(\tau)
\label{step3}
\eeq
\vspace{-4.5ex}
\item insert (\ref{step1}), (\ref{step2}) and (\ref{step3}) into (\ref{step0}) and match the result with the form (\ref{eq2.34}) 
of the differential equation  to obtain the first row of the $R_{\vec{\eta}}(\ep_k)$
\item repeat the above steps for permutations $ \partial_\tau  Y^\tau_{\vec{\eta}}(\sigma| 1,\rho(2,3,\ldots,n) ) 
,\ \rho \in {\cal S}_{n-1}$ with appropriate relabelling of $s_{ij}$ and $\eta_j$
to generate the remaining rows of $R_{\vec{\eta}}(\ep_k)$
\end{itemize}


\section{Two-point results}
\label{appBgen}


\subsection{\texorpdfstring{$\alpha'$-expansions of component integrals}{alpha'-expansions of component integrals}}
\label{appB.1}

In this appendix, we collect further representative examples of $\ap$-expansions
of two-point component integrals (\ref{eq2.14}) to order 10. Their Laurent polynomials at the cusp 
are generated by (\ref{eq4.1}) and yield the component results in (\ref{eq4.ex}) and
\begin{subequations}
\label{morelaur}
\begin{align}
Y^\tau_{(0|2)} \, \big|_{q^0 \bar q^0} &=
s_{12} \Big(
\frac{ y}{180 } {-} \frac{ \zeta_3}{8 y^2 }
\Big)
+
s^2_{12} \Big(
\frac{y^2}{3780} {-} \frac{ \zeta_5}{16 y^3} \Big)
+
s^3_{12} \Big(
\frac{y^3}{22680} {+} \frac{ \zeta_3}{144 }
 {-} \frac{ 5 \zeta_5}{ 48 y^2} {+}  \frac{ \zeta_3^2}{16 y^3 }
 {-}  \frac{ 9 \zeta_7}{128 y^4} \Big) \notag \\
&\quad+ s^4_{12} \Big(
\frac{y^4}{449064 } + \frac{ y \zeta_3 }{ 648  } - \frac{ \zeta_3^2}{48 y^2}
 + \frac{ 9 \zeta_3 \zeta_5}{64 y^4} - \frac{7 \zeta_7}{64 y^3}
  - \frac{ 15 \zeta_9}{128 y^5} \Big)
+ {\cal O}(s_{12}^5)\,,
\label{eq4.555}
\\
Y^\tau_{(4|0)} \, \big|_{q^0 \bar q^0} &=
{-}s_{12} \Big( \frac{ 32 y^5}{945} + 2 \zeta_5 \Big) + 
 s_{12}^2 \Big(-\frac{ 16 y^6}{14175} - \frac{ 8 y^3 \zeta_3}{ 45} + 2 \zeta_3^2
  - \frac{  9 \zeta_7}{2 y} \Big) \notag \\
  &\quad + 
 s_{12}^3 \Big({-} \frac{ 16 y^7}{66825} - \frac{ 8 y^4 \zeta_3  }{135} + \frac{  y^2 \zeta_5 }{ 15} 
  - \frac{ 7 \zeta_7}{2} + \frac{  9 \zeta_3 \zeta_5}{y} - \frac{ 45 \zeta_9}{4 y^2} \Big)
 + {\cal O}(s_{12}^4)\,,
  \label{eq4.5}
\\
Y^\tau_{(3|5)} \, \big|_{q^0 \bar q^0} &=
-\frac{ y^3}{18900} + \frac{ 15 \zeta_7}{128 y^4} + 
 s_{12} \Big( \frac{ y^4}{1871100} + \frac{  \zeta_5}{480 y} - \frac{ 15 \zeta_3 \zeta_5}{  64 y^4} 
 + \frac{ 91 \zeta_9}{128 y^5} \Big)
 + {\cal O}(s_{12}^2)\,.
  \label{eq4.7}  
\end{align}
\end{subequations}
For generic $\tau$, the $\ap$-expansion (\ref{eqYctp2pt}) of the two-point
generating series yields components
\begin{subequations}
\begin{align}
& Y^\tau_{(4|0)}=
s_{12} ( 5 \bsvBR{4}{6}{\tau} - 2 \zeta_5 )
+ s_{12}^2 \Big( 63 \bsvBR{5}{8}{\tau} + 9 \bsvBR{2& 2}{4& 4}{\tau} - 6  \zeta_3 \bsvBR{2}{4}{\tau} + 
 2 \zeta_3^2 - \frac{9 \zeta_7}{2 y} \Big) \notag \\
 &\quad\quad  + s_{12}^3 \Big( 810 \bsvBR{6}{10}{\tau} - 15 \bsvBR{1& 4}{4& 6}{\tau} + 
 150 \bsvBR{2& 3}{4& 6}{\tau} + 30 \bsvBR{3& 2}{6& 4}{\tau}  \! \! \label{eq4.0}  \\
 &\quad\quad \ \ + 
 105 \bsvBR{4& 1}{6& 4}{\tau} - 100  \zeta_3 \bsvBR{3}{6}{\tau} + \frac{5   \zeta_3}{2 y}  \bsvBR{4}{6}{\tau} 
   - 42 \zeta_5 \bsvBR{1}{4}{\tau}  
 - \frac{3   \zeta_5}{y}  \bsvBR{2}{4}{\tau}  
  \nn\\
  &\quad\quad  \ \ 
    - \frac{7 \zeta_7}{2}  + \frac{9 \zeta_3 \zeta_5}{y}
  - \frac{45 \zeta_9}{4 y^2}\Big) +{\cal O}(s_{12}^4)\,,
\notag \\
 &Y^\tau_{(4|2)} =
-20 \bsvBR{3}{6}{\tau} + \frac{2 \zeta_5}{y}   + 
 s_{12} \Big({-}  18 \bsvBR{1& 2}{4& 4}{\tau} - 
    18 \bsvBR{2& 1}{4& 4}{\tau}   \label{Yexpand.20} \\
&\quad \ \  -315 \bsvBR{4}{8}{\tau}+ 12  \zeta_3 \bsvBR{1}{4}{\tau} 
    + \frac{    3  \zeta_3}{y} \bsvBR{2}{4}{\tau} - \frac{2 \zeta_3^2}{y} 
    + \frac{45 \zeta_7}{8 y^2}\Big) \notag \\
 &\quad  +  s_{12}^2 \Big( {-}4860 \bsvBR{5}{10}{\tau} + 15 \bsvBR{0& 4}{4& 6}{\tau} - 
    240 \bsvBR{1& 3}{4& 6}{\tau} - 450 \bsvBR{2& 2}{4& 6}{\tau} - 
    90 \bsvBR{2& 2}{6& 4}{\tau} \notag \\
&\quad \ \ - 480 \bsvBR{3& 1}{6& 4}{\tau} - 
    105 \bsvBR{4& 0}{6& 4}{\tau} + 300 \zeta_3 \bsvBR{2}{6}{\tau}  
    + \frac{  40  \zeta_3}{y} \bsvBR{3}{6}{\tau} - \frac{5 \zeta_3}{ 8 y^2} \bsvBR{4}{6}{\tau}  \notag\\
&\quad \ \  + 42  \zeta_5 \bsvBR{0}{4}{\tau} + \frac{ 48  \zeta_5}{y} \bsvBR{1}{4}{\tau}
    + \frac{9  \zeta_5}{ 4 y^2} \bsvBR{2}{4}{\tau} + \frac{7 \zeta_7}{2 y} - \frac{45 \zeta_3 \zeta_5}{4 y^2} 
     + \frac{ 135 \zeta_9}{8 y^3} \Big) + {\cal O}(s_{12}^3) \,,
    \notag \\
 &Y^\tau_{(3|5)} =
-105 \bsvBR{2}{8}{\tau} + \frac{15 \zeta_7}{128 y^4}  \label{Yexpand.26} \\
&\quad +  s_{12} \Big({-}3276 \bsvBR{3}{10}{\tau} - 90 \bsvBR{0& 2}{4& 6}{\tau} - 
    15 \bsvBR{0& 2}{6& 4}{\tau} - 120 \bsvBR{1& 1}{4& 6}{\tau} - 
    120 \bsvBR{1& 1}{6& 4}{\tau}  \notag \\
&\quad \ \ - 15 \bsvBR{2& 0}{4& 6}{\tau} - 
    90 \bsvBR{2& 0}{6& 4}{\tau} + 10 \zeta_3 \bsvBR{0}{6}{\tau}  + 
    \frac{  20  \zeta_3}{y} \bsvBR{1}{6}{\tau} + \frac{15  \zeta_3}{ 4 y^2} \bsvBR{2}{6}{\tau} \notag \\
&\quad \ \  + \frac{9  \zeta_5}{4 y^2} \bsvBR{0}{4}{\tau} + \frac{  3  \zeta_5}{4 y^3} \bsvBR{1}{4}{\tau}
    + \frac{3  \zeta_5}{ 128 y^4} \bsvBR{2}{4}{\tau} - \frac{15 \zeta_3 \zeta_5}{64 y^4} 
    + \frac{91 \zeta_9}{ 128 y^5} \Big) + {\cal O}(s_{12}^2)   \, .  \notag
\end{align}
\end{subequations}
Consistency with the Laurent polynomials in (\ref{eq4.ex}) and 
the asymptotics (\ref{more.bsv}) of $\beta^{\rm sv}$.


\subsection{\texorpdfstring{All $\beta^{\rm sv}$ at depth one involving ${\rm G}_8$ and ${\rm G}_{10}$}{All betasv at depth one involving G8 and G10}}
\label{appBd1}

The MGF-representations of the $\bsvBRno{j}{8}$ and $\bsvBRno{j}{10} $ in terms of and Cauchy--Riemann derivatives of ${\rm E}_4, {\rm E}_5$, complementing the discussion in section \ref{sec4.2b}.
 are given by
\vspace{-1ex}
 \begin{align}
& &  \bsvBRno{0}{10}  &= -  \frac{  ( \pi \overline{ \nabla} )^4 {\rm E}_5}{3870720 y^8} 
+ \frac{  \zeta_9}{  294912 y^8}\,,
\nn\\
 \bsvBRno{0}{8} & =  \frac{ ( \pi \overline{ \nabla} )^3 {\rm E}_4}{53760 y^6} +  \frac{  \zeta_7}{14336 y^6} \, , 
& \bsvBRno{1}{10} &=  \frac{  ( \pi \overline{ \nabla} )^3 {\rm E}_5}{967680 y^6} 
+  \frac{  \zeta_9}{73728 y^7} \,,
\nn\\
 \bsvBRno{1}{8}  &= -  \frac{  ( \pi \overline{ \nabla} )^2 {\rm E}_4}{13440 y^4} +  \frac{  \zeta_7}{3584 y^5} \, ,
& \bsvBRno{2}{10}  &= -  \frac{  ( \pi \overline{ \nabla} )^2 {\rm E}_5}{120960 y^4} 
+  \frac{  \zeta_9}{  18432 y^6}\,,
\nn\\
 \bsvBRno{2}{8}  &=  \frac{ \pi \overline{ \nabla} {\rm E}_4}{1680 y^2} +  \frac{  \zeta_7}{896 y^4} \, ,
&  \bsvBRno{3}{10}  &=  \frac{  \pi \overline{ \nabla} {\rm E}_5}{10080 y^2} +  \frac{  \zeta_9}{4608 y^5}\,,
 \label{eq4.19app} 
\\
  \bsvBRno{3}{8} &= -\frac{1}{140} {\rm E}_4 + \frac{ \zeta_7}{224 y^3} \, ,
 & \bsvBRno{4}{10} &= - \frac{1}{630} {\rm E}_5 + \frac{ \zeta_9}{1152 y^4}\,,
\nn\\
  \bsvBRno{4}{8} &= \frac{1}{105} \pi \nabla{\rm E}_4 + \frac{ \zeta_7}{56 y^2} \, ,
 & \bsvBRno{5}{10} &= \frac{1}{630} \pi \nabla {\rm E}_5 + \frac{ \zeta_9}{288 y^3}\,,
\nn\\
 \bsvBRno{5}{8} &= -\frac{2}{105} (\pi \nabla)^2{\rm E}_4 + \frac{\zeta_7}{14 y} \, ,
& \bsvBRno{6}{10}& = - \frac{2}{945} (\pi \nabla)^2 {\rm E}_5 + \frac{ \zeta_9}{72 y^2}\,,
\nn\\
  \bsvBRno{6}{8} &= \frac{8}{105} (\pi \nabla)^3{\rm E}_4 + \frac{ 2\zeta_7 }{7}   \, ,
&  \bsvBRno{7}{10} &= \frac{4}{945} (\pi \nabla)^3 {\rm E}_5 +  \frac{\zeta_9}{18 y}\,,
\nn\\
&&  \bsvBRno{8}{10} &= - \frac{16}{945} (\pi \nabla)^4 {\rm E}_5 +  \frac{2 \zeta_9}{9} \, ,\nn
\end{align}
consistent with the closed formulae (\ref{eq:eksol.1}) and (\ref{eq:bsv1}).
The inverse relations are
\vspace{-1ex}
\begin{align}
&&\frac{(\pi \overline{\nabla} )^4  {\rm E}_{5} }{y^8} &=
-3870720 \betasv{0\\10} + \frac{105 \zeta_9}{8 y^8}\,,
\notag \\
\frac{(\pi \overline{\nabla} )^3  {\rm E}_{4} }{y^6} &=
53760 \betasv{0\\8} - \frac{15 \zeta_7}{4 y^6} \, ,
&\frac{(\pi \overline{\nabla} )^3  {\rm E}_{5} }{y^6} &=
967680 \betasv{1\\10} - \frac{105 \zeta_9}{8 y^7}\,,
\notag \\
\frac{(\pi \overline{\nabla} )^2  {\rm E}_{4} }{y^4} &=
-13440 \betasv{1\\8} + \frac{15 \zeta_7}{4 y^5} \, ,
&\frac{(\pi \overline{\nabla} )^2  {\rm E}_{5} }{y^4} &=
-120960 \betasv{2\\10} + \frac{105 \zeta_9}{16 y^6}\,,
\notag \\
\frac{\pi \overline{\nabla}   {\rm E}_{4} }{y^2} &=
1680 \betasv{2\\8} - \frac{15 \zeta_7}{8 y^4} \, ,
&\frac{\pi \overline{\nabla}   {\rm E}_{5} }{y^2} &=
10080 \betasv{3\\10} - \frac{35 \zeta_9}{16 y^5}\,,
\notag \\
 {\rm E}_{4} &=
-140 \betasv{3\\8} + \frac{5 \zeta_7}{8 y^3} \, ,
 &{\rm E}_{5} &=
-630 \betasv{4\\10} + \frac{35 \zeta_9}{64 y^4}\,,
 \label{eq4.17app}  \\
\pi \nabla {\rm E}_{4} &=
105 \betasv{4\\8} - \frac{15 \zeta_7}{8 y^2} \, ,
&\pi \nabla {\rm E}_{5} &=
630 \betasv{5\\10} - \frac{35 \zeta_9}{16 y^3}\,,
\notag \\
(\pi \nabla)^2 {\rm E}_{4} &=
-\frac{105}{2} \betasv{5\\8} + \frac{15 \zeta_7}{4 y} \, ,
&(\pi \nabla)^2 {\rm E}_{5} &=
-\frac{945}{2} \betasv{6\\10} + \frac{105 \zeta_9}{16 y^2}\,,
\notag \\
(\pi \nabla)^3 {\rm E}_{4} &=
\frac{105}{8} \betasv{6\\8} - \frac{15 \zeta_7}{4} \, ,
&(\pi \nabla)^3 {\rm E}_{5} &=
\frac{945}{4} \betasv{7\\10} - \frac{105 \zeta_9}{8 y}\,,
\notag \\
&&(\pi \nabla)^4 {\rm E}_{5} &=
-\frac{945}{16} \betasv{8\\10} + \frac{105 \zeta_9}{8}\,,
\notag 
\end{align}
consistent with the closed formulae (\ref{eq:eksol.2}) and (\ref{eq:bsv0}).


\subsection{\texorpdfstring{Component integrals $Y^{\tau}_{(a|b)}$ at leading order}{Component integrals Yab at leading order}}
\label{closedform}

In this appendix, we derive both the closed depth-one formulae (\ref{eq:eksol.2}), (\ref{eq:bsv0})
relating non-holomorphic Eisenstein series to the $\bsv$ and the reality properties (\ref{eq4.14}) of the latter.
For this purpose, we investigate the $s_{12}^0$-order of the two-point component integrals
$Y_{(a|b)}^\tau$ in (\ref{eq2.14}) with $a{+}b \geq 4$, where the $(s_{12}\rightarrow 0)$-limit
can be performed at the level of the integrand. By the lattice-sum representations (\ref{eq2.10}) of
the $f^{(w)}$, this limit vanishes if $a=0$ or $b= 0$ and otherwise yields MGFs
\beq
Y_{(a|b)}^\tau = \frac{ (\tau{-}\bar \tau)^a }{(2\pi i)^b} \cform{a &0\\b &0} + {\cal O}(s_{12}) \, , \ \ \ \ \ \
a,b\neq 0 \, , \ \ \    \ \ \ a+b \geq 4 \, .
\label{newapp.1}
\eeq
Once the MGFs are expressed in terms of non-holomorphic Eisenstein series via (\ref{eq2.24}) and
(\ref{eq2.26}), the $s_{12}^0$-orders of the component integrals can be rewritten as ($k\geq 2, \ m<k$)
\begin{align}
Y_{(k|k)}^\tau &=  {\rm E}_k + {\cal O}(s_{12}) \,,
\notag \\
Y_{(k+m|k-m)}^\tau &= \frac{ (-4)^m (k{-}1)! (\pi \nabla)^m  {\rm E}_k}{(k{+}m{-}1)!} + {\cal O}(s_{12}) \,,
\label{newapp.2} \\
Y_{(k-m|k+m)}^\tau &= \frac{  (k{-}1)! (\pi \overline \nabla)^m  {\rm E}_k}{ (-4)^m (k{+}m{-}1)! y^{2m}} 
+ {\cal O}(s_{12}) \, .
\notag 
\end{align}
These results will now be compared with the $\ap$-expansion (\ref{eq3.21}) in terms of $\bsv$ and initial 
values. The latter can be inferred from the Laurent polynomials (\ref{eq4.1}) by acting with 
the two-point derivation $R_\eta(\ep_0)$ in (\ref{eq2.33}), and one obtains
\beq
\exp \Big(  {-} \frac{ R_{\vec{\eta}}(\ep_0) }{ 4 y} \Big)\Yhat^{i \infty}_\eta = \frac{1}{\eta \bar \eta} - \frac{2\pi i }{s_{12}} + 4\pi i \sum_{k=1}^\infty \zeta_{2k+1} \Big( \eta + \frac{ i\pi  \bar \eta }{2y} \Big)^{2k} + {\cal O}(s_{12}) \, .
\label{newapp.3}
\eeq
The $s_{12}^0$-order of the generating series $Y^\tau_{\eta}$ receives additional contributions
when the $\eta$-independent kinematic pole of (\ref{newapp.3}) is combined with one power of 
$s_{12}$ from the derivations $R_\eta(\ep_k)$. This order exclusively stems from the depth-one 
part of the series (\ref{eq3.21}) in $\bsv$,
\begin{align}
&\sum_{k=4}^{\infty} \sum_{j=0}^{k-2} \frac{(-1)^j (k{-}1) }{(k{-}j{-}2)!} \betasv{j\\k} R_\eta\big( {\rm ad}_{\ep_0}^{k-j-2}(\ep_k) \big)  \label{newapp.4} \\
& \ = s_{12} \sum_{k=4}^{\infty} \sum_{j=0}^{k-2} (2\pi i)^{k-j-2} \frac{ (k{-}1)!}{j! (k{-}j{-}2)!}   \eta^j \bar \eta^{k-j-2}   \betasv{j\\k} +{\cal O}(s_{12}^2,\partial_\eta)\, ,
\notag
\end{align}
where we have inserted $R_\eta(\ep_k)= s_{12} \eta^{k-2}$ and $R_\eta(\ep_0)= - 2\pi i \bar \eta \partial_\eta + {\cal O}(s_{12})$. In view of (\ref{newapp.3}) and (\ref{newapp.4}), the overall $(s_{12} \rightarrow 0)$-limit of the generating series is given by
\begin{align}
Y^\tau_\eta &= \frac{1}{\eta \bar \eta} - \frac{2\pi i }{s_{12}} + 4\pi i \sum_{k=1}^\infty \zeta_{2k+1} \Big( \eta + \frac{ i\pi  \bar \eta }{2y} \Big)^{2k}  \label{newapp.5}  \\
& \quad\quad- \sum_{k=4}^{\infty} \sum_{j=0}^{k-2} (2\pi i)^{k-j-1} \frac{ (k{-}1)!}{j! (k{-}j{-}2)!}   \eta^j \bar \eta^{k-j-2}   \betasv{j\\k} 
+{\cal O}(s_{12})\, .
\notag
\end{align}
By extracting the coefficients of $\eta^{a-1} \bar \eta^{b-1}$, we arrive at the following
leading orders of the component integrals (\ref{eq2.14})
\beq
Y^\tau_{(a|b)} =  \frac{ (a{+}b{-}2)! }{(a{-}1)!(b{-}1)!} \Big\{ \frac{ 2 \zeta_{a+b-1} }{ (4y)^{b-1} } - (a{+}b{-}1) \betasv{a-1\\a+b}  \Big\}
+ {\cal O}(s_{12})  \, , \ \ \ \ \
a,b\neq 0 \, , \ \ \  \ \ a+b \geq 4\, .
 \label{newapp.6} 
\eeq
Upon comparison with the earlier expression (\ref{newapp.2}) for the $s_{12}^0$-orders
in terms of non-holomor\-phic Eisenstein series, one can read off (\ref{eq:eksol.2}) 
by setting $(a,b)=(k,k)$ and (\ref{eq:bsv0}) by setting $(a,b)=(k{\pm} m,k{\mp} m)$
with $m<k$. Moreover, irrespective of the relation (\ref{newapp.2}) with ${\rm E}_k$, the
reality properties (\ref{eq2.18}) of the $Y^\tau_{(a|b)}$ enforce the $\bsv$ in (\ref{newapp.6})
to obey 
\beq
\overline{\betasv{a-1\\a+b}}  = (4y)^{a-b}  \betasv{b-1\\a+b} \, .
 \label{newapp.7} 
\eeq
This is equivalent to (\ref{eq4.14}), i.e.\ we have derived the reality properties of the $\bsv$ at
depth one from those of $Y^\tau_{(a|b)}$ and the explicit form of 
their $(s_{12}\rightarrow 0)$-limits (\ref{newapp.6}).


\subsection{Banana graphs}
\label{appB}

In this appendix, we list higher-order examples of the banana graph functions
\begin{align}
{\rm D}_n(\tau) = n! Y^\tau_{(0|0)} \, \big|_{s_{12}^n} = \int \frac{\dd^2z}{\Im\tau} (G(z,\tau))^n = \Big( \frac{ \Im \tau}{\pi} \Big)^n
{\cal C}[ \underbrace{\begin{smallmatrix} 1 &1 &\ldots &1 \\1&1&\ldots &1 \end{smallmatrix} }_{n}](\tau)\,,
\end{align} 
discussed in section \ref{sec4.2banana}, see (\ref{banana5}) for ${\rm D}_{n\leq 5}$
in terms of $\bsv$. By computing the $\ap$-expansion of the
component integral $Y_{(0|0)}^\tau$ in terms of the initial data and the $\bsv$, cf.\ (\ref{eq3.21}),
one finds the following new representations of ${\rm D}_6$ and ${\rm D}_7$
\begin{subequations}
\label{4.ban}
\begin{align}
{\rm D}_6 &=     -19440 \betasv{1&1&1\\4&4&4}+38880 \betasv{1&2&0\\4&4&4}+38880 \betasv{2&0&1\\4&4&4} 
 -116640 \betasv{2&1&0\\4&4&4} \nn\\
&\quad +45360 \betasv{1&3\\4&8}-272160 \betasv{2&2\\4&8}
 +45360 \betasv{3&1\\8&4}-272160 \betasv{4&0\\8&4}\nn\\
&\quad  
+18000 \betasv{2&2\\6&6} -144000 \betasv{3&1\\6&6}
-144000 \betasv{4&0\\6&6}-831600 \betasv{5\\12} \nn\\
&\quad 
+\frac{3240\zeta_3}{y} \betasv{1&1\\4&4}
-25920 \zeta_3\betasv{0&1\\4&4}
+77760 \zeta_3\betasv{1&0\\4&4}
-\frac{6480\zeta_3}{y} \betasv{2&0\\4&4}
 \\
&\quad 
+181440 \zeta_3\betasv{2\\8}
-\frac{7560\zeta_3}{y} \betasv{3\\8}
-600 \zeta_3\betasv{2\\6}
+57600\zeta_5 \betasv{0\\6}
   \nn\\
&\quad  
+\frac{14400\zeta_5}{y} \betasv{1\\6} 
-\frac{450\zeta_5}{y^2} \betasv{2\\6}
 +7200 \zeta_5\betasv{0\\4}
   -\frac{900\zeta_5}{y} \betasv{1\\4}
  \nn\\
&\quad
 -\frac{4320\zeta_3^2}{y} \betasv{0\\4} 
 +\frac{270\zeta_3^2}{y^2} \betasv{1\\4} 
 -\frac{405\zeta_7}{2y^3} \betasv{1\\4} 
+\frac{4860\zeta_7}{y^2} \betasv{0\\4}
\nn\\
&\quad 
+34 \zeta_3^{2}
+\frac{483 \zeta_7}{2y}
-\frac{135 \zeta_3 \zeta_5}{y^2}
+\frac{45 \zeta_3^{3}}{y^3}
+\frac{405 \zeta_9}{2y^3}
 -\frac{675 \zeta_3 \zeta_7}{4y^4} 
-\frac{675 \zeta_5^{2}}{8 y^4}
 +\frac{4725 \zeta_{11}}{32y^5}\,,\nn\\
{\rm D}_7  &=  
-226800 \betasv{1&1&2\\4&4&6}+907200 \betasv{1&2&1\\4&4&6}-226800 \betasv{1&2&1\\4&6&4}+907200 \betasv{1&3&0\\4&6&4} \nn\\
&\quad+453600 \betasv{2&0&2\\4&4&6}-2721600 \betasv{2&1&1\\4&4&6}+907200 \betasv{2&1&1\\4&6&4}
 -226800 \betasv{2&1&1\\6&4&4}\nn\\
&\quad -1814400 \betasv{2&2&0\\4&4&6}-4989600 \betasv{2&2&0\\4&6&4}+453600 \betasv{2&2&0\\6&4&4} +907200 \betasv{3&0&1\\6&4&4}\nn\\
&\quad -2721600 \betasv{3&1&0\\6&4&4} -1814400 \betasv{4&0&0\\6&4&4}   +2041200 \betasv{1&4\\4&10} -16329600 \betasv{2&3\\4&10}  \nn\\
&\quad   +2041200 \betasv{4&1\\10&4} -16329600 \betasv{5&0\\10&4} +529200 \betasv{2&3\\6&8} -6350400 \betasv{3&2\\6&8} \nn\\
&\quad   -12700800 \betasv{4&1\\6&8} +529200 \betasv{3&2\\8&6} -6350400 \betasv{4&1\\8&6} -12700800 \betasv{5&0\\8&6} \nn\\
&\quad-61916400 \betasv{6\\14} +1209600 \zeta_3 \betasv{2&0\\4&6} -302400 \zeta_3 \betasv{0&2\\4&6}
+1814400\zeta_3 \betasv{1&1\\4&6}  \nn\\
&\quad   -604800\zeta_3 \betasv{1&1\\6&4} +3326400 \zeta_3 \betasv{2&0\\6&4} 
+\frac{37800\zeta_3}{y} \betasv{1&2\\4&6} -\frac{151200\zeta_3}{y} \betasv{2&1\\4&6}  \nn\\
 &\quad +\frac{37800\zeta_3}{y} \betasv{2&1\\6&4}  -\frac{151200\zeta_3}{y} \betasv{3&0\\6&4} +10886400\zeta_3 \betasv{3\\10}   -\frac{340200\zeta_3}{y} \betasv{4\\10} \nn\\
&\quad +7560\zeta_3 \betasv{1&1\\4&4} 
 -15120\zeta_3 \betasv{2&0\\4&4}
+725760\zeta_5 \betasv{0&0\\4&4}
-\frac{90720\zeta_5}{y} \betasv{0&1\\4&4}
\\
&\quad  +\frac{272160\zeta_5}{y} \betasv{1&0\\4&4} +\frac{5670\zeta_5}{y^2} \betasv{1&1\\4&4}
 -\frac{11340\zeta_5}{y^2} \betasv{2&0\\4&4}
 -17640\zeta_3 \betasv{3\\8}
    \nn\\
&\quad  +5080320\zeta_5 \betasv{1\\8}
    +\frac{635040\zeta_5}{y} \betasv{2\\8} -\frac{13230\zeta_5}{y^2} \betasv{3\\8}
+168000\zeta_5 \betasv{1\\6}
\nn\\
&\quad  -\frac{10500 \zeta_5}{y}\betasv{2\\6} -403200 \zeta_3^2\betasv{0\\6}  
  -\frac{100800\zeta_3^2}{y} \betasv{1\\6} +\frac{3150\zeta_3^2}{y^2} \betasv{2\\6}
    \nn\\
&\quad   +\frac{907200\zeta_7}{y} \betasv{0\\6}
  +\frac{113400\zeta_7}{y^2} \betasv{1\\6}  -\frac{4725\zeta_7}{2y^3} \betasv{2\\6}
 -1806\zeta_5 \betasv{1\\4} 
 \nn\\
&\quad  +10080 \zeta_3^2\betasv{0\\4} 
-\frac{1260\zeta_3^2}{y} \betasv{1\\4}
+\frac{52920\zeta_7}{y} \betasv{0\\4}  -\frac{6615\zeta_7}{2y^2} \betasv{1\\4} \nn \\
&\quad
-\frac{68040\zeta_3\zeta_5}{y^2} \betasv{0\\4}  +\frac{2835\zeta_3\zeta_5}{y^3} \betasv{1\\4}
 +\frac{56700\zeta_9}{y^3} \betasv{0\\4}
 -\frac{14175\zeta_9}{8y^4} \betasv{1\\4}  \nn\\
&\quad +\frac{877 \zeta_7}{2}+\frac{819 \zeta_3\zeta_5}{y}
 -\frac{105 \zeta_3^{3}}{y^2} +\frac{5145\zeta_9}{4y^2} 
  -\frac{1575 \zeta_5^{2}}{2y^3} -\frac{1575 \zeta_3 \zeta_7}{y^3} +\frac{86625 \zeta_{11}}{32y^4} \nn \\
&\quad  +\frac{4725\zeta_3^2\zeta_5}{4y^4}  -\frac{33075\zeta_5\zeta_7}{16y^5}   -\frac{33075\zeta_3\zeta_9}{16y^5}+\frac{297675 \zeta_{13}}{128y^6}\,.\nn
\end{align}
 \end{subequations}
An expression for the five-loop banana function ${\rm D}_6$ in terms of simpler modular graph functions can be found in~\cite{DHoker:2016quv}. It takes the following form when expressed in terms
of the modular graph functions of \cite{Broedel:2018izr},
\begin{align}
{\rm D}_6 &=  720 {\rm E}_{2,2,2} + 48 {\rm E}_{2,4}  - 640 {\rm E}_{3,3} + 
1200 {\rm E}_{3,3}' + 300 {\rm E}_{6} \\
&\quad\quad+ 360 {\rm E}_{2} {\rm E}_{2,2} 
+ 54 {\rm E}_{2} {\rm E}_{4} + 10 {\rm E}_{3}^2  
+ 15 {\rm E}_{2}^3  + 20 \zeta_3{\rm E}_{3} + 10 \zeta_3^2 \, ,
\notag 
\end{align}
see (\ref{morebsv1}) to (\ref{morebsv4}) for $\beta^{\rm sv}$-representations of ${\rm E}_{2,2,2} , {\rm E}_{2,4} 
,{\rm E}_{3,3} $ and ${\rm E}_{3,3}' $ and (\ref{eq2.com}) for their lattice-sum representations.


\section{Initial data without MZVs for four points}
\label{app:n4}

As explained in section~\ref{sec6.1}, the MZV-free part of the initial data $\Yhat^{i\infty}_{\vec\eta}$ plays an important role in determining the leading modular behaviour~\eqref{eqSlead} of the $\beta^{\rm sv}$. The results for $n{=}2$ and $n{=}3$, given by $\Yhat^{i\infty}_{\eta} \big|_{\zeta_k{=}0}= \frac{1}{\eta \bar \eta} - \frac{ 2\pi i }{s_{12}}$ and~\eqref{3ptlow}, admit a simple generalisation to $n{=}4$ points. Based on this we conjecture that the MZV-free part at four points is given by
\begin{subequations}
\begin{align}
 \Yhat^{i\infty}_{\vec\eta}(2{,}3{,}4|2{,}3{,}4) \big|_{\zeta_{k}{=}0}
&=  
\frac{1}{\eta _4 \eta _{34} \eta _{234} \bar{\eta }_4 \bar{\eta }_{34} \bar{\eta }_{234}}
-\frac{2\pi i}{\eta _4 \eta _{34}  \bar{\eta }_4 \bar{\eta }_{34} s_{12}}
-\frac{2\pi i}{\eta _4 \eta _{234}  \bar{\eta }_4 \bar{\eta }_{234} s_{23}}\\
&\quad
-\frac{2\pi i}{\eta _{34} \eta _{234}  \bar{\eta }_{34} \bar{\eta }_{234} s_{34}}
+\frac{(2\pi i)^2}{  \eta _4 \bar{\eta }_4 s_{12} s_{123}}
+\frac{(2\pi i)^2}{ \eta _4 \bar{\eta }_4s_{23} s_{123} }
+\frac{(2\pi i)^2}{ \eta _{34}\bar{\eta }_{34} s_{12} s_{34}}\nn\\
&\quad
+\frac{(2\pi i)^2}{ \eta _{234}\bar{\eta }_{234} s_{23} s_{234}}
+\frac{(2\pi i)^2}{\eta _{234}\bar{\eta }_{234} s_{34} s_{234} }
-\frac{(2\pi i)^3}{s_{12} s_{34} s_{1234}}
-\frac{(2\pi i)^3}{s_{12} s_{123} s_{1234}}\nn\\
&\quad
-\frac{(2\pi i)^3}{s_{23} s_{123} s_{1234}}
-\frac{(2\pi i)^3}{s_{23} s_{234} s_{1234}}
-\frac{(2\pi i)^3}{s_{34} s_{234} s_{1234}}\,,\nn\\
\Yhat^{i\infty}_{\vec\eta}(2{,}3{,}4|2{,}4{,}3)\big|_{\zeta_k{=}0} 
&=  
\frac{1}{\eta _3 \eta _{34} \eta _{234} \bar{\eta }_4 \bar{\eta }_{34} \bar{\eta }_{234}}
-\frac{2\pi i}{\eta _3 \eta _{34}  \bar{\eta }_4 \bar{\eta }_{34}s_{12}}
+\frac{2\pi i}{\eta _{34} \eta _{234}  \bar{\eta }_{34} \bar{\eta }_{234}s_{34}}\\
&\quad
-\frac{(2\pi i)^2}{\eta _{34}  \bar{\eta }_{34}s_{12} s_{34}}
-\frac{(2\pi i)^2}{\eta _{234}  \bar{\eta }_{234}s_{34} s_{234}}
+\frac{(2\pi i)^3}{s_{12} s_{34} s_{1234}}
+\frac{(2\pi i)^3}{s_{34} s_{234} s_{1234}}\,,\nn\\
 \Yhat^{i\infty}_{\vec\eta}(2{,}3{,}4|3{,}2{,}4)\big|_{\zeta_k{=}0} 
&=  
\frac{1}{\eta _4 \eta _{24} \eta _{234} \bar{\eta }_4 \bar{\eta }_{34} \bar{\eta }_{234}}
+\frac{2\pi i}{\eta _4 \eta _{234}  \bar{\eta }_4 \bar{\eta }_{234}s_{23}}
-\frac{(2\pi i)^2}{\eta _4 \bar{\eta }_4  s_{23} s_{123}}
\\
&\quad 
-\frac{(2\pi i)^2}{\eta _{234}  \bar{\eta }_{234}s_{23} s_{234}}
+\frac{(2\pi i)^3}{s_{23} s_{123} s_{1234}}
+\frac{(2\pi i)^3}{s_{23} s_{234} s_{1234}}\,,\nn\\
 \Yhat^{i\infty}_{\vec\eta}(2{,}3{,}4|3{,}4{,}2)\big|_{\zeta_k{=}0} 
&=  
\frac{1}{\eta _2 \eta _{24} \eta _{234} \bar{\eta }_4 \bar{\eta }_{34} \bar{\eta }_{234}}
-\frac{2\pi i}{\eta _2 \eta _{234}  \bar{\eta }_{34} \bar{\eta }_{234}s_{34}}
-\frac{(2\pi i)^2}{\eta _{234}  \bar{\eta }_{234}s_{34} s_{234}}\\
&\quad
+\frac{(2\pi i)^3}{s_{34} s_{234} s_{1234}}\,,\nn\\
 \Yhat^{i\infty}_{\vec\eta}(2{,}3{,}4|4{,}2{,}3)\big|_{\zeta_k{=}0} 
&=  
\frac{1}{\eta _3 \eta _{23} \eta _{234} \bar{\eta }_4 \bar{\eta }_{34} \bar{\eta }_{234}}
-\frac{2\pi i}{\eta _{23} \eta _{234}  \bar{\eta }_4 \bar{\eta }_{234}s_{23}}
-\frac{(2\pi i)^2}{\eta _{234} \bar{\eta }_{234} s_{23} s_{234}}\\
&\quad
+\frac{(2\pi i)^3}{s_{23} s_{234} s_{1234}}\,,\nn\\
 \Yhat^{i\infty}_{\vec\eta}(2{,}3{,}4|4{,}3{,}2)\big|_{\zeta_k{=}0} 
&=  
\frac{1}{\eta _2 \eta _{23} \eta _{234} \bar{\eta }_4 \bar{\eta }_{34} \bar{\eta }_{234}}
+\frac{2\pi i}{\eta _{23} \eta _{234}\bar{\eta }_4 \bar{\eta }_{234} s_{23} }
+\frac{2\pi i}{\eta _2 \eta _{234} \bar{\eta }_{34} \bar{\eta }_{234}s_{34} }\\*
&\quad
+\frac{(2\pi i)^2}{\eta _{234}  \bar{\eta }_{234}s_{23} s_{234}}
+\frac{(2\pi i)^2}{\eta _{234}  \bar{\eta }_{234}s_{34} s_{234}}
-\frac{(2\pi i)^3}{s_{23} s_{234} s_{1234}}
-\frac{(2\pi i)^3}{s_{34} s_{234} s_{1234}}\,,\nn
\end{align}
\end{subequations}
where the notation $\big|_{\zeta_{k}{=}0}$ informally instructs to disregard MZVs of arbitrary depth.
We have verified that this conjectured initial data is annihilated by the four-point instance of the operator $R_{\vec\eta}(\epsilon_0)$ reviewed in appendix \ref{deriv.1}. It is moreover manifestly invariant under the scaling $\eta_j \rightarrow \bar \tau \eta_j$  and $\bar \eta_j \rightarrow \frac{ \bar \eta_j}{\bar \tau}$ which is crucial for obtaining the leading modular weight of the $\bsv$ given in~\eqref{eqSlead}. Note that the pattern of kinematic poles follows similar patterns in tree-level amplitudes of the open string~\cite{Broedel:2013tta}.


\section{\texorpdfstring{Detailed expressions for $\Esv$, $\bsv$ and modular graph forms}{Detailed expressions for Esv, betasv and modular graph forms}}
\label{appDetail}

In this appendix, we collect for reference the expressions for the $\bsv$ and $\Esv$ in terms of the basis of modular graph forms presented in table~\ref{tab:Ebasis}. Expressions for
all $\betasv{j\\k}, \EsvBRno{j}{k}$ and $\betasv{j_1&j_2\\k_1&k_2}, \EsvBRno{j_1&j_2}{k_1&k_2}$ 
with $k\leq 10$ and $k_1{+}k_2\leq 10$ in terms of MGFs
can also be found in the ancillary file within the arXiv submission of this paper.

\subsection{\texorpdfstring{Expressions for $\bsv$ in terms of modular graph forms}{Expressions for betasv in terms of modular graph forms}}
\label{appbetasv}

The expressions for $\bsv$ of weight $10$ in terms of modular graph forms can be obtained by inverting the relations~\eqref{excusp.2} and \eqref{excusp.4}--\eqref{excusp.6}.

For $\betasv{j_1&j_2\\6&4}$ with $0\leq j_1\leq 4$ and $0\leq j_2\leq 2$
we find
\begin{align}
\label{bsvwt10}
\betasv{0&0\\6&4}&= -\frac{(\pi\overline{\nabla})^3\overline{\BB}'_{2,3}}{483840y^6} + \frac{ \zeta_{5} \pi\overline{\nabla}\EE_{2}}{15360y^6} -\frac{\zeta_{3}}{907200y} 
-\frac{\zeta_{5}}{345600y^3} +\frac{\zeta_{3}\zeta_{5}}{30720y^6} \,,\nn\\
\betasv{0&1\\6&4}&= \frac{(\pi\overline{\nabla})^2\overline{\BB}'_{2,3}}{120960y^4}
+\frac{(\pi\overline{\nabla})^2\EE_{2, 3}}{7680y^4}
+\frac{\EE_{2}(\pi\overline{\nabla})^2\EE_{3}}{5760y^4}
+ \frac{ (\pi\overline{\nabla}\EE_{2}) \pi\overline{\nabla}\EE_{3}}{11520y^4}  
-\frac{ \zeta_{5}\EE_{2}}{3840y^4}
\nn\\
&\quad -\frac{\zeta_{3}}{226800}
+\frac{\zeta_{5}}{172800y^2}
-\frac{7 \zeta_{7}}{92160y^4}
 +\frac{\zeta_{3}\zeta_{5}}{7680y^5}\,,\nn\\
\betasv{0&2\\6&4}&=
\frac{\pi\overline{\nabla} \,  \overline{\BB}'_{2,3}}{15120y^2}
-\frac{\pi\overline{\nabla}\,  \overline{\BB}_{2,3}}{2160y^2}
-\frac{\pi\overline{\nabla}\EE_{2, 3}}{2880y^2}
-\frac{\EE_{2}\pi\overline{\nabla}\EE_{3}}{1440y^2}
+\frac{\EE_{3}\pi\overline{\nabla}\EE_{2}}{1440y^2}
-\frac{ (\pi\nabla \EE_{2}) (\pi\overline{\nabla})^2\EE_{3}}{1440y^4}\nn\\
&\quad 
-\frac{ \zeta_{3}  \pi\overline{\nabla}\EE_{2}}{6048y^2}+\frac{ \zeta_{5}\pi\nabla\EE_{2}}{960y^4}
-\frac{\zeta_{5}}{21600y}-\frac{\zeta_{3}^{2}}{5040y^2}
-\frac{7 \zeta_{7}}{11520y^3} +\frac{\zeta_{3}\zeta_{5}}{1920y^4}\, , \nn\\
\betasv{1&0\\6&4}&=\frac{(\pi\overline{\nabla})^2\overline{\BB}'_{2,3}}{120960y^4}+\frac{ \zeta_{5}\pi\overline{\nabla}\EE_{2}}{3840y^5}+\frac{\zeta_{3}}{907200} -\frac{\zeta_{5}}{86400y^2}+\frac{7 \zeta_{7}}{184320y^4} +\frac{\zeta_{3}\zeta_{5}}{7680y^5}\,,\nn\\
\betasv{1&1\\6&4}&= \frac{\pi\overline{\nabla} \, \overline{\BB}_{2,3}}{4320y^2}
-\frac{\pi\overline{\nabla} \, \overline{\BB}'_{2,3}}{7560y^2}
 -\frac{\pi\overline{\nabla}\EE_{2, 3}}{1152y^2}
 -\frac{\EE_{2}\pi\overline{\nabla}\EE_{3}}{2880y^2}
 -\frac{\EE_{3}\pi\overline{\nabla}\EE_{2}}{2880y^2}
 \nn\\*
&\quad -\frac{ \zeta_{3} \pi\overline{\nabla}\EE_{2}}{60480y^2}  -\frac{ \zeta_{5} \EE_{2}}{960y^3} 
+\frac{\zeta_{5}}{43200y}+\frac{\zeta_{3}^{2}}{20160y^2} -\frac{7 \zeta_{7}}{46080y^3}
+\frac{\zeta_{3}\zeta_{5}}{1920y^4}\,,\nn\\
\betasv{1&2\\6&4}&=-\frac{\BB_{2,3}'}{1260}+\frac{(\pi\overline{\nabla} \EE_{3})\pi\nabla \EE_{2}}{360y^2}
+ \frac{ \zeta_{5} \pi\nabla\EE_{2}}{240y^3}+\frac{\zeta_{5}}{43200}-\frac{\zeta_{3}^{2}}{2520y}
-\frac{7 \zeta_{7}}{3840y^2} +\frac{\zeta_{3}\zeta_{5}}{480y^3}\,,\nn\\
\betasv{2&0\\6&4}&=
-\frac{\pi\overline{\nabla}\, \overline{\BB}'_{2,3}}{15120y^2}
-\frac{\zeta_{3} \pi\overline{\nabla}\EE_{2}}{15120y^2}
+ \frac{ \zeta_{5}  \pi\overline{\nabla}\EE_{2}}{960y^4}
-\frac{\zeta_{5}}{21600y} -\frac{\zeta_{3}^{2}}{30240y^2}+\frac{7 \zeta_{7}}{23040y^3} +\frac{\zeta_{3}\zeta_{5}}{1920y^4}\,,\nn\\
\betasv{2&1\\6&4}&=
-\frac{\EE_{2, 3}}{144}
+\frac{\BB_{2,3}}{540}-\frac{\BB'_{2,3}}{756}
+\frac{\EE_{2}\EE_{3}}{360}
-\frac{\zeta_{3} \EE_{2}}{1080}-\frac{\zeta_{5}\EE_{2}}{240y^2}
 +\frac{\zeta_{5}}{21600}
+\frac{\zeta_{3}^{2}}{3780y} +\frac{\zeta_{3}\zeta_{5}}{480y^3} \, ,\nn\\
\betasv{2&2\\6&4}&=
\frac{\pi\nabla\BB'_{2,3}}{945}
-\frac{\EE_{3}\pi\nabla\EE_{2}}{45}
+\frac{ \zeta_{5} \pi\nabla\EE_{2}}{60y^2}-\frac{\zeta_{3}^{2}}{1890} -\frac{7 \zeta_{7}}{1440y} + \frac{\zeta_{3}\zeta_{5}}{120y^2}\,,\nn\\
\betasv{3&0\\6&4}&=-\frac{\EE_{2, 3}}{120}-\frac{\BB'_{2,3}}{1260}+\frac{ \zeta_{5} \pi\overline{\nabla}\EE_{2}}{240y^3}-\frac{\zeta_{5}}{4800} -\frac{\zeta_{3}^{2}}{2520y}  +\frac{7 \zeta_{7}}{3840y^2}
+\frac{\zeta_{3}\zeta_{5}}{480y^3} \,,\nn\\
\betasv{3&1\\6&4}&=
\frac{\pi\nabla\EE_{2, 3}}{72}
+\frac{2 \pi\nabla\BB'_{2,3}}{945}
-\frac{ \pi\nabla \BB_{2,3}}{270}
-\frac{\EE_{2}\pi\nabla\EE_{3}}{180}
+\frac{\EE_{3}\pi\nabla\EE_{2}}{180}\nn\\
&\quad +\frac{ \zeta_{3} \pi\nabla\EE_{2}}{540}-\frac{ \zeta_{5} \EE_{2}}{60y}
+\frac{\zeta_{3}^{2}}{1260} +\frac{7 \zeta_{7}}{2880y} +\frac{\zeta_{3}\zeta_{5}}{120y^2}\,,\\
\betasv{3&2\\6&4}&=
-\frac{2 (\pi\nabla)^2\BB'_{2,3}}{945}
+\frac{2 (\pi\nabla\EE_{2})\pi\nabla\EE_{3}}{45}
+\frac{\zeta_{5} \pi\nabla \EE_{2}}{15y}
-\frac{7 \zeta_{7}}{720}
+\frac{\zeta_{3}\zeta_{5}}{30y}\,,\nn\\
\betasv{4&0\\6&4}&= \frac{\pi\nabla\EE_{2, 3}}{180} +\frac{\pi\nabla\BB_{2,3}}{135}
-\frac{\pi\nabla\BB'_{2,3}}{945} -\frac{\EE_{3}\pi\nabla\EE_{2}}{90}
+\frac{\EE_{2}\pi\nabla\EE_{3}}{90}
\nn\\
&\quad 
-\frac{ \zeta_{3} \pi\nabla \EE_{2}}{270}
+\frac{ \zeta_{5} \pi\overline{\nabla}\EE_{2}}{60y^2}
-\frac{\zeta_{3}^{2}}{315}+\frac{7 \zeta_{7}}{720y} 
+\frac{\zeta_{3}\zeta_{5}}{120y^2}\,,\nn\\
\betasv{4&1\\6&4}&=
-\frac{(\pi\nabla)^2\EE_{2, 3}}{30}
-\frac{2 (\pi\nabla)^2\BB'_{2,3}}{945}
-\frac{(\pi\nabla \EE_{2})\pi\nabla\EE_{3}}{45}
-\frac{ \zeta_{5}\EE_{2}}{15}
+\frac{7 \zeta_{7}}{360}
+\frac{\zeta_{3}\zeta_{5}}{30y}\,,\nn\\
\betasv{4&2\\6&4}&=
\frac{8 (\pi\nabla)^3\BB'_{2,3}}{945}
-\frac{8 (\pi\nabla\EE_{2}) (\pi\nabla)^2\EE_{3}  }{45}
+\frac{4 \zeta_{5} \pi\nabla\EE_{2}}{15}
+\frac{2 \zeta_{3}\zeta_{5}}{15}\,.\nn
\end{align}
The analogous expressions for $\betasv{j_1&j_2\\4&6}$ follow from this
via the shuffle relations (\ref{eq3.shuff2}). 

\subsection{\texorpdfstring{Expressions for $\Esv$ in terms of modular graph forms}{Expressions for Esv in terms of modular graph forms}}
\label{appEsv}

The MGF expressions for the $\Esv$ can be obtained by applying formula~\eqref{eq3.23} to the expressions of the $\bsv$ in terms of MGFs given in the preceding section~\ref{appbetasv}.

At depth one, we find
\vspace{-3ex}
\begin{align}
&&\esv{0\\6}&=-\frac{\EE_{3}}{80y^2}-\frac{\Re[\pi \nabla\EE_{3}]}{120y^3}
-\frac{\Re[(\pi \nabla)^2\EE_{3}]}{480y^4}
\,\nn,\\
\esv{0\\4}&=\frac{\EE_{2}}{12y}+\frac{\Re[\pi \nabla \EE_{2}]}{12y^2} \, \ \ \ \ \ \ 
&\esv{1\\6}&=\frac{\EE_3}{40y} +\frac{\pi \overline{\nabla}\EE_{3}}{240y^2}+\frac{\pi\nabla \EE_{3}}{80y^2}+\frac{(\pi\nabla)^2\EE_{3}}{240y^3}
\,,\nn\\
\esv{1\\4}&=-\frac{\EE_{2}}{6}-\frac{\pi\nabla \EE_{2}}{6y}\, ,
&\esv{2\\6}&=-\frac{\EE_3}{30}-\frac{\pi\nabla \EE_{3}}{30y}-\frac{(\pi\nabla)^2\EE_{3}}{60y^2}\,,\\
\esv{2\\4}&=\frac{2 \pi\nabla \EE_{2}}{3}+\frac{2\zeta_3}{3}\,, 
&\esv{3\\6}&=\frac{\pi\nabla \EE_{3}}{15}+\frac{(\pi\nabla)^2\EE_{3}}{15y}\,,\nn\\
&&\esv{4\\6}&=-\frac{4(\pi\nabla)^2\EE_{3}}{15}+\frac{2\zeta_5}{5}
\,.\nn
\end{align}

Using~\eqref{eq3.23} and the closed formulae (\ref{eq:eksol.1}) and (\ref{eq:bsv1}) one can also find closed expressions for the $\Esv$ at depth one  ($0 \leq m \leq k{-}1$)
\begin{align}
\esv{k-1+m\\2k} &= - (-4)^m \frac{(k{-}1)!(k{-}1{-}m)!}{(2k{-}1)!} \sum_{p=0}^{k-1-m} \frac{(\pi\nabla)^{m+p} \EE_k}{ p!\, y^p} + \delta_{m,k-1} \frac{2\zeta_{2k-1}}{2k{-}1}\,,
\notag \\
\esv{k-1\\2k} &=  -  \frac{[(k{-}1)!]^2}{(2k{-}1)!} \sum_{p=0}^{k-1}\frac{(\pi\nabla)^p \EE_k}{p!\, y^p}\,,
 \\
\esv{k-1-m \\2k} &= - \frac{(k{-}1)!(k{-}1{+}m)!}{  (-4)^{m} (2k{-}1)!} \bigg\{ \sum_{p=0}^{m-1}\frac{ (k{-}1{-}m{+}p)!\, (\pi\overline\nabla)^{m-p} \EE_k}{(k{-}1{+}m{-}p)! \, p!\, y^{2m-p}} + \sum_{p=m}^{k-1+m}  \frac{(\pi\nabla)^{p-m}\EE_k}{p!\, y^p} \bigg\}\,.
\notag 
\end{align}

At depth two, the expressions (\ref{eq4.27}) for $\bsvBRno{j_1 &j_2}{4 &4}$ for $j_1>j_2$ are equivalent to following formulae for the $\Esv$
\begin{align}
\esv{1&0\\4&4}&=\frac{\EE_{2, 2}}{72y}+\frac{\pi\overline{\nabla}\EE_{2, 2}}{144y^2}+\frac{\pi\nabla\EE_{2, 2}}{144y^2}-\frac{\EE_{2}^{2}}{144y}-\frac{(\pi\nabla \EE_{2})^{2}}{288y^3}-\frac{\EE_{2}\pi\nabla \EE_{2}}{144y^2}-\frac{\zeta_{3}}{2160} \,,\nn\\
\esv{2&0\\4&4}&=-\frac{\EE_{2, 2}}{18}-\frac{\pi\nabla\EE_{2, 2}}{18y}+\frac{(\pi\nabla\EE_{2})^{2}}{72y^2}+\frac{\zeta_{3}\pi\overline{\nabla}\EE_{2}}{36y^2}+\frac{\zeta_3 \pi\nabla\EE_{2}}{36y^2}+\frac{\zeta_{3}\EE_{2}}{18y} \,, \label{eq:compare}\\
\esv{2&1\\4&4}&= \frac{\pi\nabla\EE_{2, 2}}{9} -\frac{(\pi\nabla \EE_{2})^{2}}{18y}-\frac{\zeta_{3} \pi\nabla \EE_{2}}{9y}-\frac{\zeta_{3}\EE_{2}}{9}+\frac{5\zeta_{5}}{108}\,,\nn
\end{align}
while the analogous expressions for $\esv{j_1&j_2\\4&4}$ with $j_1 \leq j_2$ 
follow from the shuffle relations (\ref{eq3.shuff2}) and the depth-one results.

 
 The expressions in the ${\rm G}_4{\rm G}_6$ sector for the $\Esv$ in terms of MGFs are much longer compared to the $\bsv$, showing the advantage of working with the $\bsv$. 
 At depth two we find for $\esv{j_1&j_2\\6&4}$ the expressions
\begin{align}
\label{eq:esvexpr}
\esv{0&0\\6&4}&=-\frac{\EE_{2}(\pi\overline{\nabla})^2\EE_{3}}{11520y^5}+\frac{\pi\overline{\nabla}\overline{\BB}_{2,3}}{11520y^4}-\frac{\pi\overline{\nabla}\overline{\BB}'_{2,3}}{11520y^4} -\frac{\pi\nabla \BB_{2,3}}{11520y^4}+\frac{\pi\nabla\BB'_{2,3}}{11520y^4}-\frac{(\pi\overline{\nabla})^2\EE_{2, 3}}{15360y^5}\nn\\
&\quad+\frac{(\pi\nabla)^2\EE_{2, 3}}{15360y^5}-\frac{\EE_{2}\EE_{3}}{1920y^3} -\frac{(\pi\overline{\nabla})^2\EE_{3}\pi\nabla\EE_{2}}{23040y^6}-\frac{(\pi\overline{\nabla}\EE_{2})\pi\overline{\nabla}\EE_{3}}{23040y^5}-\frac{\zeta_3\pi\overline{\nabla}\EE_{2}}{23040y^4}\nn\\
&\quad -\frac{(\pi\nabla)^2\EE_{3}\pi\nabla\EE_{2}}{23040y^6}+\frac{\zeta_{3} \pi\nabla\EE_{2}}{23040y^4}-\frac{\EE_{3}\pi\nabla \EE_{2}}{2560y^4}-\frac{\BB_{2,3}}{2880y^3}+\frac{\BB'_{2,3}}{2880y^3}\nn\\
&\quad -\frac{\EE_{2}\pi\overline{\nabla}\EE_{3}}{4608y^4}-\frac{(\pi\overline{\nabla})^3\overline{\BB}'_{2,3}}{483840y^6}+\frac{(\pi\nabla)^3\BB'_{2,3}}{483840y^6}+\frac{\zeta_{3}\EE_{2}}{5760y^3}-\frac{(\pi\overline{\nabla}\EE_{3})\pi\nabla\EE_{2}}{5760y^5}\nn\\
&\quad -\frac{\EE_{2}\pi\nabla\EE_{3}}{7680y^4}-\frac{\EE_{3}\pi\overline{\nabla}\EE_{2}}{7680y^4} -\frac{(\pi\nabla\EE_{2})\pi\nabla\EE_{3}}{7680y^5}-\frac{(\pi\overline{\nabla})^2\overline{\BB}'_{2,3}}{80640y^5}+\frac{(\pi\nabla)^2\BB'_{2,3}}{80640y^5}\nn\\
&\quad -\frac{7\pi\overline{\nabla}\EE_{2, 3}}{15360y^4}+\frac{7\pi\nabla \EE_{2, 3}}{15360y^4}+\frac{7\EE_{2,3}}{3840y^3}\,,\nn\\
\esv{0&1\\6&4}&=\frac{11\pi\overline{\nabla}\EE_{2, 3}}{11520y^3}+\frac{(\pi\overline{\nabla}\EE_{2})\pi\overline{\nabla}\EE_{3}}{11520y^4}-\frac{\pi\nabla\EE_{2, 3}}{1152y^3}+\frac{(\pi\overline{\nabla})^2\overline{\BB}'_{2,3}}{120960y^4}-\frac{(\pi\nabla)^3\BB'_{2,3}}{120960y^5}\nn\\
&\quad +\frac{\BB_{2,3}}{1440y^2}-\frac{\BB'_{2,3}}{1440y^2} +\frac{(\pi\overline{\nabla}\EE_{3})\pi\nabla \EE_{2}}{1440y^4}+\frac{\zeta_{3} \pi\overline{\nabla}\EE_{2}}{17280y^3}+\frac{\EE_{2}\pi\overline{\nabla}\EE_{3}}{1920y^3}-\frac{\zeta_3}{226800}\nn\\
&\quad -\frac{(\pi\nabla)^2\BB'_{2,3}}{24192y^4}+\frac{\EE_{2}\pi\nabla \EE_{3}}{2880y^3}-\frac{\zeta_{3}\EE_{2}}{2880y^2}-\frac{\EE_{2, 3}}{384y^2}+\frac{\pi\nabla\BB_{2,3}}{4320y^3}-\frac{\pi\nabla\BB'_{2,3}}{4320y^3}\nn\\
&\quad +\frac{\zeta_{5}}{57600y^2}+\frac{\EE_{2}(\pi\overline{\nabla})^2\EE_{3}}{5760y^4} +\frac{\EE_{3}\pi\overline{\nabla}\EE_{2}}{5760y^3}+\frac{(\pi\overline{\nabla})^2\EE_{3}\pi\nabla\EE_{2}}{5760y^5}+\frac{(\pi\nabla)^2\EE_{3}\pi\nabla\EE_{2}}{5760y^5}\nn\\
&\quad +\frac{\EE_{3}\pi\nabla\EE_{2}}{576y^3}+\frac{(\pi\overline{\nabla})^2\EE_{2, 3}}{7680y^4}-\frac{(\pi\nabla)^2\EE_{2, 3}}{7680y^4}-\frac{\pi\overline{\nabla}\overline{\BB}_{2,3}}{8640y^3}+\frac{\pi\overline{\nabla}\overline{\BB}'_{2,3}}{8640y^3}+\frac{\EE_{2}\EE_{3}}{960y^2}\nn\\
&\quad -\frac{\zeta_{3} \pi\nabla \EE_{2}}{8640y^3}+\frac{7(\pi\nabla\EE_{2})\pi\nabla\EE_{3}}{11520y^4}\,,\nn\\
\esv{0&2\\6&4}&=-\frac{\EE_{3}\pi\nabla\EE_{2}}{120y^2}+\frac{\BB'_{2,3}}{1260y}-\frac{\zeta_{5}}{14400y}-\frac{\EE_{2}\pi\overline{\nabla}\EE_{3}}{1440y^2}+\frac{\EE_{3}\pi\overline{\nabla}\EE_{2}}{1440y^2}-\frac{(\pi\overline{\nabla})^2\EE_{3}\pi\nabla\EE_{2}}{1440y^4}\nn\\*
&\quad -\frac{(\pi\nabla)^2\EE_{3}\pi\nabla\EE_{2}}{1440y^4}+\frac{\pi\overline{\nabla}\overline{\BB}'_{2,3}}{15120y^2} -\frac{\pi\overline{\nabla}\overline{\BB}_{2,3}}{2160y^2}+\frac{\pi\nabla\BB'_{2,3}}{2520y^2}-\frac{\pi\overline{\nabla}\EE_{2, 3}}{2880y^2}+\frac{(\pi\nabla)^3\BB'_{2,3}}{30240y^4}\nn\\*
&\quad -\frac{(\pi\overline{\nabla}\EE_{3})\pi\nabla\EE_{2}}{360y^3}-\frac{(\pi\nabla\EE_{2})\pi\nabla\EE_{3}}{360y^3}-\frac{\zeta_{3} \pi\overline{\nabla} \EE_{2}}{6048y^2}+\frac{(\pi\nabla)^2\BB'_{2,3}}{7560y^3}\,,\nn\\
\esv{1&0\\6&4}&=-\frac{\zeta_4}{115200y^2}+\frac{(\pi\overline{\nabla})^2\overline{\BB}'_{2,3}}{120960y^4}-\frac{(\pi\nabla)^3\BB'_{2,3}}{120960y^5}+\frac{\BB_{2,3}}{1440y^2}-\frac{\BB'_{2,3}}{1440y^2}+\frac{\EE_{3}\pi\nabla \EE_{2}}{1440y^3}\nn\\*
&\quad +\frac{\zeta_{3} \pi\overline{\nabla} \EE_{2}}{17280y^3}+\frac{\pi\overline{\nabla}\EE_{2, 3}}{2304y^3}-\frac{\EE_{2, 3}}{240y^2} -\frac{(\pi\nabla)^2\BB'_{2,3}}{24192y^4}+\frac{\EE_{2}\pi\nabla\EE_{3}}{2880y^3}-\frac{\zeta_{3}\EE_{2}}{2880y^2}\nn\\*
&\quad +\frac{(\pi\nabla\EE_{2})\pi\nabla\EE_{3}}{2880y^4} -\frac{(\pi\nabla)^2\EE_{2, 3}}{3840y^4}+\frac{(\pi\nabla)\BB_{2,3}}{4320y^3}-\frac{\pi\nabla\BB'_{2,3}}{4320y^3}+\frac{\EE_{2}\pi\overline{\nabla}\EE_{3}}{5760y^3}\nn\\*
&\quad +\frac{\EE_{3}\pi\overline{\nabla}\EE_{2}}{5760y^3}+\frac{(\pi\overline{\nabla}\EE_{3})\pi\nabla\EE_{2}}{5760y^4}+\frac{(\pi\nabla)^2\EE_{3}\pi\nabla\EE_{2}}{5760y^5}-\frac{\pi\nabla\EE_{2, 3}}{720y^3}-\frac{\pi\overline{\nabla}\overline{\BB}_{2,3}}{8640y^3}\nn\\*
&\quad +\frac{\pi\overline{\nabla}\overline{\BB}'_{2,3}}{8640y^3} -\frac{\zeta_{3} \pi\nabla \EE_{2}}{8640y^3}+\frac{\zeta_3}{907200}+\frac{\EE_2\EE_3}{960y^2}\,,\nn\\
\esv{1&1\\6&4}&=-\frac{\pi\overline{\nabla}\EE_{2, 3}}{1152y^2}+\frac{\zeta_{3}\EE_{2}}{1440y}-\frac{(\pi\overline{\nabla}\EE_{3})\pi\nabla\EE_{2}}{1440y^3} -\frac{(\pi\nabla)^2\EE_{3}\pi\nabla\EE_{2}}{1440y^4}-\frac{\pi\nabla\BB_{2,3}}{1440y^2}+\frac{\pi\nabla\BB'_{2,3}}{1680y^2}\nn\\
&\quad+\frac{(\pi\nabla)^2\EE_{2, 3}}{1920y^3} +\frac{\EE_{2,3}}{192y}-\frac{\EE_{2}\pi\overline{\nabla}\EE_{3}}{2880y^2}-\frac{\EE_{3}\pi\overline{\nabla}\EE_{2}}{2880y^2}+\frac{\zeta_{3} \pi\nabla \EE_{2}}{2880y^2}+\frac{(\pi\nabla)^3\BB'_{2,3}}{30240y^4}\nn\\
&\quad -\frac{\EE_{3}\pi\nabla\EE_{2}}{320y^2}+\frac{\pi\nabla\EE_{2, 3}}{384y^2}+\frac{\pi\overline{\nabla}\overline{\BB}_{2,3}}{4320y^2}-\frac{\EE_{2}\EE_{3}}{480y}-\frac{\zeta_{5}}{57600y}-\frac{(\pi\nabla\EE_{2})\pi\nabla\EE_{3}}{576y^3}\nn\\
&\quad -\frac{\zeta_{3}\pi\overline{\nabla}\EE_{2}}{60480y^2} -\frac{\BB_{2,3}}{720y}-\frac{\pi\overline{\nabla}\overline{\BB}'_{2,3}}{7560y^2}+\frac{(\pi\nabla)^2\BB'_{2,3}}{7560y^3}+\frac{\BB'_{2,3}}{840y}-\frac{\EE_{2}\pi\nabla\EE_{3}}{960y^2}\,,\nn\\
\esv{1&2\\6&4}&=\frac{(\pi\nabla\EE_{2})\pi\nabla\EE_{3}}{120y^2}-\frac{\BB'_{2,3}}{1260}-\frac{\pi\nabla\BB'_{2,3}}{1260y}-\frac{(\pi\nabla)^2\BB'_{2,3}}{2520y^2}+\frac{(\pi\overline{\nabla}\EE_{3})\pi\nabla\EE_{2}}{360y^2}\nn\\*
&\quad+\frac{(\pi\nabla)^2\EE_{3}\pi\nabla\EE_{2}}{360y^3}+\frac{\zeta_5}{43200}+\frac{\EE_{3}\pi\nabla\EE_{2}}{60y} -\frac{(\pi\nabla)^3\BB'_{2,3}}{7560y^3}\,,\\
\esv{2&0\\6&4}&=-\frac{\BB_{2,3}}{1080y}+\frac{11\EE_{2,3}}{1440y}+\frac{11\pi\nabla\EE_{2, 3}}{2880y^2}-\frac{\EE_{2}\pi\nabla\EE_{3}}{1440y^2}-\frac{\EE_{3}\pi\nabla\EE_{2}}{1440y^2}-\frac{(\pi\nabla)^2\EE_{3}\pi\nabla\EE_{2}}{1440y^4}\nn\\
&\quad -\frac{(\pi\nabla\EE_{2})\pi\nabla\EE_{3}}{1440y^3}-\frac{\zeta_{3} \pi\overline{\nabla} \EE_{2}}{15120y^2}-\frac{\pi\overline{\nabla}\overline{\BB}'_{2,3}}{15120y^2}+\frac{\pi\nabla\BB'_{2,3}}{1890y^2}+\frac{\zeta_3\EE_2}{2160y}+\frac{(\pi\nabla)^2\EE_{2, 3}}{960y^3}\nn\\
&\quad+\frac{\BB'_{2,3}}{945y} -\frac{\pi\nabla\BB_{2,3}}{2160y^2}+\frac{\zeta_{5}}{28800y} +\frac{(\pi\nabla)^3\BB'_{2,3}}{30240y^4}+\frac{\zeta_{3} \pi\nabla \EE_{2}}{4320y^2}-\frac{\EE_{2}\EE_{3}}{720y}+\frac{(\pi\nabla)^2\BB'_{2,3}}{7560y^3}\,,\nn\\
\esv{2&1\\6&4}&=-\frac{\zeta_{3}\EE_{2}}{1080}-\frac{\zeta_{3} \pi\nabla\EE_{2}}{1080y}-\frac{\EE_{2,3}}{144}-\frac{\pi\nabla\EE_{2, 3}}{144y}+\frac{\zeta_5}{21600}+\frac{(\pi\nabla\EE_{2})\pi\nabla\EE_{3}}{240y^2}+\frac{\EE_{2}\EE_{3}}{360}\nn\\*
&\quad -\frac{(\pi\nabla)^2\BB'_{2,3}}{2520y^2}+\frac{\EE_{2}\pi\nabla\EE_{3}}{360y} +\frac{\EE_{3}\pi\nabla\EE_{2}}{360y}+\frac{(\pi\nabla)^2\EE_{3}\pi\nabla\EE_{2}}{360y^3}-\frac{(\pi\nabla)^2\EE_{2, 3}}{480y^2}+\frac{\BB_{2,3}}{540}\nn\\*
&\quad +\frac{\pi\nabla\BB_{2,3}}{540y}-\frac{(\pi\nabla)^3\BB'_{2,3}}{7560y^3}-\frac{\BB'_{2,3}}{756}-\frac{\pi\nabla\BB'_{2,3}}{756y}\,,\nn\\
\esv{2&2\\6&4}&=\frac{(\pi\nabla)^3\BB'_{2,3}}{1890y^2}-\frac{\zeta_{3}^{2}}{1890}-\frac{\EE_{3}\pi\nabla\EE_{2}}{45}-\frac{(\pi\nabla\EE_{2})\pi\nabla\EE_{3}}{45y}
 -\frac{(\pi\nabla)^2\EE_{3}\pi\nabla\EE_{2}}{90y^2}\nn\\
 &\quad +\frac{(\pi\nabla)^2\BB'_{2,3}}{945y}+\frac{\pi\nabla\BB'_{2,3}}{945}\,,\nn\\
\esv{3&0\\6&4}&=-\frac{\EE_{2, 3}}{120}-\frac{\pi\nabla\EE_{2, 3}}{120y}-\frac{\BB'_{2,3}}{1260}-\frac{\pi\nabla\BB'_{2,3}}{1260y}-\frac{(\pi\nabla)^2\EE_{2, 3}}{240y^2}+\frac{(\pi\nabla)^2\EE_{3}\pi\nabla\EE_{2}}{360y^3}-\frac{\zeta_5}{4800}\nn\\
&\quad -\frac{(\pi\nabla)^2\BB'_{2,3}}{2520y^2}-\frac{(\pi\nabla)^3\BB'_{2,3}}{7560y^3}\,,\nn\\
\esv{3&1\\6&4}&=\frac{(\pi\nabla)^2\EE_{2, 3}}{120y}+\frac{\zeta_3^2}{1260}-\frac{\EE_{2}\pi\nabla\EE_{3}}{180}+\frac{\EE_{3}\pi\nabla\EE_{2}}{180}-\frac{(\pi\nabla\EE_{2})\pi\nabla\EE_{3}}{180y} +\frac{(\pi\nabla)^3\BB'_{2,3}}{1890y^2}\nn\\*
&\quad-\frac{\pi\nabla\BB_{2,3}}{270}+\frac{\zeta_{3} \pi\nabla\EE_{2}}{540}+\frac{\pi\nabla\EE_{2, 3}}{72} -\frac{(\pi\nabla)^2\EE_{3}\pi\nabla\EE_{2}}{90y^2}+\frac{(\pi\nabla)^2\BB'_{2,3}}{945y}+\frac{2\pi\nabla\BB'_{2,3}}{945}\,,\nn\\
\esv{3&2\\6&4}&=\frac{2(\pi\nabla)^2\EE_{3}\pi\nabla\EE_{2}}{45y}+\frac{2(\pi\nabla\EE_{2})\pi\nabla\EE_{3}}{45}-\frac{2(\pi\nabla)^2\BB'_{2,3}}{945}-\frac{2(\pi\nabla)^3\BB'_{2,3}}{945y}-\frac{7\zeta_7}{720}\,,\nn\\
\esv{4&0\\6&4}&=\frac{\pi\nabla\BB_{2,3}}{135}+\frac{\pi\nabla\EE_{2, 3}}{180}+\frac{(\pi\nabla)^3\BB'_{2,3}}{1890y^2}-\frac{\zeta_{3} \pi\nabla \EE_{2}}{270}+\frac{\zeta_{5}\EE_{2}}{30y} -\frac{\zeta_{3}^{2}}{315}+\frac{\zeta_{5} \pi\overline{\nabla} \EE_{2}}{60y^2}\nn\\
&\quad+\frac{(\pi\nabla)^2\EE_{2, 3}}{60y}+\frac{\zeta_{5} \pi\nabla \EE_{2}}{60}+\frac{\EE_{2}\pi\nabla\EE_{3}}{90}-\frac{\EE_{3}\pi\nabla\EE_{2}}{90}-\frac{(\pi\nabla)^2\EE_{3}\pi\nabla\EE_{2}}{90y^2}\nn\\
&\quad +\frac{(\pi\nabla \EE_{2})\pi\nabla\EE_{3}}{90y}+\frac{(\pi\nabla)^2\BB'_{2,3}}{945y} -\frac{\pi\nabla\BB'_{2,3}}{945}\,,\nn\\
\esv{4&1\\6&4}&=-\frac{\zeta_{5}\EE_{2}}{15}-\frac{\zeta_{5} \pi\nabla\EE_{2}}{15y}-\frac{(\pi\nabla)^2\EE_{2, 3}}{30}-\frac{(\pi\nabla\EE_{2})\pi\nabla\EE_{3}}{45} +\frac{2(\pi\nabla)^2\EE_{3}\pi\nabla\EE_{2}}{45y}\nn\\*
&\quad-\frac{2(\pi\nabla)^2\BB'_{2,3}}{945}-\frac{2(\pi\nabla)^3\BB'_{2,3}}{945y}+\frac{7\zeta_7}{360}\,,\nn\\
\esv{4&2\\6&4}&=\frac{2\zeta_{3}\zeta_{5}}{15}+\frac{4\zeta_{5}\pi\nabla\EE_{2}}{15}-\frac{8(\pi\nabla)^2\EE_{3}\pi\nabla \EE_{2}}{45}+\frac{8(\pi\nabla)^3\BB'_{2,3}}{945}\,.\nn
\end{align}
The results for  $\esv{j_1&j_2\\4&6}$ follow by shuffle relations and the depth-one results.


\section{\texorpdfstring{$S$-modular transformations of the $\bsv$}{S-modular transformations of the betasv}}
\label{appSbeta}

In this appendix, we display various modular $S$ transformations of the $\bsv$ that follow
from (\ref{modtrfY.1}) and (\ref{Yexpand.7x}). The $\betasvStau{j\\4}$ have already been displayed in 
(\ref{Sexampl}), the next case is
\begin{align}
\betasvStau{0\\6}&=\bar\tau^{4} \Big\{ \betasvtau{0\\6}+\frac{\zeta_5 (\tau^4{-}1)}{640y^4}
\Big\}\,,\nn\\
\betasvStau{1\\6}&=\bar\tau^{2} \Big\{\betasvtau{1\\6}+\frac{\zeta_5 ( \tau^3 \bar \tau {-} 1) }{160y^3}
\Big\}\,,\nn\\
\betasvStau{2\\6}&=\betasvtau{2\\6}+\frac{\zeta_5 ( |\tau|^4{-}1)}{40y^2} \,,\\
\betasvStau{3\\6}&=\frac{1}{{\bar\tau}^2} \Big\{ \betasvtau{3\\6} + \frac{\zeta_5( \tau \bar \tau^3 {-} 1)}{10 y} \Big\} \,,\nn\\
\betasvStau{4\\6}&=\frac{1}{{\bar\tau}^4} \Big\{ \betasvtau{4\\6}+\frac{2\zeta_{5} ( \bar \tau^4{-}1)}{5} \Big\} \,,\nn
\end{align}
and $\betasvStau{j\\k}$ with $k\geq 8$ follow from the closed formula (\ref{allSbsv}).

For depth two in the $(4,4)$ sector, the complete set of shuffle irreducibles is
\begin{align}
\betasvStau{1&0\\4&4}&=\bar\tau^{2} \Big\{ \betasvtau{1&0\\4&4}
+\frac{\zeta_3( |\tau|^2 {-} 1)}{6y}\betasvtau{0\\4}
+\frac{\zeta_3(1{-}\bar\tau^{-2} )}{2160} 
\nn\\
&\quad \quad 
+\frac{5\zeta_5( \tau^2 {-} 1)}{1728y^2}
+\frac{\zeta_3^2 ( \tau^3 \bar \tau {-} 2 |\tau|^2 {+} 1)}{288y^3} \Big\}
\,,\\
\betasvStau{2&0\\4&4}&=
\betasvtau{2&0\\4&4} + \frac{ 2 \zeta_3}{3}  (\bar \tau^2{-}1) \betasvtau{0\\4} 
+\frac{ 5 \zeta_5}{216 y} \big(|\tau|^2{-}1 \big)
+\frac{ \zeta_3^2 }{72y^2} \big(1{-} 2 \bar \tau^2 {+} | \tau |^4 \big) 
\,,\nn\\
\betasvStau{2&1\\4&4}&=\frac{1}{{\bar\tau}^2} \Big\{\betasvtau{2&1\\4&4} 
+\frac{2\zeta_3(\bar\tau^2 {-}1)}{3 }\betasvtau{1\\4} 
+\frac{5\zeta_5(\bar\tau^2 {-}1)}{108 }
+\frac{\zeta_3^2(1- 2\bar\tau^2 + \tau \bar\tau^3  )}{18  y} \Big\}
 \, ,\nn
\end{align}
while the expressions for $\betasvStau{j&j\\4&4}$ as well as $\betasvStau{0&1\\4&4},\betasvStau{0&2\\4&4},\betasvStau{1&2\\4&4}$ follow from shuffle relations (\ref{eq3.shuff2}).

In the $(6,4)$ sector we have
\begin{align}
\betasvStau{0&0\\6&4}&=\bar\tau^{6} \Big\{ \betasvtau{0&0\\6&4}+
\frac{ \zeta_5 (\tau^4{-}1)}{640y^4} \betasvtau{0\\4}
 + \frac{ \zeta_3 (\bar \tau^5{-}\tau)}{907200 \bar \tau^5 y} 
\nn\\
&\quad \quad  - \frac{ \zeta_5 (\tau^3 {-}\bar \tau^3)}{345600 \bar \tau^3 y^3}
 + \frac{ \zeta_3 \zeta_5 (\tau^6{-} 2 \tau^4   {+}1)}{30720} \Big\}\,,\nn\\
\betasvStau{0&1\\6&4}&=\bar\tau^{4} \Big\{ \betasvtau{0&1\\6&4}
+ \frac{ \zeta_5 (\tau^4{-}1)}{640y^4} \betasvtau{1\\4}
 + \frac{ \zeta_3 (\bar \tau^4 {-}1) }{226800 \bar \tau^4} 
+ \frac{ \zeta_5(\tau^2{-} \bar \tau^2) }{172800 \bar \tau^2 y^2}\nn\\
&\quad \quad 
- \frac{ \zeta_7(\tau^4{-}1) }{92160 y^4}
+ \frac{ \zeta_3 \zeta_5 ( \tau^5 \bar \tau{-}2 \tau^4 {+}1) }{7680 y^5} \Big\}
\,,\nn\\
\betasvStau{0&2\\6&4}&=\bar\tau^{2} \Big\{\betasvtau{0&2\\6&4}+
\frac{ \zeta_5 (\tau^4{-}1)}{640y^4} \betasvtau{2\\4}
- \frac{ \zeta_5(\tau{-}\bar \tau) }{21600 \bar \tau y}
- \frac{ \zeta_3^2(\tau^2{-}1) }{5040y^2} \nn\\
&\quad \quad 
- \frac{ 7 \zeta_7 (\tau^3 \bar\tau{-}1)}{11520 y^3}
+ \frac{ \zeta_3 \zeta_5 (\tau^4 \bar \tau^2{-}2 \tau^4{+}1) }{1920 y^4} \Big\}
\,,\nn\\
\betasvStau{1&0\\6&4}&=\bar\tau^{4} \Big\{\betasvtau{1&0\\6&4}
+\frac{ \zeta_5 (\tau^3 \bar \tau{-}1)}{160 y^3} \betasvtau{0\\4}
- \frac{ \zeta_3(\bar \tau^4{-}1) }{907 200 \bar \tau^4}
- \frac{ \zeta_5(\tau^2 {-}\bar \tau^2)}{86400 \bar \tau^2 y^2}\nn\\
&\quad \quad 
+ \frac{ 7 \zeta_7 (\tau^4{-}1) }{184320 y^4}
+ \frac{ \zeta_3 \zeta_5 (\tau^5 \bar \tau{-}2 \tau^3 \bar \tau{+}1) }{7680 y^5}
\Big\}
\,,\nn\\
\betasvStau{1&1\\6&4}&=\bar\tau^{2} \Big\{ \betasvtau{1&1\\6&4}
+\frac{ \zeta_5 (\tau^3 \bar \tau{-}1)}{160 y^3} \betasvtau{1\\4}
+\frac{ \zeta_5(\tau{-}\bar \tau) }{43200 \bar \tau y} + \frac{ \zeta_3^2(\tau^2{-}1) }{20160 y^2}\nn\\
&\quad \quad 
- \frac{ 7 \zeta_7 (\tau^3 \bar \tau{-}1) }{46080 y^3}
+ \frac{ \zeta_3 \zeta_5 (\tau^4 \bar \tau^2{-}2 \tau^3 \bar \tau{+}1) }{1920 y^4} \Big\}
\,,\nn\\
\betasvStau{1&2\\6&4}&=\betasvtau{1&2\\6&4}+
\frac{ \zeta_5 (\tau^3 \bar \tau{-}1)}{160 y^3} \betasvtau{2\\4}
- \frac{ \zeta_3^2 (\tau \bar \tau{-}1) }{2520 y}\nn\\
&\quad \quad  - \frac{ 7 \zeta_7 (\tau^2 \bar \tau^2{-}1) }{3840 y^2}
+ \frac{ \zeta_3 \zeta_5 (\tau^3 \bar \tau^3{-}2 \tau^3 \bar \tau{+}1) }{480 y^3}\,,\\
\betasvStau{2&0\\6&4}&=\bar\tau^{2} \Big\{ \betasvtau{2&0\\6&4}+
\frac{ \zeta_5 (\tau^2 \bar \tau^2{-}1)}{40 y^2} \betasvtau{0\\4}
- \frac{ \zeta_5 (\tau{-}\bar \tau) }{21600 \bar\tau y} - \frac{ \zeta_3^2(\tau^2{-}1) }{30240 y^2}\nn\\
&\quad \quad 
+ \frac{ 7 \zeta_7(\tau^3\bar\tau{-}1) }{23040 y^3}
+\frac{ \zeta_3 \zeta_5 (\tau^4 \bar \tau^2{-}2\tau^2 \bar \tau^2{+}1) }{1920 y^4}
\Big\}\,,\nn\\
\betasvStau{2&1\\6&4}&=\betasvtau{2&1\\6&4}+
\frac{ \zeta_5 (\tau^2 \bar \tau^2{-}1)}{40 y^2} \betasvtau{1\\4}
+ \frac{ \zeta_3^2 (\tau \bar \tau{-}1) }{3780 y} + \frac{ \zeta_3 \zeta_5 (\tau^3 \bar \tau^3{-}2 \tau^2 \bar \tau^2{+}1) }{480 y^3}\,,\nn\\
\betasvStau{2&2\\6&4}&= \frac{1}{{\bar\tau}^2} \Big\{ \betasvtau{2&2\\6&4} + 
\frac{ \zeta_5 (\tau^2 \bar \tau^2{-}1)}{40 y^2} \betasvtau{2\\4}
- \frac{ \zeta_3^2 (\bar \tau^2{-}1)}{1890} \nn\\
&\quad \quad  - \frac{ 7 \zeta_7 (\tau \bar \tau^3{-}1) }{1440y} 
+ \frac{ \zeta_3 \zeta_5 (\tau^2 \bar \tau^4{-}2 \tau^2 \bar \tau^2{+}1) }{120y^2} 
\Big\}\,,\nn\\
\betasvStau{3&0\\6&4}&=\betasvtau{3&0\\6&4}
+ \frac{ \zeta_5 (\tau \bar \tau^3{-}1)}{10 y} \betasvtau{0\\4}
- \frac{ \zeta_3^2 (\tau \bar \tau {-}1) }{2520y}  \nn\\
&\quad \quad + \frac{ 7 \zeta_7 (\tau^2 \bar \tau^2{-}1) }{3840 y^2}
+ \frac{ \zeta_3 \zeta_5 (\tau^3 \bar \tau^3 {-}2 \tau \bar \tau^3 {+}1) }{480 y^3}\,,\nn\\
\betasvStau{3&1\\6&4}&= \frac{1}{{\bar\tau}^2} \Big\{\betasvtau{3&1\\6&4}
+ \frac{ \zeta_5 (\tau \bar \tau^3{-}1)}{10 y} \betasvtau{1\\4}
+ \frac{ \zeta_3^2(\bar \tau^2{-}1) }{1260} \nn\\
&\quad \quad + \frac{ 7 \zeta_7 (\tau \bar \tau^3{-}1) }{2880 y}
+ \frac{ \zeta_3 \zeta_5 (\tau^2 \bar \tau^4{-}2 \tau \bar \tau^3{+}1) }{120y^2}
\Big\}
\,,\nn\\
\betasvStau{3&2\\6&4}&= \frac{1}{{\bar\tau}^4} \Big\{\betasvtau{3&2\\6&4}
+ \frac{ \zeta_5 (\tau \bar \tau^3{-}1)}{10 y} \betasvtau{2\\4}
- \frac{7 \zeta_7 (\bar \tau^4{-}1) }{720} + \frac{ \zeta_3 \zeta_5(\tau \bar \tau^5{-}2 \tau \bar \tau^3{+}1)}{30y}
\Big\}
\,,\nn\\
\betasvStau{4&0\\6&4}&=
\frac{1}{{\bar\tau}^2} \Big\{\betasvtau{4&0\\6&4}
+ \frac{2 \zeta_5 ( \bar \tau^4{-}1)}{5} \betasvtau{0\\4}
- \frac{ \zeta_3^2 (\bar \tau^2{-}1) }{315} \nn\\
&\quad \quad  + \frac{ 7 \zeta_7(\tau \bar \tau^3{-}1) }{720y}
+ \frac{ \zeta_3 \zeta_5(\tau^2 \bar \tau^4{-}2 \bar \tau^4{+}1) }{120y^2}
\Big\}
\,,\nn\\
\betasvStau{4&1\\6&4}&= \frac{1}{{\bar\tau}^4} \Big\{ \betasvtau{4&1\\6&4}
+
 \frac{2 \zeta_5 ( \bar \tau^4{-}1)}{5} \betasvtau{1\\4}
+ \frac{ 7 \zeta_7 (\bar \tau^4{-}1) }{360} + \frac{ \zeta_3 \zeta_5 (\tau \bar \tau^5{-}2 \bar \tau^4{+}1) }{30y}
\Big\}
\,,\nn\\
\betasvStau{4&2\\6&4}&=\frac{1}{{\bar\tau}^6} \Big\{
\betasvtau{4&2\\6&4} +
 \frac{2 \zeta_5 ( \bar \tau^4{-}1)}{5} \betasvtau{2\\4}
+ \frac{2 \zeta_3 \zeta_5(\bar \tau^6{-}2 \bar \tau^4{+}1) }{15} \Big\}\,,\nn
\end{align}
and similar expressions for the $\betasvStau{j_1&j_2\\4&6}$ follow from shuffle
relations. Expressions for
all $\betasvStau{j\\k}$ and $\betasvStau{j_1&j_2\\k_1&k_2}$ 
with $k\leq 10$ and $k_1{+}k_2\leq 10$ in machine-readable form
can be found in the ancillary file within the arXiv submission of this paper.


\section{\texorpdfstring{$T$-invariance and convergent iterated Eisenstein integrals ${\cal E}_0$}{T-invariance and convergent iterated Eisenstein integrals E0}}
\label{appC}

In this appendix, we will rewrite various expressions from the main text in terms
of convergent versions ${\cal E}_0(\ldots)$ of 
iterated Eisenstein integrals employed in \cite{Broedel:2018izr, Broedel:2019vjc}. 
Since the ${\cal E}_0(k_1,\ldots)$
with $k_1\neq 0$ are invariant under the modular $T$-transformation $\tau \rightarrow \tau{+}1$,
the rewritings in this appendix will manifest the $T$-invariance
of certain antiholomorphic integration constants and imaginary cusp forms.


\subsection{\texorpdfstring{Definitions and $q$-expansions}{Definitions and q-expansions}}
\label{secC.0}

Once we subtract the zero-mode of the holomorphic Eisenstein series
 \begin{equation}
 {\rm G}^0_{k}(\tau)= {\rm G}_{k}(\tau) - 2
\zeta_{k} \, , \ \ k \neq 0 \,, \ \ \ \ \ \ 
{\rm G}_{0}(\tau)={\rm G}^0_{0}(\tau)=-1 \, ,
\label{GGk0}
\end{equation}
one can obtain convergent and $T$-invariant iterated Eisenstein integrals 
from $ {\cal E}_0(;\tau)=1$ and~\cite{Broedel:2015hia}
\begin{align}
  {\cal E}_0(k_1,k_2,\ldots,k_r;\tau) & :=
  2\pi i \int^{i\infty}_{\tau} \dd \tau_r \  \frac{{\rm G}^0_{k_r}(\tau_r)}{(2\pi i)^{k_r}}  {\cal E}_0(k_1,k_2,\ldots,k_{r-1};\tau_r)  \label{eq11.3} \, ,
\end{align}
provided that $k_1\neq 0$. Their $q$-expansions straightforwardly follow from those of
${\rm G}_k$ and were given in \cite{Broedel:2015hia}
\begin{align}
&{\cal E}_0(k_1,0^{p_1-1},k_2,0^{p_2-1},\ldots,k_r,0^{p_r-1};\tau) =(-2)^r
  \bigg(  \prod_{j=1}^{r} \frac{ 1 }{(k_j-1)!} \bigg)
\label{qgamma1}\\
& \ \ \ \ \ \ \ \times \sum_{m_i,n_i=1}^{\infty} \frac{m_1^{k_1-1} m_2^{k_2-1} \ldots m_r^{k_r-1}  q^{m_1n_1+m_2n_2+\ldots +m_rn_r}}{(m_1 n_1)^{p_1} (m_1n_1+m_2n_2)^{p_2} \ldots (m_1n_1+m_2n_2+\ldots +m_rn_r)^{p_r}}  \,, \notag
\end{align}
where $k_i\neq0$ and $0^p =0,0,\ldots,0$ denotes a sequence of $p$ zeros. The number $r$ of non-zero 
entries $k_i\neq0$ is referred to as the depth of the ${\cal E}_0$ in (\ref{qgamma1}). 
Divergent instances of (\ref{eq11.3}) with $k_1=0$ can be shuffle-regularised based on
${\cal E}_0(0;\tau)= 2\pi i \tau$. The dictionary between Brown's iterated Eisenstein integrals (\ref{eq3.11})
and (\ref{eq11.3}) has been discussed in section 3.3 of \cite{Broedel:2018izr}, and it specialises
as follows at depth $\leq 2$
\begin{align}
\EBR{j_1}{k_1}{\tau} &=  j_1 ! \,  {\cal E}_0(0^{j_1},k_1;\tau) +  \frac{ B_{k_1} }{k_1!} \frac{ (2 \pi i \tau)^{j_1+1} }{j_1+1} \,,
\notag \\
\EBR{j_1 &j_2}{k_1 &k_2}{\tau} &=   j_2! \sum_{a=0}^{j_2} \frac{(j_1{+}a)!}{a!} \Big\{ {\cal E}_0(0^{j_1+a},k_1,0^{j_2-a},k_2;\tau)
+\frac{B_{k_2} }{k_2!}  {\cal E}_0(0^{j_1+a},k_1,0^{j_2-a+1};\tau)  \Big\} 
\notag \\
& \ \ \ \
+\frac{B_{k_1} }{k_1!}     \frac{(j_1{+}j_2{+}1)!}{j_1{+}1}
{\cal E}_0(0^{j_1+j_2+1},k_2;\tau)
 +  \frac{ B_{k_1} B_{k_2} }{k_1! k_2!} \frac{ (2 \pi i \tau)^{j_1+j_2+2} }{(j_1{+}1)(j_1{+}j_2{+}2)}
 \, , 
 \label{more.02ext}
\end{align}
see (\ref{more.02}) for the power-behaved terms $\sim \tau^{j_1+1}$ and $\sim \tau^{j_1+j_2+2}$.
The $q$-expansion of (\ref{more.02ext}) is available from (\ref{qgamma1}) once we enforce a
non-zero first entry via shuffle-relations, e.g.\ \cite{Broedel:2018izr}
\begin{align}
 {\cal E}_0(0^{p_0},k_1;\tau) &= \sum_{r=0}^{p_0} \frac{ (-1)^{p_0-r} }{r!} (2\pi i \tau)^r {\cal E}_0(k_1,0^{p_0-r};\tau)\,,
 \label{Eqexp} \\
  {\cal E}_0(0^{p_0},k_1,0^{p_1},k_2;\tau) &= \sum_{r=0}^{p_0} \frac{ (-1)^{p_0-r} }{r!} (2\pi i \tau)^r
  \sum_{s=0}^{p_0-r} \binom{p_1{+}s}{s}
  {\cal E}_0(k_1,0^{p_1+s},k_2,0^{p_0-r-s};\tau) \, ,\notag 
\end{align}
where $k_1,k_2\neq 0$. As before, we will drop the reference to the argument $\tau$
of ${\cal E}[\ldots]$ and ${\cal E}_0(\ldots)$ in the rest of this appendix.


\subsection{Integration constants at depth two}
\label{secC.1}

We will now rewrite the expressions for $ \alphaBRno{j_1& j_2}{4& 4} $
and $ \alphaBRno{j_1& j_2}{6& 4} $ in (\ref{eq4.24})
and (\ref{G4G6alpha}) in terms of convergent iterated Eisenstein integrals
(\ref{eq11.3}) at depth $\ell=1$. In this way, the power-behaved terms $\sim \tau^m$
conspire with the $ \EBRno{j}{k}$ to yield manifestly $T$-invariant combinations such as
\beq
\EBRno{0}{4} + \frac{ i \pi \tau}{360} = {\cal E}_0(4) \, , \ \ \ \ \ \ 
2 \pi i \tau  \EBRno{0}{4} -   \EBRno{1}{4}  
 - \frac{ \pi^2 \tau^2}{360} =  {\cal E}_0(4,0)\, .
 \label{eq4.25}
\eeq
We have used the depth-one instances of (\ref{more.02ext}) and (\ref{Eqexp}) 
to derive (\ref{eq4.25}) and the expressions (recall that $ \alphaBRno{1& 0}{4& 4}= \alphaBRno{0& 1}{4& 4}=0)$
\beq
 \alphaBRno{2& 0}{4& 4} =
\frac{2  \zeta_3}{3}  {\cal E}_0(4)  = - \alphaBRno{0& 2}{4& 4}  \, , \ \ \ \ \ \ 
 \alphaBRno{2& 1}{4& 4} =
\frac{2 \zeta_3}{3}   {\cal E}_0(4, 0) = -  \alphaBRno{1& 2}{4& 4}
\label{G4G4app}
\eeq
as well as ($\alphaBRno{0& 0}{6& 4}  =  
\alphaBRno{1& 0}{6& 4}  =
\alphaBRno{0& 1}{6& 4}  = 0$)  
\begin{align}
\alphaBRno{2& 0}{6& 4}  &= 
-\frac{ \zeta_3}{630} {\cal E}_0(4)\, ,
&\alphaBRno{0& 2}{6& 4}  &=
-\frac{\zeta_3}{105} {\cal E}_0(4)  - \frac{2 \zeta_3}{3} {\cal E}_0(6) 
 \notag\,,\\
 \alphaBRno{1& 1}{6& 4}  &=
\frac{\zeta_3}{420} {\cal E}_0(4) \, ,
&\alphaBRno{1& 2}{6& 4}  &= 
-\frac{\zeta_3}{210} {\cal E}_0(4, 0)  - \frac{2 \zeta_3}{3} {\cal E}_0(6, 0) 
 \notag \,,\\
\alphaBRno{3& 0}{6& 4}  &=
-\frac{\zeta_3}{210} {\cal E}_0(4, 0) \, ,
&\alphaBRno{4& 0}{6& 4}  &=
-\frac{2 \zeta_3}{105} {\cal E}_0(4, 0, 0)  + \frac{2 \zeta_5}{5} {\cal E}_0(4) 
\label{G4G6app} \,,\\
 \alphaBRno{2& 1}{6& 4}  &=
\frac{\zeta_3}{315} {\cal E}_0(4, 0) \, ,
&\alphaBRno{2& 2}{6& 4}  &= 
-\frac{\zeta_3}{315} {\cal E}_0(4, 0, 0)  -\frac{ 4 \zeta_3}{3} {\cal E}_0(6, 0, 0) 
 \notag\,, \\
 \alphaBRno{3& 1}{6& 4}  &=
\frac{ \zeta_3}{210} {\cal E}_0(4, 0, 0)\, ,\ \ \
& \alphaBRno{3& 2}{6& 4}  &=
-4  \zeta_3{\cal E}_0(6, 0, 0, 0) 
\notag \,,\\
 \alphaBRno{4& 1}{6& 4}  &=
\frac{2  \zeta_5}{5} {\cal E}_0(4, 0) \, ,
&\alphaBRno{4& 2}{6& 4}  &= 
-16  \zeta_3 {\cal E}_0(6, 0, 0, 0, 0) + \frac{4 \zeta_5}{5} {\cal E}_0(4, 0, 0)  \, .
\notag
\end{align}
%


\subsection{\texorpdfstring{Cusp forms in terms of ${\cal E}_0$}{Cusp forms in terms of E0}}
\label{secC.3}

The imaginary cusp forms at weight five can be easily expanded in terms of  ${\cal E}_0$
by combining their $\beta^{\rm sv}$-representations in (\ref{excusp.2}) and (\ref{excusp.3})
with the rearrangements (\ref{more.02ext}) and (\ref{Eqexp}) of iterated Eisenstein integrals. We arrive
at the following new expressions that manifest their $q$-expansion and $T$-invariance
(see (\ref{DHKcusp}) for ${\cal A}_{1,2;5} = \left(\frac{\Im\tau}{\pi}\right)^5\aform{0&2&3\\3&0&2} $): \small
\begin{align}
i {\cal A}_{1,2;5} &=
 \Big(  \frac{ 8y^2}{3} + \frac{ 120 \zeta_3  }{y } \Big) \Im[{\cal E}_0(6, 0^2)] + 
\Big( \frac{ 4y}{3}  + \frac{  240  \zeta_3 }{y^2 } \Big)\Im[{\cal E}_0(6, 0^3)] -
\Big(  \frac{  70}{3} -150 \zeta_3 \Big) \frac{   \Im[{\cal E}_0(6, 0^4)] }{y^3} 
 \notag \\
 & \! \! \! \! \! \! \! \!
 - \Big( \frac{ 8y^3}{315}+   9 \zeta_5 \Big) \frac{ \Im[{\cal E}_0(4, 0)] }{y^2}
 - \frac{ 15  \zeta_5\Im[{\cal E}_0(4, 0^2)] }{2 y^3}
   + 1920 \Im[{\cal E}_0(6, 0, 4, 0^2)]
+ 3600 \Im[{\cal E}_0(6, 4, 0^3)]
  \notag \\
 & \! \! \! \! \! \! \! \!  
  + 720 \Im[{\cal E}_0(6, 0^2, 4, 0)] 
 -  720 \Im[{\cal E}_0(4, 0, 6, 0^2)] - 3600 \Im[{\cal E}_0(4, 6, 0^3)]
 - \frac{ \Im[{\cal E}_0(4, 0^4)]}{3}   - 
\frac{  \Im[{\cal E}_0(4, 0^5)]}{y }  \notag \\
 & \! \! \! \! \! \! \! \!  - \frac{ 360 \Im[{\cal E}_0(4, 0^2, 6, 0^2)]}{y} - 
\frac{  2880 \Im[{\cal E}_0(4, 0, 6, 0^3)]}{y} - 
\frac{  10800 \Im[{\cal E}_0(4, 6, 0^4)]}{y} - 
\frac{  70 \Im[{\cal E}_0(6, 0^5)]}{y}   \notag \\
 & \! \! \! \! \! \! \! \! + 
\frac{  1080 \Im[{\cal E}_0(6, 0^3, 4, 0)]}{y} + 
\frac{  3240 \Im[{\cal E}_0(6, 0^2, 4, 0^2)]}{y} + 
\frac{  6480 \Im[{\cal E}_0(6, 0, 4, 0^3)]}{y} + 
\frac{  10800 \Im[{\cal E}_0(6, 4, 0^4)]}{y}  \notag \\
 & \! \! \! \! \! \! \! \! - \frac{ 5 \Im[{\cal E}_0(4, 0^6)]}{ 4 y^2} - 
\frac{ 720 \Im[{\cal E}_0(4, 0^2, 6, 0^3)]}{y^2} - 
\frac{  4140 \Im[{\cal E}_0(4, 0, 6, 0^4)]}{y^2} - 
\frac{  13500 \Im[{\cal E}_0(4, 6, 0^5)]}{y^2} \notag \\
 & \! \! \! \! \! \! \! \!  - 
\frac{  175 \Im[{\cal E}_0(6, 0^6)]}{2 y^2} + 
\frac{  540 \Im[{\cal E}_0(6, 0^4, 4, 0)]}{y^2} + 
\frac{  2160 \Im[{\cal E}_0(6, 0^3, 4, 0^2)]}{y^2} + 
\frac{  4860 \Im[{\cal E}_0(6, 0^2, 4, 0^3)]}{y^2 } \notag \\
 & \! \! \! \! \! \! \! \! + 
\frac{  8640 \Im[{\cal E}_0(6, 0, 4, 0^4)]}{y^2} + 
\frac{  13500 \Im[{\cal E}_0(6, 4, 0^5)]}{y^2} - 
\frac{  5 \Im[{\cal E}_0(4, 0^7)]}{8 y^3} - 
\frac{  450 \Im[{\cal E}_0(4, 0^2, 6, 0^4)]}{y^3} \notag \\
 & \! \! \! \! \! \! \! \!  - 
\frac{  2250 \Im[{\cal E}_0(4, 0, 6, 0^5)]}{y^3} - 
\frac{  6750 \Im[{\cal E}_0(4, 6, 0^6)]}{y^3} - 
\frac{  175 \Im[{\cal E}_0(6, 0^7)]}{4 y^3} + 
\frac{  450 \Im[{\cal E}_0(6, 0^4, 4, 0^2)]}{y^3}  \notag \\
 & \! \! \! \! \! \! \! \!  + 
\frac{  1350 \Im[{\cal E}_0(6, 0^3, 4, 0^3)]}{y^3} + 
\frac{  2700 \Im[{\cal E}_0(6, 0^2, 4, 0^4)]}{y^3} + 
\frac{  4500 \Im[{\cal E}_0(6, 0, 4, 0^5)]}{y^3} + 
\frac{  6750 \Im[{\cal E}_0(6, 4, 0^6)]}{y^3} \notag \\
 & \! \! \! \! \! \! \! \!  - 
 720 \Im[{\cal E}_0(6, 0^2)] \Re[{\cal E}_0(4, 0)] - 
\frac{  1080 \Im[{\cal E}_0(6, 0^3)] \Re[{\cal E}_0(4, 0)]}{y} - 
\frac{  540 \Im[{\cal E}_0(6, 0^4)] \Re[{\cal E}_0(4, 0)]}{y^2} \notag \\
 & \! \! \! \! \! \! \! \! - 
\frac{  360 \Im[{\cal E}_0(6, 0^2)] \Re[{\cal E}_0(4, 0^2)]}{y}  - 
\frac{  720 \Im[{\cal E}_0(6, 0^3)] \Re[{\cal E}_0(4, 0^2)]}{y^2} - 
\frac{  450 \Im[{\cal E}_0(6, 0^4)] \Re[{\cal E}_0(4, 0^2)]}{y^3} \notag \\
 & \! \! \! \! \! \! \! \!  + 
 720 \Im[{\cal E}_0(4, 0)] \Re[{\cal E}_0(6, 0^2)] + 
\frac{  360 \Im[{\cal E}_0(4, 0^2)] \Re[{\cal E}_0(6, 0^2)]}{y} + 
\frac{  1080 \Im[{\cal E}_0(4, 0)] \Re[{\cal E}_0(6, 0^3)]}{y} \notag \\
& \! \! \! \! \! \! \! \!  + 
\frac{  720 \Im[{\cal E}_0(4, 0^2)] \Re[{\cal E}_0(6, 0^3)]}{y^2} + 
\frac{  540 \Im[{\cal E}_0(4, 0)] \Re[{\cal E}_0(6, 0^4)]}{y^2} + 
\frac{  450 \Im[{\cal E}_0(4, 0^2)] \Re[{\cal E}_0(6, 0^4)]}{y^3}
\notag \\
i {\rm B}_{2,3} &= 
\Big( \frac{ 4y^3}{105 } +  6 \zeta_3 + \frac{ 27 \zeta_5}{ 2 y^2}  \Big) \Im[{\cal E}_0(4, 0)]  
+\Big( \frac{ y^2}{15 } + \frac{ 3 \zeta_3  }{y} - \frac{ 9  \zeta_5}{2 y^3} \Big) \Im[{\cal E}_0(4, 0^2)]    \\
& \! \! \! \! \! \! \! \! - \Big( 4 y^2+  \frac{  180 \zeta_3  }{y} \Big) \Im[{\cal E}_0(6, 0^2)]
 - \Big( 16 y + \frac{  45 \zeta_3  }{y^2} \Big) \Im[{\cal E}_0(6, 0^3)] 
-\Big(35- \frac{  90  \zeta_3}{y^3}  \Big) \Im[{\cal E}_0(6, 0^4)]\,,
   \notag \\
 & \! \! \! \! \! \! \! \! + 1080 \Im[{\cal E}_0(4, 0, 6, 0^2)] - 
 2160 \Im[{\cal E}_0(4, 6, 0^3)] 
 - 1080 \Im[{\cal E}_0(6, 0^2, 4, 0)] - 360 \Im[{\cal E}_0(6, 0, 4, 0^2)] \notag \\
 & \! \! \! \! \! \! \! \!  +  2160 \Im[{\cal E}_0(6, 4, 0^3)]   -  \frac{ 1}{4} \Im[{\cal E}_0(4, 0^4)] + 
\frac{  540 \Im[{\cal E}_0(4, 0^2, 6, 0^2)]}{y} + 
\frac{  540 \Im[{\cal E}_0(4, 0, 6, 0^3)]}{y} \notag \\
 & \! \! \! \! \! \! \! \!  - 
\frac{  6480 \Im[{\cal E}_0(4, 6, 0^4)]}{y} - 
\frac{ 105 \Im[{\cal E}_0(6, 0^5)]}{ 2 y} - \frac{ 1620 \Im[{\cal E}_0(6, 0^3, 4, 0)]}{y} 
- \frac{  1080 \Im[{\cal E}_0(6, 0^2, 4, 0^2)]}{y}  \notag \\
 & \! \! \! \! \! \! \! \! + 
\frac{  1620 \Im[{\cal E}_0(6, 0, 4, 0^3)]}{y} + 
\frac{  6480 \Im[{\cal E}_0(6, 4, 0^4)]}{y}
- \frac{ 5 \Im[{\cal E}_0(4, 0^5)]}{8 y} 
+ \frac{ 135 \Im[{\cal E}_0(4, 0^2, 6, 0^3)]}{y^2} \notag \\
& \! \! \! \! \! \! \! \!  - 
\frac{  1350 \Im[{\cal E}_0(4, 0, 6, 0^4)]}{y^2} - 
\frac{  8100 \Im[{\cal E}_0(4, 6, 0^5)]}{y^2} - 
\frac{  105 \Im[{\cal E}_0(6, 0^6)]}{2 y^2} - 
\frac{  810 \Im[{\cal E}_0(6, 0^4, 4, 0)]}{y^2} \notag \\
& \! \! \! \! \! \! \! \! - \frac{  405 \Im[{\cal E}_0(6, 0^3, 4, 0^2)]}{y^2} + 
\frac{  1215 \Im[{\cal E}_0(6, 0^2, 4, 0^3)]}{y^2} + 
\frac{  4050 \Im[{\cal E}_0(6, 0, 4, 0^4)]}{y^2} + 
\frac{  8100 \Im[{\cal E}_0(6, 4, 0^5)]}{y^2} \notag \\
& \! \! \! \! \! \! \! \! - \frac{ 3 \Im[{\cal E}_0(4, 0^6)]}{ 4 y^2}  - 
\frac{  3 \Im[{\cal E}_0(4, 0^7)]}{8 y^3} - 
\frac{  270 \Im[{\cal E}_0(4, 0^2, 6, 0^4)]}{y^3} - 
\frac{  1350 \Im[{\cal E}_0(4, 0, 6, 0^5)]}{y^3} \notag \\
& \! \! \! \! \! \! \! \! - \frac{  4050 \Im[{\cal E}_0(4, 6, 0^6)]}{y^3} -
\frac{  105 \Im[{\cal E}_0(6, 0^7)]}{4 y^3} + 
\frac{  270 \Im[{\cal E}_0(6, 0^4, 4, 0^2)]}{y^3} + 
\frac{  810 \Im[{\cal E}_0(6, 0^3, 4, 0^3)]}{y^3} \notag \\
& \! \! \! \! \! \! \! \! + \frac{  1620 \Im[{\cal E}_0(6, 0^2, 4, 0^4)]}{y^3} + 
\frac{  2700 \Im[{\cal E}_0(6, 0, 4, 0^5)]}{y^3} + 
\frac{  4050 \Im[{\cal E}_0(6, 4, 0^6)]}{y^3} \notag \\
& \! \! \! \! \! \! \! \!  + 
 1080 \Im[{\cal E}_0(6, 0^2)] \Re[{\cal E}_0(4, 0)] + 
\frac{  1620 \Im[{\cal E}_0(6, 0^3)] \Re[{\cal E}_0(4, 0)]}{y} + 
\frac{  810 \Im[{\cal E}_0(6, 0^4)] \Re[{\cal E}_0(4, 0)]}{y^2} \notag \\
& \! \! \! \! \! \! \! \!  + 
\frac{  540 \Im[{\cal E}_0(6, 0^2)] \Re[{\cal E}_0(4, 0^2)]}{y} + 
\frac{  135 \Im[{\cal E}_0(6, 0^3)] \Re[{\cal E}_0(4, 0^2)]}{y^2} - 
\frac{  270 \Im[{\cal E}_0(6, 0^4)] \Re[{\cal E}_0(4, 0^2)]}{y^3} \notag \\
& \! \! \! \! \! \! \! \!  - 
 1080 \Im[{\cal E}_0(4, 0)] \Re[{\cal E}_0(6, 0^2)] - 
\frac{  540 \Im[{\cal E}_0(4, 0^2)] \Re[{\cal E}_0(6, 0^2)]}{y} - 
\frac{  1620 \Im[{\cal E}_0(4, 0)] \Re[{\cal E}_0(6, 0^3)]}{y} \notag \\
& \! \! \! \! \! \! \! \!  - 
\frac{  135 \Im[{\cal E}_0(4, 0^2)] \Re[{\cal E}_0(6, 0^3)]}{y^2} - 
\frac{  810 \Im[{\cal E}_0(4, 0)] \Re[{\cal E}_0(6, 0^4)]}{y^2} + 
\frac{  270 \Im[{\cal E}_0(4, 0^2)] \Re[{\cal E}_0(6, 0^4)]}{y^3} \,.
\notag
\end{align} \normalsize
The leading orders of the $q,\bar q$-expansion of ${\cal A}_{1,2;5}$ have been 
checked\footnote{We are grateful to Eric D'Hoker
and Justin Kaidi for their help in performing this check.} to line up with
the all-order results of \cite{DHoker:2019txf} on the Fourier-expansion of two-loop MGFs.


\renewcommand{\baselinestretch}{1.15}

\providecommand{\href}[2]{#2}\begingroup\raggedright\endgroup

\end{document}